\documentclass[12pt,a4paper,twoside,normalheadings]{scrbook}

\newcommand{\captionfonts}{\small}
\makeatletter  
\long\def\@makecaption#1#2{%
  \vskip\abovecaptionskip
  \sbox\@tempboxa{{\captionfonts #1: #2}}%
  \ifdim \wd\@tempboxa >\hsize
    {\captionfonts #1: #2\par}
  \else
    \hbox to\hsize{\hfil\box\@tempboxa\hfil}%
  \fi
  \vskip\belowcaptionskip}
\makeatother   

\makeatletter                                  
\newcommand{\thefigurename}{Fig.}              
\def\fnum@figure{\textit{\textbf{              
                \thefigurename\ \thefigure}}}
\newcommand{\thetablename}{Tab.}               
\def\fnum@table{\textit{\textbf{               
                \thetablename\ \thetable}}}

\newcommand{\pinholeindex}{$\;\!^{\mathrm{p}}$}
\newcommand{\openindex}{$\;\!^{\mathrm{b}}$}
\newcommand{\+}{$\,$}
\newcommand{\AMS}{\mbox{AMS-01}}

\def\eV{\ifmmode {\mathrm{\ e\kern -0.1em V}}\else\textrm{e\kern -0.1em V}\fi}%
\def\keV{\ifmmode {\mathrm{\ ke\kern -0.1em V}}\else\textrm{ke\kern -0.1em V}\fi}%
\def\MeV{\ifmmode {\mathrm{\ Me\kern -0.1em V}}\else\textrm{Me\kern -0.1em V}\fi}%
\def\GeV{\ifmmode {\mathrm{\ Ge\kern -0.1em V}}\else\textrm{Ge\kern -0.1em V}\fi}%
\def\TeV{\ifmmode {\mathrm{\ Te\kern -0.1em V}}\else\textrm{Te\kern -0.1em V}\fi}%
\def\PeV{\ifmmode {\mathrm{\ Pe\kern -0.1em V}}\else\textrm{Pe\kern -0.1em V}\fi}%

\newbox\slashbox \setbox\slashbox=\hbox{$/$}
\newbox\Slashbox \setbox\Slashbox=\hbox{\large$/$}
\def\pFMslash#1{\setbox\@tempboxa=\hbox{$#1$}
  \@tempdima=0.5\wd\slashbox \advance\@tempdima 0.5\wd\@tempboxa
  \copy\slashbox \kern-\@tempdima \box\@tempboxa}
\def\pFMSlash#1{\setbox\@tempboxa=\hbox{$#1$}
  \@tempdima=0.5\wd\Slashbox \advance\@tempdima 0.5\wd\@tempboxa
  \copy\Slashbox \kern-\@tempdima \box\@tempboxa}
\def\FMslash{\protect\pFMslash}

\newcommand{\etmiss}{\mbox{$\FMslash{E}_T$}} 

\newlength{\offset}
\newcommand{\STRUT}{\rule[-1ex]{0ex}{3.5ex}}
\newcommand{\diff}[2]{\frac{\mathrm{d}#1}{\mathrm {d}#2}}
\newcommand{\LL}[1]{\smash{\lower 1.5ex \hbox{#1}}}

\setcapindent{0mm}

\usepackage[english]{babel}
\usepackage{booktabs}
\usepackage{here}
\usepackage{textcomp}
\usepackage{amssymb}
\usepackage{picinpar} 
\usepackage{graphicx}
\usepackage{picins}
\usepackage{setspace}
\usepackage{color}
\usepackage{wrapfig}
\usepackage{rotating}
\definecolor{DarkBlue}{rgb}{0,0,0.5}

\begin{document}

\pagestyle{empty}
\begin{titlepage} 
\vspace*{10mm}          

{\centering \sf \huge \bfseries
Signatures of SUSY Dark Matter\\
at the LHC and in\\
the Spectra of Cosmic Rays\\
}

\vspace{2.0cm}

{\centering \normalsize
Von der Fakult\"at f\"ur Mathematik, Informatik und Naturwissenschaften \\
der Rheinisch-Westf\"alischen Technischen Hochschule Aachen \\
zur Erlangung des akademischen Grades eines Doktors der Naturwissenschaften \\
genehmigte Dissertation \\
}

\vspace{1.5cm}
 
{\centering vorgelegt von \\ }

\vspace{1.5cm}

{\centering 
Diplom--Physiker\\
Jan Olzem\\
\vspace{1cm}
aus Krefeld \\
\vspace{1cm}
} 
 
\vspace{15mm}
\begin{center}
\begin{tabbing} 
\hspace{35mm}  \= Berichter: \= Universit\"atsprofessor Dr. St.~Schael \\
               \>            \> Universit\"atsprofessorin Dr. L.~Baudis \\
\end{tabbing}

Tag der m\"undlichen Pr\"ufung: 27. Februar 2007\\
\vspace{15mm}
Diese Dissertation ist auf den Internetseiten der Hochschulbibliothek online verf\"ugbar.
\end{center}

\end{titlepage}

\cleardoublepage
\section*{Abstract}
Over the past decades it has become evident that luminous matter
amounts only to a small fraction of the energy density in the
universe. More than 75\+\% is accounted for by what is called the dark
energy, and about 20\+\% must exist in the form of some kind of
non-relativistic dark matter. Among the candidates for the
constituents of dark matter, the supersymmetric (SUSY) neutralino is
one of the most promising. This thesis discusses the search for
supersymmetry at the future Large Hadron Collider (LHC) and the
ongoing construction of one of the four large LHC experiments, the
Compact Muon Solenoid (CMS), and focuses on the detection of signals
from the annihilation of supersymmetric dark matter in the spectra of
cosmic rays.

CMS relies on the excellent performance of its components and thus
requires strict quality control before their assembly. The final steps
of assembly of 1061 silicon microstrip detector modules for the CMS
tracker endcaps are performed at the 1. Physikalisches Institut B at
the RWTH Aachen. A laser test facility for these modules was developed
and is described in this thesis. In contrast to test procedures based
only on the evaluation of pedestal and noise data, the test facility
relies on the generation of signals in the silicon sensors by infrared
laser illumination. Subsequent analysis of the signals allows reliable
detection of module defects. The fully automatic test facility
provides high throughput and easy operation for the series production
of the modules. Its performance is validated by investigating a
reference module with artificially prepared defects of three types:
open wirebonds, short-circuited strips and pinholes. It is shown that
all defects are clearly detected. In addition to defect detection, an
indication for the type of defect is provided. In a further validation
step, nine modules from a prototype series are investigated with the
laser test facility. Confirming the earlier results on the reference
module, defective strips are reliably identified. The results are in
agreement with those from other test facilities using different
techniques.

Measurements of cosmic ray antiparticles, such as positrons, can
impose strong constraints on the nature of new physics beyond the
Standard Model. However, cosmic ray positron measurements are
experimentally very challenging due to the vast proton
background. This thesis describes a novel approach of positron
identification with the space-borne \AMS{} experiment, namely through
the detection of bremsstrahlung conversion in a silicon microstrip
detector. In contrast to earlier single-track analyses, this approach
involves the selection and reconstruction of multi-track
events. Subsequent to an introduction to cosmic ray physics and a
description of the \AMS{} experiment, the discussion of the signal
process shows that bremsstrahlung from protons is suppressed by a
factor of more than $3\cdot 10^6$ with respect to positrons. The
background to the positron sample can largely be suppressed using the
topological and geometrical properties of the events.

In order to obtain the highest positron selection efficiency possible,
novel combinatorial track finding algorithms were developed,
particularly optimized for the signature of converted bremsstrahlung.
By applying restrictions on the invariant mass of particles the
background to the positron sample is largely eliminated. The remaining
background contamination is determined from large samples of Monte
Carlo data taking into account the effects of the geomagnetic
field. It amounts to 26\+\% of the positron counts and is corrected
for. In order to remove atmospheric secondaries from the
positron and electron samples, a precise method involving trajectory
backtracing in the magnetic field of the Earth was developed and is
applied individually to all positron and electron candidates.

The results of the positron measurement show that the bremsstrahlung
approach extends the sensitivity range of \AMS{} to positron momenta
up to 50\+\GeV{}/c, which is far beyond the original scope of the
experiment. The precision of the positron measurement is statistically
limited by the small number of particle counts. The positron fraction
$e^+ / (e^+ + e^-)$ is calculated for particle momenta in the range
from 1 to 50\+\GeV{}/c. For momenta up to 8\+\GeV{} it is found to be
in good agreement with model predictions for background from purely
secondary positron production, while at higher momenta there is
indication for a positron overabundance. Therefore, the \AMS{} data
lend further weight to the hints of a positron overabundance seen in
the data from earlier experiments.

In addition to the positron fraction, the absolute fluxes of positrons
and electrons are calculated from the event samples of the present
analysis. For this purpose, a method was developed which allows the
determination of the geomagnetic transmission as a function of
momentum and direction of incidence with high accuracy. The results of
the flux calculation are found to be in very good agreement with
earlier data, confirming the good performance of electron and positron
selection with the bremsstrahlung approach.

Finally, the positron fraction results from this analysis have been
combined with results from earlier experiments. In the combined
results, the significance of the positron overabundance with respect
to the background expectation for purely secondary positron production
increases to 5.3 standard deviations, which would be reduced to 4.2
standard deviations without the results of the present analysis.
Therefore, a statistical fluctuation causing the positron
overabundance in the data with respect to the background-only
expectation can now be excluded.

\section*{Zusammenfassung} 
Es ist eine der weitreichendsten Entdeckungen der letzten Jahrzehnte,
da\ss\ die sichtbare Materie im Universum lediglich einen sehr kleinen
Teil zu dessen Gesamtenergiedichte beitr\"agt. Zu mehr als 75\+\% ist
es von einer vollkommen unbekannten Form von Energie dominiert, f\"ur
die der Name {\it Dunkle Energie} gepr\"agt worden ist, w\"ahrend etwa
20\+\% als gravitierende nichtrelativistische {\it Dunkle Materie}
existieren mu\ss. Unter den zahlreichen Kandidaten f\"ur die
Konstituenten der Dunklen Materie ist das supersymmetrische (SUSY)
Neutralino das vielversprechendste. Die vorliegende Arbeit befa\ss t
sich mit der Suche nach supersymmetrischen Teilchen am zuk\"unftigen
Large Hadron Collider (LHC) und dem Bau eines der vier gro\ss en
LHC-Experimente, dem Compact Muon Solenoid (CMS), und konzentriert
sich auf den Nachweis von Signalen aus der Annihilation
supersymmetrischer Dunkler Materie in den Spektren der Kosmischen
Strahlung.

Die einwandfreie Funktion seiner Komponenten ist eine wesentliche
Voraussetzung f\"ur den Erfolg eines technisch so anspruchsvollen
Gro\ss experimentes wie CMS und macht eine strenge
Qualit\"atskontrolle w\"ahrend ihrer Fertigung erforderlich. Die
Endmontage von 1061 Siliziumstreifendetektormodulen f\"ur den
Spurdetektor von CMS wird im 1. Physikalischen Institut B der RWTH
Aachen durchgef\"uhrt. Die vorliegende Arbeit beschreibt die
Entwicklung eines vollst\"andig automatisierten Testverfahrens zur
Qualit\"atskontrolle dieser Module. Es basiert auf mittels eines
Infrarotlasers in den Siliziumsensoren angeregten Signalen, deren
Auswertung eine zuverl\"assige Erkennung von Defekten erlaubt. Um die
Anwendbarkeit des Testverfahrens zu best\"atigen, wurde ein
Referenzmodul mit k\"unstlich pr\"aparierten Defekten untersucht. Wie
gezeigt wird, k\"onnen alle Defekte einwandfrei detektiert werden, und
es ergibt sich aus der Messung ein Hinweis auf den jeweiligen
Defekttyp. Dar\"uberhinaus wurde das Testverfahren zur Untersuchung
von neun Modulen einer Prototypserie verwendet. In
\"Ubereinstimmung mit den Ergebnissen aus der Untersuchung des
Referenzmoduls konnten Moduldefekte zuverl\"assig identifiziert
werden. Die Ergebnisse werden von denjenigen anderer Testverfahren,
die auf abweichenden Nachweistechniken beruhen, best\"atigt.

Die Energiespektren von Antiteilchen in der Kosmischen Strahlung, wie
etwa von Positronen, k\"onnen wichtige Hinweise auf die Natur der
Dunklen Materie liefern. Die Identifizierung von Positronen in der
Kosmischen Strahlung stellt experimentell eine erhebliche
Herausforderung dar aufgrund des gro\ss en von Protonen verursachten
Untergrundes. Die vorliegende Arbeit besch\"aftigt sich mit einem
neuen Ansatz zur Identifikation von Positronen auf der Grundlage des
Nachweises konvertierter Bremsstrahlungsphotonen in einem
Siliziumstreifendetektor. Die Wahrscheinlichkeit der Emission von
Bremsstrahlung von Protonen ist gegen\"uber Positronen um einen Faktor
von mehr als $3\cdot 10^6$ unterdr\"uckt und erlaubt somit eine
Unterscheidung beider Teilchensorten. Dieses Verfahren wurde anhand
der Daten des weltraumgest\"utzten \AMS{}-Experimentes realisiert. Im
Gegensatz zu fr\"uheren auf Einzelspurmessungen basierenden Analysen
erfordert es die Rekonstruktion von Mehrspurereignissen.

Um eine bestm\"ogliche Selektionseffizienz f\"ur Positronen zu
errreichen, wurden speziell f\"ur die Signatur konvertierter
Bremsstrahlungsphotonen optimierte Spurfindungsalgorithmen
entwickelt. Schnitte auf die invariante Masse der gemessenen Teilchen
eliminieren weitgehend den Untergrund zu den Positronkandidaten. Der
verbleibende irreduzible Untergrund wurde mit Hilfe einer gro\ss en
Anzahl simulierter Ereignisse unter Ber\"ucksichtigung des Einflusses
des Erdmagnetfeldes bestimmt. Die Zahl der Positronen wurde um diesen
Untergrund, der 26\+\% der Signalereignisse entspricht,
korrigiert. Zur Unterdr\"uckung atmosph\"arischer Sekund\"arteilchen
unter den Positron- und Elektronkandidaten wurde eine Methode
entwickelt, die auf der R\"uckverfolgung von Teilchenspuren im
Erdmagnetfeld basiert.

Die Ergebnisse dieser Positronmessung zeigen, da\ss\ der
Bremsstrahlungsansatz den Me\ss bereich von \AMS{} bis hin zu
Positronimpulsen von 50\+\GeV{}/c erweitert, was weit jenseits der
urspr\"unglich angestrebten Reichweite des Experimentes von
3,5\+\GeV{}/c liegt. Die Genauigkeit der Positronmessung ist hierbei
statistisch limitiert durch die Anzahl der gemessenen
Positronen. Weiterhin wurde der Positronenanteil $e^+ / (e^+ + e^-)$
in der Kosmischen Strahlung f\"ur Teilchenimpulse im Bereich von 1 bis
50\+GeV{}/c berechnet. F\"ur Impulse bis hin zu 8\+\GeV{}/c ist dieser
in guter \"Ubereinstimmung mit den Ergebnissen von Modellrechnungen
f\"ur den Untergrund von ausschlie\ss lich sekund\"ar produzierten
Positronen. Bei h\"oheren Impulsen hingegen zeichnet sich ein
\"Uberschu\ss\ von Positronen ab. Somit liefern auch die Daten des
\AMS{}-Experimentes Hinweise auf einen Positronen\"uberschu\ss, wie er 
bei fr\"uheren Experimenten bereits beobachtet wurde.

Zus\"atzlich zum Positronenanteil wurden auch die absoluten Fl\"usse
von Elektronen und Positronen aus der Anzahl der Kandidaten dieser
Analyse errechnet. Hierzu wurde ein Verfahren entwickelt, das es
erlaubt, die Durchl\"assigkeit des Erdmagnetfeldes f\"ur Elektronen
und Positronen als Funktion des Impulses und der Flugrichtung der
Teilchen mit hoher Genauigkeit zu ermitteln. Die Ergebnisse der Flu\ss
berechnung zeigen eine sehr gute \"Ubereinstimmung mit denjenigen
fr\"uherer Messungen.

Abschlie\ss end wurde der im Rahmen dieser Arbeit ermittelte
Positronenanteil mit den Ergebnissen anderer Experimente statistisch
kombiniert. Aufgrund der Ergebnisse der vorliegenden Arbeit steigt die
Signifikanz des Positronen\"uberschusses gegen\"uber dem erwarteten
Untergrund von 4,2 auf 5,3 Standardabweichungen. Somit kann nun eine
statistische Fluktuation als Ursache f\"ur den \"Uberschu\ss\ von
Positronen gegen\"uber dem erwarteten Untergrund weitgehend
ausgeschlossen werden.

\pagestyle{empty}
\renewcommand*{\chapterpagestyle}{empty}
\tableofcontents
\cleardoublepage
\pagestyle{headings}
\renewcommand*{\chapterpagestyle}{plain}

\pagenumbering{arabic}
\setcounter{page}{1}
\chapter{Introduction}
\parpic(7.5cm,7.5cm)[r]{\includegraphics[width=7cm]{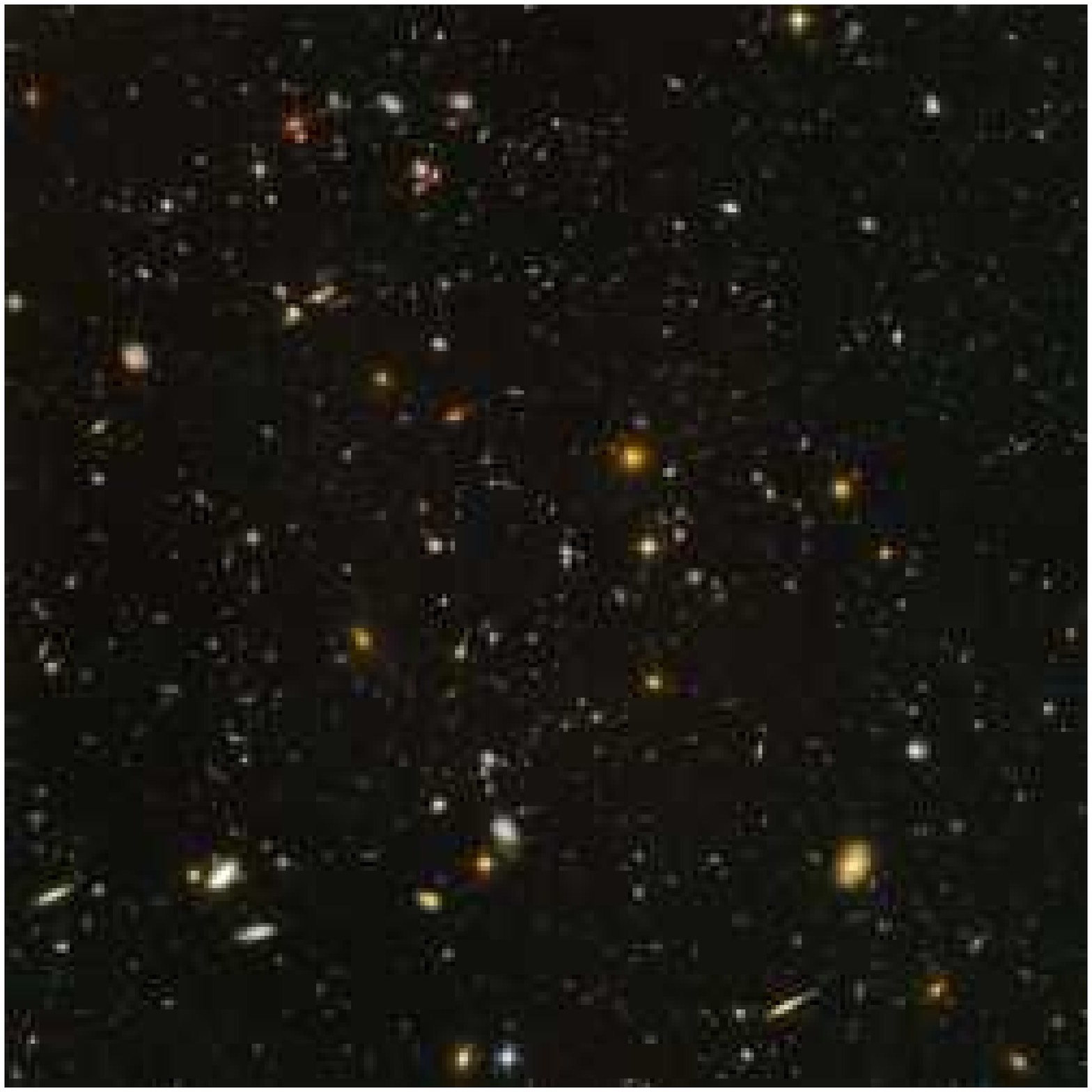}}
\noindent The picture shown here is known as the {\sl Hubble Ultra Deep Field}
and was taken by the Hubble Space Telescope in
2004~\cite{beckwith06a}. It is the result of an ultra-long exposure --
more than 11 days effectively -- to the light coming from a tiny area
of the sky, as small as 3 by 3 arcminutes. In other words, the image
represents far less than one millionth of the sphere that surrounds
the Earth. An area of this size could easily be obscured by a grain of
sand at the distance of an arm.

Apart from a few faint stars which happened to be in the field of
view, about 10,000 galaxies appear in the image. With the coarse but
reasonable assumption that a typical galaxy contains about one hundred
billion stars and that galaxies are uniformly distributed over large
scales, one concludes that the total number of stars in the universe
must be roughly of the order of 10$^{22}$, or
10,000,000,000,000,000,000,000. What an unimaginable amount of mass
and energy fills the universe!

Nevertheless, in the last few decades it became apparent that luminous
matter and all the invisible dust and gas inbetween, whose elementary
composition is in principle similar to that of matter in our familiar
surroundings, amounts only to a surprisingly small fraction of the
energy density in the universe. In reality, the universe must be
regarded as essentially dominated by exotic and novel forms of matter,
and an even bigger reservoir of strange non-material energy, whose
nature yet remains a complete mystery.  Consequently, the assertion of
basically knowing what fills space must be abandoned and the terms
{\sl dark matter} and {\sl dark energy} have been coined for the
unknown bulk ingredients of the universe.

At the same time the Standard Model of particle physics, which
structures the knowledge about the elementary constituents and
interactions of matter, has matured. It has been and still is
extraordinarily successful in explaining practically all experimental
observations in particle physics and has received continuous
confirmation with highest precision for more than two decades. Despite
its success, it is widely assumed that the Standard Model must merely
be a part of a superior and more comprehensive concept, i.e. it must
be a low-energy effective theory truly valid only in the energy regime
open to current experiments. Thus a large diversity of hypotheses has
emerged to anticipate whatever may appear when the present
experimental limits are exceeded. Among such hypotheses, the idea of
supersymmetry is one of the most cogent and intriguing.

It is becoming apparent that cosmology, astrophysics and particle
physics are simultaneously approaching the same problem from different
points of view and that their synergy may finally lead to a
fundamentally new understanding of basic physics. In a manner of
speaking, the universe acts as a huge autonomous reactor for all its
constituents, even the possibly unknown, and offers by far the most
powerful particle accelerators. Therefore, the study of cosmic ray
particles can deliver strong constraints on those new phenomena which
are as yet beyond the reach of any laboratory experiment and can point
the way towards better theoretical and experimental approaches. With
the upcoming collider facilities, such as the LHC, it will thus be
possible to target new discoveries with greater precision.

This thesis discusses the search for supersymmetry at the future
LHC collider and the ongoing construction of one of the four large LHC
experiments, the Compact Muon Solenoid (CMS), and focuses on the
detection of signals from supersymmetric dark matter candidates in the
spectra of cosmic rays. The {\sl second chapter\/} introduces the
question of the nature of dark matter together with an overview of the
history and current state of the universe, as well as the evidence for
the existence of as yet invisible new forms of matter. The
supersymmetric extension of the Standard Model of particle physics
leads to the presently most promising dark matter candidate, namely
the hypothetical neutralino.

The possible discovery of supersymmetry at the LHC is the subject of
{\sl chapter three\/}. It starts with an overview of the LHC facility
and of the four large collider ring experiments. Subsequently, the
discovery potential for supersymmetry of the LHC experiments is
discussed, including the production mechanisms and the decay channels
of supersymmetric particles, as well as the experimental signatures
which would allow their detection and the determination of
supersymmetry parameters. The CMS experiment at the LHC is currently
under construction and will feature the largest silicon microstrip
tracker ever built. Such a complex detector relies on the excellent
performance of its components and thus requires strict quality control
before their assembly. Following an overview of the experiment and in
particular of the tracker and its components, a laser test facility
for silicon microstrip detector modules is described. The
investigation of artificially created module defects establishes the
applicability of the test facility and the results of tests of a
prototype series of detector modules are presented.

The {\sl fourth chapter\/} addresses cosmic rays as possible
messengers for signals from reactions of supersymmetric dark matter
particles.  Subsequent to a description of the composition and
characteristics of cosmic rays, their origin and propagation in the
interstellar medium, as well as the modulation of cosmic ray spectra
in the vicinity of the Earth are discussed. As it turns out, the
annihilation of neutralinos from dark matter may leave its traces in
cosmic rays in the form of an overabundance of particles, particularly
of positrons, antiprotons, and gamma rays. The processes involved are
therefore described and a characterization of the expected signatures
in the corresponding particle spectra is given. The chapter ends with
an overview of the evidence for neutralino annihilation which has so
far been found in cosmic ray data.

Such evidence still inconclusive and more experimental results are
necessary to further investigate the hypothesis of supersymmetric dark
matter. While new cosmic ray experiments are under construction or
have recently started taking data, existing data from past experiments
can be re-examined using new analysis techniques. The second part of
this thesis reanalyzes the data taken by the space-borne \AMS{}
detector in 1998 with the aim of determining the spectrum of cosmic
ray positrons up to energies of 50\+\GeV, far beyond the original scope
of the experiment. After a detailed description of \AMS{} and its
Space Shuttle flight in {\sl chapter five\/}, {\sl chapter sixth\/}
introduces the challenge of cosmic ray positron measurements and the
novel approach of positron identification through the detection of
converted bremsstrahlung photons in silicon microstrip detectors. An
in-depth description of the analysis includes discussion of the signal
and background processes, methods of particle track and vertex
reconstruction, background suppression and correction for the
background remaining among the signal candidates, as well as the
complications which arise due to the circumstances under which \AMS{}
was operated, namely the influence on particle detection of the
spacecraft and of the magnetic field of the Earth.

The results of the positron measurement are the subject of {\sl
chapter seven\/}, where the positron fraction -- the ratio of positron
to electron counts \mbox{$e^+/(e^{+}+e^-)$} in which several
systematic uncertainties cancel -- is presented as a function of
particle momentum. It is shown that by using the conversion of
bremsstrahlung photons to tag positrons and electrons in cosmic rays
the \AMS{} data substantially improve the knowledge of the cosmic ray
positron fraction up to the highest momenta so far accessible to
experiments. To validate the tagging method, the chapter ends with the
calculation of the absolute fluxes of positrons and electrons based on
the results of this analysis and the comparison of these fluxes to the
existing data~\cite{alcaraz00d,duvernois01a}.

The {\sl concluding chapter\/} summarizes the results of this
thesis. Finally, the positron fraction measurements from the most
recent experiments~\cite{golden96a,boezio00a,alcaraz00a,beatty04a} are
combined with the results of this analysis, giving the most precise
positron fraction data yet available. The positron fraction measured
in this analysis lends further weight to the hints seen in the data
of~\cite{beatty04a} of a positron overabundance for momenta above
6\+\GeV.

\chapter{Dark Matter in the Universe}\label{chapter:darkmatter} 
Over the past decades, numerous pioneering ideas and seminal
discoveries have significantly broadened the understanding of the
structure and evolution of the universe. The theoretical fundaments of
space-time in an homogeneous and isotropic world based on General
Relativity were already known in 1922 to contradict a static
universe~\cite{friedman22a}. Seven years later it was indeed
discovered that the recessional velocity of distant galaxies is
proportional to their distance, a fact which is nowadays referred to
as {\sl Hubble's law}~\cite{hubble29a}. Hence, the universe was found
to be continuously expanding. The manifest idea of the {\sl Hot Big
  Bang} -- postulating that the universe primordially emerged from a
tremendously dense and hot state -- became increasingly popular,
especially when it was shown that it could explain the observed
abundances of the lightest nuclei~\cite{gamow46a,gamow48a}. However,
the strongest confirmation for the theory came with the discovery of
the cosmic microwave background (CMB) in 1965~\cite{penzias65a}, an
omnipresent electromagnetic radiation field with an almost perfect
thermal 2.7\+K black-body spectrum, which had been predicted before as
a Hot Big Bang relic~\cite{alpher49a}. Recently, accurate measurements
of the fluctuations in the CMB led to a precise determination of the
cosmological parameters~\cite{spergel03a}.

The first sky surveys and galaxy mappings surprisingly revealed that
there is structure in the distribution of matter at very large scales
in the order of 100\+Mpc~\cite{delapparent86a}. Using the Big Bang
model and assumptions about the types of matter that make up the
universe, it is possible to predict the matter distribution.
Furthermore, the survey of Type Ia supernovae yields a relation
between the host galaxies' redshift and their distance, so that
supernovae of this species are regarded as standard candles, allowing
to investigate the expansion history of the universe~\cite{tonry03a}.
The large-scale structure and supernova data as well as the CMB
results clearly favor a class of models for which the term
$\Lambda$CDM was coined\footnote{CDM -- cold dark matter; $\Lambda$
  denotes the cosmological constant.}. The basic statement of the
latter is that ordinary matter (also referred to as {\sl baryonic\/}
matter) constitutes only a small fraction of the energy density of the
universe~\cite{spergel06a}. More than 75\+\% is accounted for by what
is called the {\sl dark energy\/}, and about 20\+\% must exist in form
of some kind of non-relativistic {\sl dark matter\/}. The nature of
these two most dominant forms of energy in the universe is one of the
cardinal unsolved questions in modern cosmology, and the answer will
doubtlessly give rise to fundamentally new physical concepts. While
cosmology is presently far from a distinct approach to solving the
dark energy problem, the nature of dark matter is possibly now close
to being resolved.

\section{The History of the Universe}
\begin{figure}[bt]
\begin{center}
\includegraphics[width=16cm]{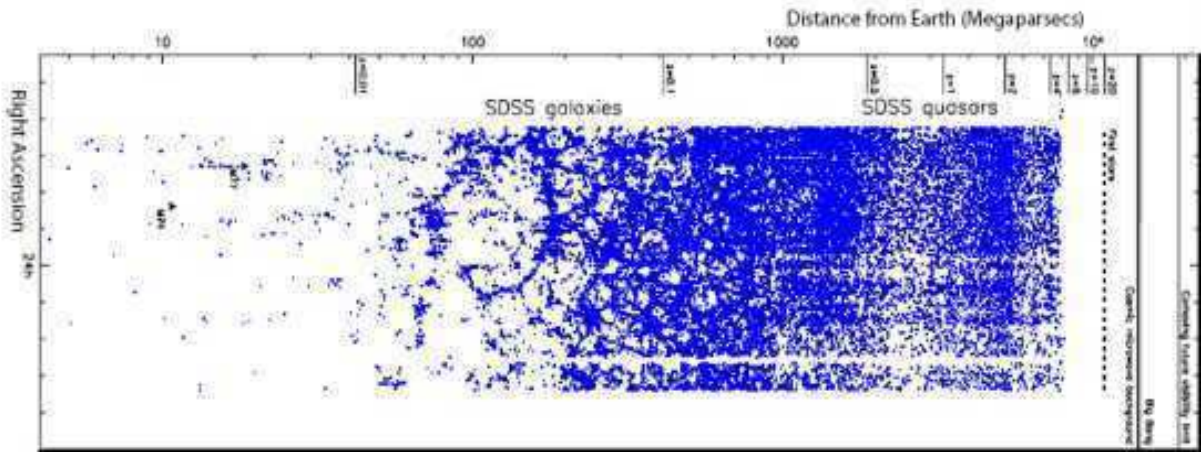}
\caption{\label{fig:sdss} A map of the universe: the distribution of 
distant galaxies exhibits structure on large scales~\cite{gott05a}.}
\end{center} 
\end{figure}
The standard model of the universe is the Hot Big Bang
model. According to this theory, the universe came into existence 13.7
billion years ago~\cite{spergel06a}, starting in an enormously
compressed and very hot state. Very little is known about the first
$10^{-43}$ s, the so called {\sl Planck Epoch}, in which the three
forces we know today (see \S\+\ref{section:sm}) were presumably
unified. After the end of this period, the universe entered an
approximately $10^{-35}$\+s long phase of exponentially quick expansion
{\sl (inflation)\/}, during which its size increased by about 25
orders of magnitude~\cite{guth05a}.  The driving force of this process
is believed to be a yet unknown scalar field, called the {\sl
inflaton\/}, whose initial quantum fluctuations from the Planck Epoch
were blown up to macroscopic scales. These are nowadays represented by
the large-scale structure (Figure~\ref{fig:sdss}). Simultaneously, a
possible curvature of space-time in the early universe was stretched
out by inflation, giving the universe its Euclidean geometry which is
observed today~\cite{spergel06a}.

After the end of inflation, the energy in the universe was dominated
by radiation and a sea of relativistic particle-antiparticle pairs
near thermal equilibrium~\cite{turner99a}. As the universe further
expanded and cooled, particle species dropped out of equilibrium once
the thermal energy $k_{\mathrm{B}}T$ fell below their mass, and
vanished through annihilation. According to {\sl Sakharov's
conditions\/}~\cite{sakharov67a}, the existence of CP-violating (see
\S\+\ref{section:sm}) and baryon number violating processes out of
thermal equilibrium may have lead to a tiny surplus of particles with
respect to antiparticles which remained. At $k_{\mathrm{B}}T\approx
200$\+\MeV{}, $10^{-6}$\+s after the inflation, a phase transition
occured from a quark-gluon plasma to protons, neutrons, and pions,
along with the leptons, antileptons, and photons {\sl
(baryogenesis)\/}. When the universe was about 100\+s old and
$k_{\mathrm{B}}T\approx 1$\+\MeV{}, the temperature was low enough to
allow for the creation of light nuclei such as D, $^3$He, $^4$He, and
$^7$Li {\sl (nucleosynthesis)\/}.  During this period, the last
particle-antiparticle pairs, the electrons and positrons, annihilated.

It took 370,000 years~\cite{spergel03a} until the temperature was low
enough ($k_{\mathrm{B}}T\approx 0.3$\+\eV{}) for stable hydrogen atoms to
be formed from the remaining protons and electrons {\sl
(recombination)\/}. At that time, the universe became transparent for
radiation, which since then propagates almost freely. Today, it has
cooled down to 2.7\+K and is observable as the remarkably isotropic
CMB. With the recombination and until the emergence of the first stars
several hundred million years later, the {\sl Dark Age\/} of the
universe had begun.

The present universe can be characterized as spatially flat --
Euclidean -- and homogenous and isotropic on the largest scales. It is
expanding so that objects move away from each other with a speed of
73\+km/s per Mpc distance~\cite{spergel06a}. There is strong evidence
that the expansion is presently accelerated~\cite{tonry03a}, which can
be explained by the presence of non-material dark energy with
particular properties. More than 75\+\% of the energy density of the
universe exists as dark energy, another 20\+\% as an unknown form of
gravitating matter which otherwise does not observably participate in
electromagnetic interactions. Energy densities are usually stated as
fractions $\Omega=\rho / \rho_{\mathrm{c}}$ of the {\sl critical
density\/} $\rho_{\mathrm{c}}$ which leads to a spatially flat
universe. The value of $\rho_{\mathrm{c}}$ corresponds to about six
hydrogen atoms per cubic meter. In these terms,
$\Omega_{\mathrm{\Lambda}}=0.76$ for dark energy,
$\Omega_{\mathrm{m}}=0.24$ for the total matter density, and
$\Omega_{\mathrm{b}}=0.04$ for baryonic matter
only~\cite{spergel06a}. The total energy density of the universe is
$\Omega = 1.003\pm 0.015$~\cite{spergel03a}. Matter is concentrated in
galaxies, whose spatial distribution exhibits structure on scales of
1-100\+Mpc in form of clusters, walls and voids. Since the hot and
highly pressurized baryonic matter would have quickly washed out such
structure, its formation can best be explained by the presence of dark
matter.

\section{Evidence for Dark Matter}
\begin{figure}[tb]
\begin{center}
\begin{tabular}{lr}
\begin{minipage}{7.5cm}
\begin{center}
\includegraphics[width=7.5cm]{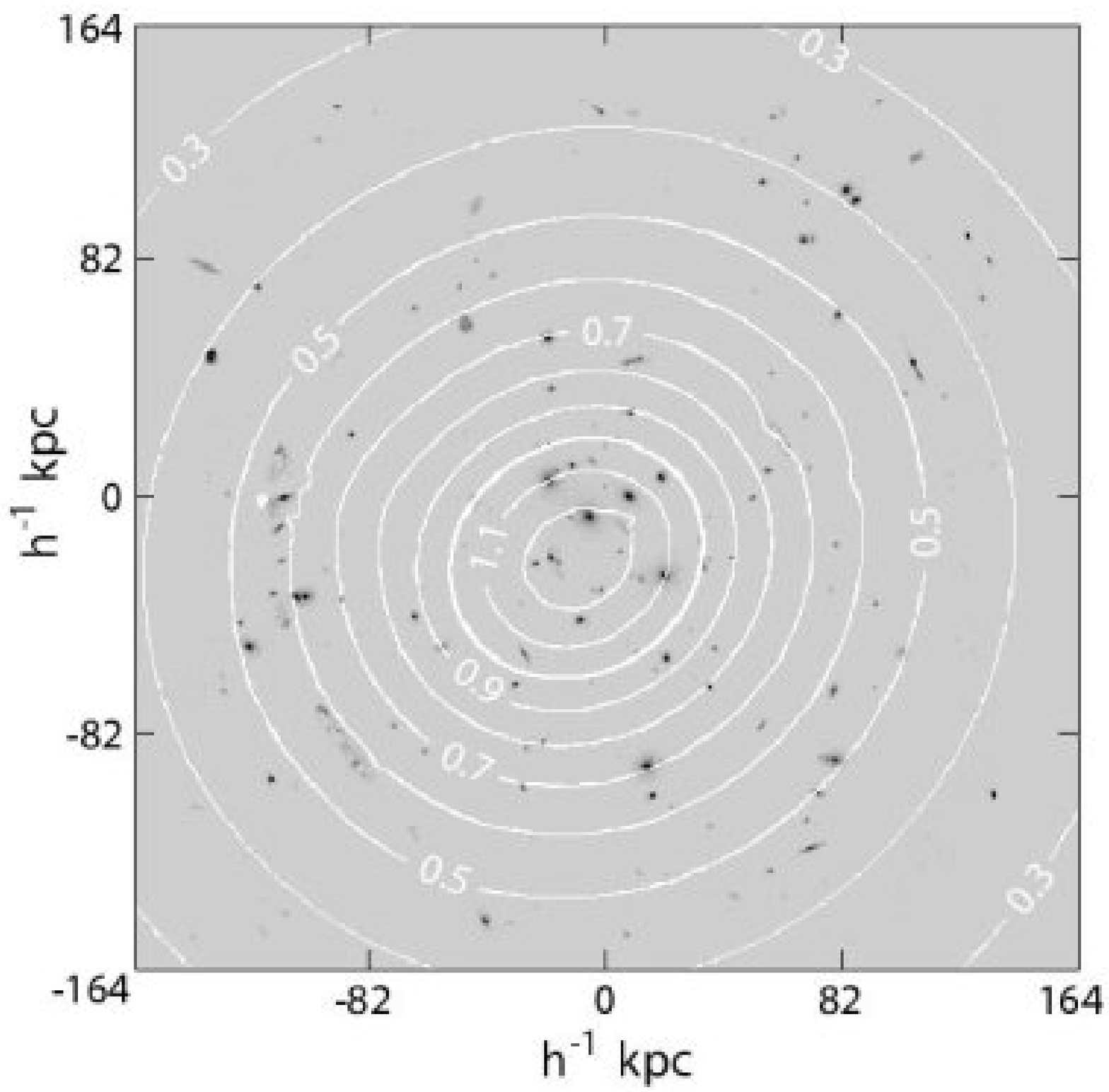}
\end{center}
\end{minipage}
&
\begin{minipage}{7.5cm}
\begin{center}
\includegraphics[width=7.5cm]{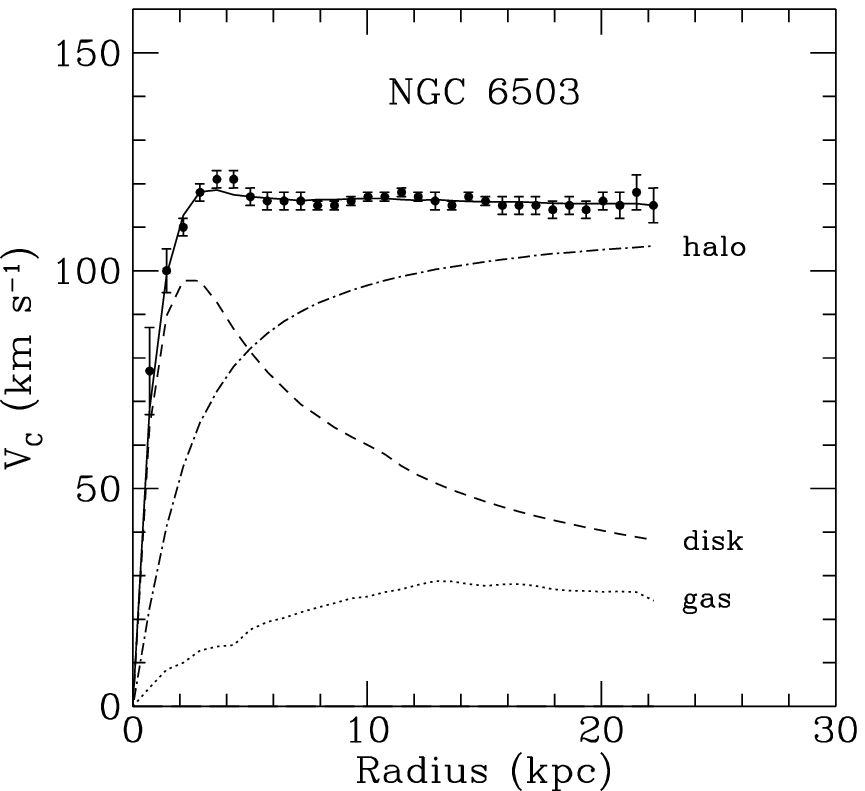}
\end{center}
\end{minipage}
\end{tabular} 
\end{center} 
\caption{\label{fig:cl0024-ngc6503} The reconstructed surface mass 
density not associated with visible galaxies in the galaxy cluster
CL0024 (white contours, in units of $2.1\cdot
10^3\,\mathrm{M}_{\odot}\, \mathrm{pc}^{-2}$) superposed on an image
by the Hubble Space Telescope~\cite{tyson98a} {\sl (left)\/}; rotation
curve of the galaxy NGC6503~\cite{begeman91a} with contributions from
stars (disk), interstellar gas and the dark matter halo {\sl
(right)\/}.}
\end{figure}
Evidence for the presence of significant amounts of invisible matter
is numerous and apparent from small to the largest scales. First
indication was found as early as 1933 in the Coma galaxy
cluster~\cite{zwicky33a}, which revealed a mass-to-light ratio two
orders of magnitude larger than in the solar neighborhood. The mass of
a cluster can be determined via several methods, including application
of the virial theorem to the observed distribution of radial
velocities, by weak gravitational lensing\footnote{A gravitational
lens distorts the image of a background light source. In contrast to
strong lensing, where there are easily visible distortions such as
multiple images of the light source, weak lensing only leads to faint
distortions on the percent level.} and by studying the profile of
X-ray emission that traces the distribution of hot emitting
gas~\cite{bertone04a}. Figure~\ref{fig:cl0024-ngc6503} {\sl (left)\/}
shows the reconstructed distribution of non-luminous -- dark -- excess
matter in the galaxy cluster CL0024~\cite{tyson98a}, which apparently
concentrates toward the cluster's center. Recent observations in this
field are consistent with a contribution of 20--30\+\% from matter to
the total energy density (see e.g.~\cite{carlberg98a}). This is way
above what can be expected from the amount of visible matter alone.

The most direct evidence for dark matter on galactic scales comes from
observations on the rotation curves of galaxies. The radial velocity of a
star at distance $r$ from the center of the galaxy is (classically)
given by:
\begin{equation}
v(r) = \sqrt{\frac{G\cdot M(r)}{r}}\, ,
\end{equation}
with $G$ being the gravitational constant, and $M(r)$ = $\int\rho
(r)r^2dr$ the integrated matter density. Thus, the velocity should be
falling $\propto 1/\sqrt{r}$ outside the luminous galactic disk, as
indicated by the dashed rotation curve in
Figure~\ref{fig:cl0024-ngc6503} {\sl (right)\/} for the galaxy
NGC6503~\cite{begeman91a}. However, the data show that the actual
radial velocity is largely a constant function of the distance from
the center. This could be explained with the assumption that the
galactic disk is embedded in a {\sl galactic halo\/} of non-luminous
dark matter, represented by the dash-dotted line in
Figure~\ref{fig:cl0024-ngc6503} {\sl (right)\/}. Basically all galaxies
studied up to now share this behavior~\cite{sofue01a}.

\begin{figure}[bt]
\begin{center}
\includegraphics[width=10cm]{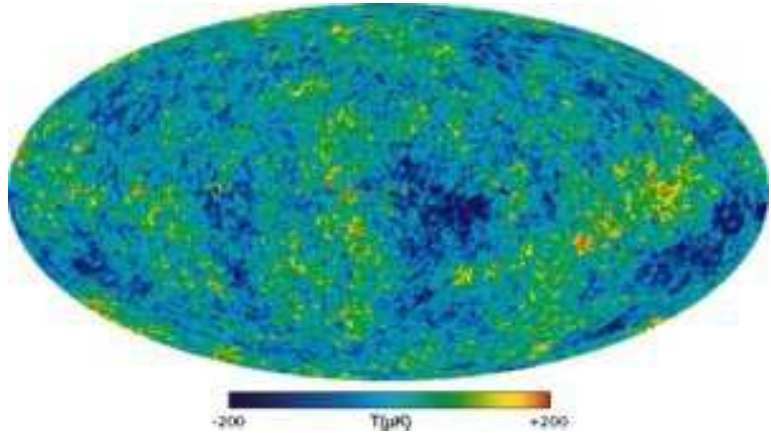}
\caption{\label{fig:cmbmap}Map of CMB temperature fluctuations by WMAP~\cite{hinshaw06a}.}
\end{center} 
\end{figure}
\begin{figure}[bt]
\begin{center}
\includegraphics[width=10cm]{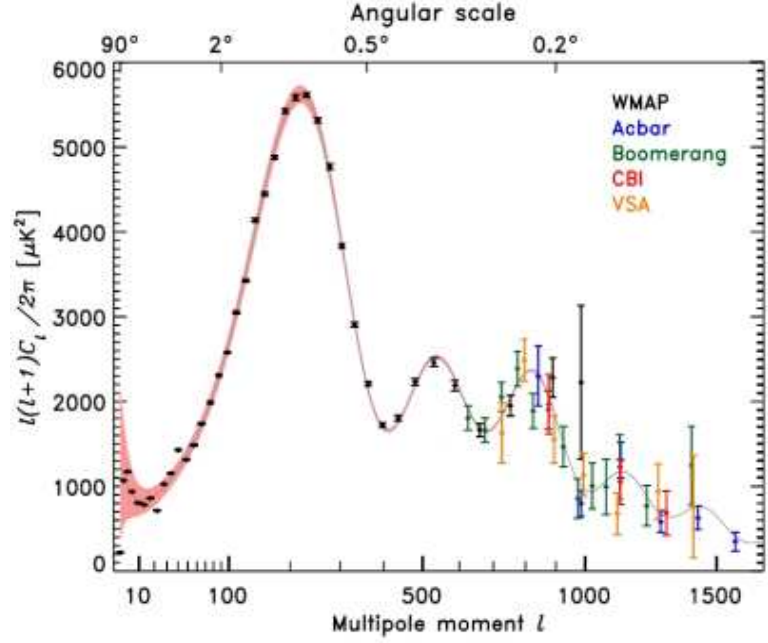}
\caption{\label{fig:wmapcmb}A compilation of CMB power spectrum data and the best
$\Lambda$CDM fit~\cite{hinshaw06a}.}
\end{center} 
\end{figure}

\begin{figure}[bt]
\begin{center}
\includegraphics[width=13cm]{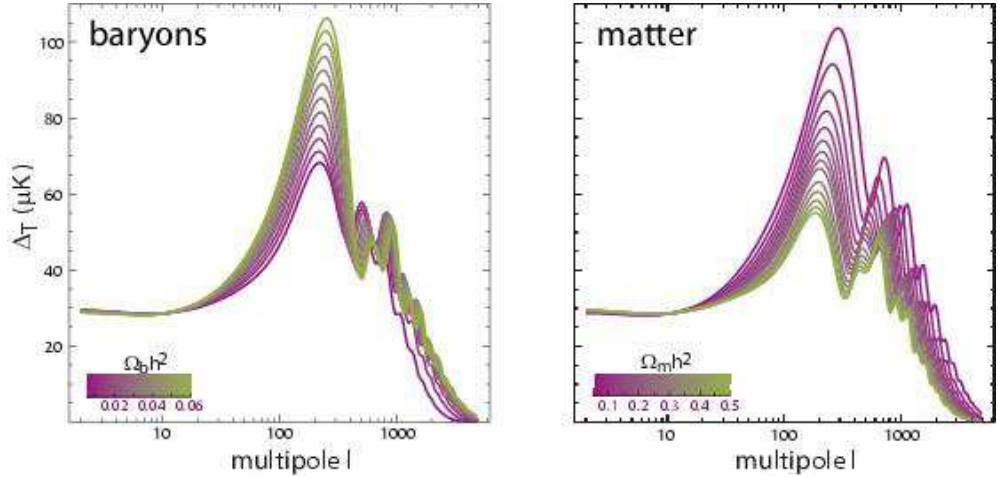}
\caption{\label{fig:cmbmodel} Sensitivity of the CMB power spectrum
to the baryonic and all-matter energy density contributions
$\Omega_{\mathrm{b}}$ {\sl (left)\/} and $\Omega_{\mathrm{m}}$ {\sl
(right)\/}~\cite{hu03a}. The parameter $h$ equals
about~0.74~\cite{spergel06a}.}
\end{center} 
\end{figure}
The spectacular recent CMB measurements from the Wilkinson Microwave
Anisotropy Probe (WMAP)~\cite{spergel03a} and other experiments allow
the determination of the cosmological parameters with unprecedented
accuracy. The observed temperature anisotropies in the sky (see
Figure~\ref{fig:cmbmap}) are usually expanded in spherical harmonics
$Y_{\ell m}\left( \theta,\phi\right)$ as
\begin{equation}
\frac{\delta T}{T}\left( \theta,\phi\right) = \sum^{+\infty}_{\ell =2}\sum^{\ell}_{m=-\ell}
a_{\ell m}Y_{\ell m}\left( \theta,\phi\right)
\end{equation}
with the variance $C_{\ell}\equiv \langle|a_{\ell
  m}|^2\rangle$~\cite{bertone04a}. Plotted as a function of $\ell$ in
a power spectrum (see Figure~\ref{fig:wmapcmb}), the variance reveals a
noticeable structure with a series of distinct peaks. They roughly
correspond to resonances of acoustic oscillations in the photon-baryon
plasma at the time of recombination. The oscillations occured around
the equilibrium of radiation pressure and gravity. Since dark matter
gravitates but does not interact with photons, it had a significant
impact on the resonances. As a consequence, besides a large set of
additional cosmological parameters, the CMB power spectrum is
sensitive to the dark matter content of the universe.
Figure~\ref{fig:cmbmodel} illustrates the effect of varying values of
$\Omega_{\mathrm{b}}$ and $\Omega_{\mathrm{m}}$ on the power spectrum.
The analysis of the data in the framework of the $\Lambda$CDM
model~\cite{spergel06a} gives $\Omega_{\mathrm{b}}h^2=0.0223\pm
0.0008$ and $\Omega_{\mathrm{m}}h^2=0.126\pm 0.009$ with $h=0.74\pm
0.03$, which confirms that only about one fifth of the matter content
of the universe is comprised of ordinary matter.

As already stated, dark matter has left traces in the large-scale
distribution of matter in the universe. Data from recent galaxy
surveys~\cite{cole05a,tegmark04b} yield the spatial density
distribution of large numbers of galaxies. The squared Fourier
transform of this distribution -- the power spectrum $P(k)$ --
constrains large-scale structure formation and cosmological
parameters, such as the ratio
$\Omega_{\mathrm{b}}/\Omega_{\mathrm{m}}=0.185\pm
0.016$~\cite{cole05a}, which is in excellent agreement with the CMB
results. Furthermore, it can be inferred from the data that most of
the dark matter particles must be nonrelativistic {\sl (''cold'')\/} and
thus quite heavy, since relativistic particles would have washed out
structure on medium scales due to their large free-streaming length.

Altogether, there is clear evidence for the existence of an unknown
non-baryonic contribution to the matter content of the universe with
an average density corresponding to roughly one hydrogen atom per cubic
meter. Hence, the question for the nature of this dark matter
arises. Is the current understanding of particle physics sufficient to
explain its presence, or will new theoretical approaches be
necessary?

\section{The Standard Model of Particle Physics}\label{section:sm}
The Standard Model of Particle Physics (SM) has, for many years,
accounted for all observed particles and interactions (see
e.g.~\cite{yao06a} for reviews). In the SM, the fundamental
constituents of matter are particles with half-integer spin
(fermions): quarks and leptons. Their interactions are mediated by
integer spin particles called gauge bosons. Strong interactions are
mediated by gluons $g$, electroweak interactions by $W^{\pm}$, $Z^0$
and $\gamma$. The masses of the fermions and the $W^{\pm}$ and $Z^0$
are generated through the Higgs mechanism~\cite{higgs64a} via
couplings to the Higgs boson $H^0$. Gravitation is presently not
covered by the SM, since General Relativity and the SM alone cannot be
combined in one quantum theory. The left-handed leptons and quarks are
arranged into three generations of $SU(2)_L$ doublets:
\begin{equation}
\begin{array}{c}
\left(\begin{array}{c}\nu_e\\e^-\end{array}\right)_L \\ \\
\left(\begin{array}{c}u\\d'\end{array}\right)_L
\end{array}
\begin{array}{c}
\left(\begin{array}{c}\nu_{\mu}\\ \mu^-\end{array}\right)_L \\ \\
\left(\begin{array}{c}c\\s'\end{array}\right)_L
\end{array}
\begin{array}{c}
\left(\begin{array}{c}\nu_{\tau}\\ \tau^-\end{array}\right)_L \\ \\
\left(\begin{array}{c}t\\b'\end{array}\right)_L
\end{array}
\end{equation}
with the corresponding right-handed fields transforming as singlets
under $SU(2)_L$. Each generation contains two flavors of quarks with
baryon number $B = 1/3$ and lepton number $L = 0$ and two leptons with
$B=0$ and $L=1$. Each particle also has a corresponding antiparticle
with the same mass and opposite quantum numbers. The primed quarks are
weak eigenstates related to mass eigenstates by the unitary
Cabibbo-Kobayashi-Maskawa (CKM) matrix
$V_{\mathrm{CKM}}$~\cite{cabibbo63a,kobayashi73a} through
\begin{equation}
\left(\begin{array}{c}d'\\s'\\b'\end{array}\right) = 
\left(\begin{array}{ccc}
V_{ud}&V_{us}&V_{ub} \\
V_{cd}&V_{cs}&V_{cb} \\
V_{td}&V_{ts}&V_{tb} 
\end{array}\right)
\left(\begin{array}{c}d\\s\\b\end{array}\right)
= V_{\mathrm{CKM}} \left(\begin{array}{c}d\\s\\b\end{array}\right)
\end{equation}
with complex parameters $V_{ij}$. Under the assumption that exactly
three quark generations exist and that the CKM matrix is unitary, the
parameters can be constrained and partially absorbed into arbitrary
phases. Four independent parameters remain, one of which can be
interpreted as an imaginary phase. In the SM, this phase is
responsible for the violation of the CP symmetry, which transforms
particles into antiparticles with regard to their handedness.

Local gauge symmetries play a fundamental role in particle physics. It
is in fact in terms of symmetries and using the formalism of gauge
theories that electroweak and strong interactions are described. The
SM is based on the $SU(3)_C\otimes SU(2)_L\otimes U(1)_Y$ gauge
theory, which undergoes the spontaneous breakdown
\begin{equation}
SU(3)_C\otimes SU(2)_L\otimes U(1)_Y \rightarrow SU(3)_C\otimes U(1)_Q\, ,
\end{equation}
where $Y$ and $Q$ denote the  weak hypercharge and the electric charge
generators, respectively,  and $SU(3)_C$ describes  the strong (color)
interaction, known as  Quantum ChromoDynamics  (QCD). This spontaneous
symmetry breaking results in the  generation of the massive  $W^{\pm}$
and $Z$ gauge bosons as well as a massive scalar Higgs field. 

The only viable dark matter candidate from the SM particle table is
the neutrino $\nu$. The neutrino number density is known to be about
3/11 of the CMB photon density~\cite{overduin04a}, thus yielding
$\Omega_{\nu}=\sum m_{\nu}c^2 / (h^2\cdot 93.8\,\mathrm{\eV{}})$ with
the sum running over the three generations. With an upper limit on the
sum of the neutrino masses of 0.72\+\eV{}/c$^2$ (95\+\% C.L.) from
large-scale structure and CMB data~\cite{spergel06a}, the energy
density contribution from neutrinos can be constrained to
$\Omega_{\nu}<0.014$. Hence, neutrinos seem unlikely to constitute a
relevant fraction of the dark matter in the universe. Further
candidates from conventional physics include primordial black holes
(see e.g.~\cite{overduin04a}) or massive objects\footnote{referred to
as MACHOS -- massive compact halo objects} such as cold dwarf
stars~\cite{paczynski86a}. Up to now there has been no evidence that
one of these is a good candidate, and in particular baryonic matter
cannot constitute more than a small fraction of the total dark matter,
thus it is likely that new physics needs to be introduced in order to
approach the nature of dark matter. There is a wide range of
candidates from non-Standard Model theories, including, amongst many
others, axions and Kaluza-Klein states (see
e.g.~\cite{bertone04a,overduin04a} for reviews). However, the best
motivated are provided by {\sl supersymmetric extensions} to the SM.

\section{Supersymmetric Dark Matter}\label{section:susy}
Supersymmetry (SUSY)~\cite{martin97a} is a generalization of the
space-time symmetries of quantum field theory that transforms fermions
into bosons and vice versa. In other words, the generators $Q$ of
supersymmetry act on particle fields as
\begin{equation}
Q|\mathrm{fermion}\rangle = |\mathrm{boson}\rangle\, ,\quad 
Q|\mathrm{boson}\rangle = |\mathrm{fermion}\rangle\, ,
\end{equation}
such that there exist superpartners of the SM fields which mix or
directly form supersymmetric particles {\sl (''sparticles'')\/}.
Supersymmetry was originally postulated as a solution for the problem
of quadratically divergent renormalization corrections to the Higgs
boson mass, as such divergences cancel out with the introduction of
equally heavy superpartners of the SM fields.  Additionally,
supersymmetry at \TeV{} energies could modify the running coupling
constants of electroweak and strong interactions in a way that they
naturally converge to a common value at some very high energy scale
$\sim 2\cdot 10^{16}$\+\GeV{} (GUT\footnote{GUT -- grand unified
theories} scale)~\cite{amaldi91a}. This physically attractive feature
of SUSY is not present in the SM framework alone. Moreover, in
contrast to the SM, the local extension of supersymmetry {\sl
(supergravity)\/} may provide a framework for the unification of the
known forces with gravity.

\begin{table}[bt]
\small 
\begin{center}
\begin{tabular}{lllllll} \hline 
  \multicolumn{2}{c}{Standard Model particles and fields} & \multicolumn{5}{c}{Supersymmetric partners} \\
  & & \multicolumn{3}{l}{Interaction eigenstates} & \multicolumn{2}{l}{Mass 
  eigenstates} \\
  Symbol & Name & Symbol & Name & & Symbol & Name \\ \hline
  $q=d,c,b,u,s,t$ & quark & $\tilde{q}_{L}$, $\tilde{q}_{R}$ & 
  squark & & $\tilde{q}_{1}$, $\tilde{q}_{2}$ & squark \\
  $l=e,\mu,\tau$ & lepton & $\tilde{l}_{L}$, $\tilde{l}_{R}$ & slepton & 
  & $\tilde{l}_{1}$, $\tilde{l}_{2}$ & slepton \\
  $\nu = \nu_{e}, \nu_{\mu}, \nu_{\tau}$ & neutrino & $\tilde{\nu}$ & 
  sneutrino & & $\tilde{\nu}$ & sneutrino \\
  $g$ & gluon & $\tilde{g}$ & gluino & & $\tilde{g}$ & gluino \\
  $W^\pm$ & $W$-boson & $\tilde{W}^\pm$ & wino & & & \\
  $H^-$ & Higgs boson & $\tilde{H}_{1}^-$ & higgsino & 
  \raisebox{-.25ex}[0ex][0ex]{$\left. \raisebox{0ex}[-3.3ex][3.3ex]{}
  \right\}$} &  $\tilde{\chi}_{1,2}^\pm$ & chargino \\
  $H^+$ & Higgs boson & $\tilde{H}_{2}^+$ & higgsino & & & \\
  $B$ & $B$-field & $\tilde{B}$ & bino & & & \\
  $W^3$ & $W^3$-field & $\tilde{W}^3$ & wino & & & \\
  $H_{1}^0$ & Higgs boson & 
  \raisebox{-1.75ex}[0ex][0ex]{$\tilde{H}_{1}^0$} & 
  \raisebox{-1.75ex}[0ex][0ex]{higgsino} & 
  \raisebox{.25ex}[0ex][0ex]{$\left. \raisebox{0ex}[-5.25ex][5.25ex]{}
  \right\}$} & \raisebox{0.5ex}[0ex][0ex]{$\tilde{\chi}_{1,2,3,4}^0$} & 
  \raisebox{.5ex}[0ex][0ex]{neutralino} \\[0.5ex]
  $H_{2}^0$ & Higgs boson & 
  \raisebox{-1.75ex}[0ex][0ex]{$\tilde{H}_{2}^0$} & 
  \raisebox{-1.75ex}[0ex][0ex]{higgsino} & & & \\[0.5ex]
  $H_{3}^0$ & Higgs boson & & & & & \\[0.5ex] \hline
\end{tabular}
\label{tab:susyparticles}
\caption[Standard Model particles and their superpartners in the MSSM]
{\label{tab:susy}Standard Model particles and their superpartners in
  the MSSM~\cite{bertone04a}.}
\end{center}
\end{table}
The {\sl Minimal Supersymmetric Standard Model} (MSSM) has the
smallest possible field content necessary to give rise to all the
particles of the SM (see Table~\ref{tab:susy}). In the MSSM, unlike
the SM, two Higgs doublets, corresponding to five physical Higgs
bosons, are needed to give masses to all the fermions. The four {\sl
neutralinos} $\tilde{\chi}_{1,2,3,4}^0$ are linear combinations of the
superpartners of the gauge bosons $B$ and $W^3$ and the neutral
components of the Higgs doublets (higgsinos), as
\begin{equation}\label{eq:neutralinoComposition}
\tilde{\chi}_i^0 = c_{1i}\tilde{B} + c_{2i}\tilde{W}^3 + c_{3i}\tilde{H}_1^0 + c_{4i}\tilde{H}_2^0\, ,
\end{equation} 
while the two {\sl charginos} $\tilde{\chi}_{1,2}^\pm$ represent
superpositions of the charged components $\tilde{H}^{\pm}$ and
$\tilde{W}^{\pm}$ of the wino and the higgsinos. One additional
ingredient usually used to constrain the MSSM is the conservation of
{\sl R-parity}. $R$ is a multiplicative quantum number defined as
$R\equiv (-1)^{3B+L+2S}$ for a particle with spin $S$, baryon number
$B$ and lepton number $L$. All SM particles have $R=1$ and all their
superpartners have $R=-1$. As a consequence of $R$-parity
conservation, SUSY particles can only decay into an odd number of SUSY
particles, plus SM particles. Therefore, the {\sl lightest
  supersymmetric particle} (LSP) is stable.

If SUSY were an exact symmetry of nature, then particles and their
superpartners would have identical mass. Since sparticles have not yet
been observed, SUSY must be a broken symmetry. The simplest symmetry
breaking mechanism, spontaneous symmetry breaking, is experimentally
ruled out at a mass scale below 1\+\TeV{}~\cite{chung05a}. Thus SUSY
is thought to be broken spontaneously in a {\sl hidden sector\/},
which consists of sparticles which do not directly interact with the
{\sl visible\/} SM particles or their superpartners. Symmetry breaking
is then communicated to the visible sector via superheavy messenger
particles. The cancellation of quadratic divergences mentioned above
cannot be maintained in a broken SUSY, however the remaining
divergences are only logarithmic and also proportional to the mass
difference of particles and their superpartners. As a consequence, the
sparticle masses are constrained and at least several of them should
not significantly exceed 1\+\TeV{}/c$^2$~\cite{martin97a}. Thus
sparticles are in principle accessible to high energy colliders such
as the LHC, which will be the subject of the next chapter.

\begin{figure}[bt]
\begin{center}
\begin{tabular}{lr}
\begin{minipage}{7.9cm}
\begin{center}
\includegraphics[width=7.5cm]{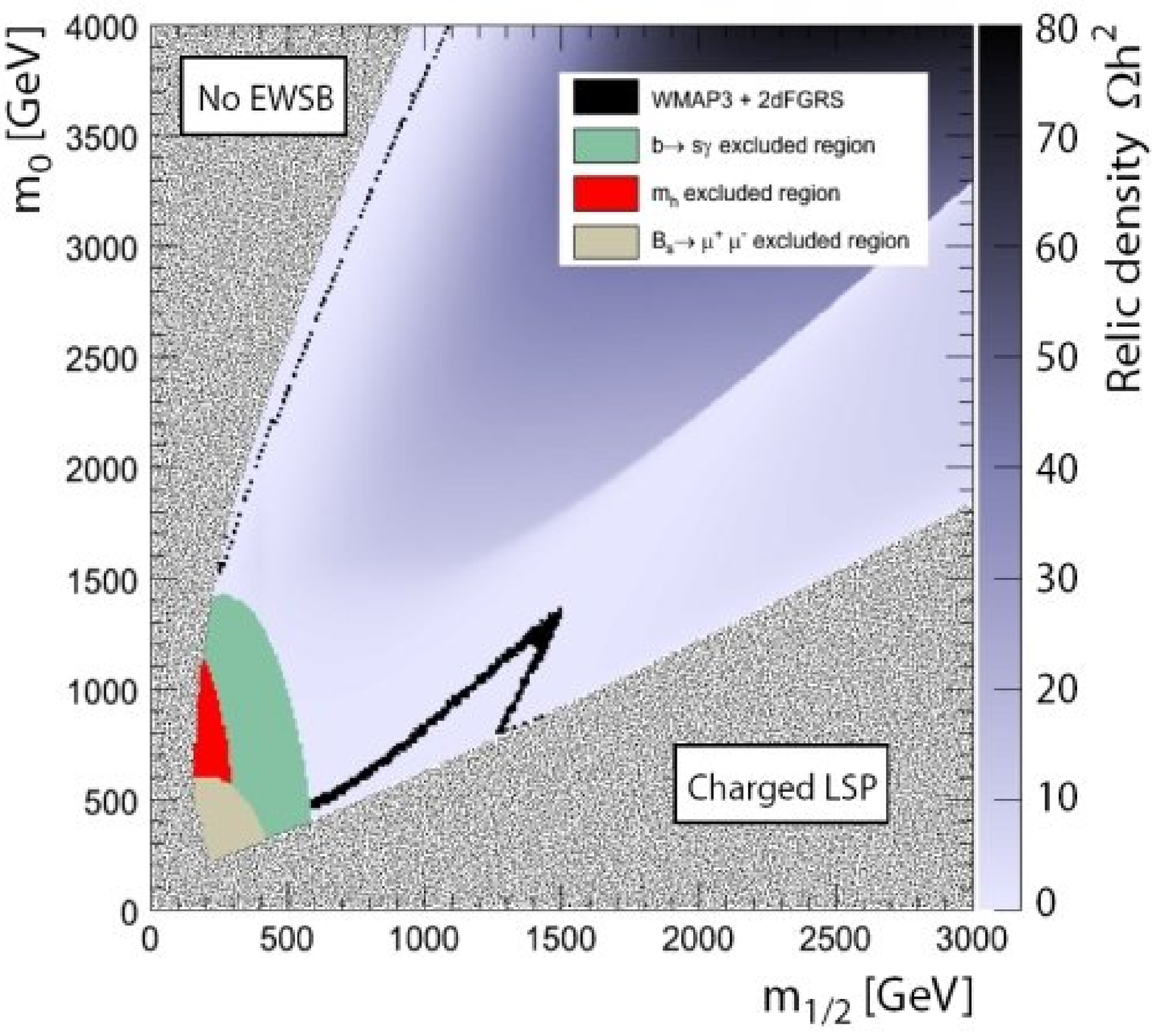}
\end{center}
\end{minipage}
&
\begin{minipage}{7.2cm}
\begin{center}
\includegraphics[width=7.2cm]{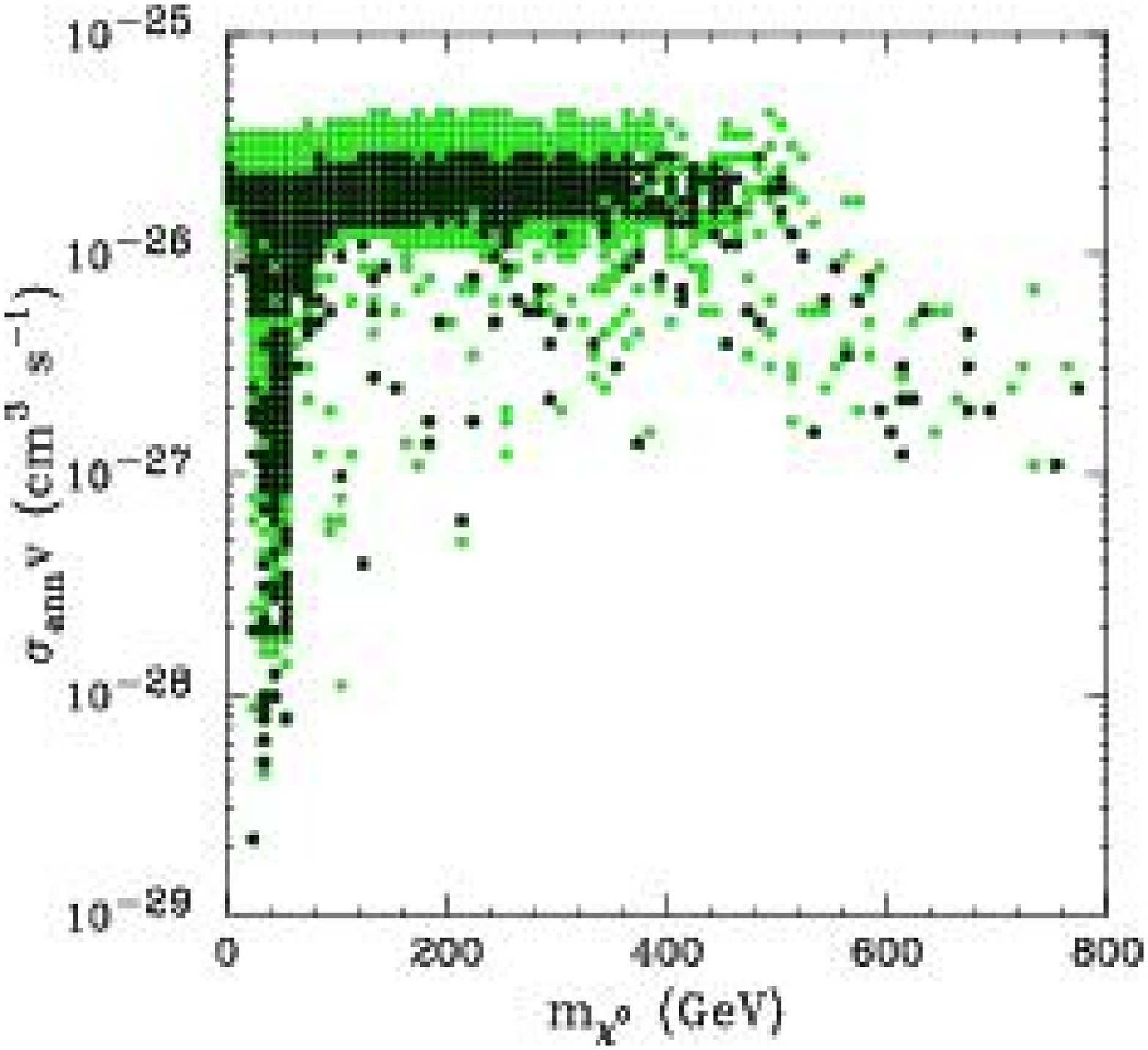}
\end{center}
\end{minipage}
\end{tabular} 
\end{center} 
\caption{\label{fig:msugra}Constraints on the mSUGRA parameters $m_0$ and
$m_{1/2}$ and the neutralino relic density in the allowed regions for
$\tan\beta=52$, $A_0=0$, $\mu>0$. The grey dotted and solid colored
regions are excluded. The black contour denotes the cosmologically
allowed region~\cite{chung07a} {\sl (left)\/}; neutralino annihilation cross
sections for parameters in the MSSM which are compatible with
$\Omega_{\mathrm{CDM}}h^2=0.095-0.129$ (dark points) or
$\Omega_{\mathrm{CDM}}h^2=0.06-0.16$ (light green points) for
$\tan\beta=50$~\cite{hooper05a} {\sl (right)\/}.}
\end{figure}

The MSSM has 124 free parameters, compared to 19 of the
SM\footnote{Here, the neutrinos are considered to be
massless.}~\cite{chung05a}. In the so called {\sl minimal
supergravity\/} (mSUGRA) models, gravitation is additionally taken
into account, so that the spin-2 graviton has a spin-3/2 fermion
superpartner called {\sl gravitino\/}. Gravitational interactions then
communicate the symmetry breaking from the hidden to the visible
sector. In the mSUGRA case the number of parameters can be
significantly reduced by assuming that the model obeys a set of well
motivated boundary conditions at the highest energies, thus
eliminating all but 5 parameters:
\begin{itemize}\setlength{\itemsep}{-0.2mm}
\item $\tan\beta$, the ratio of the vacuum expectation values of the
two Higgs doublets,
\item $m_{1/2}$, the common gaugino mass at the GUT scale,
\item $m_0$, the common scalar mass at the GUT scale,
\item $A_0$ denoting the trilinear couplings of Higgs bosons to the sfermions, and
\item $\mathrm{sign}(\mu )$, the sign of the Higgs boson mass parameter.
\end{itemize}\setlength{\itemsep}{+0.2mm}
This small number of parameters makes mSUGRA a fruitful field for
supersymmetry phenomenology. Figure~\ref{fig:msugra} {\sl (left)\/}
shows the plane spanned by the parameters $m_0$ and $m_{1/2}$ for
fixed values of $\tan\beta$, $A_0$ and $\mathrm{sign}(\mu )$. Large
regions can already be excluded on a theoretical basis, for example
the region $m_0\gg m_{1/2}$, where the model does not feature
electroweak symmetry breaking (see previous section), or at low $m_0$,
where the LSP would be a charged sparticle. Further constraints are
imposed by a wide range of data from e.g. WMAP, collider and cosmic
ray experiments~\cite{chung07a}. For the allowed regions, the
corresponding neutralino relic density $\Omega_{\chi}h^2$ is
given. The black contour denotes the region consistent with results
from WMAP and galaxy surveys. The allowed regions are typically
located close to the borders of the excluded areas.

In mSUGRA and in a multitude of other SUSY models, the LSP is the
lightest of the four neutralinos $\tilde{\chi}_1^0$. The mass of the
$\tilde{\chi}_1^0$ is currently constrained by collider experiments to
be larger than 46\+\GeV{}/c$^2$ (95\+\%~C.L.)~\cite{yao06a}. Neutralinos are
Majorana particles, thus they can annihilate pairwise into SM
particle states (see \S\+\ref{sec:susyspectra}). The product of the
annihilation cross section times the particle velocity
$\sigma_{\mathrm{ann}}v$ as a function of the $\tilde{\chi}_1^0$ mass
is displayed in Figure~\ref{fig:msugra} {\sl (right)\/} for those sets
of MSSM parameters which are consistent with the WMAP observations with
$\tan\beta=50$. Apparently, the data prefer neutralino masses of few
hundred \GeV{} and $\sigma_{\mathrm{ann}}v\approx
10^{-26}\,\mathrm{cm}^3\,\mathrm{s}^{-1}$. Given these numbers, it can
be calculated that neutralinos must have left the primordial
equilibrium at non-relativistic energies and that the cross section
is indeed close to the rough expectation of
$\sigma_{\mathrm{ann}}v\approx
10^{-25}\,\mathrm{cm}^3\,\mathrm{s}^{-1}$ for particles in this mass
range annihilating through weak interactions~\cite{jungman96a}. Taking
all this into account, the neutralino appears as one of the most
favorable candidates for particles which constitute large parts
of cold dark matter in the universe: it has a high mass but does not carry charge,
interacts only weakly with ordinary matter, has the correct
abundance and propagates at non-relativistic speeds. Particles with
these properties are generally referred to as {\sl weakly interacting
  massive particles\/} (WIMPs).  The fact that the neutralino does
interact at all and, moreover, is embedded in a framework of other
sparticles, in principle opens the possibility to identify it as the
constituent of dark matter by experiment.

\section{Approaches to Understanding Dark  Matter}\label{sec:understanding_dm} 
It has been pointed out in this chapter that evidence for the
existence of dark matter is widespread in astrophysics and cosmology.
The fact that particle physics can provide reliable models for dark
matter candidates has established a new creative interplay of these
domains in the recent past. A number of collider and non-accelerator
experiments are currently operating or will soon be deployed that will
shed light on dark matter. In addition, there exist both direct and
indirect non-accelerator dark matter search experiments that are
ongoing or proposed~\cite{baer04a}. Prospects for detecting dark
matter and determining its properties are particularly bright in the
case of the supersymmetric neutralino.

From the view of particle physics, the dark matter question is
strongly connected to those new physical phenomena which inevitably
wait at energies beyond the reach of present experiments. Therefore,
the particle physics program includes searching for supersymmetric
particles at new collider facilities, particularly the LHC, to
possibly identify the LSP and determine its properties. The LHC and
its experiments are currently under construction, so that their
completion is naturally among the primary tasks regarding the
search for dark matter. Since the field of supersymmetry contains an
enormous variety of ideas, models and parameters, input from the
domain of astronomy essentially helps to put constraints on
supersymmetry and to focus the theoretical as well as the experimental
work on realistic scenarios.

From the view of astrophysics and cosmology, particle physics offers
checkable predictions about the behavior of dark matter and its
relation to fundamental concepts of cosmology and the evolution of the
universe. However, evidence for the existence of dark matter and
insight in its properties come from various domains in astrophysics
and cosmology, and these can deliver data at energies which are
currently unreachable for collider experiments. The so called {\sl
  direct detection\/} of WIMPs is based on the measurement of nuclear
recoils in elastic WIMP scattering processes~\cite{gaitskell04a},
while the {\sl indirect search\/} for dark matter is concerned with
the detection of WIMPs through the analysis of their annihilation
products in cosmic rays. At present, the experimental data show that
the cosmic ray fluxes of positrons~\cite{beatty04a},
antiprotons~\cite{aguilar02a}, and gamma rays~\cite{deboer06a} are not
quite in agreement with expectations, and that these discrepancies may
well be interpreted as a signal of WIMP annihilation (see
chapter~\ref{chapter:cosmicrays}). However, the data are still too
sparse to make convincing statements. Hence, while waiting for new
experiments to be deployed in the near future, putting more effort
into the analysis of existing data can yield results with increased
significance.

\chapter{SUSY Detection at the LHC}\label{chapter:lhc}
\section{The LHC and its Experiments}
\begin{figure}[bt]
\begin{center}
\includegraphics[width=12cm]{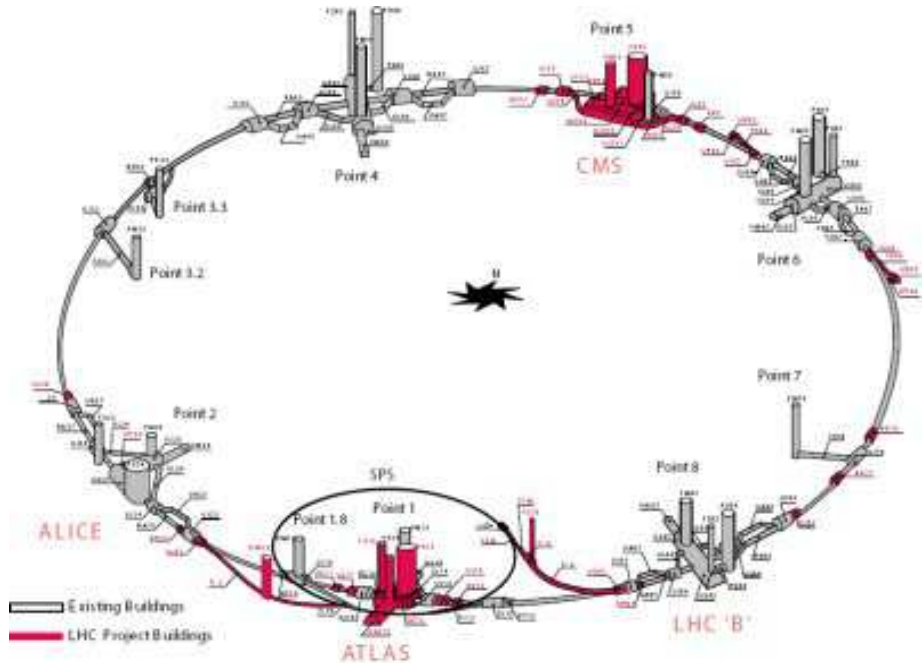}
\caption{\label{fig:lhc}The underground structures and experiments of the LHC~\cite{bruening04b}.}
\end{center} 
\end{figure}
The {\sl Large Hadron Collider\/}
(LHC)~\cite{bruening04a,bruening04b,benedikt04a} is currently under
construction at the CERN particle physics laboratory in the same
underground tunnel where the LEP collider~\cite{brandt00a,aleph06a}
was previously housed.  It will provide proton-proton collisions at a
center-of-mass energy of 14\+\TeV{}, as well as heavy ion collisions
(lead-lead) up to 1.15\+\PeV{}. Two separated beam pipes will be used to
circulate the protons or ions. The particles are forced on their
trajectories by an 8.3\+T magnetic field produced by superconducting
magnets and are accelerated by superconducting radio frequency
cavities. First operation of the LHC is scheduled for 2007 with beam
energies of 450\+\GeV~\cite{lamont06a}.

A certain process, which is characterized by its cross section
$\sigma$, occurs with a rate $\dot{N} = \mathcal{L}\sigma$, where
$\mathcal{L}$ is the luminosity of the accelerator. The LHC is
designed for a final luminosity $\mathcal{L}\approx
10^{34}\+\mathrm{cm}^{-2}\mathrm{s}^{-1}$ in the proton-proton mode,
however it is planned to run at $\mathcal{L}\approx
10^{33}\+\mathrm{cm}^{-2}\mathrm{s}^{-1}$ during the {\sl low
luminosity phase\/} for three years~\cite{lamont05a}, which sums
to approximately 30\+fb$^{-1}$ integrated luminosity. To achieve the
design luminosity the two beams will guide up to 2808 bunches of about
$1.15\times 10^{11}$ protons each. The bunch spacing will be 25\+ns,
which is equivalent to a bunch crossing rate of 40\+MHz.  Within the
total circumference of about 26.7\+km, there will be eight arcs and
straight sections. Each straight section is approximately 528\+m long
and can serve as an experimental or utility insertion.
As shown in Figure~\ref{fig:lhc}, four large experiments will be
situated in the four beam crossing sites along the collider:
\begin{itemize}\setlength{\itemsep}{-0.2mm}
\item ATLAS~\cite{atlas99}, which is characterized by its
  sophisticated superconducting air-core toroid muon spectrometer, and
  features liquid argon calorimetry and silicon pixel and microstrip
  tracking, as well as transition radiation detectors, within a 2\+T
  solenoid field;
\item CMS~\cite{dellanegra06a,dellanegra06b}, which is described in
  more detail below;
\item LHCb~\cite{lhcb98a,antunesnobrega03a}, an experiment dedicated
  to studying phenomena related to CP-violation and rare B-meson
  decays at low luminosity, and
\item ALICE~\cite{alice95a}, which is customized for the investigation
  of ion-ion collisions.
\end{itemize}\setlength{\itemsep}{+0.2mm}
ATLAS and CMS are multi-purpose experiments and were optimized to
perform high quality measurements in pp collisions of leptons,
hadronic jets and high energy photons. The two detectors have been
designed following similar guiding principles and differ mainly in the
choice of detector technologies. While ATLAS has invested a large
fraction of its resources into superconducting toroid magnets and into
a set of precise muon chambers, CMS has put emphasis on the highest
possible magnetic field combined with an inner tracker consisting
solely of silicon pixel and silicon microstrip detectors, which
provide high granularity at all radii. Main topics of the experiments'
physics programs are the investigation of the mechanism responsible
for electroweak symmetry breaking, which involves finding the Higgs
boson, top quark physics and B-meson studies. However, since the LHC
is often referred to as a {\sl discovery machine\/}, a very important
task is to search for new physics at high energies, such as SUSY.

\section{Discovery Potential of the LHC Experiments for Supersymmetry}
Assuming that supersymmetry is in fact realized in nature, its
discovery at the LHC will be relatively
straightforward~\cite{abdullin02a}. There are many possibilities to
create superpartners at a hadron collider: besides the quark-antiquark
annihilation channel, there are numerous processes of gluon fusion,
quark-antiquark and quark-gluon scattering. Gluon fusion leads to the
largest cross sections, of the order of a few picobarn. SUSY would be
revealed by an excess of events with a number of characteristic
signatures over the standard model expectations. With $R$-parity
conserved, the final state in the decay chains of sparticles always
contains LSPs, which escape the experiment undetected. Hence,
signatures of SUSY are characterized by typically large values of
missing transverse energy~$\etmiss$.
\begin{table}[htb]\begin{center}
\begin{tabular}{c|l|l} \hline
initial state & main decay modes & signature\\
\hline
$\tilde{g}\tilde{g}, \tilde{q}\tilde{q}, \tilde{g}\tilde{q}$
&$\left.\begin{array}{l} \tilde{g} \to q\bar q \tilde{\chi}^0_1 \\
 q\bar q\,' \tilde{\chi}^\pm_1  \\
 g\tilde{\chi}^0_1 \end{array} \right\}\rule[-1ex]{0ex}{6.5ex} 
 m_{\tilde{q}}>m_{\tilde{g}}$ &
$\begin{array}{c} \etmiss
+ \mbox{~multijets}~(+\mbox{leptons}) \end{array}$ \\
& $\left.\begin{array}{l}\tilde{q} \to q \tilde{\chi}^0_i \\
    \tilde{q} \to q\,' \tilde{\chi}^\pm_i \end{array} \right\}\rule[-2.9ex]{0ex}{1ex} 
  m_{\tilde{g}}>m_{\tilde{q}}$ & \\ \hline
$\tilde{\chi}^\pm_1\tilde{\chi}^0_2$ & $\tilde{\chi}^\pm_1
\to \tilde{\chi}^0_1 \ell^\pm \nu, \ \tilde{\chi}^0_2 \to
 \tilde{\chi}^0_1 \ell\ell$ \STRUT &
$\mbox{~three leptons} + \etmiss$ \\
 & $\tilde{\chi}^\pm_1 \to \tilde{\chi}^0_1 q \bar q\,',
\tilde{\chi}^0_2 \to \tilde{\chi}^0_1 \ell\ell,$ &
$\mbox{~two leptons + jet} + \etmiss$ \rule[-1.3ex]{0ex}{1ex} \\
\hline
$
\tilde{\chi}^+_1\tilde{\chi}^-_1$ \STRUT &
$\tilde{\chi}^{\pm}_1 \to \tilde{\chi}^0_1 \ell^\pm \nu$ &
$\mbox{~two leptons} + \etmiss$ \rule[-1.3ex]{0ex}{1ex} \\
\hline
$
\tilde{\chi}^0_i\tilde{\chi}^0_i$ \STRUT &
$\tilde{\chi}^0_i \to \tilde{\chi}^0_1 X, \tilde{\chi}^0_i \to \tilde{\chi}^0_1 X'$ &
$\mbox{~two leptons + jet} + \etmiss$ \rule[-1.3ex]{0ex}{1ex} \\
\hline
$
\tilde{t}_1\tilde{t}_1$ \STRUT &
$\tilde{t}_1 \to c \tilde{\chi}^0_1$ &
$\mbox{~2 noncollinear jets}+ \etmiss $\\ 
 & $\tilde{t}_1 \to b \tilde{\chi}^\pm_1, \tilde{\chi}^\pm_1 \to \tilde{\chi}^0_1 q\bar q\,'$ &
$\mbox{~single lepton} + \etmiss + b$ \\
 &$\tilde{t}_1 \to b \tilde{\chi}^\pm_1,\tilde{\chi}^\pm_1 \to \tilde{\chi}^0_1 \ell^\pm\nu,$ &
$\mbox{~two leptons} + \etmiss + b $ \rule[-1.3ex]{0ex}{1ex} \\
\hline
$\tilde{l}\tilde{l},\tilde{l}\tilde{\nu},\tilde{\nu}\tilde{\nu}$ \rule[-1ex]{0ex}{4ex} &
 $\tilde{\ell}^\pm \to \ell^\pm \tilde{\chi}^0_i,\tilde{\ell}^\pm \to \nu_\ell \tilde{\chi}^\pm_i$ &
 $\mbox{~two leptons}+ \etmiss$ \\
& $\tilde{\nu} \to \nu \tilde{\chi}^0_1$ & $\mbox{~single~lepton} + \etmiss$  
\rule[-1.3ex]{0ex}{1ex} \\ \hline
\end{tabular}\end{center}
\caption{Compilation of initial sparticle states, main decay modes, and their signatures at the LHC~\cite{gladyshev06a}.} \label{tab:modes}
\end{table}

\begin{figure}[!h]
\begin{center}
\begin{tabular}[b]{ll}
\begin{minipage}[b]{8cm}
\includegraphics[width=8cm]{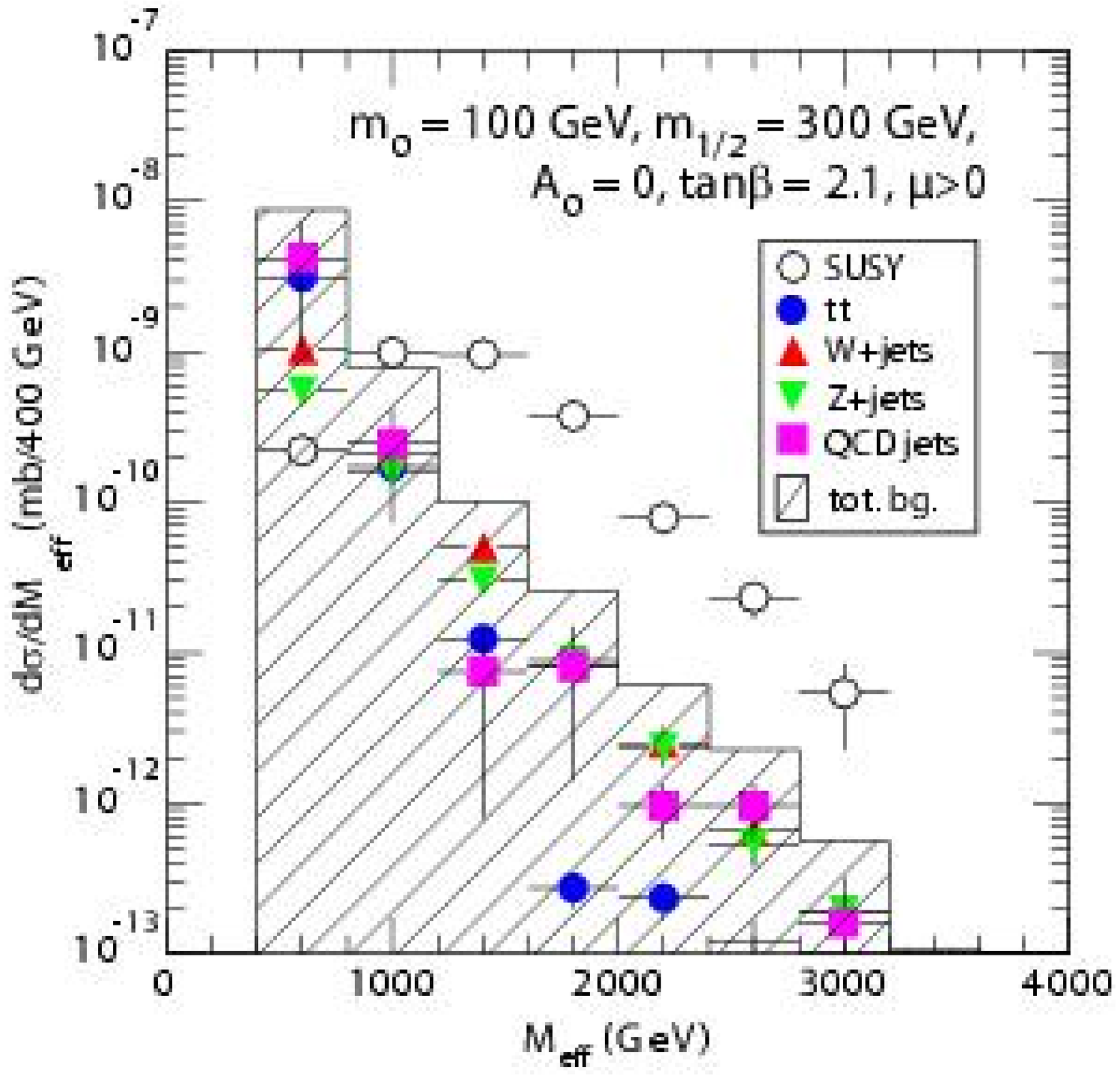}
\end{minipage}
&
\begin{minipage}[b]{7cm}
\begin{center}
\includegraphics[width=7cm]{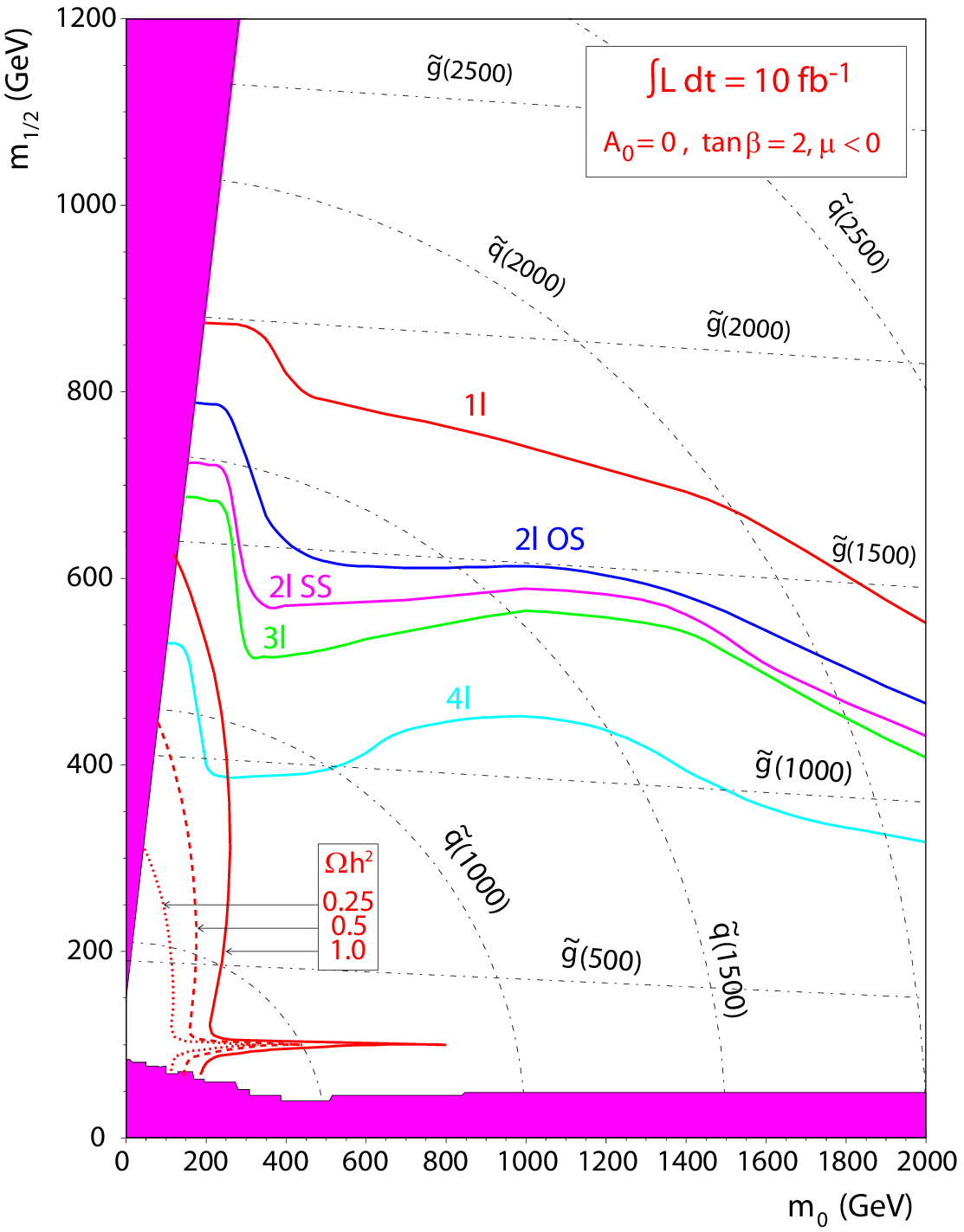}
\end{center}
\end{minipage}
\end{tabular} 
\end{center} 
\caption{\label{fig:squarkGluinoSignal} Distribution of $M_{\mathrm{eff}}$ 
for the SUSY signal and the expected background for an integrated
luminosity of 10\+fb${}^{-1}$ {\sl (left)\/}~\cite{hinchliffe97a};
reach (5$\sigma$ contours) of the gluino/squark search in various
lepton multiplicity final states (OS -- opposite charge sign; SS --
same sign) for 10\+fb${}^{-1}$. The neutralino dark matter density
contours of $\Omega h^2 = 0.25$, 0.5 and 1.0 are also
shown. Dash-dotted lines denote squark/gluino isomass contours {\sl
(right)\/}~\cite{abdullin02a}.}
\end{figure}

Table~\ref{tab:modes} gives an overview of the most common initial
sparticle states and their expected signatures at the LHC. At LHC
energies, the total sparticle production cross section is dominated by
strongly interacting gluinos and squarks. They initiate decay cascades
which lead to the occurrence of final states with numerous jets and
leptons and missing energy due to the existence of at least two LSPs
and possibly neutrinos~\cite{abdullin02a}. As a consequence of LSP
production, a complete mass reconstruction of gluinos and squarks is
possible only in long decay chains. So SUSY signal observability is
based on an excess of events of a given topology over known or
expected backgrounds. Figure~\ref{fig:squarkGluinoSignal} {\sl (left)\/}
shows the distribution of the quantity $M_{\mathrm{eff}}$ -- i.e. the
sum of $\etmiss$ and the transverse momenta $p_{\mathrm{T}}$ of the
four highest energetic jets -- for events with $\etmiss > 100$\+\GeV{} and
at least four jets with $p_{\mathrm{T}} > 50$\+\GeV{}/c, compared to the
SM background. With an integrated luminosity of 10\+fb${}^{-1}$, the
signal over background ratio is approximately 10 for large values of
$M_{\mathrm{eff}}$~\cite{atlas99}.

The LHC can discover squarks and gluinos up to masses well in excess
of 2\+\TeV{}/c$^2$, which covers the entire parameter space over which
supersymmetry can plausibly be relevant to electroweak symmetry
breaking.  In Figure~\ref{fig:squarkGluinoSignal} {\sl (right)\/} the
reach of the gluino/squark search is displayed for 10\+fb${}^{-1}$ of
data. The solid lines show the $5\sigma$ contours for final states
with different lepton multiplicity. Additionally, contour lines for
the resulting neutralino dark matter densities $\Omega h^2 = 0.25$,
0.5 and 1.0 are given.  The cosmologically preferred density of
$\Omega h^2\approx 0.1$ (see chapter~\ref{chapter:darkmatter}) is well
within the reach of the LHC.

Sleptons can be produced in pairs, for example, directly via
$q\bar{q}$ annihilation. If sleptons are more massive than the
lightest charginos and neutralinos, they decay directly into these via
$\tilde{\ell}^{\pm}\to\ell^{\pm}\tilde{\chi}^{0}_{1,2}$ or
$\tilde{\ell}^{\pm}\to\nu_{\ell}\tilde{\chi}^{\pm}_1$. Similar decay
channels exist for sneutrinos. On the contrary, light sleptons can
indirectly originate from charginos and neutralinos via
e.g. $\tilde{\chi}_2^0\to\tilde{\ell}^{\pm}\ell^{\mp}$ or
$\tilde{\chi}_1^{\pm}\to\tilde{\nu}_{\ell}\ell^{\pm}$, where the
charginos and neutralinos themselves dominantly stem from gluino and
squark pairs. This leads to an experimental signature characterized by
2 isolated leptons, missing transverse energy, and the absence of
jets, which allows to detect sleptons up to masses of about
350\+\GeV{}/c$^2$~\cite{abdullin02a}.

\begin{figure}[bt]
\begin{center}
\includegraphics[width=14cm]{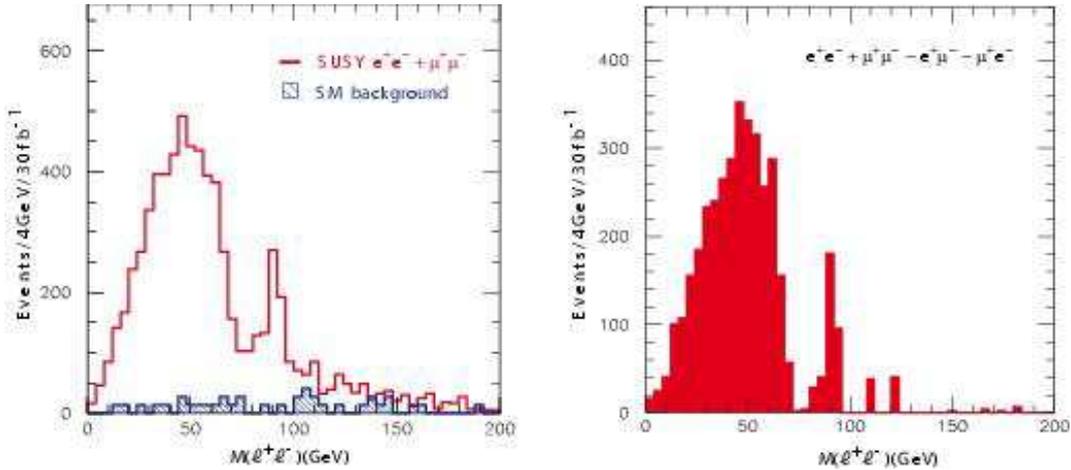}
\caption{\label{fig:dilepton}Distribution of the dilepton invariant mass for $m_0=800$\+\GeV{}/c$^2$,
$m_{1/2}=200$\+\GeV{}/c$^2$, $A_0=0$, $\tan\beta=10$, $\mu>0$, and the SM
background, for $\mathcal{L}_{\mathrm{int}}=30\+\mathrm{fb}^{-1}$ {\sl
(left)\/}; same with mixed-flavor pair distribution subtracted {\sl
(right)\/}. Taken from~\cite{atlas99}.}
\end{center} 
\end{figure}

Following the possible detection of SUSY signals in channels such
as those described above, the most important tasks would be to measure
the sparticle masses and to narrow down the range of model parameters.
Leptonic decays of the second-lightest neutralino, $\tilde{\chi}_2^0$,
have a useful kinematical feature: the two-lepton invariant mass
spectrum has an edge near the kinematical upper limit with a
maximum value of
$m_{\ell^+\ell^-}^{\mathrm{max}}=m_{\tilde{\chi}_2^0}-m_{\tilde{\chi}_1^0}$
in the case of direct three-body decays via
$\tilde{\chi}_2^0\to\ell^+\ell^-\tilde{\chi}_1^0$, and
\begin{equation}\label{eq:dilepton}
m_{\ell^+\ell^-}^{\mathrm{max}}=\sqrt{\left(m_{\tilde{\chi}_2^0}^2-m_{\tilde{\ell}}^2\right)
\left(m_{\tilde{\ell}}^2-m_{\tilde{\chi}_1^0}^2\right)}/m_{\tilde{\ell}}
\end{equation}
in the case of two-body cascade decays
$\tilde{\chi}_2^0\to\ell^{\pm}\tilde{\ell}^{\mp}\to\ell^+\ell^-\tilde{\chi}^0_1$,
thus allowing a measurement of the mass difference of the two
neutralinos.  The $\tilde{\chi}_2^0$ dominantly stem from gluinos,
which decay into heavy quark-squark pairs, as for example
$\tilde{g}\to\tilde{b}b$, $\tilde{b}\to\tilde{\chi}_2^0 b$. Thus the
experimental signature is characterized by the occurrence of isolated
same-flavor opposite-charge lepton pairs in conjunction with at least
four jets. Figure~\ref{fig:dilepton} {\sl (left)\/} shows the
invariant mass of lepton pairs from events which match these
requirements for an integrated luminosity of $30\+\mathrm{fb}^{-1}$
and a particular choice of parameters. The background expected from SM
processes is also shown. While the narrow peak around 90\+\GeV{}/c$^2$
stems from $Z^0$ production in the decays of heavy neutralinos, the
edge at lower invariant masses indicates the mass difference $\Delta
m_{\chi} = m_{\tilde{\chi}_2^0}-m_{\tilde{\chi}_1^0}$, which in this
case approximately equals 68\+\GeV{}/c$^2$. The background is mostly
due to $t\bar{t}$ or $WW$+jet production. However, in that case the
final state contains as many same-flavor leptons as different-flavor
ones and with identical distributions. Hence, by subtracting the
different-flavor distribution, the SM background can be canceled up to
statistical fluctuations (Figure~\ref{fig:dilepton}, {\sl right\/}).
$\Delta m_{\chi}$ can then be determined with a statistical accuracy
of about 50\+\MeV{}/c$^2$~\cite{atlas99}.

As eq.~(\ref{eq:dilepton}) suggests, the dilepton edge is also
sensitive to the intermediate slepton mass. Through analysis of
particular kinematical distributions, the masses of the two lightest
neutralinos as well as the slepton mass can be
determined~\cite{abdullin02a}. Additionally, events near the edge can
be used to reconstruct the $\tilde{\chi}^0_2$ momentum vector. This
would permit the search for resonance structures in the distribution
of the $\tilde{\chi}^0_2$+jet invariant mass and possibly provide
access to the mass of the decaying squarks. Apart from these event
topologies, a variety of other signatures would facilitate the
determination of sparticle masses and SUSY model
parameters~\cite{atlas99,dellanegra06b,abdullin02a,hinchliffe97a,kitano06a,denegri05a}.
The ultimate goal of such studies would be to use very many
measurements to make an overconstrained fit to the model, rather in
the same way that current data are used to test the SM.

\section{Silicon Microstrip Detectors for CMS}
\subsection{The CMS Experiment}
\begin{figure}[tb]
\begin{center}
\includegraphics[width=14cm]{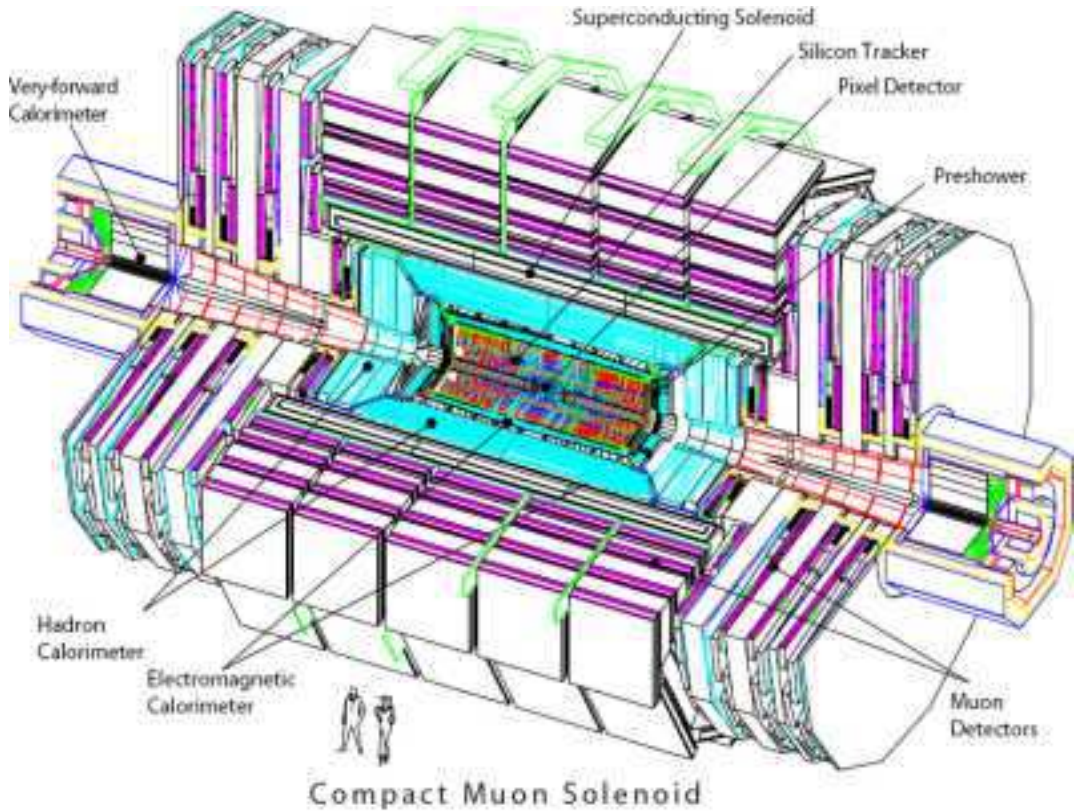}
\caption{\label{fig:cms}The CMS experiment (from~\cite{dellanegra06a}).}
\end{center} 
\end{figure}

CMS is a general purpose high energy physics experiment which is
particularly optimized for the detection of particles stemming from
proton-proton collisions. Its design goals comprise, amongst others,
providing a high-quality and redundant muon system, best possible
electromagnetic calorimetry and excellent tracking in a strong
magnetic field. Its layout is that of a classical hermetic detector
with a barrel region and two endcaps, covering practically the full
$4\pi$ solid angle. CMS will have a length of 21.6\+m, a diameter of
14.6\+m, and 14500\+t of weight~\cite{cms94a}. Figure~\ref{fig:cms} gives
an overview of the experiment and its subdetectors, which will briefly
be discussed in the following.

{\bf The Magnet System}~\cite{cms97b} consists of a superconducting
solenoid coil in conjunction with a saturated iron yoke for flux
return. The 12.5\+m long solenoid will provide an homogeneous magnetic
field of 4\+T in the central region, which decreases to about 1.7\+T in
the outermost parts of the experiment.

{\bf The Muon System}~\cite{cms97a} is integrated in the return yoke
of the magnet system and will be used to identify muons and measure
their charge sign and momentum. Furthermore, it will play an important
role for triggering the CMS experiment. The Muon System consists of
three subdetectors: drift tube chambers in the barrel region, cathode
strip chambers in the endcaps, and fast resistive plate chambers in
both the barrel and endcap. When combined with data from the tracker,
the expected momentum resolution
$\sigma_{p_{\mathrm{T}}}/p_{\mathrm{T}}$ for muons ranges from 1\+\%
for $p_{\mathrm{T}} < 10$\+\GeV{}/c to about 6--17\+\% for
$p_{\mathrm{T}} = 1$\+\TeV{}/c, depending on the direction of the
emerging particle. The Muon System also provides a track
reconstruction efficiency of better than 90\+\% in the transverse
momentum range up to 100\+\GeV{}/c, decreasing to $\gtrsim 70$\+\% at
$p_{\mathrm{T}} = 1$\+\TeV{}/c.

{\bf The Calorimeters}~\cite{cms97c,cms97d} of CMS will measure the
energy and direction of electrons, photons and jets and, moreover,
deliver information crucial for the trigger system. They are divided
into two parts, the electromagnetic calorimeter (ECAL) and the hadron
calorimeter (HCAL). The complete ECAL and most parts of the HCAL are
located within the magnet coil. The ECAL is composed of 82728
scintillating PbWO$_4$ crystals with a depth of 25.8 (barrel) and 24.7
(endcap) radiation lengths, read out by avalanche photodiodes and
vacuum phototriodes, respectively. In the barrel part, the ECAL
granularity is $2.2\times2.2$\+cm$^2$, providing an excellent
separation power for close-to-collinear particles. The endcaps have
less fine granularity and a silicon strip preshower detector will be
installed in front of them to help with $\pi^0$-$\gamma$ separation.

The HCAL consists of a barrel and two endcap (HC) sections inside the
magnet coil and a tail catcher outside the magnet. It is a sampling
calorimeter composed of copper absorber plates interleaved with 4\+mm
(front part) or 8\+mm (back part) plastic scintillator tiles read out
by hybrid photodiodes. In addition two forward quartz fiber
calorimeters (HF) are placed around the beam pipe beyond the endcap
magnet yokes to provide hermetic closure of the detector. The energy
resolution $\sigma_E/E$ of the calorimeters is given by
\begin{equation}
\left(\frac{\sigma_E}{E}\right)^2=\left(\frac{a}{\sqrt{E}}\right)^2+
\left(\frac{\sigma_n}{E}\right)^2+b^2\, ,
\end{equation}
with $E$ in units of \GeV{} and $a=2.7\+\%$, $\sigma_n=155$\+\MeV{},
$b=0.5$\+\% for the ECAL (barrel, $\eta=0$), and $a=100\+\%$,
$b=4.5$\+\% with negligible $\sigma_n$ for the HCAL with the ECAL in
front.

\begin{figure}[tb]
\begin{center}
\includegraphics[width=15cm]{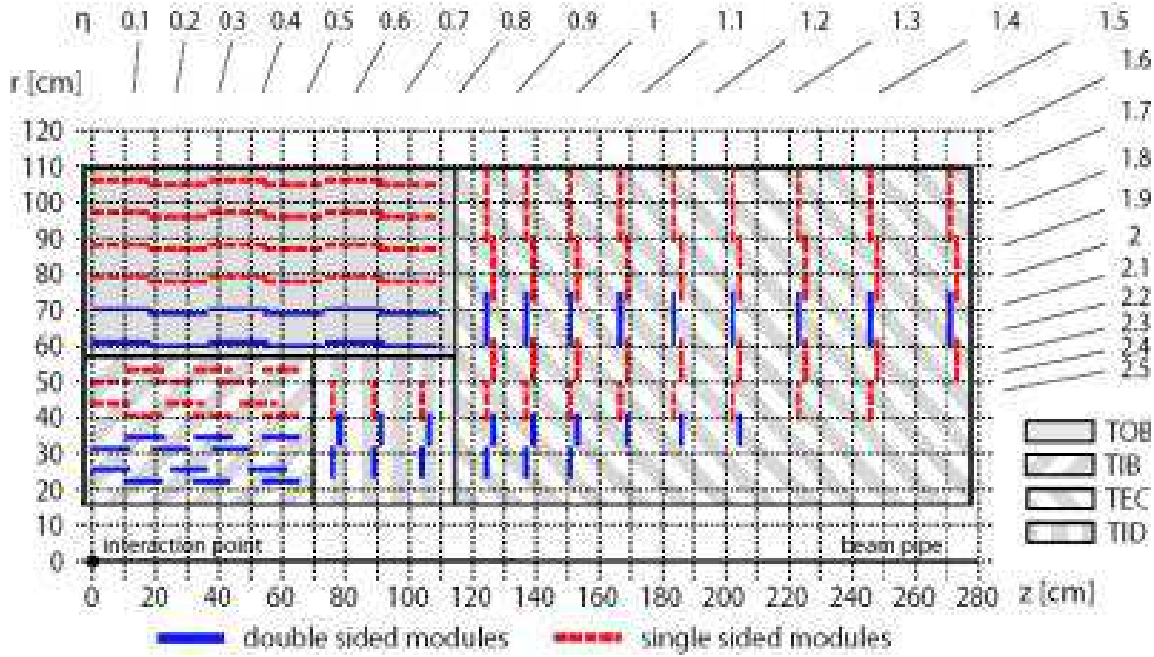}
\caption{\label{fig:cmsTracker}Cross section through one octant of the CMS 
silicon strip tracker~\cite{brauer05a}. The pixel detector in not shown.}
\end{center} 
\end{figure} 

{\bf The Tracking System}~\cite{cms98a,cms00a} consists of two
subsystems, the pixel detector and the silicon strip detector, with an
overall transverse momentum resolution of $\delta
p_{\mathrm{T}}/p_{\mathrm{T}}\lesssim (15\cdot
p_{\mathrm{T}}/\mathrm{\TeV{}}\oplus0.5)$\+\%. The pixel detector
forms the innermost part of the tracker, close to the beam pipe. It
consists of three 53\+cm long barrel layers and two endcap disks on
each side of the barrel. The barrel layers are placed at distances of
4.4, 7.3 and 10.2\+cm from the beam line, while the two disks are
situated at 34.5 and 46.5\+cm from the interaction point. To achieve
an equally good hit resolution of 15\+$\mu$m in the transverse and the
longitudinal planes, a design with a square pixel shape of dimensions
$150\times 150$\+$\mu$m$^2$ and thickness 300\+$\mu$m is
used~\cite{cucciarelli05a}.

The pixel detector is enclosed by the silicon strip detector (SST).
With more than 15000 silicon modules, adding up to a total active
silicon area of about 200\+m$^2$, and roughly 10 million electronics
channels, it will be the largest silicon detector ever
built~\cite{brauer05a}. Figure~\ref{fig:cmsTracker} displays a cross
section through one octant of the SST and shows its division into four
subsystems: the {\sl Tracker Inner Barrel\/} (TIB), the {\sl Tracker
  Inner Disks\/} (TID), the {\sl Tracker Outer Barrel\/} (TOB) and the
{\sl Tracker Endcaps\/} (TEC). The TEC and their components are
described in the next section. 

\subsection{Silicon Microstrip Detectors} 
\begin{figure}[tb]
\begin{center}
\includegraphics[width=12cm]{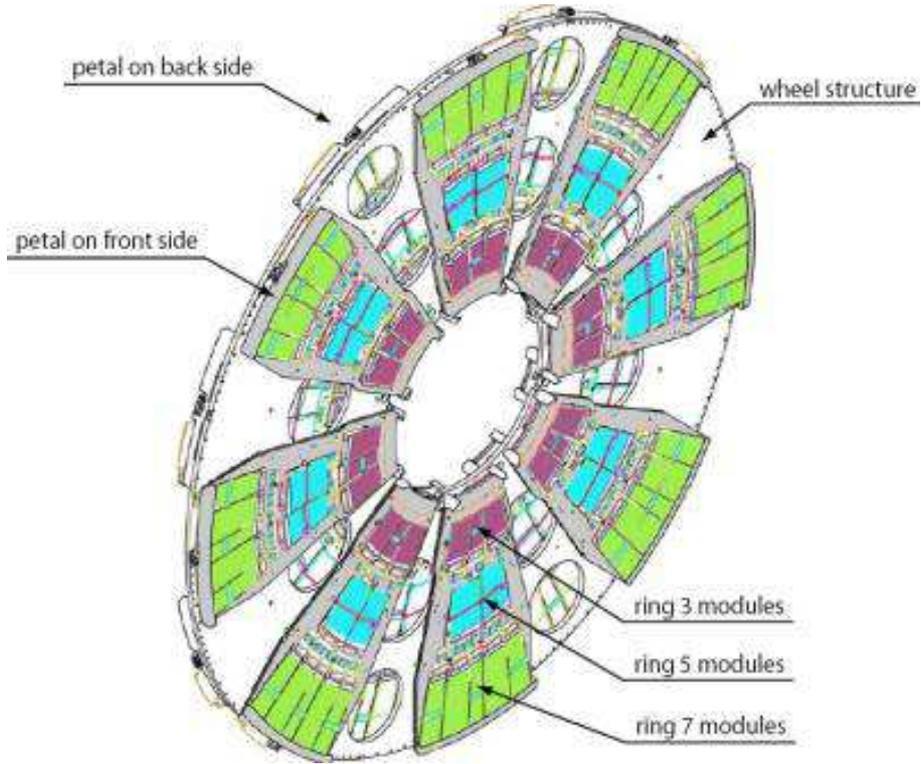}
\caption{\label{fig:tecdisk}Schematic view of the fourth TEC disk 
($z\approx 170$\+cm). It carries 16 petals -- eight on the front and
eight on the back side -- on which the silicon strip modules are
mounted. The petals on this disk do not carry modules of ring 1
(compare Figure~\ref{fig:cmsTracker}).}
\end{center} 
\end{figure}
\begin{figure}[t]
\begin{center}
\includegraphics[width=16cm]{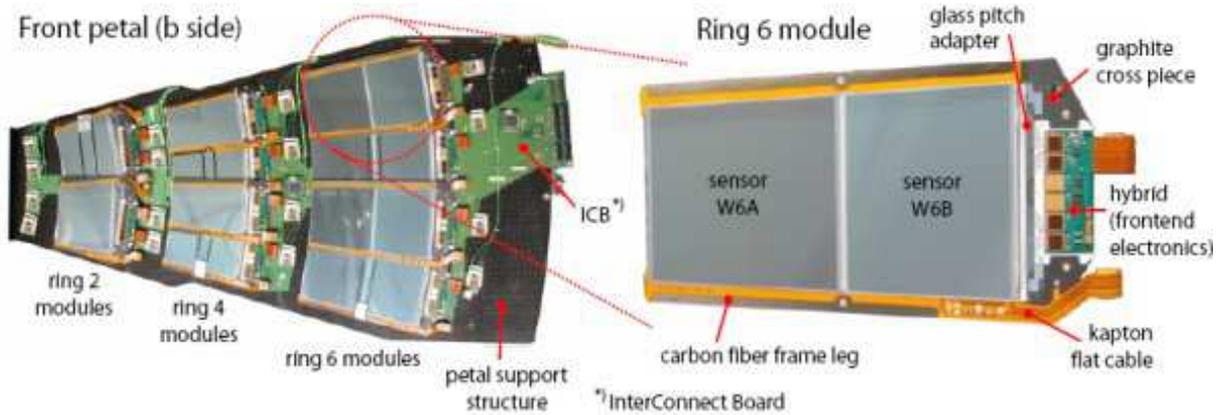}
\caption{\label{fig:petalAndModule}Photograph of a fully assembled TEC petal
(b-side), carrying modules of rings 2, 4 and 6, and a view of
a 22.5\+cm long ring 6 module~\cite{brauer05a}.}
\end{center} 
\end{figure}

Each TEC consists of nine circular disks perpendicular to the beam
direction, on which the the silicon sensors are mounted via modular
support structures {\sl (petals)\/} as shown in
Figure~\ref{fig:tecdisk}. One or two silicon sensors are integrated in
{\sl modules\/} together with the frontend readout electronics. The
modules are arranged on both sides of the support structures forming
up to seven concentric {\sl rings\/} around the beam pipe with the
sensitive strips oriented radially. Modules belonging to odd-numbered
rings (1, 3, 5 and 7) are mounted on one petal side, while those of
even-numbered rings (2, 4 and 6) are mounted on the other side, so
that there is an overlap between adjacent rings in $r$ and
$\phi$. Rings 3, 4, 6 and 7 have single sided modules consisting of
one single silicon plane. Modules of the rings 1, 2 and 5 are double
sided {\sl (stereo)\/} with two coplanar silicon planes mounted back
to back and tilted at an angle of 100\+mrad with respect to each
other. Since at a given $z$ position the silicon microstrip sensors
are sensitive only to one of the coordinates spanning the sensor
plane, single sided modules yield measurements of only the $\phi$ and
$z$ coordinates of particle hits, while double sided modules measure
all coordinates simultaneously. The spatial resolution of single hit
measurements in a TEC sensor is about $30\+\mu$m.
Figure~\ref{fig:petalAndModule} shows a fully assembled petal
(b-side), carrying modules of rings 2, 4 and 6. Due to the overall
circular geometry of the TEC the modules have a trapezoidal shape and
increase in size towards the outer edge of the disks.

For the modules of ring 6, the Physikalisches Institut 1B at the RWTH
Aachen is involved in the final assembly steps of wirebonding. Such a
ring 6 module is shown in the enlarged view on the right side of
Figure~\ref{fig:petalAndModule}. Two carbon fiber frame legs are
connected via a graphite plate {\sl (cross piece)\/}, serving as the
mechanical support structure for the two trapezoidal silicon sensors
named W6A and W6B. The latter are electrically insulated from the
support structure with bands of kapton foil, which at the same time
carry electrical circuitry for the application of high voltage to the
sensors. The frontend electronics, assembled on a kapton/copper board
laminated on thin ceramics {\sl (hybrid)\/}, are glued on the cross
piece.  A thin glass plate with strip conductors, the {\sl pitch
adapter\/}, guides the signals from the sensors to the readout
electronics. Each of the single sided (double sided) sensors has 512
(768) readout strips. For ring 6 the strip pitch varies from 163 to
205\+$\mu$m. The inter-strip connections from sensor to sensor, sensor
to pitch adapter, and pitch adapter to hybrid are established via
25\+$\mu$m aluminum wirebonds.

\begin{figure}[tb]
\begin{center}
\includegraphics[width=13cm]{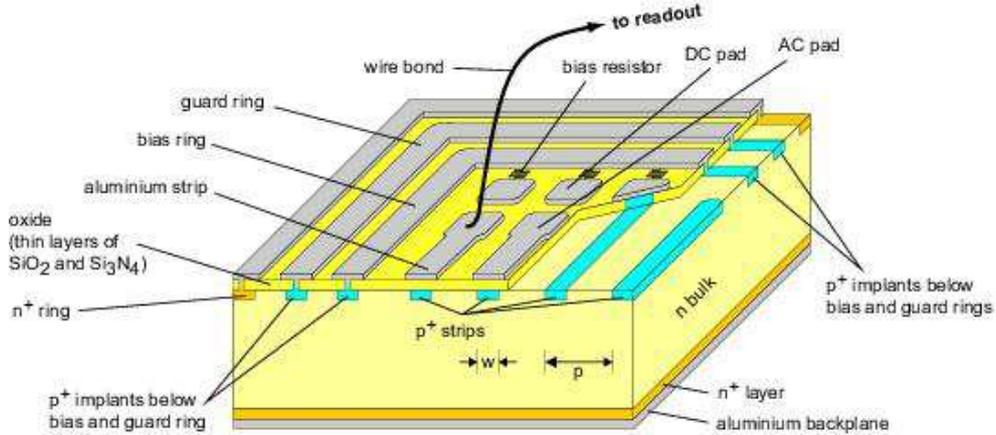}
\caption{\label{fig:cmsSensor}Schematic cross section of a silicon microstrip sensor~\cite{franke05a}.}
\end{center} 
\end{figure}
The structure and working principle of a silicon microstrip sensor is
illustrated in Figure~\ref{fig:cmsSensor}. Strips of p$^+$ doped silicon
are implanted into the surface of the 500\+$\mu$m thick\footnote{For
  rings 5--7; sensors of the rings 1--4 have a thickness of 300\+$\mu$m
  instead.} n$^+$ doped silicon bulk. They are insulated via a
continuous SiO$_2$/Si$_3$N$_4$ layer, and are covered with aluminum
metallization for readout purposes {\sl (strips)\/}. The strip width
varies between 53 and 67\+$\mu$m for ring 6 modules. In this
configuration, the implants form a series of p-n junctions, with the
aluminum strips capacitively coupled to them.  On both ends of each
strip, the aluminum broadens and forms two {\sl AC pads\/} to house
the wirebond. In addition to the AC pads, {\sl DC pads\/} are located
beyond the ends of each strip. These provide a direct connection to
the implants and are thus not used for wirebonding. The DC pads are
connected to the {\sl bias ring\/} via 1.5\+M$\Omega$ polysilicon bias
resistors, hence providing all implants with a common potential. The
opposite surface of the sensors is finished with a thin n$^+$ silicon
layer and finally coated with an aluminum backplane.

The frontend electronics on the hybrid are equipped with {\sl
APV25-S1}\footnote{APV -- Analogue Pipeline (Voltage
mode)}~\cite{french01a} integrated circuits (IC). The APV25-S1 is a
128 channel analog pipeline with 192 columns of analog storage. Each
strip is connected to one single electronics channel. Strip signals
are amplified into 50\+ns shaped pulses of magnitude 100\+mV per 25000
electrons. These are sampled at a rate of 40\+MHz and stored in the
pipeline. Useful data are marked after a programmable latency, and
held in the pipeline until such a time that they can be read out. If
no trigger is received within 4.8\+$\mu$s the pipeline cells are
overwritten. A set of 512 strip signals is referred to as an {\sl
event\/} in the following. The APV25-S1 can be operated in two
different readout modes; in {\sl peak mode} only one sample per
channel is read from the pipeline while in {\sl deconvolution mode}
three samples are read sequentially, and the final output is the
weighted sum of all three.  In the latter case, the shaping constant
is reduced from 50\+ns to 25\+ns at the expense of slightly higher
noise. Besides a set of other parameters, the gain of the amplifier
stage of the APV25-S1 can be varied. In addition to the APV25-S1, the
hybrid houses an IC to multiplex a pair of two APV25-S1 analog outputs
onto a single differential line {\sl (MUX)\/}, the {\sl detector
control unit (DCU)\/} IC for slow control purposes and an IC for
processing the trigger signal supplied by the CMS trigger electronics
{\sl (TPLL)\/}~\cite{axer03a}.

With a voltage $V$ applied to the sensor backplane and the bias ring
grounded, the p-n junctions are reverse biased, and {\sl depletion
zones\/} form at their locations. The width $d$ of the depletion zones
can be calculated in good approximation as
\begin{equation}
d\approx\sqrt{\frac{2\epsilon\epsilon_0}{qN_{\mathrm{D}}} V}\, ,
\end{equation}
where $N_{\mathrm{D}}$ is the effective doping concentration and $q$
is the elementary charge. Full depletion is reached if $d$ equals the
sensor thickness $d_{\mathrm{s}}$, hence the corresponding depletion
voltage $V_{\mathrm{dep}}$ is found to be
\begin{equation}
V_{\mathrm{dep}}\approx\frac{qN_{\mathrm{D}}}{\epsilon\epsilon_0} \,\frac{d_{\mathrm{s}}^2}{2}\, ,
\end{equation}
equal to about 100\+V for $N_{\mathrm{D}}\approx 0.5\cdot
10^{12}$/cm$^{-3}$~\cite{cms98a} in a 500\+$\mu$m thick ring 6
sensor. However, it must be pointed out that $N_{\mathrm{D}}$ and
consequently $V_{\mathrm{dep}}$ will strongly vary as the silicon
is irradiated during operation of the CMS experiment, essentially
leading to a type inversion from the initially n-type silicon to
p-type silicon~\cite{braibant02a}.

Charged particles traversing the sensor ionize the silicon and create
electron-hole pairs. A minimum-ionizing particle (MIP) deposits a most
probable energy of 260\+\eV{} within 1\+$\mu$m of material along its
flight path~\cite{yao06a}; with an energy of 3.6\+\eV{} required to
create one electron-hole pair in silicon~\cite{scholze00a}, it follows
that a MIP releases about 36000 pairs or 5.8\+fC per charge
sign when traversing a ring 6 sensor. The charge carriers are
separated in the electric field of the depletion zones and move
towards the electrodes. The signal is then created on the aluminum
strips by induction and carried to the frontend electronics through
the wirebonds.

Irradiation with light also generates electron-hole pairs in silicon.
The penetration depth $a$ -- i.e. the depth at which the intensity $I$
of the incident light has decreased to $1/e$ of its initial value
$I_0$ -- strongly depends on the wavelength $\lambda$. In the near
infrared domain, it is a monotonously increasing function of
$\lambda$, and reaches a value of about 400\+$\mu$m at $\lambda
=1050$\+nm in the case of pure silicon~\cite{abt99a}. The mean depth
$\bar{x}$ of energy loss of light in a silicon sensor of thickness $b$
as the expectation value of the intensity distribution
$I(x)=I_0\cdot\exp(-x/a)$ along $x$ is
\begin{equation}
\bar{x}=a-b\cdot\frac{e^{-b/a}}{1-e^{-b/a}}\, .
\end{equation}
For $\lambda =1050$\+nm and $b=500\+\mu$m, $\bar{x}$ equals
200\+$\mu$m, which is close to $\bar{x}=250\+\mu$m obtained for the
uniform energy loss attributed to a MIP. Although the silicon of the
sensors is not pure but doped, the effect of the doping concentration
on $\bar{x}$ is negligible for photons with energies above the
1.12\+\eV{} band gap of silicon, equivalent to a wavelength of
1107\+nm.

There exist several sources of noise superimposed on the strip
signals. Besides noise due to thermal excitations in the silicon, to
the readout electronics and low-frequency voltage noise, the module
output suffers to a small extent from the so called {\sl common mode
  noise\/} (CMN), which is basically caused by electromagnetic pick-up
at the APV25-S1 preamplifier inputs. One characteristic feature of the
CMN is that it simultaneously affects groups of strips, and thus
appears as a shift of their common baseline randomly varying with
time. The APV25-S1 features a facility for CMN reduction; it has an
external resistor in the power supply line common to all 128 channels,
which collectively drives down their output in case of a CMN pulse.
The remaining CMN can be determined and subsequently subtracted for
each of these groups separately within the data of one event.

During the intended operating time of ten years, the components of the
CMS tracker will be exposed to an aggressive environment. The
radiation level within the tracker volume will be extremely high, such
that some components will encounter a fluence of up to $1.6\cdot
10^{14}$ \MeV{}-equivalent neutrons per cm$^2$~\cite{dierlamm03a}. As a
consequence, the silicon has to be protected from radiation damage by
cooling it to a temperature of -10$^{\circ}$C.

\section{A Laser Test Facility for Silicon Microstrip Detectors}\label{sec:MTF}
To provide a stable and proper operation of the CMS tracker according
to its performance specifications, care has to be taken that all
assembled components fulfill strict quality requirements. Regarding
the silicon microstrip detector modules, these requirements are
summarized within a grading scheme as follows~\cite{meschini04a}.
Modules are classified as {\sl grade A} if they have less than 1\+\%
defective strips, as {\sl grade B} if they have between 1 and 2\+\%
defective strips, or {\sl grade C} otherwise. Modules of grades A and
B are used for tracker assembly with grade A preferred. In addition to
the requirement regarding defective strips, modules may also be
disqualified due to deviations in their electric properties, the
response of the frontend electronics and the mechanical precision
of assembly.

\subsection{Possible Module Defects} 
After completing the ring 6 module assembly with the final steps of
sensor to sensor and sensor to pitch adapter wirebonding, there are
three main categories of flaws which may be responsible for the
occurrence of a defective strip:
\begin{itemize}
\item {\bf Open} or {\bf missing wirebonds}, which are caused by
  flawed calibration of the wirebonding facilities or improper
  handling of the modules. In principle, this category also includes
  interruption of any signal lines, such as the pitch adapter
  conductors or the aluminum strips themselves. Open or missing
  wirebonds can largely be repaired.
\item {\bf Inter-strip short-circuits} -- ohmic contacts between
  adjacent strips -- which almost exclusively result from
  manufacturing defects.  They degrade the detector resolution and are
  in most cases irreparable defects. Hence, inter-strip short-circuits
  are considered in the module grading and must therefore be
  identified.
\item The so called {\bf pinholes}, which represent ohmic contacts
  between the aluminum strips and the underlying p$^+$ implants
  through the insulating oxide/nitride layer. Pinholes may result from
  manufacturing defects or from flawed wirebonding. One particular
  issue about pinholes is that already a single one allows current to
  flow from the sensor bulk into the frontend electronics, possibly
  leading to a breakdown of the corresponding 128 APV25-S1 channels at
  a time. Hence, strips affected by pinholes must be detected and
  detached from the hybrid by removing the wirebonds.
\end{itemize}
A test facility for silicon microstrip detector modules must reliably
detect defects from these three categories. It has to provide easy
handling and, once the module is mounted for testing, fully automatic
operation and high throughput are required. In total, 1061 ring 6
modules (including spares) are produced for the SST, all of which are
wirebonded at the Physikalisches Institut 1B at the RWTH Aachen. A
test facility for these modules, which meets the above requirements,
has been developed and is described in the following.

\subsection{The Setup of the Module Test Facility}
The basic principle of the module test facility (MTF) is the
generation of signals in the silicon sensors by laser illumination and
their subsequent analysis.  As will be shown, each defect leaves a
signature in the signal of the affected strip, characteristic of each
of the above categories and allowing their detection and tagging.
Since the implants are capacitively coupled to the readout strips, the
illuminating laser must be pulsed.

The modules are read out via the ARC\footnote{ARC -- APV readout
controller} system~\cite{axer03a,axer02a}. ARC is a multicomponent
readout system developed for module testing purposes and comprises the
ARC controller board, two frontend hybrid adapters, an ISA card
interface for personal computers, and the operating software package
ARCS. The ARC board represents the core of the system and performs
clock and trigger generation and distribution, data sampling and
buffering, as well as slow control and voltage control for the IC on
the hybrids. Two modules can be read out in parallel; their data are
sampled by a group of three 8-bit analog-to-digital converters at
optionally 20 or 40\+MHz. The ARC board additionally accepts external
trigger signals. Of the two frontend adapters, one houses amplifiers,
line drivers and communication circuits while the other serves as a
voltage regulator for the hybrids. The operating software ARCS is
arranged in three levels: a hardware device driver, a package of
communication routines written in the high-level programming language
C++, and a graphical user interface. In the configuration implemented
in the MTF, the ARC system can deliver data at a maximum rate of about
200\+Hz. A detailed description of ARC is given in~\cite{axer03a}.

\begin{figure}[htb]
\begin{center}
\includegraphics[width=10cm]{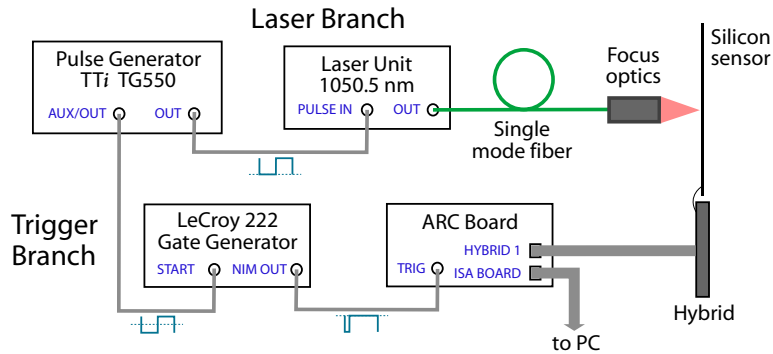}
\caption{\label{fig:teststandScheme}Schematic view of the laser test 
facility's optical and electrical circuitry.}
\end{center} 
\end{figure}
The electronic and optical signal lines of the MTF are depicted in
Figure~\ref{fig:teststandScheme}. They are grouped in two branches, a
laser or signal generation branch and a branch for trigger generation
and readout. Both branches start at the pulse generator's output
jacks, which deliver two separate square-wave signals with a duty
cycle of 50\+\% and a frequency of 180\+Hz. The first signal with
positive voltage and an amplitude of 0.8\+V is fed in the laser unit,
which emits at a central wavelength of 1050.5\+nm with an amplitude of
0.43\+mW. This infrared light is guided via a single-mode optical fiber
to the laser optics, which focus the beam on the silicon sensor
surface at a distance of 50\+mm. The resulting laser spot has a nearly
Gaussian intensity profile with a minimum achievable width of
10\+$\mu$m. For the purpose of module testing, the spot width has been
fixed close to this minimum value, so that the strips can be
targeted individually.

Simultaneously, a second signal leaving the pulse generator with
alternating voltage sign is fed into the trigger unit (gate
generator), which transforms it into a series of standardized logic
NIM\footnote{NIM -- Nuclear Instrumentation Modules} signals with
voltages of 0\+V and $-0.8$\+V for logic 0 and 1 respectively. The
output is connected to the trigger input jack of the ARC board. On the
arrival of a logic 1 pulse, the ARC board executes the readout
sequence of the APV25-S1 at a constant phase with respect to the laser
pulse inducing signals in the silicon sensor.  The phase -- i.e. the
pipeline slot to be read out -- is determined by the difference of the
total delay times for signals traveling along the two branch lines and
has been optimized by sampling tests.

\begin{figure}[htb]
\begin{center}
\includegraphics[width=11cm]{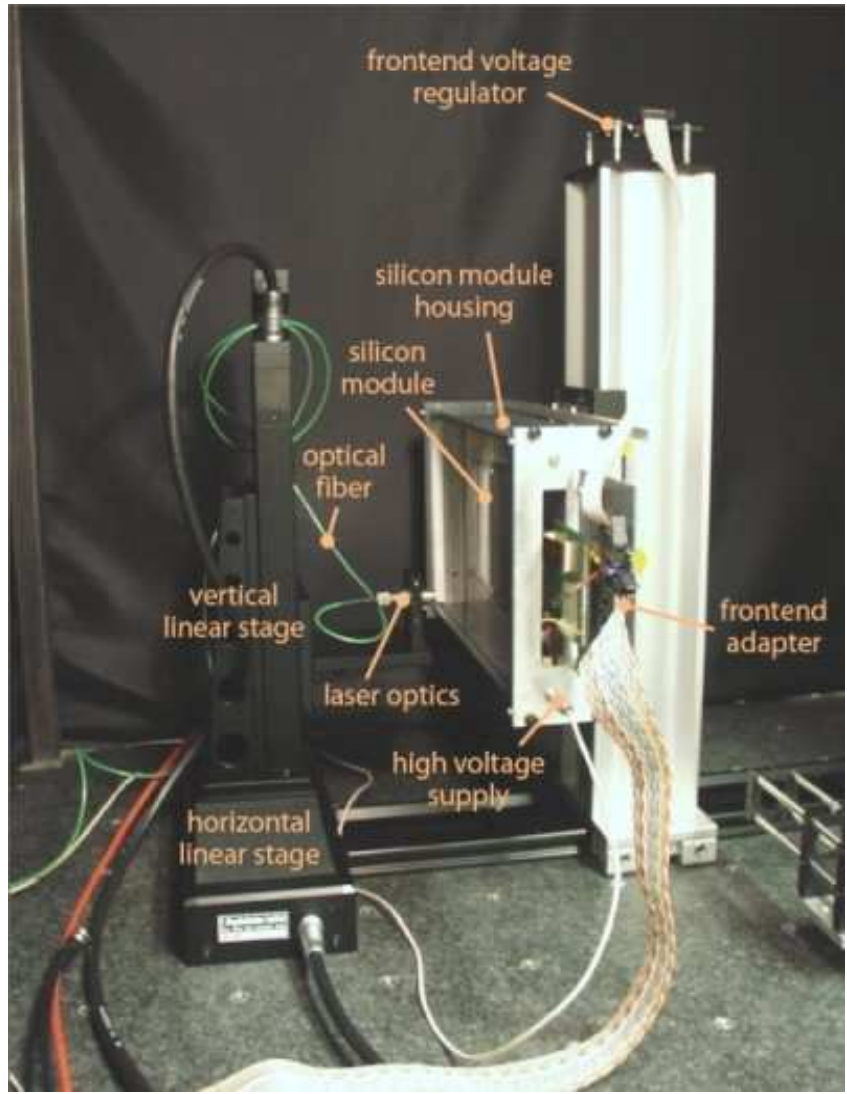}
\caption{\label{fig:teststandPhoto}Photograph of the module test facility.}
\end{center} 
\end{figure}
\begin{table}[hb]\begin{center}\small
\begin{tabular}{p{0.19\textwidth}|p{0.72\textwidth}}\hline
  controller & \STRUT Intel P3 866\+MHz personal computer \\ \hline
  linear stages & \STRUT OWIS Limes 90 (vertical) and Limes 120
  (horizontal) DC servo stages with 150\+mm and 200\+mm travel,
  respectively, and $<$16\+$\mu$m positioning accuracy; OWIS DSC~1000
  servo amplifier actuated by a Galil DMC~1040 motion controller for ISA
  bus
  \\ \hline
  pulse generator & TT{\it i} TG550 function generator \STRUT \\ \hline
  laser source & \STRUT Custom built single-mode diode laser unit, center
  wavelength $\lambda = 1050.5$\+nm, with output power $< 1$\+mW and 1\+ns
  rise time\\ \hline
  trigger unit & \STRUT LeCroy model 222 gate generator \\ \hline
  laser optics & \STRUT Sch\"after+Kirchhoff micro focus ($f=50$\+mm)
  mounted on a fiber collimator ($f=18.4$\+mm); 1.5\+m long single-mode
  fiber with $7.5\+\mu$m core diameter; 8$^{\circ}$ angled polish
  connectors \\ \hline
  module readout & \STRUT ARC system including ARC board, frontend
  adapters, ISA computer interface and cabling \\ \hline
  depletion voltage supply & \STRUT ISEG SHQ~222m power supply
  with RS~232 interface \\ \hline
  software & \STRUT Linux kernel 2.2.14 operating system, custom driver
  for linear stages, ARCS driver, custom software for module test
  operation \\ \hline
\end{tabular}\end{center}
\caption{\label{tab:teststandComponents}Details of the components of the laser test facility.} 
\end{table}
Figure~\ref{fig:teststandPhoto} shows a photograph of the mechanical
layout of the MTF. During the final steps of their assembly, the
modules are mounted on standardized aluminum support plates for easy
handling and interim storage. For laser testing, a module is placed in
an aluminum/plastic composite box which is equipped with a slide-in
slot for quick insertion of the support plate. The box itself is
mounted on a rigid aluminum pole and carries the electrical connectors
for the readout system and the depletion voltage supply.  Across from
the module housing box, a motorized linear stage is mounted
horizontally, itself carrying a second vertically oriented motorized
stage, to which the laser optics are attached. By these means, the
optics can be moved in a plane parallel to the module and illuminate
any point on the silicon sensors. Both stages are driven by servo
motors and offer a positioning accuracy of less than 16\+$\mu$m.  The
entire setup is housed in a lighttight cabinet and is mounted on an
air-cushioned granite table to protect it from vibrations.
Table~\ref{tab:teststandComponents} summarizes the components of the
MTF and their properties.

\subsection{The Test Procedure}\label{sec:testProcedure} 
\begin{figure}[htb]
\begin{center}
\includegraphics[width=10cm]{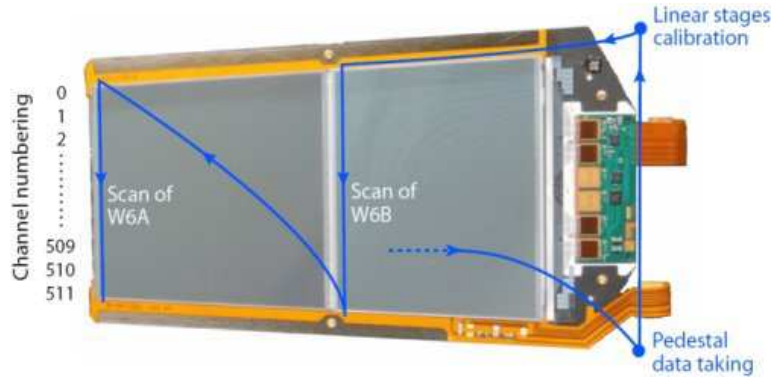}
\caption{\label{fig:laserPath}Schematic view of the path which the laser spot
follows during the test procedure in the plane of the module surface
(solid blue line).}
\end{center} 
\end{figure}

A personal computer (PC) is deployed for controlling the course of a
module test session with the MTF. It addresses the two linear stages,
operates and monitors the high voltage power supply, and stores the
silicon module data from the ARC board. The custom software package
for the MTF is written in C++ and makes use of the ARCS hardware
device driver and hybrid communication routine level, as well as a
custom device driver for the linear stage controller board.
Figure~\ref{fig:laserPath} illustrates the path followed by the laser
spot in the plane of the module surface during the course of the fully
automated testing procedure. In a first step, after insertion of the
module into the MTF box, the depletion voltage for the module is
slowly ramped up to 100\+V to avoid high currents flowing into the
capacitance of the growing depletion zones around the p-n junctions.
The ramp time is set to 30\+s. Subsequently, the laser spot is moved to
a position away from the silicon sensor surfaces in order to take
pedestal data without laser illumination.

The pedestal $P_i$ of a sensitive detector element $i$ is defined as
its average output level if no signal is present. Due to manufacturing
tolerances of the hybrid IC, the pedestals of the individual strips
differ slightly from each other and must therefore be determined for
later subtraction from the signal data. For this purpose, the
module is read out a certain amount of times $n$ with the laser in the
offside position and the resulting data $D_{i,n}$ are stored for each
strip $i$. The pedestals are then calculated according to
\begin{equation}
P_i = \frac{1}{n}\sum^n_{k=1} D_{i,k}\, .
\end{equation}
In the case of the MTF, a value of $n=10000$ has been chosen to determine
the pedestals with high accuracy.

Since at the beginning of a test session the state and position of the
linear stages must be assumed to be unknown, an absolute position
calibration is performed as the next step. For this, the two stages
are moved towards the upper (vertical stage) or right (horizontal
stage) boundary of their range, until they are stopped by their
internal limit switches. This position is reproducible within the
16\+$\mu$m accuracy of the servomotors and is then defined as the
reference point by resetting the position encoders to zero.

After calibration, the MTF is ready for laser testing. The laser spot
is moved successively over both sensors in a direction perpendicular
to the strips at a constant speed of 200\+$\mu$m/s. During the scans
the individual strip signals are read out and stored with the APV25-S1
operated in peak mode. To allow checking the full strip length for
possible defects, the sensors are scanned at the end far from the
readout electronics. Scanning both sensors separately gives additional
information about the identity of the sensor or wirebond which is
affected by a particular defect. The channel numbering used in the
following is indicated on the left side of Figure~\ref{fig:laserPath}
and starts at zero. The laser moves along the direction of increasing
strip numbers in both cases. The final output of the procedure is an
array of 512 values for each scan -- the set of the {\sl maximum
signal\/} delivered by each of the 512 strips. From these numbers, the
corresponding pedestal values are subtracted. Including pedestal data
collection and calibration of the linear stages, the total time needed
for a test session with the MTF amounts to approximately 900 seconds
per ring 6 module.

\subsection{Investigation of Artificially Bonded Defects}
To investigate the performance of the MTF and to assure its
applicability for reliable detection of module defects, a {\sl
  reference module\/} has been prepared with artificially bonded
defects. Two pairs of short-circuited strips were simulated by placing
a wirebond between the AC pads of the strip pair. Additionally, three
pinholes resulted by wirebonding the AC and DC pads of a strip, thus
establishing a perfect short-circuit between the surface aluminum and
the underlying implant. The wirebonding was performed at the far end
of sensor W6A, at maximum distance from the hybrid. Furthermore, one
open wirebond between the sensor W6B and the pitch adapter appears on
the reference module.  Table~\ref{tab:refModDefStrips} summarizes the
artificial defects and gives the number of the affected strips on the
module.
\begin{table}[htb]
\begin{center}
\begin{tabular}{r|c|l}
\hline
defect type & number & strip numbers \\ \hline
inter-strip short-circuits & 2 & 269--270, 288--289 \\
pinholes & 3 & 309, 329, 349 \\
open wirebond & 1 & 411 \\ \hline
\end{tabular}
\caption{\label{tab:refModDefStrips}The type and number of
  artificially bonded defects on the reference module and the affected
  strip numbers.}
\end{center}
\end{table}

\begin{figure}[htb]
\begin{center}
\includegraphics[width=15.9cm]{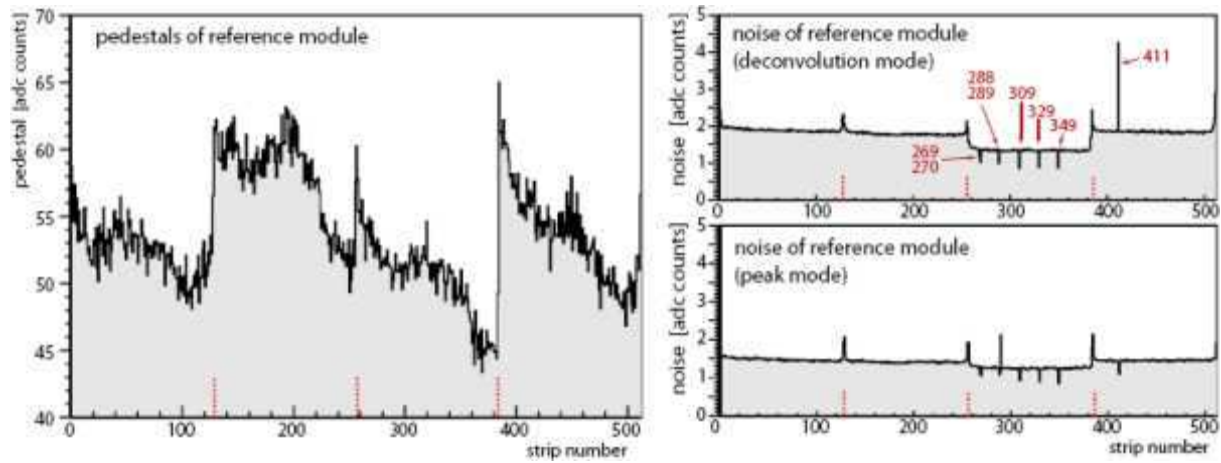}
\caption{\label{fig:refModNoisePed}The pedestals {\sl (left)\/} and
  the common mode subtracted noise for the two APV25-S1 readout modes
  {\sl (right)\/} for the MTF reference module. The APV25-S1 border
  channels are indicated by red dashed marks.}
\end{center} 
\end{figure}

Prior to the laser testing procedure, the pedestals and the CMN
corrected noise of the reference module have been determined with the
ARCS software~\cite{franke05a}. Since the analog inputs of the
APV25-S1 are bundled in four blocks of 32 channels each, the CMN
correction algorithms implemented in ARCS are based on this group
definition~\cite{axer03a}.  The results are depicted in
Figure~\ref{fig:refModNoisePed}. The left part of the figure shows the
strip pedestals, which exhibit a typical structure with sharp edges at
the APV25-S1 borders and a slope within the range of the individual
APV25-S1, which is probably due to a drop of supply voltage across the
channels. No hint for any defective strip can be observed in these
data.

The right side of Figure~\ref{fig:refModNoisePed} displays the noise
of the strips of the reference module for the two APV25-S1 readout
modes.  In both cases, the noise data show a largely flat common
baseline with a series of spikes of varying amplitude in both
directions and a lower plateau in the strip range of one APV25-S1. The
artificially bonded defects, indicated by the corresponding strip
numbers, are clearly visible as distinct peaks or dips and can thus in
principle be tagged as faults. However, a number of effects complicate
the identification of defective strips in the noise data.

First, the channels on the APV25-S1 borders and especially the
outermost strips close to the edge of the module show an increased
noise level. This behavior is mostly due to their position which
affects the capacitive couplings to their neighbors and other circuit
paths on the module (e.g. to the guard or bias ring, see
Figure~\ref{fig:cmsSensor}). Despite the peculiarities shown by border
and edge strips in the noise data, they are not to be considered as
defective, thus making it necessary to distinguish them from the truly
faulty channels. Furthermore, the signature of the different
categories of defects is not unique in the noise data. Obviously, as
apparent for example in the behavior of the pair of short-circuited
strips 288--289 and the open wirebond (number 411), the signature
varies with the readout mode, as the channels exhibit a lower or
higher than average noise level in the two cases, respectively. The
reason for this behavior lies in the CMN correction facility of the
APV25-S1, which applies also to channels with lower noise as for
example those affected by an open wirebond. The correction may then
lead to a higher apparent noise level, depending on the readout mode.
In addition, the presence of multiple pinholes in the strip range of a
single APV25-S1 leads to a significant lowering of the noise baseline
due to an increased current flow between the silicon and the APV25-S1,
which complicates the detection of defects. Finally, the signatures of
defects in the noise data are highly sensitive to the environmental
circumstances under which the module is operated, such as grounding
and electromagnetic shielding, making them generally unpredictable and
unreliable for testing purposes.

\begin{figure}[ht]
\begin{center}
\includegraphics[width=16cm]{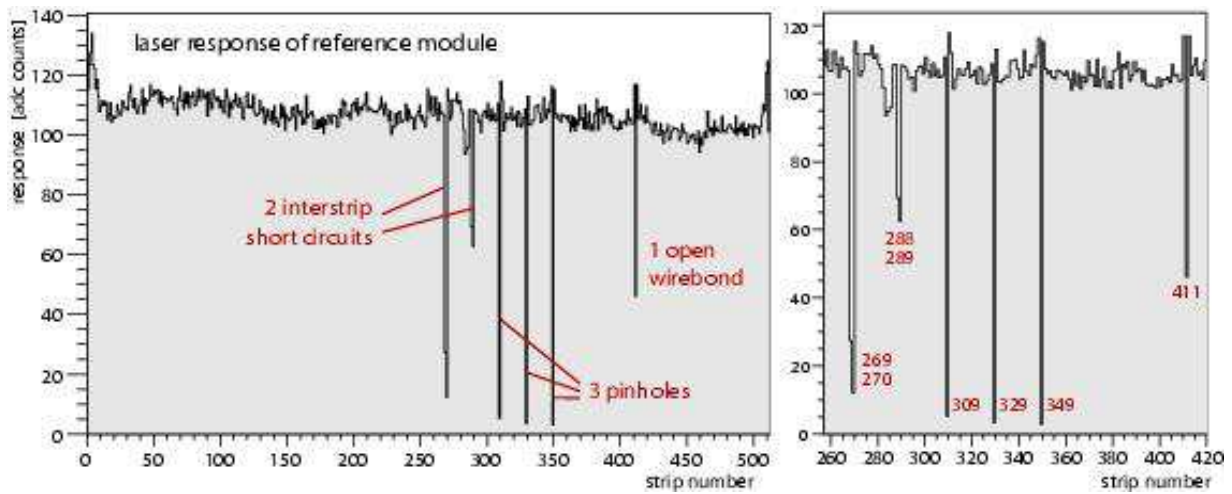}
\caption{\label{fig:refModLaserScan}Laser scan results from sensor W6A of 
the reference module {\sl (left)\/}; enlarged view of the strip
interval exhibiting conspicuous strip signals from artificially bonded
defects. The strip numbers are indicated {\sl (right)\/}.}
\end{center} 
\end{figure}
Figure~\ref{fig:refModLaserScan} displays the result of the laser
testing procedure as it was applied to sensor W6A of the reference
module. The overview histogram on the left side contains the maximum
laser scan signal for each of the 512 channels; the right side shows
an enlarged view of the strip range with bonded defects. As apparent
in the overview, the normal channels show a nearly constant response
to the laser illumination of about 110 adc counts. However, a slightly
negative slope remains, which is caused by interference effects from
light reflected at the metalized backplane of the sensors. Such
effects are due to slight variations of the sensor thickness of the
order of a few micrometers and have already been observed
previously~\cite{wittmer02a}. All artificially bonded defective
channels clearly appear as distinct negative spikes from the common
baseline and consistently show a lower response, independent of the
readout mode. An indication for the type of defect can be obtained
from the laser scan signature in the following way.
\begin{itemize}
\item Short-circuited strips are recognizable through their pairwise
  occurrence. Besides, since they are connected, the charge induced on
  one strip is shared among the two readout channels and leads to a
  reduced signal response for both. As a consequence of the increased
  strip capacitance at each channel input, the noise level is higher,
  so that the resulting laser response of two particular pairs of
  short-circuited strips may significantly differ.
\item Pinholes represent direct connections from the aluminum strips
  to the underlying p$^+$ implants, permanently forcing the channel
  inputs of the affected strips to the implant potential.  Therefore,
  no signal is observed at any time and the noise level is very
  small. Pinholes are thus identifiable through their
  particularly small response to laser illumination.
\item Open wirebonds, similar to pinholes, prohibit the observation of
  any signal in response to laser illumination. However, due to the
  noisy behavior of the affected channels, resulting from the CMN
  correction described above, their mean observed response is
  typically higher than in the case of a pinhole.
\end{itemize}

\begin{figure}[b]
\begin{center}
\includegraphics[width=10cm]{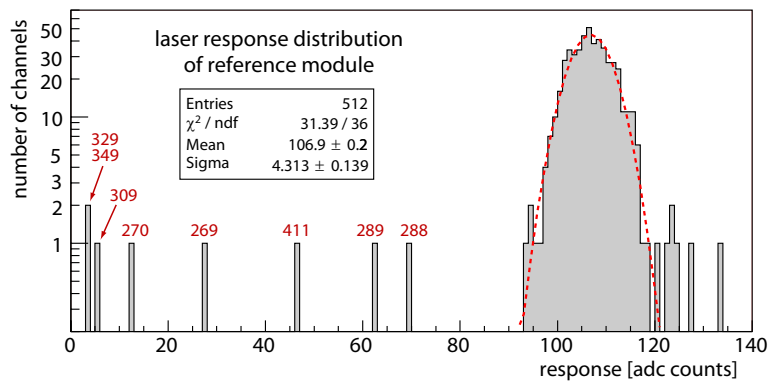}
\caption{\label{fig:refModRespDist}Distribution of the laser response
of the reference module channels. The red dashed line represents a
Gaussian fit to the bulk of faultless strips, whose parameters are
given in the legend. Defective channels are indicated by their
strip numbers.}
\end{center} 
\end{figure}
It must be pointed out that, as a consequence of the statistical
character of signal noise, the above criteria for identification of
the type of defect must not be considered to be unambiguous in every
case. The main purpose of the MTF is reliable defect detection.
Figure~\ref{fig:refModRespDist} displays the distribution of the
response for the channels of the reference module. The bulk of
flawless strips show an approximately Gaussian distribution with a
mean of 107 and a standard deviation of 4.3 adc counts. A small tail
on the right side is caused by the slightly higher response of the
module's edge strips. The region populated by the defective channels
on the left side is clearly separated from the bulk by more than 8
standard deviations.  Hence, all artificially bonded defects are
reliably identifiable through their low response to laser
illumination.

\subsection{Results from the TEC W6 Prototype Series}
The TEC collaboration produced a prototype series -- the {\sl
express-line 2} -- of nine ring 6 modules.  The main purpose of the
express-line modules was to allow the development, improvement and
validation of module test procedures prior to series production.
Contrary to the final module design, the express-line modules are
equipped with ceramic hybrids instead of the kapton/copper ones,
which, however, has no effect on their readout properties in this
procedure. Individual modules are distinguishable through
identification numbers on the support frame, which are listed in
Table~\ref{tab:expresslineResult}.

\begin{figure}[hbt]
\begin{center}
\includegraphics[width=20cm,angle=90]{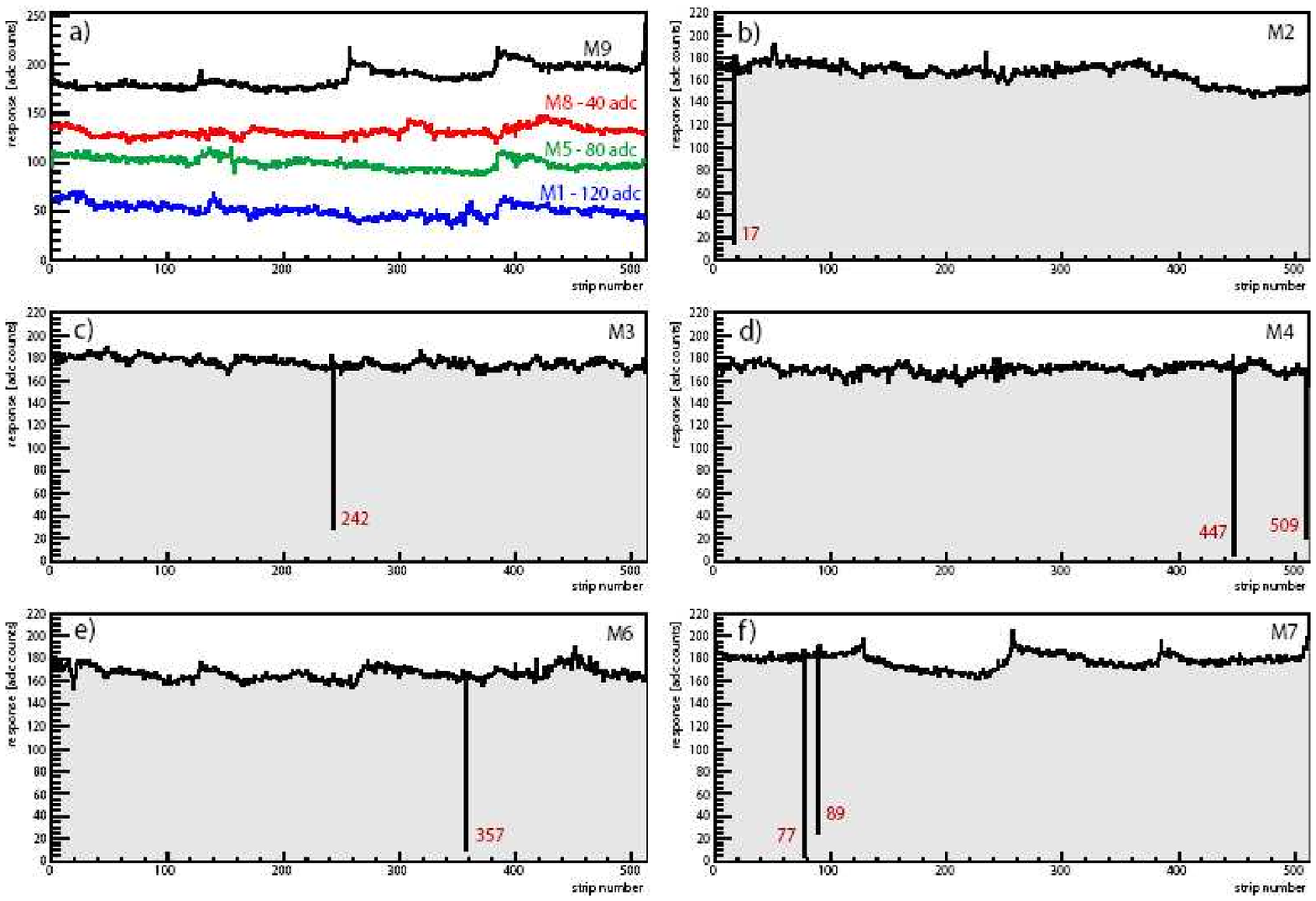}
\caption{\label{fig:exprLaserScans}Laser scan results from W6A sensors
  of the express-line modules. Display a) combines the scans of the
  four modules without defects M1, M5, M8 and M9. For the sake of
  clarity, the common baselines of the channels have been shifted
  downward by 120, 80, 40 or 0~adc counts, respectively. The remaining
  histograms b) to f) show the individual results for modules M2, M3,
  M4, M6 and M7, with the defective strips indicated by their strip
  numbers.}
\end{center} 
\end{figure}

All nine modules of the express-line have been tested with the MTF
according to the procedure described in
\S\+\ref{sec:testProcedure}. The results of the individual scans
of the W6A sensors are depicted in Figure~\ref{fig:exprLaserScans}.
Confirming the earlier results on the reference module, defective
strips are clearly visible through their strongly reduced response to
laser light. Four of the modules turned out to have no faulty channels
at all; their laser scan results are summarized in panel a) of the
figure. Three of the remaining modules (M2, M3 and M6) have one
defective channel each, while modules M4 and M7 each have two faulty
channels. According to the grading scheme introduced in
\S\+\ref{sec:MTF}, based on the number of defective strips,
all modules must be considered as grade~A. Nevertheless, modules
M5, M6 and M7 were later classified as grade~C due to their
high leakage current~\cite{axer04a}.
\begin{table}[htb]
\begin{center}
\begin{tabular}{c|c|l}
\hline
module & frame id number & faulty channels \\ \hline
M1 & 30200020000501 & --  \\
M2 & 30200020000502 & 17\openindex  \\
M3 & 30200020000641 & 242\openindex \\
M4 & 30200020000643 & 447\pinholeindex, 509\pinholeindex \\
M5 & 30200020000503 & -- \\
M6 & 30200020000644 & 357\pinholeindex \\
M7 & 30200020000640 & 77\pinholeindex, 89\openindex \\
M8 & 30200020000642 & -- \\
M9 & 30200020000645 & -- \\ \hline
\end{tabular}
\caption{\label{tab:expresslineResult}The results of the express-line
  module tests with the MTF. Faulty channels are indicated by the type
  of defect: \pinholeindex~for a pinhole and \openindex~for an open
  wirebond.}
\end{center}
\end{table}

Table~\ref{tab:expresslineResult} gives an overview of the modules and
the defective channels detected. From the seven defects listed, two
have been identified as open wirebonds by visual inspection, and four
represent pinholes. The failure of channel 242 on M3 is probably due
to a broken pitch adapter line~\cite{axer04a}. No short-circuited
strips were observed. The results of the express-line module tests
with the MTF turn out to be consistent with those from other
facilities using different testing techniques~\cite{axer03a}.

Following the expess-line module tests the MTF is established as a
practicable and reliable facility for the detection of defective
strips on the TEC modules. Furthermore, with its high positioning
accuracy and stability of the laser intensity, the MTF is well suited
as a precision tool for the in-depth investigation of silicon
microstrip sensor properties. However, the disadvantage of the
facility are the high costs of the hardware. In order to provide all
laboratories involved in module testing with identical test
facilities, the decision was made to use an extension to the ARC
system~\cite{franke05a} for testing during the series production of
the modules. Besides test procedures based on measurements of the
sensor leakage current, APV25-S1 amplifier gain linearity, noise
distributions and frontend electronics performance, the extended ARC
can perform illumination of the silicon sensors with light-emitting
diodes. 

The assembly of the CMS Tracker Endcaps was finished in November 2006
and first operation of the LHC is scheduled for 2007. The following
chapters describe the search for signatures of supersymmetry in the
cosmic ray spectra and the analysis of data from the \AMS{}
experiment, which bases on the reconstruction of multi-particle events
in a large-scale application of a silicon microstrip detector which
was operated in space. The subject is introduced with a discussion of
cosmic rays in the next chapter.

\chapter{Cosmic Rays as Tracers of Dark
Matter}\label{chapter:cosmicrays} The Earth is continuously exposed to
a flux of particles of extraterrestric origin, referred to as cosmic
rays. This was discovered in 1912 by V.~Hess, who observed an
increasing discharge rate of electrometers towards higher altitude
during a series of balloon flights~\cite{hess12a}. Historically, the
investigation of cosmic rays has played an important role in the
development of particle physics, as they were the only source of high
energy particles before the emergence of accelerators. Thus, the
positron~\cite{anderson33a}, the muon~\cite{neddermeyer37a}, the
pion~\cite{lattes47a}, the kaon~\cite{rochester47a} and other strange
mesons have been discovered in cosmic rays.
\begin{figure}[b]
\begin{center}
\begin{tabular}{ll}
\begin{minipage}{7.5cm}
\includegraphics[width=7.32cm]{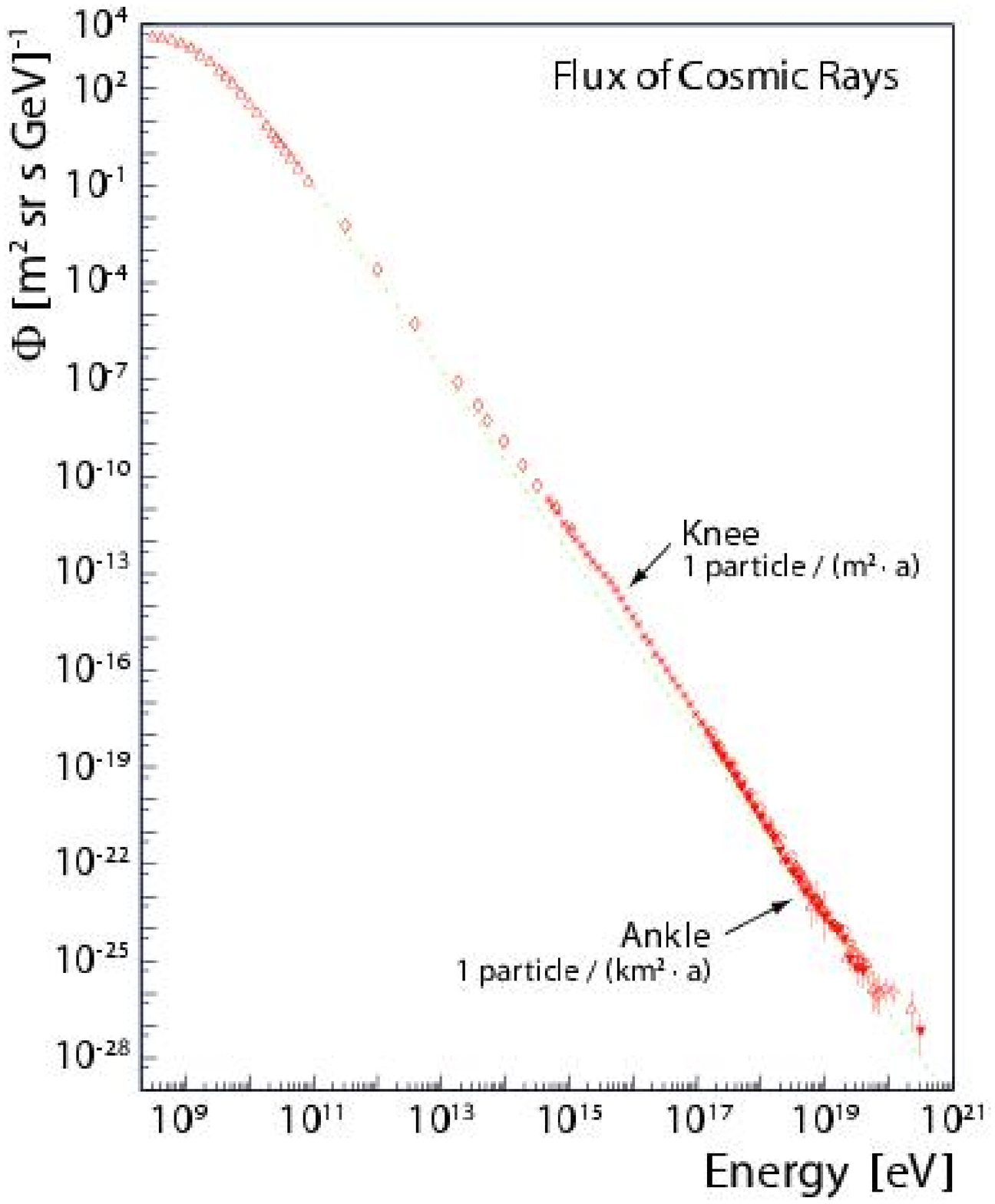}
\end{minipage}
&
\begin{minipage}{8cm}
\begin{center}
\includegraphics[width=8cm]{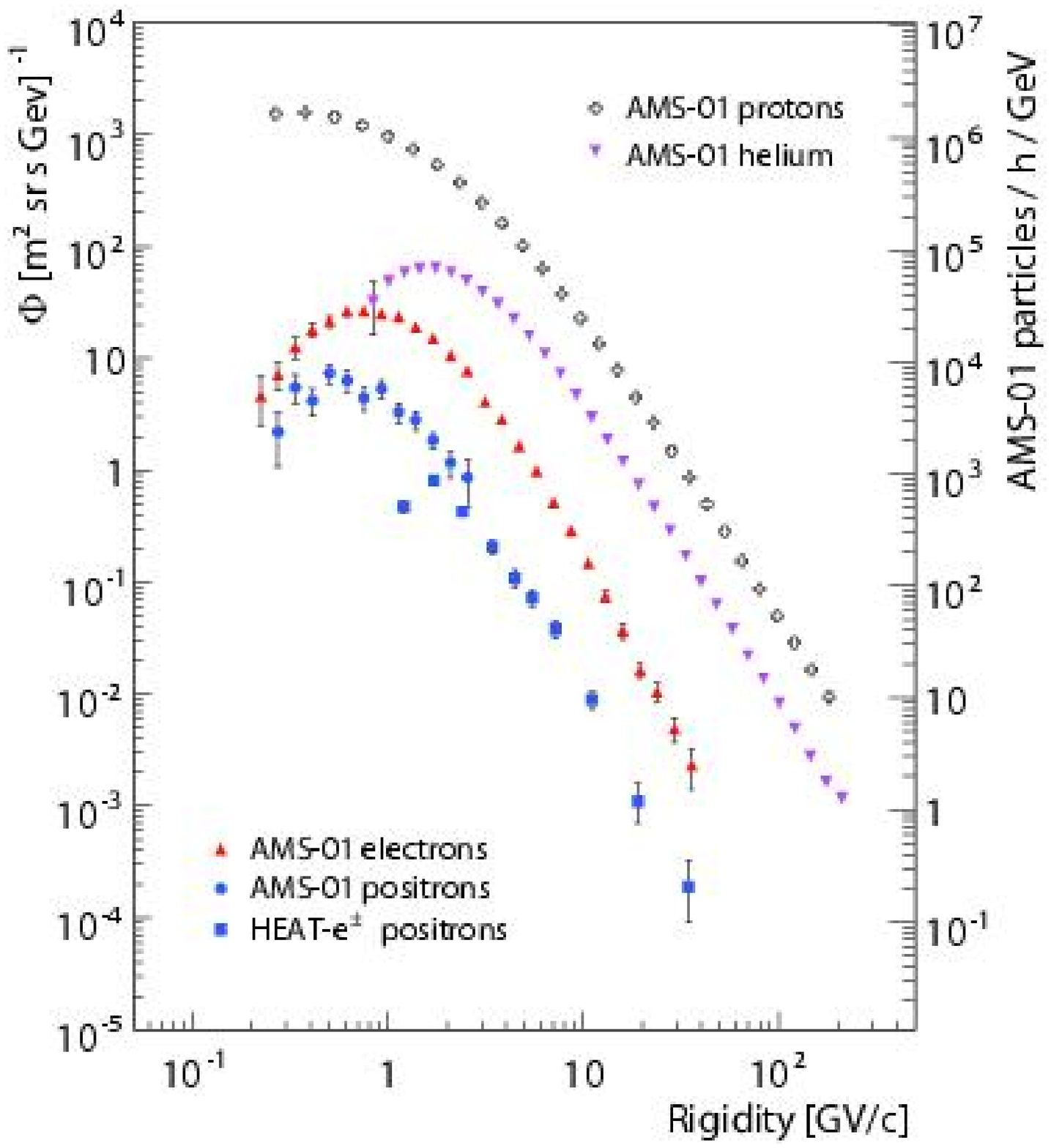}
\end{center}
\end{minipage}
\end{tabular} 
\end{center} 
\caption{\label{fig:spectra}Total cosmic ray flux as a function of energy~\cite{cronin97a} {\sl (left)};
fluxes of some of the most abundant cosmic ray components as measured by 
\AMS{}~\cite{alcaraz00d,alcaraz00b,alcaraz00c} and HEAT-e$^{\pm}$~\cite{duvernois01a} {\sl (right)}.}
\end{figure}

Experimental knowledge about cosmic rays spans a very wide energy
range. Particles from the solar wind typically propagate with a few
\keV{}, while, on the other hand, cosmic rays have been observed with
energies higher than $10^{20}$\+\eV{}~\cite{bird94a}. Moreover, the
abundance of cosmic rays strongly varies as a function of energy. Over
large energy intervals the spectra follow a steep power law with a
spectral index of roughly 3, meaning that the intensity of cosmic rays
decreases by a factor of 1000 for each decade in
energy~\cite{cronin99a}.

Regarding the search for dark matter, one of the most important
properties of the spectral shapes of cosmic rays is their smoothness
on small scales. As will be shown in this chapter, the cosmic ray
background produced in conventional high energy processes in the
universe is expected to lack any structure. Reaction chains involving
dark matter may be additional sources of standard model particles and
would cause an excess within limited energy ranges. Of particular
interest are the fluxes cosmic ray particles of low abundance, such as
the positron, since exotic sources would lead to significant excesses.

\section{Cosmic Ray Composition and Spectra} 
The particles reaching the top of the Earth's atmosphere are mainly
electrons and fully ionized light atomic nuclei, such as hydrogen and
helium. Nuclei with higher charge and other particles amount to less
than 1\+\% of the total intensity. Table~\ref{table:abundances} gives
an overview of the approximate abundance of the most dominant cosmic
ray components at 1\+\GeV{} energy. Antinuclei have not been observed
yet~\cite{alcaraz99a}.  The chemical composition of cosmic rays is
similar to that of the sun, which indicates that they are
predominantly of stellar origin.  Exceptions are the nuclei of
elements from Li to B and from Sc to Mn, which are overabundant as the
products of spallation of carbon and oxygen and of iron, respectively.
\begin{table}[!h]
\begin{center}
\begin{tabular}{lc|lc}
\hline
Component & Abundance [\%] & Component & Abundance [\%]\\
\hline
Hydrogen & 90 & Positrons & 0.4\\
Helium & 4.6 & Carbon & 0.2\\
Electrons & 2.3 & Oxygen & 0.2\\
Deuterons & 1.7 & Other nuclei & 0.3\\
\hline
\end{tabular}
\caption{\label{table:abundances} Approximate abundances of the most dominant
components of cosmic rays at 1\+\GeV{} (or 1\+\GeV{}/nucleon for
nuclei) energy, in percent of the total intensity. Based on data from
\cite{alcaraz00d,aguilar02a,alcaraz00b,alcaraz00c,orito00a,lezniak78a}.}
\end{center}
\end{table}

Due to the deflection of charged particles in the galactic and solar
magnetic fields, the arrival directions of cosmic rays are randomized
at detection, so that the fluxes are widely isotropic. Their energy
spectra above 10\+\GeV{} can be described by a segmented power law of the
form
\begin{equation}
\frac{dN}{dE}\propto E^{-\gamma}\, ,
\end{equation}
with the {\sl spectral index} $\gamma$ as an essential parameter. Up
to energies of $10^{16}$\+\eV{} $\gamma$ is equal to 2.7 and then steepens
to 3.0 for higher energies (see Figure \ref{fig:spectra}, {\sl left}).
The break in the spectra at $10^{16}$\+\eV{} is commonly referred to as
the {\sl knee}. At approximately $10^{19}$\+\eV{} -- the {\sl ankle} --
and above, the spectra seem to flatten again. This behavior is
independent of the incident direction, i.e. the spectral shapes are
also isotropic. Below 10\+\GeV{}, the spectra are modulated by the solar and
geomagnetic influence, which is the subject of
\S\+\ref{section:earthsVicinity}.

The reason for the two breaks in the spectra has not been fully
understood as yet~\cite{cronin99a}. The conventional explanation is
that they represent the power limits of different acceleration
mechanisms at successive energy scales. Obviously, these mechanisms
must be based on a common principle, since the spectral shape is
universal over large energy ranges. It was proposed in 1949 by
E.~Fermi~\cite{fermi49a}, that the spectral shape may originate from
repeated finite energy gains of particles in collisions with moving
objects, e.g. magnetic fields in shock waves of supernovae. Let a
particle of initial energy $E_0$ undergo such a repeated acceleration
and gain a fraction $\xi E$ of its energy $E$ each time. Then, after
$n$ collisions, it will have
\begin{equation}
E_n = E_0\left(1+\xi\right)^n,
\end{equation}
giving for $n$
\begin{equation}\label{eq:fermi:n}
n = \ln(E/E_0) / \ln(1+\xi)\, .
\end{equation}
Assuming a probability $p_w$ for a particle to escape the acceleration
mechanism, the number of particles with an energy larger than $E$ is
\begin{equation}
N(\geq E) \propto \sum^{\infty}_{m=n} \left(1-p_w\right)^m = \frac{\left(1-p_w\right)^n}{p_w}\, .
\end{equation}
Substituting $n$ from eq. \ref{eq:fermi:n} yields
\begin{equation}
N(\geq E) \propto \frac{1}{p_w} \left(\frac{E}{E_0}\right)^{-\alpha} \,,\, \alpha\approx p_w/\xi\, ,
\end{equation}
which leads to the observed form of the spectra~\cite{gaisser90a}.
This is the principle on which the so called {\sl second order Fermi
  acceleration mechanism} is based. However, in this simple form, it
cannot sufficiently explain the isotropy of the observed spectra.
Furthermore, it requires large initial energies and is very
inefficient, so that cosmic rays of highest energy are unlikely to
have been accelerated in this way. Consequently, Fermi's theory has
undergone fundamental further development in the past
decades~\cite{blandford78a}.

Figure~\ref{fig:spectra} {\sl (right)} shows the spectra of protons,
helium nuclei, electrons and positrons with energies up to
approximately 100\+\GeV{}. It is apparent that in this energy regime
protons are by far the most abundant cosmic ray component. In
particular, their flux exceeds that of positrons by a factor of
$10^3-10^4$.  The ratio of the positron to electron flux varies
from roughly 1:7 to 1:20. Moreover, the energy spectra of the leptons
($\gamma\approx 3.4$) are slightly steeper than those of nuclei
($\gamma\approx 2.7$).

\section{Origin and Propagation of Cosmic Rays}\label{sec:originAndProp}
Cosmic rays can be classified in two main categories of different
origin, {\sl solar} and {\sl galactic} cosmic rays. As obvious from
these terms, solar cosmic rays have their origin and are accelerated
in the vicinity of the sun, while galactic cosmic rays are produced
somewhere in the galactic volume. However, it is often assumed that
cosmic rays of highest energies may even originate from extragalactic
sources. On their way to Earth, cosmic ray particles are involved in a
plenitude of processes, including elastic and inelastic scattering off
the interstellar medium, deflection by magnetic fields and
deceleration through emission of radiation.

Solar cosmic rays mostly consist of ionized coronal material which is
accelerated by shock waves associated with the solar wind
\cite{gloeckler84a}. The abundance of these particles decreases
steeply with energy. At low energies, below 100\+\MeV{}, they constitute
the major part of the observable cosmic ray flux, but in the GeV
energy range their persistent contribution is negligible. Only in the
rare case of intense solar flares may particles be accelerated up to
1\+\GeV{} or even more \cite{steinacker89a}. These events have typical
durations of the order of minutes or hours. The composition of cosmic
rays produced in solar flares is quite similar to that of galactic
cosmic rays \cite{blandford87a}.

Supernovae provide the dominant power input to the galaxy. An average
supernova explosion releases between 1 and 10 solar masses of material
and a total energy of $10^{44}$\+J. Taking into account a supernova
explosion frequency in the galaxy of roughly $3\times 10^4$\+Ma$^{-1}$,
it follows that each supernova would only have to contribute $\sim
3$\+\% of its output power to the total cosmic ray power\footnote{i.e.
  the estimated total power leaving the galactic disk in form of
  cosmic rays} of $10^{34}$~W~\cite{blandford87a}. For this reason
supernovae are believed to be the dominant accelerators of cosmic rays
to energies up to $10^{16}$\+\eV{}. Since particles are scattered by
interstellar gas and therefore their energy declines inversely with
the cube root of the number density, they cannot be accelerated too
soon after the explosion. Rather than that, the acceleration takes
place when the interstellar gas is passed by the shock waves emerging
from the supernova. Evidence comes from obervations of supernova
remnants in the keV X-ray band~\cite{uchiyama02a}, yielding a flat
spectrum which can be interpreted in terms of bremsstrahlung emission
from protons or electrons. Furthermore, recent results from imaging
atmospheric \v Cerenkov telescopes reveal that these remnants also
emit gamma rays with up to TeV energies~\cite{aharonian04a}, in
agreement with the X-ray measurements regarding the spatial location
of the sources. Besides supernovae, other candidates for sources of
cosmic rays are pulsars, close binary systems and stellar
winds~\cite{moskalenko04a}. Particles which are produced and
accelerated in the vicinity of these sources are, with respect to
their origin, commonly referred to as {\sl primary cosmic rays}.

Between its acceleration and its detection a galactic cosmic ray
particle traverses an average amount of interstellar matter of
10\+g$\,$cm$^{-2}$~\cite{stephens98a} within typically 20\+Ma of
time~\cite{maurin01a}. The propagation volume is permeated by the
galactic magnetic field, whose regular (large scale) component has a
strength in the order of 5\+$\mu$G~\cite{vallee94a}, however with
turbulent local variations. As a consequence, in addition to their
convective thermodynamic motion, charged particles diffuse due to
interactions with the magnetic field inhomogeneities. While
propagating through the galaxy, cosmic rays also lose energy. For
nuclei, the dominant energy loss processes are ionization in the
interstellar medium and scattering off thermal electrons in
plasmas~\cite{maurin01a}. Electrons and positrons additionally undergo
bremsstrahlung, synchrotron radiation and inverse Compton scattering
with starlight or with the cosmic microwave
background~\cite{moskalenko98a}. Assuming the presence of a galactic
wind, i.e. a constant flow of magnetic irregularities, particles may
also gain energy while propagating from shock acceleration {\sl
  (diffuse reacceleration)}.
\begin{figure}[ht]
\begin{center}
\begin{tabular}{ll}
\begin{minipage}{8cm}
\includegraphics[width=8cm]{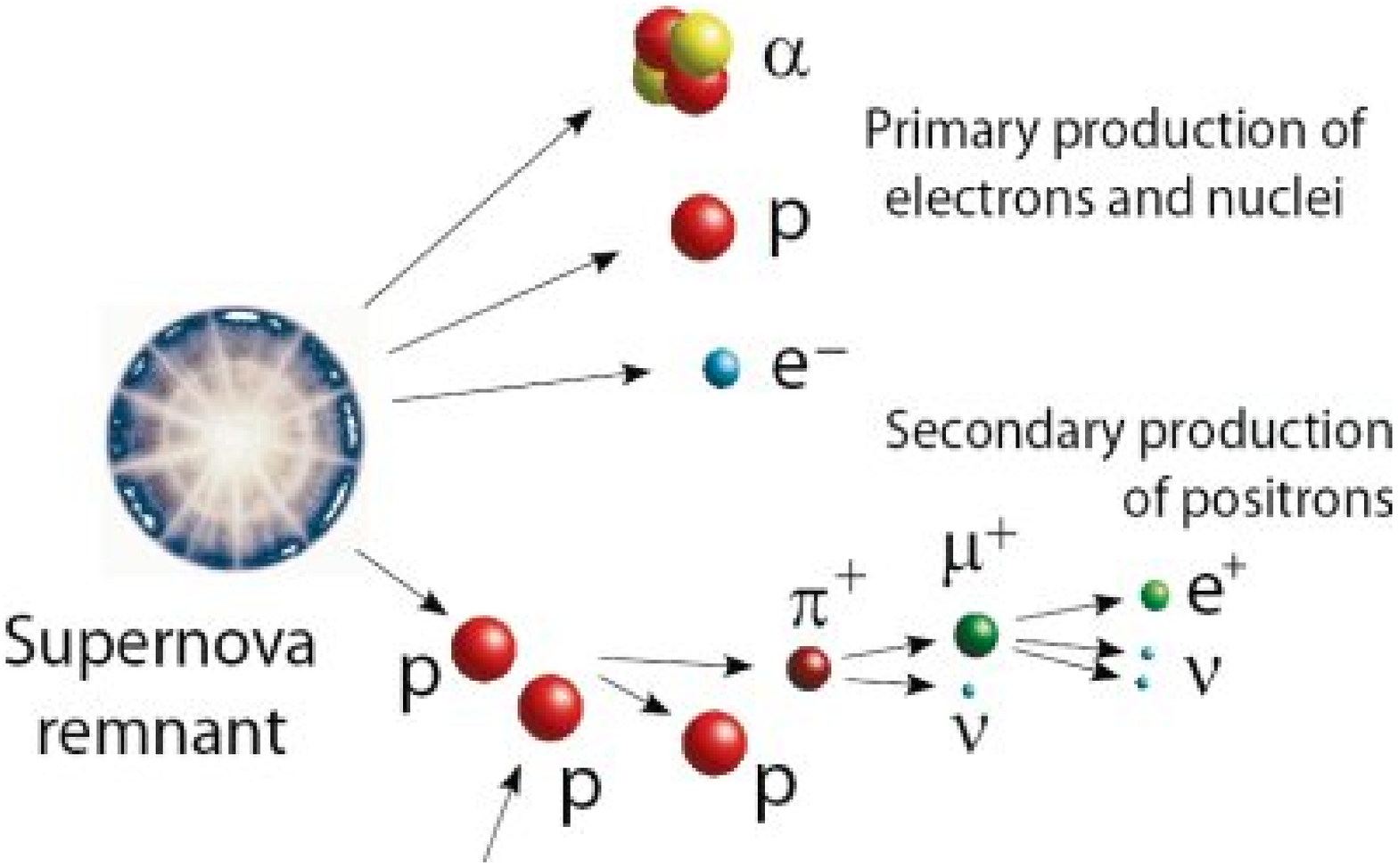}
\end{minipage}
&
\begin{minipage}{7cm}
\begin{center}
\includegraphics[width=7cm]{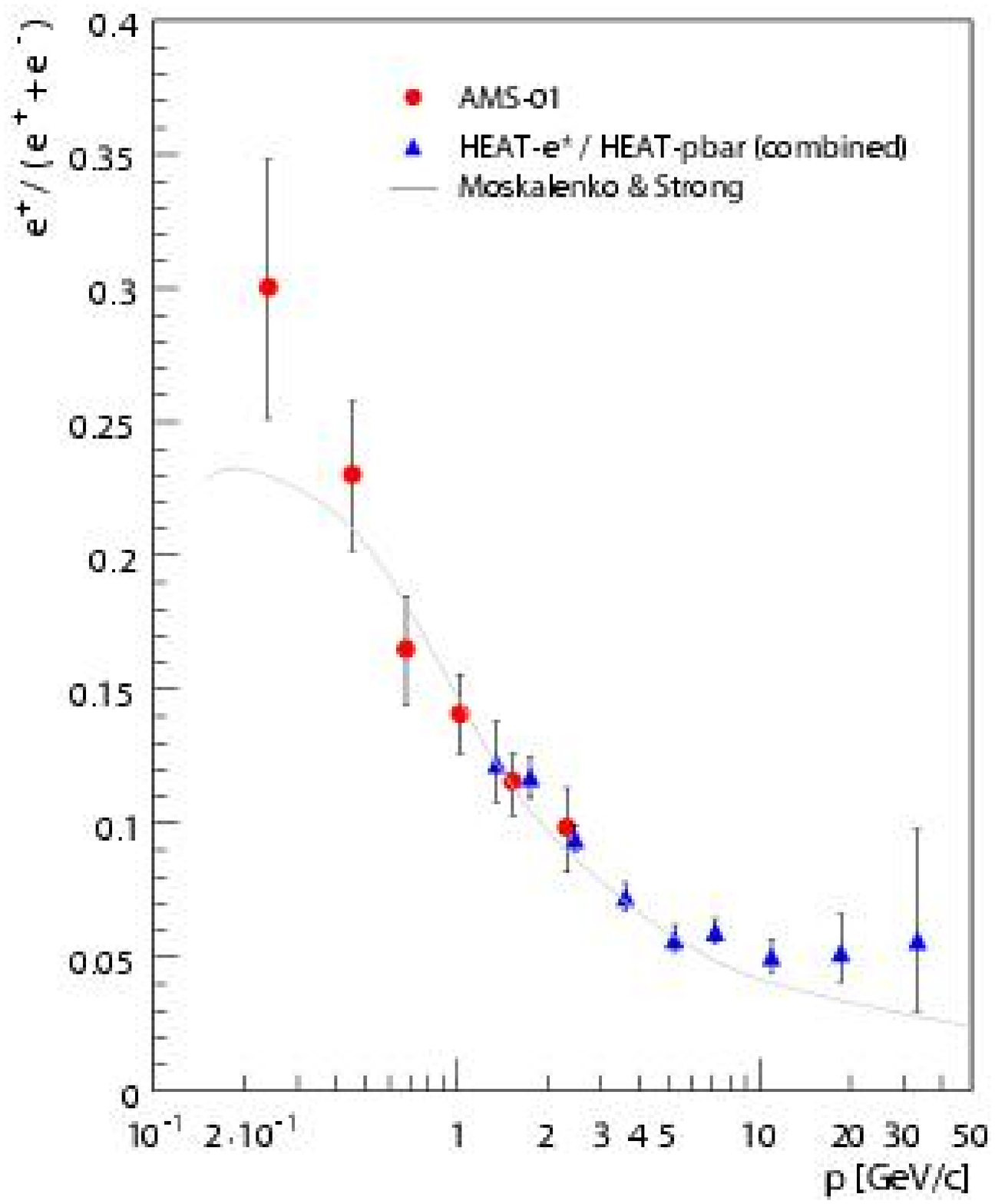}
\end{center}
\end{minipage}
\end{tabular} 
\end{center} 
\caption{\label{fig:fullfraction} Principle of primary and secondary cosmic ray production {\sl (left)\/};
model calculation of the positron fraction $e^+/(e^+ + e^-)$ for purely secondary positron production from
\protect\cite{moskalenko98a} (solid line) together with recent data from \AMS{}~\protect\cite{alcaraz00d},
HEAT-$e^{\pm}$ and HEAT-pbar~\protect\cite{beatty04a} {\sl (right)}.}
\end{figure}

The destruction of primary nuclei via spallation in the interstellar
medium gives rise to secondary nuclei and rare isotopes. Photons and
antiparticles are also rare in cosmic rays. Apparently, they are not
directly produced in the cosmic ray sources, but emerge as secondary
products from processes involving primary cosmic rays, and are thus
called {\sl secondary cosmic rays} (see Figure~\ref{fig:fullfraction},
{\sl left}). Antiprotons largely stem from proton-proton,
proton-nucleus and nucleus-nucleus interactions \cite{moskalenko98b}.
Photons are decay products of neutral pions from such processes and
also result from synchrotron radiation of high energetic particles or
inverse Compton scattering. Positrons are mainly created by protons
in reactions of the type $pp\to\pi^++X$ or $pp\to K^++X$, where the
pions and kaons decay into positrons via muons. Furthermore, the
$\gamma e^+e^-$ decay channel of neutral pions from reactions of the
type $pp\to \pi^{0}+X$ contributes to the positron flux. In addition
to protons, light nuclei may also be involved in these interactions
\cite{murphy87a}.

The cross sections of the processes which contribute to the positron
flux are strongly energy dependent and the spectra of their individual
contributions are smeared by energy losses during particle
propagation. The ratio of the positron and electron fluxes, commonly
written in terms of the {\sl positron fraction} $e^+/(e^++e^-)$, is
thus a smooth function of momentum. Figure~\ref{fig:fullfraction} {\sl
  (right)} shows a model calculation of the positron fraction for
purely secondary positron production (without reacceleration) together
with recent experimental data. In the range from a few hundred \MeV{}/c
to 50\+\GeV{}/c it decreases monotonously from 0.2 to roughly 0.03 and
then flattens out. In particular, for purely secondary positron
production, the positron fraction does not exhibit any small scale
features, peaks or dips.

\section{Cosmic Rays in the Earth's  Vicinity}\label{section:earthsVicinity} 
\begin{figure}[ht]
\begin{center}
\includegraphics[width=11cm]{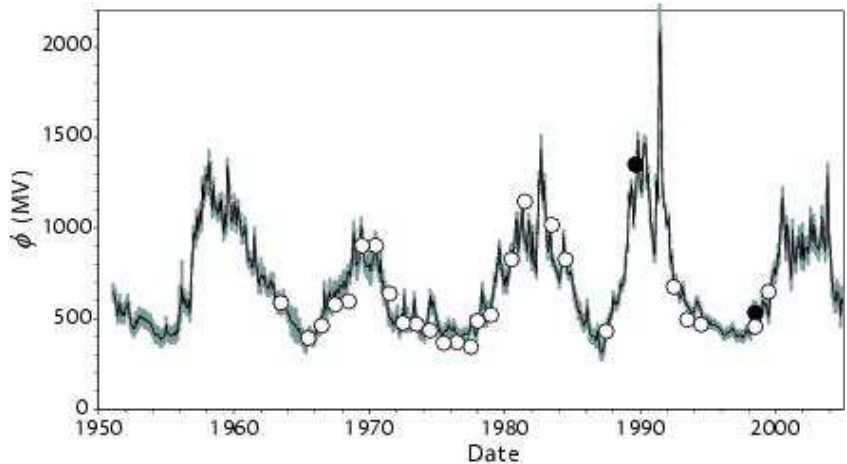}
\caption{\label{fig:solarmodulation} The solar modulation parameter $\phi$
(solid line) with 68\+\% confidence intervals (grey area) as a function
of time~\cite{usoskin05a}. The circles denote reference data.}
\end{center} 
\end{figure}

Within the heliosphere, cosmic rays are influenced by the solar wind,
a continuous stream of charged particles emanating from the sun. The
magnetic field associated with this stream has irregularities which
scatter particles in a frame of reference moving outwards of the solar
system. As a result, cosmic rays are effectively decelerated (or even
deflected) in the vicinity of the sun, an effect which is referred to
as {\sl solar modulation}. According to the so called force field
approximation \cite{gleeson68a}, the influence of solar modulation can
be parameterized with a single parameter $\phi$, which represents an
effective potential and is given in units of volts. Thus a particle
with charge $Ze$ experiences an energy loss $Ze\phi$ when approaching
the Earth from outside the heliosphere. As displayed in
Figure~\ref{fig:solarmodulation}, the solar modulation parameter is
time dependent and follows the 11 year solar cycle, typically varying
from 300\+MV to 1300\+MV. As can be seen e.g. in
Figure~\ref{fig:spectra} {\sl (right)}, the solar modulation is
apparent as a decrease with respect to the power law expectation of
the cosmic ray fluxes below approximately 10\+\GeV{}. There is
evidence that solar modulation is charge sign dependent at low
particle energies~\cite{clem96a}.

\begin{figure}[ht]
\begin{center}
\begin{tabular}{ll}
\begin{minipage}{8.5cm}
\includegraphics[width=8.5cm]{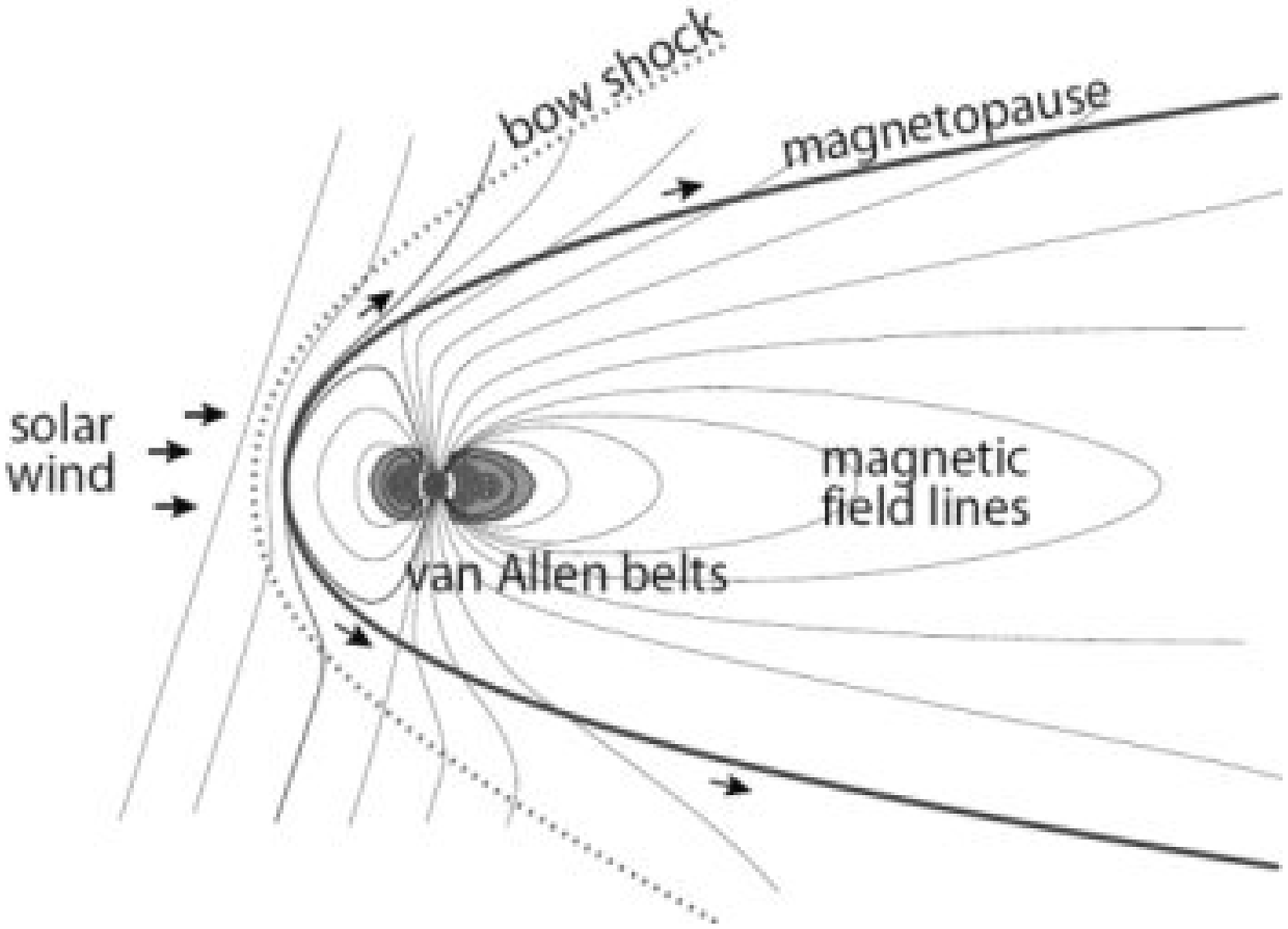}
\end{minipage}
\hfill
\begin{minipage}{7cm}
\begin{center}
\includegraphics[width=7cm]{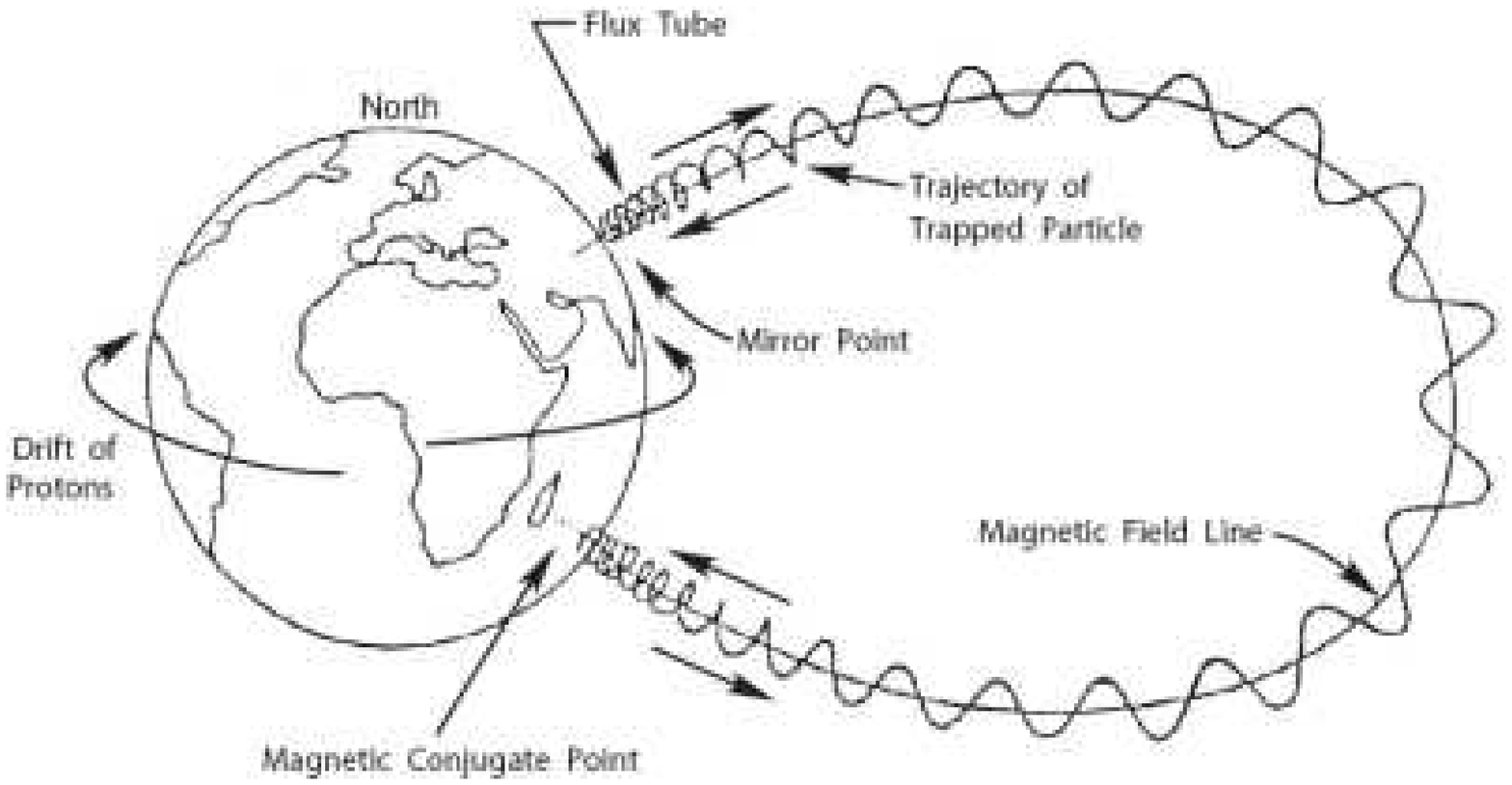}
\end{center}
\end{minipage}
\end{tabular} 
\end{center} 
\caption{\label{fig:gyromotion} Schematic view of the Earth's
  magnetosphere and the van Allen belts, as distorted by the solar
  wind {\sl (left)}; trajectory of a charged particle in the
  geomagnetic field {\sl (right)} \cite{russell91a}.}
\end{figure}

Immersed in the heliospheric particle streams is the magnetosphere of
the Earth. The rather complex magnetic field generated by the Earth's
body {\sl (internal field)\/} is commonly treated as the derivative of
a scalar potential, $\mathbf{B}=-\mbox{grad}\,V$, with $V$ expanded in
terms of spherical harmonics,
\begin{equation}\label{equation:sphericalharmonics}
V=a\sum_{n=1}^{\infty}\sum_{m=0}^{n}\left( g_n^m \cos m\phi + h_n^m
\sin m\phi\right)\left( \frac{a}{r}\right)^{n+1} P_n^m \left( \cos\theta\right)\, .
\end{equation}
Here, $a$ is the mean radius of the Earth, $(r,\phi,\theta)$ are
geocentric coordinates, $P_n^m \left( \cos\theta\right)$ are the
associated Legendre functions, and $\left(g_n^m, h_n^m\right)$ are the
Gauss coefficients describing the field contribution of the separate
terms in the sum~\cite{olsen00a}. Additional terms may also
incorporate external magnetic field sources, such as particle streams
generated by interactions with the solar wind. The dominant terms in
eq.~(\ref{equation:sphericalharmonics}) are related to $n=1$, i.e. the
geomagnetic field is in first order that of a dipole, tilted by about
168.5$^{\circ}$ with respect to the rotation axis of the
Earth\footnote{The geomagnetic north pole lies in the southern
  geographic hemisphere.} and displaced approximately 400\+km from its
center. The field strength on the Earth's surface varies between 20
$\mu$T and 70 $\mu$T. As shown in Figure~\ref{fig:gyromotion} {\sl
  (left)}, at distances of a few Earth radii, the geomagnetic field is
distorted by the solar wind, a highly dynamic process which gives rise
to a shock front {\sl (bow shock)} along the border between
interplanetary magnetic fields and the Earth's magnetosphere ({\sl
  magnetopause}).

Low energy charged particles can be trapped in the geomagnetic field.
The path which such a particle travels is schematically illustrated in
Figure~\ref{fig:gyromotion} {\sl (right)}. The basic trajectory is a
helix following the field lines, with an increasing radius of
curvature of its gyromotion ({\sl gyroradius}) as the particle moves
further away from the Earth. Near the Earth's surface, where the field
strength is larger, the gyroradius decreases. The particle's energy
then becomes dominated by the gyromotion contribution at the expense
of the forward motion along the field line. Consequently, the particle
is deflected back along its path, only to repeat the process at the
other end. As the trapped particle bounces back and forth along the
magnetic field, it also drifts around the Earth because, in part, its
gyroradius is larger on the outer part of its trajectory than on the
inner part. This effect causes negatively charged particles to drift
in the direction of the Earth's rotation (eastward) and positively
charged ones to drift in the opposite direction~\cite{russell91a}.

It follows that, in the dipole approximation, the region which can be
occupied by trapped particles forms a nearly toroid-shaped volume
symmetrical about the dipole axis. The {\sl Van Allen radiation belts}
surround the Earth at a distance of roughly 1.5 and 4-5 Earth radii
from its center~\cite{li01a}. While the inner belt mostly consists of
protons with energies larger than 100\+\MeV{} and is relatively stable,
the outer belt is largely populated with electrons ranging in energy
from 400\+\keV{} to above 15\+\MeV{}, and dynamically deformed by the solar
wind. Protons from the inner belt can interact with atmospheric
molecules, giving rise to secondary particles {\sl (atmospheric
  secondaries)} such as, amongst others, positrons.

\begin{figure}[ht]
\begin{center}
\includegraphics[width=12cm]{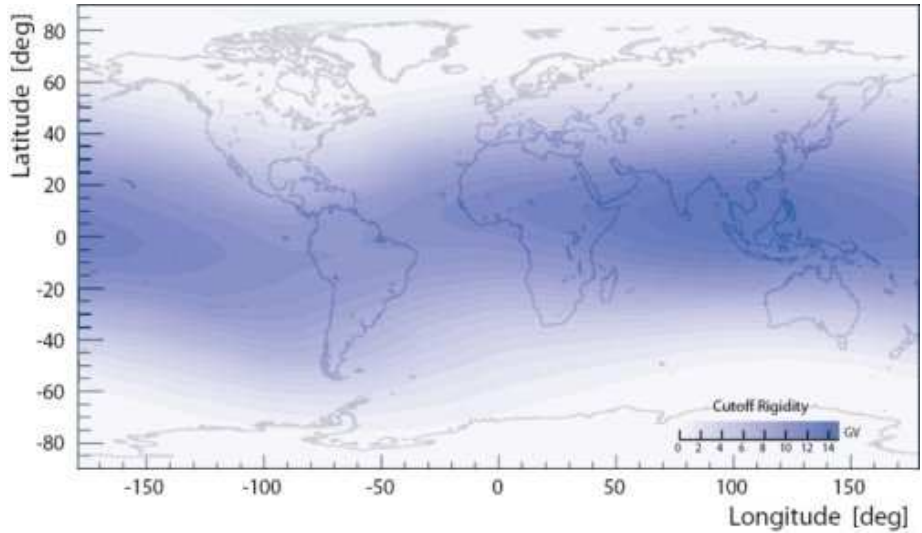}
\caption{\label{fig:cutoff}The vertical cutoff rigidity at an altitude of 450\+km under magnetically quiet conditions. 
Based on data from \cite{smart99a}.}
\end{center} 
\end{figure}
The ability of a cosmic ray particle to penetrate the Earth's
magnetosphere is uniquely determined by its {\sl rigidity}, the
momentum divided by the charge, given in units of volts. Cosmic ray
particles with rigidities below a certain {\sl geomagnetic cutoff} are
deflected by the geomagnetic field and cannot reach the vicinity of
the Earth. The value of this cutoff not only depends on the
geomagnetic coordinates of the particle's entry point into the
magnetosphere, but also on its incident direction.

Figure~\ref{fig:cutoff} shows the cutoff rigidity for vertical incidence
at an altitude of 450\+km above the Earth's surface under magnetically
quiet conditions \cite{smart99a}. It ranges from a few MV in the
polar regions up to approximately 15\+GV near the equator. Similar to
the solar modulation effect, the geomagnetic cutoff modulates the
spectra of cosmic rays having rigidities in this range. Particles
detected in an experiment with rigidities below their cutoff cannot be
of galactic origin, but must stem from within the magnetosphere. They
are mostly produced as atmospheric secondaries and are trapped in the
Earth's radiation belts.

\section{Signatures of SUSY Dark Matter in the Cosmic Ray Spectra}\label{sec:susyspectra} 
As stated already in sec.~\ref{section:susy}, neutralinos as Majorana
particles could annihilate pairwise into SM particles, which then
propagate as cosmic rays and are thus accessible to experiments. This
process would constitute an additional primary source of cosmic rays
with a unique spectral shape, and may thus appear as an excess of
particles over conventional expectations. However, since the
corresponding source strength is assumed to be small compared to that
of supernovae, the dark matter annihilation signal should be
sufficiently significant only in the spectra of cosmic ray particles
which are exclusively produced through secondary processes at low
rates, such as positrons, antiprotons or gamma rays (see \S\+\ref{sec:originAndProp}).

\begin{figure}[t]
\begin{center}
\includegraphics[width=14cm]{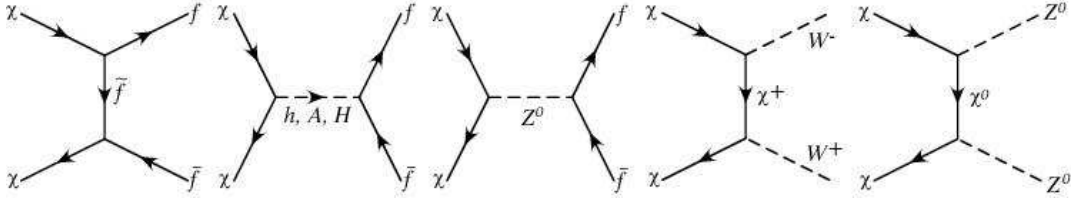}
\caption{\label{fig:wimpAnnFeynman}The dominant annihilation channels for the 
lightest neutralino $\chi$~\cite{deboer06a}.}
\end{center} 
\end{figure}

Figure~\ref{fig:wimpAnnFeynman} shows the Feynman diagrams for the
dominant neutralino annihilation channels. The process can proceed
through the t-channel exchange of sfermions, charginos or neutralinos,
or through s-channel annihilation via $Z^{0}$ or pseudoscalar Higgs
bosons. The final states are either fermions or gauge bosons. The
total cross section for neutralino annihilation strongly depends on
the individual channel amplitudes, and thus on the choice of
supersymmetric parameters. For example, in the case of fermion final
state diagrams, the amplitude is proportional to the fermion mass in
the low velocity limit and inversely proportional to the squared mass
of the sfermion $\tilde{f}$. Furthermore, the amplitude for Higgs
exchange is proportional to $\tan\beta$ for down-type
quarks~\cite{deboer06a}. This implies that at values of
$\tan\beta\gtrsim 5$ neutralino annihilation is dominated by s-channel
Higgs exchange with $b\bar{b}$ pairs as the final state.

\begin{figure}[b]
\begin{center}
\includegraphics[width=14cm]{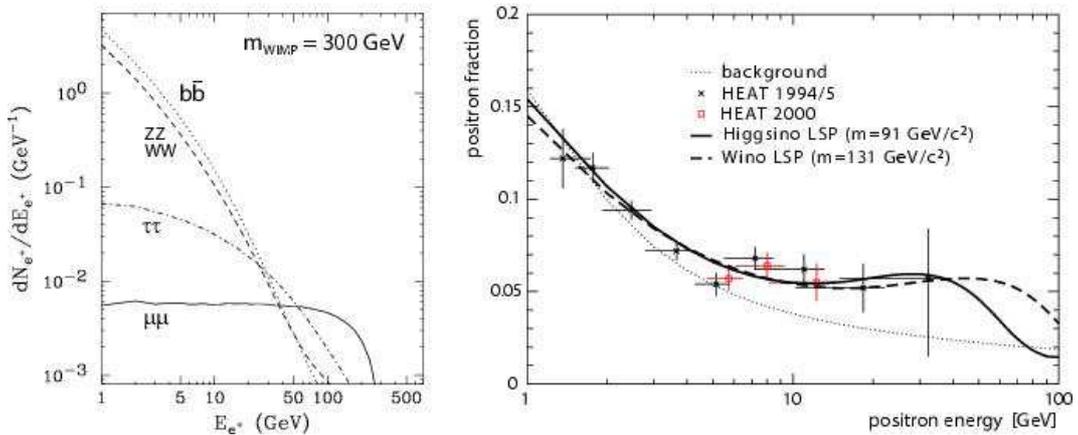}
\caption{\label{fig:positronSignals}The positron spectrum from
  neutralino annihilations, prior to propagation, for selected
  annihilation modes and a neutralino mass of
  300\+\GeV{}/c$^2$~\cite{hooper05a} {\sl (left)}; positron fraction data
  and simulation results for purely secondary positron production
  (background) and with additional positrons from neutralino
  annihilation for two mSUGRA parameter sets, resulting in either a
  higgsino- or wino-like neutralino~\cite{kane02a} {\sl (right)}.}
\end{center} 
\end{figure}
Subsequent to their generation, $b$-quarks hadronize and give rise to
decay cascades in which, amongst others, positrons and antiprotons are
produced, as well as typically 30--40 photons per
annihilation~\cite{deboer05a}. Additional positrons originate from the
decay of gauge bosons and of $\tau$ and $\mu$. Direct annihilation
into electron positron pairs is helicity suppressed~\cite{jungman96a}.
As an example, Figure~\ref{fig:positronSignals} {\sl (left)} shows the
raw spectra of positrons (prior to their propagation through the
interstellar medium) from the different annihilation modes of
neutralinos with a mass of 300\+\GeV{}/c$^2$. The spectral shapes differ
significantly, since annihilations to heavy quarks and gauge bosons
produce a much softer spectrum than the $\tau$ or $\mu$ channels. The
shape of the overall resulting spectrum is a mixture of the above and
is determined by the cross section of the individual modes, which
depend on the supersymmetric parameters. The positrons then propagate
through the galactic volume and experience energy loss and deflection
due to scattering and interactions with magnetic fields, which finally
result in a smearing of the initial raw spectrum when observed near
Earth.

Quark hadronization functions and branching ratios of the decay modes
involved in positron production from neutralino annihilations are
known to a high degree of precision~\cite{yao06a}. Using models of
particle propagation in the interstellar medium, the spectra of
secondary particles near Earth can be simulated and compared to flux
measurements. Figure~\ref{fig:positronSignals} {\sl (right)} displays
the positron fraction as a function of the particle energy together
with data from HEAT-$e^{\pm}$~\cite{duvernois01a} and
HEAT-pbar~\cite{coutu01a}. When compared to the background
expectation, the data show an indication for an excess number of
positrons at energies above 6\+\GeV{}. The solid and dashed lines
represent simulations of the positron fraction including contributions
from neutralino annihilation for two particular sets of supersymmetric
parameters~\cite{kane02a}. The spectral shape differs mainly due to
the choice of neutralino mass. In the one case, the neutralino with a
mass of 91\+\GeV{}/$c^2$ is {\sl Higgsino-like\/}, meaning that in the
linear combination of eq.~(\ref{eq:neutralinoComposition}) the
coefficients $c_{31}$ and $c_{41}$ are large compared to the bino and
wino contributions. In the other scenario, the 131\+\GeV{}/c$^2$
neutralino is {\sl wino-like\/} with a dominant contribution from
$c_{21}$. In contrast to the {\sl bino-like\/} neutralino, which
largely annihilates to heavy fermions, gauge boson final states are
preferred in the two cases above~\cite{hooper05a}.

\begin{figure}[b]
\begin{center}
\includegraphics[width=14cm]{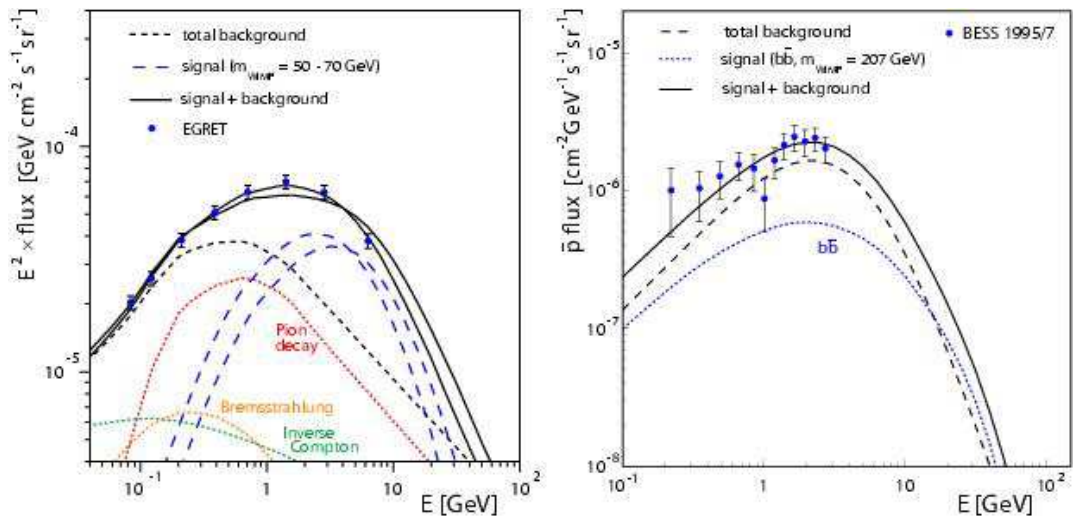}
\caption{\label{fig:gammaAndAntiprotons}Data of the diffuse gamma ray
  energy spectrum fitted with the simulated background from different
  processes and an annihilation signal from a 50--70\+\GeV{}/c$^2$
  neutralino {\sl (left)}~\cite{deboer06a}; same for antiprotons with
  an assumed neutralino mass of 207\+\GeV{}/c$^2$ {\sl
    (right)}~\cite{deboer04a}.}
\end{center} 
\end{figure}
Both results involving dark matter annihilation clearly improve the
agreement with the data compared to the background-only prediction.
Clearly, in order to further constrain the mass range of the
neutralino, better data is needed, especially at higher energies in
order to capture the trailing edge of the dark matter signal. The
simulated positron flux resulting from annihilation is almost always
too small to produce a visible signal. However, it may be increased by
a {\sl boost factor\/} $B$, which can be explained by assuming a
clumpy nature of the dark matter distribution. In particular, since
the neutralino annihilation rate is proportional to the square of its
density, a signal enhancement is expected in high density regions like
the center of the Galaxy. The boost factor is thus proportional to the
mean deviation of the squared density $\rho$, $B\propto
\left<\rho^2\right>/\left<\rho\right>^2$.

Figure~\ref{fig:gammaAndAntiprotons} shows the simulated background
spectra of gamma rays and antiprotons compared to experimental
results. In both cases the background alone turns out to be
inconsistent with the measured fluxes, so that an additional source of
particles is assumed to fit the data. In the gamma ray spectrum on the
left side of Figure~\ref{fig:gammaAndAntiprotons}, a signal from the
annihilation of neutralinos with masses of 50--70\+\GeV{}/c$^2$ is found
to explain the observed excess well~\cite{deboer06a}. The right side of
Figure~\ref{fig:gammaAndAntiprotons} shows the measured spectrum of
antiprotons, which can again be fitted with the assumption of
neutralino annihilation, however with a significantly different
neutralino mass of 207\+\GeV{}/c$^2$~\cite{deboer04a}.

Despite the hints for the presence of a dark matter signal in the
spectra of secondary particles, the neutralino hypothesis remains
controversial. It has been pointed out recently that the excess of
gamma rays, if explained as a signal of neutralino annihilation as
shown in Figure~\ref{fig:gammaAndAntiprotons} {\sl (left)\/}, may not
be compatible with the observed abundance of
antiprotons~\cite{bergstrom06a}. Other possible contributions to the
cosmic ray positron flux have been proposed, such as electron-positron
pairs from annihilations of hypothetical Kaluza-Klein particle
states~\cite{hooper05a}, or from conversion of synchroton photons
emitted by galactic pulsars~\cite{zhang01a}. Consequently, more data
is necessary to allow further development of these models. While new
experiments are yet under construction, one possible approach is the
reanalysis of data from past experiments such as \AMS{} with new
techniques. The description of the AMS-01 detector is subject of the
next chapter.

\chapter{The AMS-01 Experiment}
\section{AMS-01 Detector Construction}
As a prototype for the \mbox{AMS-02} experiment, the \AMS{} detector was
flown on board the space shuttle {\sl Discovery} in a near Earth orbit
during the {\sl STS-91} mission from June 2nd to 12th, 1998. Its main
purpose was to verify the feasibility of engineering a space borne
high-energy particle detector, launching it into orbit, and operating
it safely under space environment conditions. Furthermore, the main
design principles of the future AMS-02 experiment were tested. During
its flight, \AMS{} has recorded a large amount of data for the
determination of particle fluxes in the Earth's vicinity, leading to
precise proton~\cite{alcaraz00a,alcaraz00b} and
helium~\cite{alcaraz00c} flux spectra, measurements of the
characteristics of leptons in near earth orbit~\cite{alcaraz00d}, and
to a significant improvement of the upper limit on the relative flux
of antihelium to helium~\cite{alcaraz99a} in cosmic rays.
\begin{figure}[!h]
\begin{center}
\begin{tabular}{lr}
\begin{minipage}{7cm}
\begin{center}
\includegraphics[width=7cm]{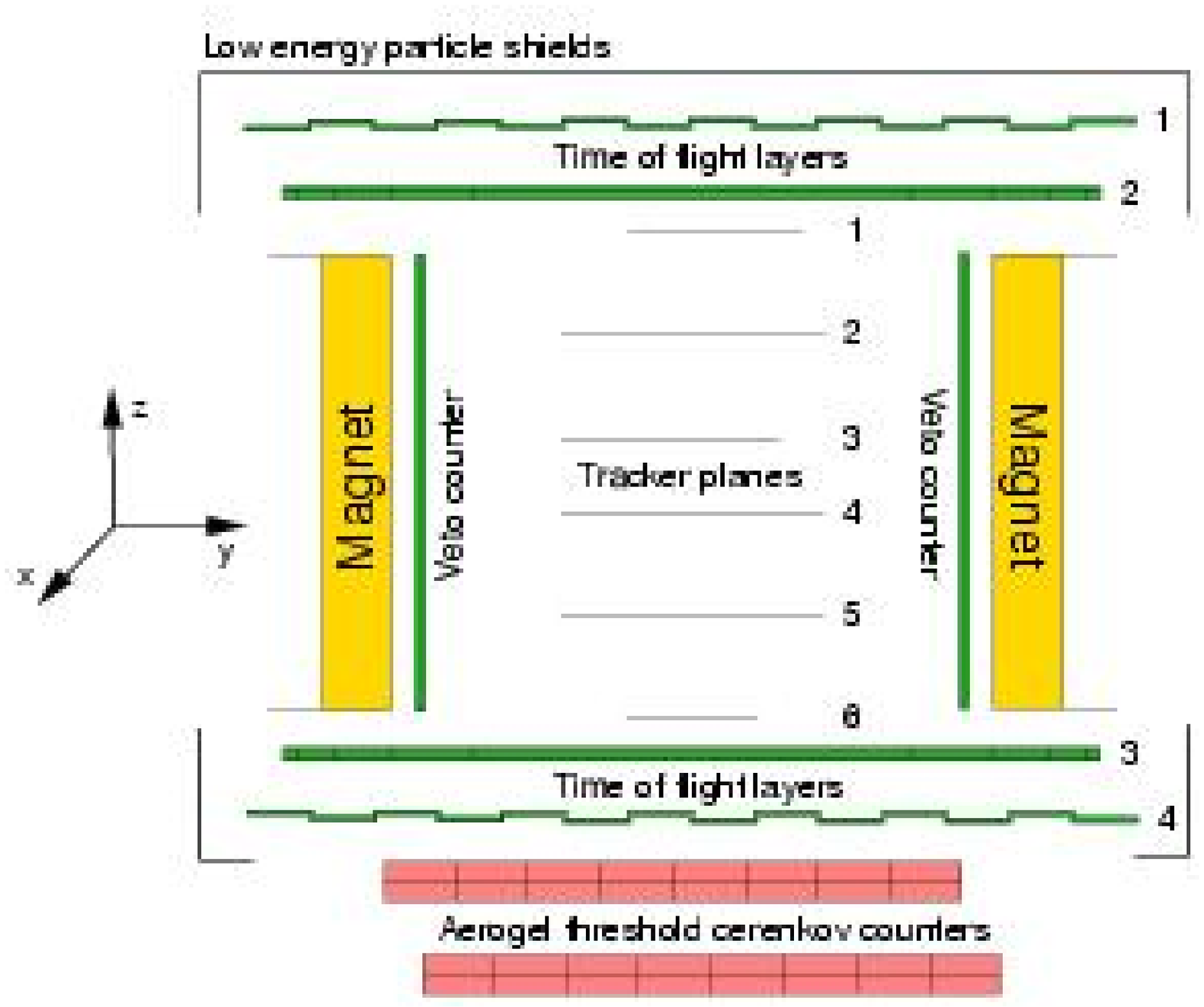}
\end{center}
\end{minipage}
&
\begin{minipage}{8cm}
\begin{center}
\includegraphics[width=8cm]{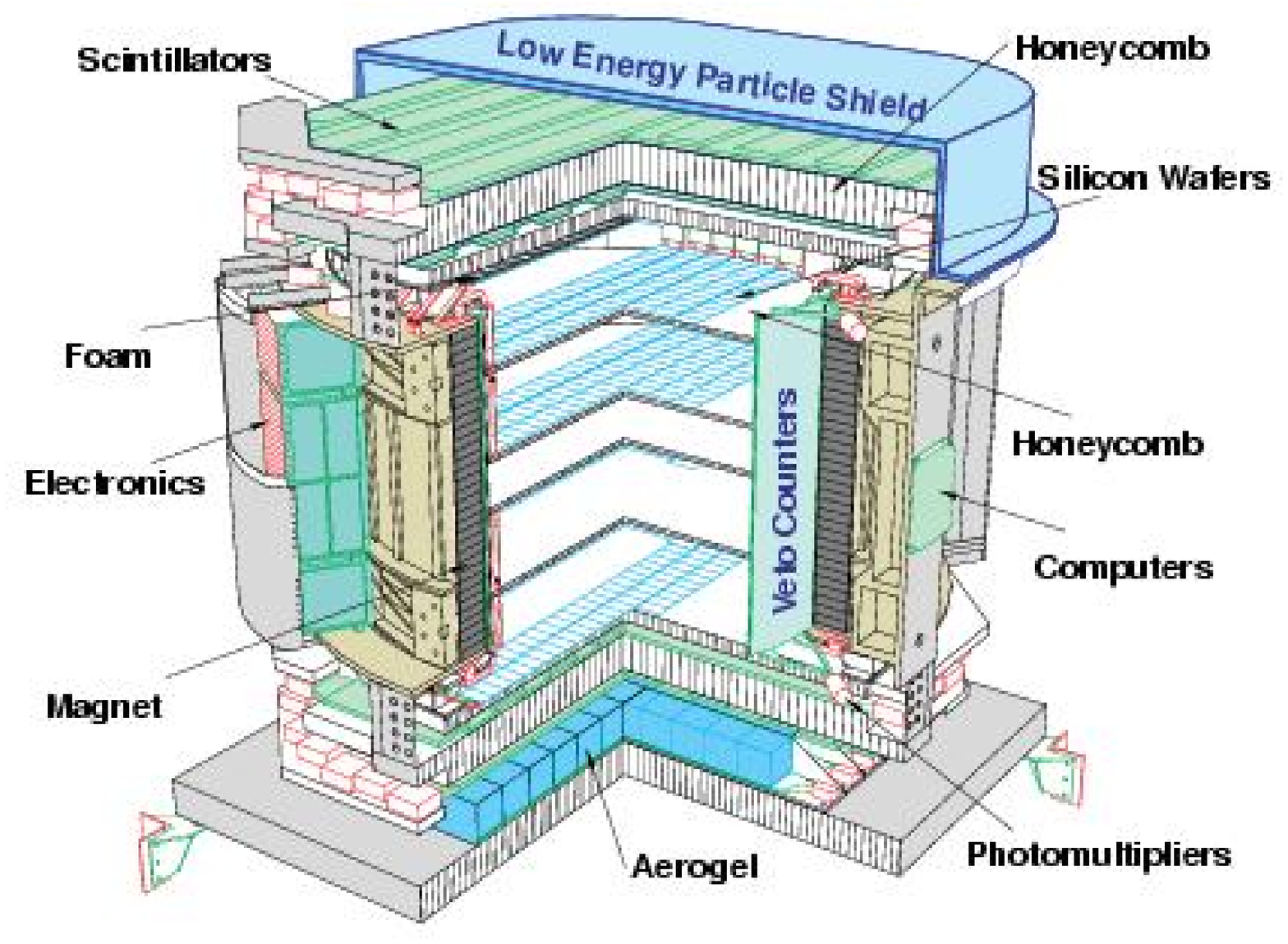}
\end{center}
\end{minipage}
\end{tabular} 
\end{center} 
\caption{\label{fig:ams:3d}The \AMS{} detector: schematic cross section in the y-z plane {\sl (left)\/} and
three dimensional illustration {\sl (right)\/}. }
\end{figure}

The structure of \AMS{} is illustrated in
Figure~\protect\ref{fig:ams:3d}. Six layers of double sided silicon
strip sensors enclosed in a cylindrical permanent magnet formed the
experiment's tracking chamber. The particle flight direction and
velocity was measured with plastic scintillator paddles arranged in
two double planes, one above and one below the magnet (time of flight
hodoscopes, TOF). Additionally, the inner surface of the magnet was
covered with scintillators to reject background from particles which
traverse the magnet wall or the support structure (anticoincidence or
veto counters). As a complement to the velocity measurement with the
hodoscopes, two layers of threshold \u Cerenkov counters were situated
at the lower end of the detector. In order to keep very low energy
particles (up to several \MeV{}) from entering, \AMS{} was shielded by
thin carbon fiber walls with a thickness of 10\+mm above or 6\+mm
underneath the experiment (low energy particle shield, LEPS). The top
shielding was covered with a blanket of Beta
cloth\protect\footnote{Beta cloth is a fabric woven from fine quartz
  threads and impregnated with teflon or silicone based
  materials~\cite{nasa99b}.} and a Nomex/Aluminum/Mylar composite with
a thickness of approximately 1.7\+mm. By these means \AMS{} was
protected against mechanical damage and excessive warming from
exposure to direct sunlight.

Figure~\ref{fig:ams:3d} {\sl (left)\/} shows the definition of the
\AMS{} coordinate system. Its origin lies in the center of the magnet
bore. The z-axis is perpendicular to the tracker and TOF planes and
points upward, with the \u Cerenkov counter at negative values of z.
The x-axis points into the direction of the magnetic field's dominant
component\protect\footnote{Details about the connection of the \AMS{}
  coordinate system to the Shuttle and the Earth frames of reference
  are given in~\cite{suter98a}.}.

\subsection{The Magnet}
\begin{figure}[!h]
\begin{center}
\includegraphics[width=12cm]{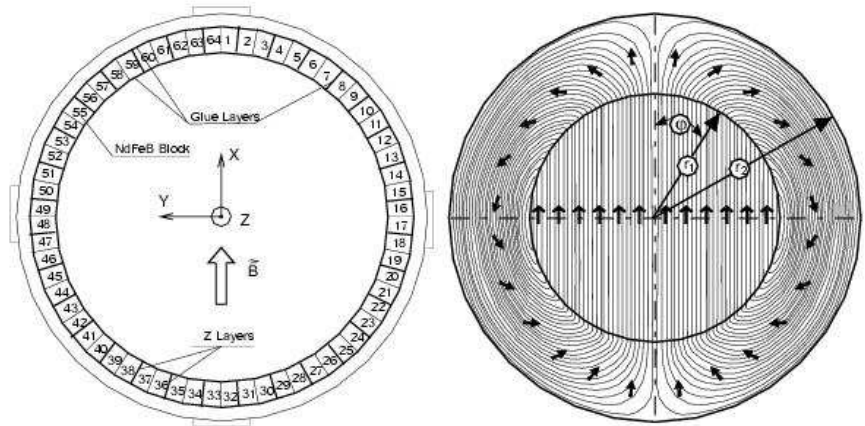}
\end{center} 
\caption{\label{fig:magnet:blocks}x-y cross section through the \AMS{} permanent magnet: arrangement of block sectors and AMS coordinate 
system definition {\sl (left)\/}; magnetization of blocks and resulting field lines {\sl (right)\/}. }
\end{figure}

A permanent magnet with a nominal bending power of $BL^2 =
0.14$\+Tm$^2$ was chosen for \AMS{} because of its compact and rigid
structure and the low amount of maintenance and operating expenses.
The magnet had a cylindrical shape with an inner radius of 558\+mm,
95\+mm wall thickness and a length of 800\+mm along its cylinder axis,
resulting in a free area of 0.98\+m$^2$ and a geometrical acceptance of
0.82\+m$^2$sr. It consisted of 6400 blocks of \mbox{Nd-Fe-B} alloy,
sized $2\times2\times1$ inches each, arranged in 64 sectors. The
blocks were magnetized in such a way that the resulting dipole field
was to first order perpendicular to the longitudinal axis (z-axis) and
parallel to the positive x-axis as shown in
Figure~\ref{fig:magnet:blocks}.

\begin{figure}[!h]
\begin{center}
\begin{tabular}{ll}
\begin{minipage}{7cm}
\includegraphics[width=7cm]{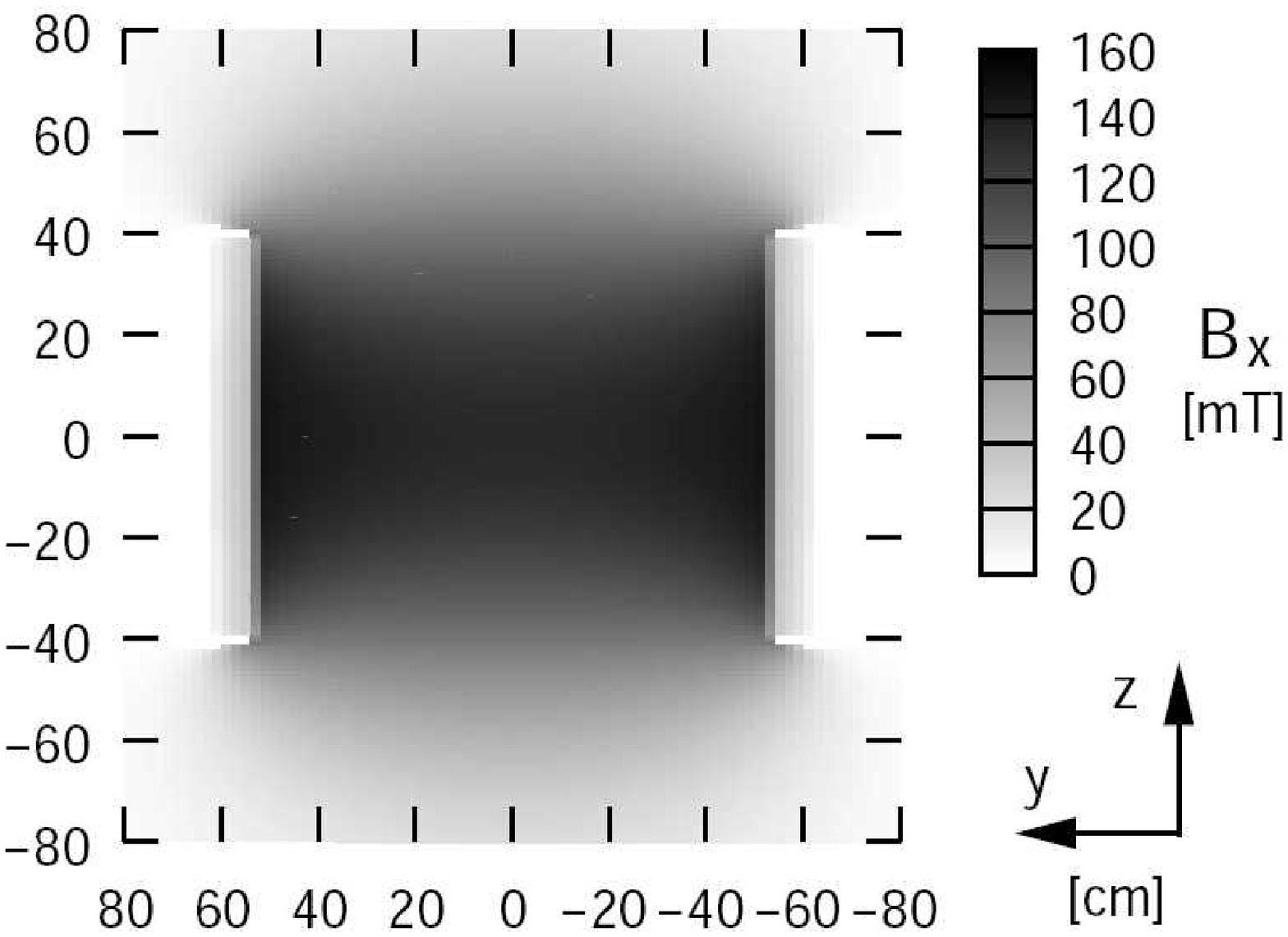}
\end{minipage}
&
\begin{minipage}{7cm}
\begin{center}
\includegraphics[width=7cm]{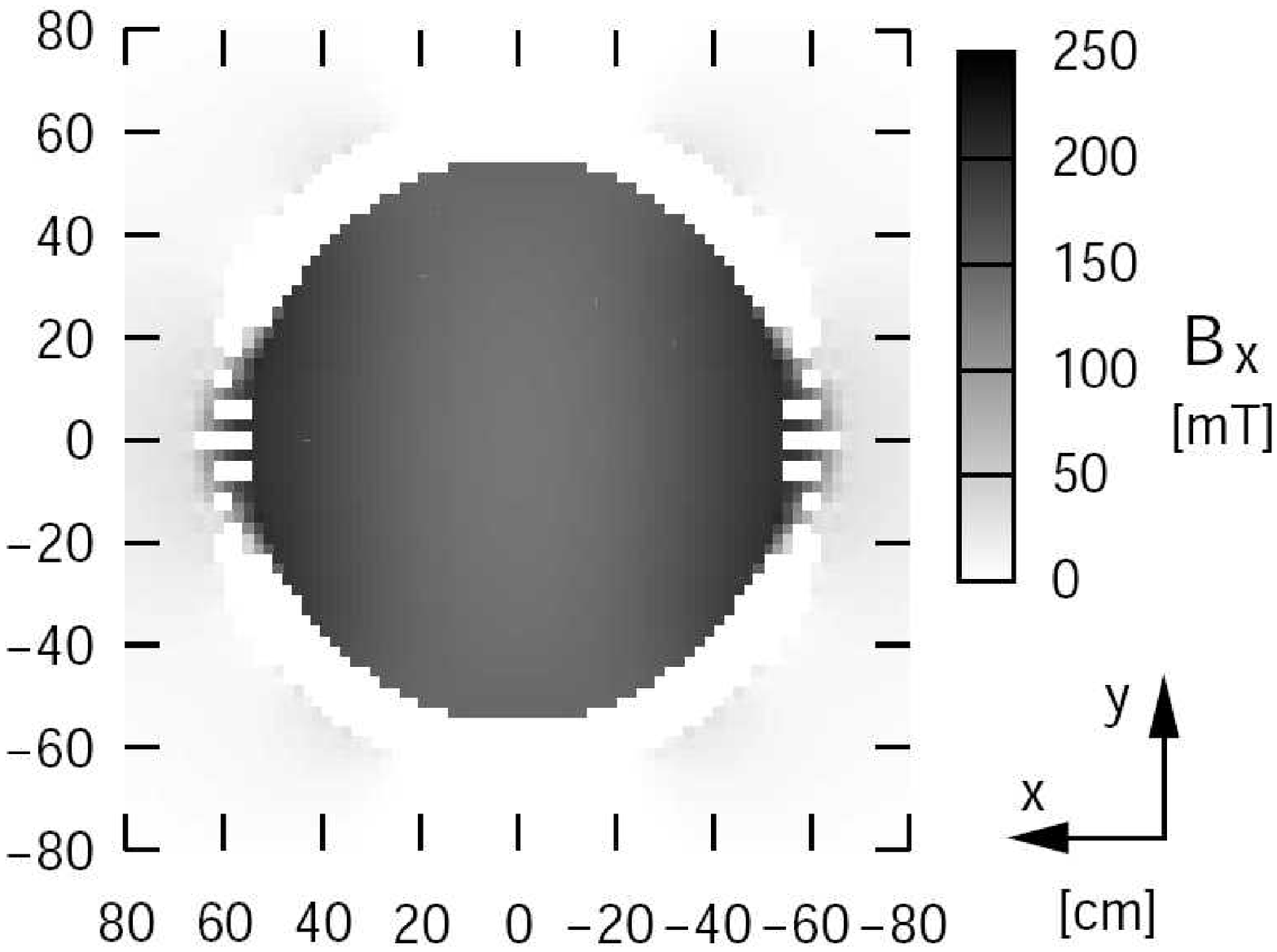}
\end{center}
\end{minipage}
\end{tabular} 
\end{center} 
\caption{\label{fig:magnet:field} The strength of the x-component
  $B_x$ of the magnetic field: vertical cross section at $x=0$ ({\sl
    left}) and horizontal cross section at $z=0$ ({\sl right}). White
  areas denote negative values.}
\end{figure}

The magnetic field was not homogenous. As seen in
Figure~\ref{fig:magnet:field} {\sl (left)\/}, its dominant component
$B_x$ reached a flat maximum of 0.14\+T at the center of the magnet
and dropped quickly with increasing values of $\left| z \right|$. At
the upper and lower edges of the magnet its strength decreased to
about 64\+\% of its maximum, while it was down to 14.9\+\% at $z=83$\+cm,
the upper end of the detector.  Furthermore, $B_x$ was not constant
within a cross section in the x-y-plane (Figure~\ref{fig:magnet:field},
{\sl right\/}), but grew with increasing values of $\left| x \right|$
up to approximately 250\+mT. The magnet had very small flux leakage
outside its volume and a vanishing dipole moment, since the latter, in
combination with the geomagnetic field, would have implied a torque on
the spacecraft. The strength of the magnetic field as a function of
the space coordinates was measured with high accuracy and is available
for the offline analysis of \AMS{} data.

\subsection{The Time of Flight Hodoscopes}
The time of flight scintillation counter hodoscopes (TOF) of \AMS{} basically had three different tasks:
\begin{itemize}\addtolength{\itemsep}{-3mm}
\item delivering a fast trigger signal to the data acquisition
  system,
\item measuring the value and direction of an incident particle's
    velocity,
\item determining the absolute charge of a particle from its
  energy loss in the scintillators.
\end{itemize}
The TOF system consisted of four planes of Bicron BC-408 scintillator
paddles, each plane covering a total area of roughly 1.6\+m$^2$. The
planes were grouped in two stations, i.e. double layers of
scintillator paddles oriented perpendicularly to each other, allowing
two space points to be measured on the particle trajectory, one above
and one below the magnet. The distance of two paired planes was
12\+cm, the distance from the upper to the lower station approximately
130\+cm.  Thus a particle with $\beta \approx 1$ traverses the time of
flight system in about 4.3\+ns.

One TOF plane was composed of 14 rectangular scintillator panels, 1\+cm
thick and 11\+cm wide with a length between 72 and 136\+cm to account
for the circular shape of the plane. In order to avoid dead space,
adjacent paddles were mounted with 0.5\+cm overlap. Each paddle was
wrapped in aluminized mylar foil and encased in a 0.6\+mm thick carbon
fiber housing. Each scintillator double layer was supported by a 10\+cm
thick aluminum honeycomb panel glued between two aluminum skins of 5\+mm
thickness~\cite{alvisi99a}.
\begin{figure}[!h]
\begin{center}
\includegraphics[width=10cm]{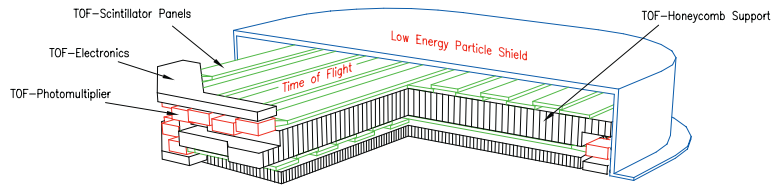}
\end{center} 
\caption{\label{fig:tof:scheme} Schematic view of the upper TOF planes and the LEPS.}
\end{figure}

A total of six photomultipliers (Hamamatsu R5900), grouped in two
triplets, was used to read out a single paddle. Both sides of the
paddles were terminated with one triplet of photomultipliers via 5\+cm
long trapezoidal light guides. The anode signals from a triplet, as
well as those from the next to last dynodes, were summed to form
two output signals for each side of a paddle.

Prior to the launch of the experiment the time resolution of a single
scintillator plane was determined from cosmic ray
tests~\cite{alvisi99a} to be 126\+ps, resulting in an error of 178\+ps
on the total flight time. The position of passage of a particle
through the scintillators can be determined with an accuracy of
1.8\+cm along a panel.

The charge measurement is performed using the time information, based
on a time-over-threshold method. Though not optimized for this task,
the TOF system was thus capable of distinguishing particles with charge
$|Z|=e$, $|Z|=2e$ and $|Z|>2e$ with a purity better than 1\+\%
~\cite{baldini01a}.

\subsection{The Silicon Strip Tracker}\label{section:tracker}
For the measurement of position and energy loss of particles \AMS{}
featured a high quality tracking device based on silicon microstrip
sensor technology~\cite{alcaraz99b}. It was designed to provide a
position resolution of 10\+$\mu$m in the bending plane and 30\+$\mu$m in
the non-bending plane of the magnetic field.
\begin{figure}[!ht]
\begin{center}
\begin{tabular}{ll}
\begin{minipage}{8cm}
\includegraphics[width=8cm]{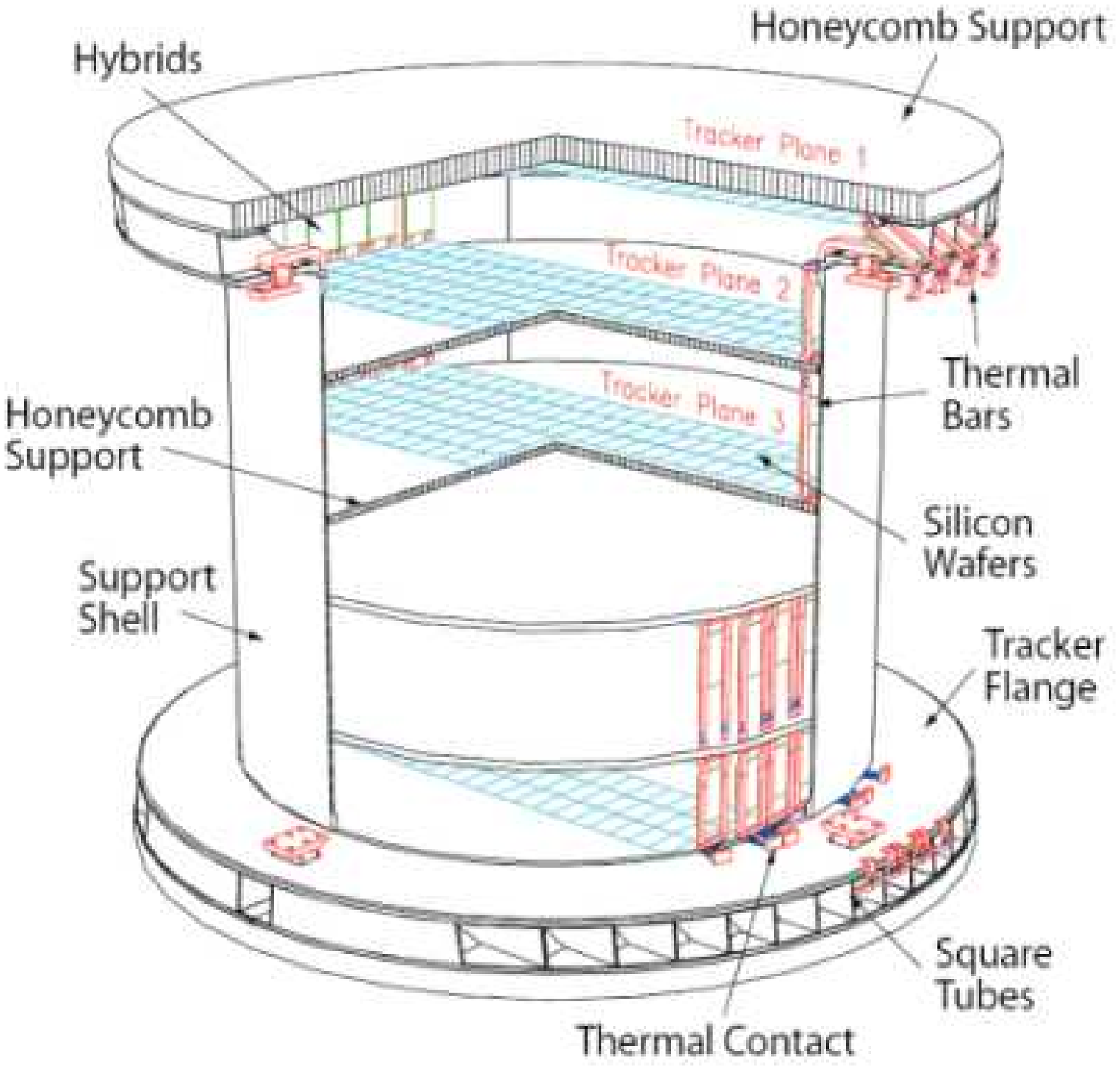}
\end{minipage}
&
\begin{minipage}{7cm}
\begin{center}
\includegraphics[width=7cm]{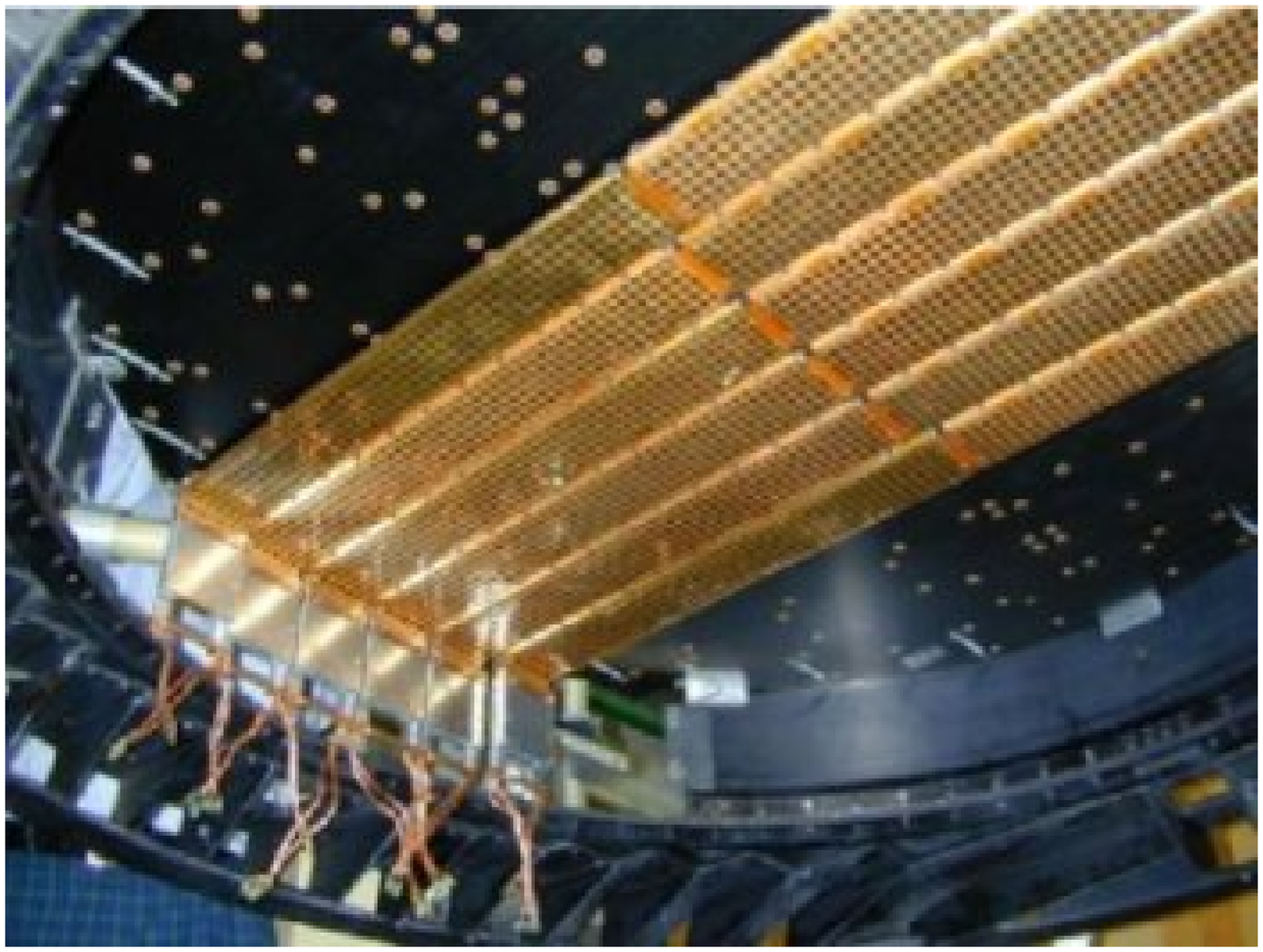}
\end{center}
\end{minipage}
\end{tabular} 
\end{center} 
\caption{\label{fig:tracker:scheme} Schematic view of the silicon strip 
tracker and its support structure ({\sl left}); picture of the
fifth tracker layer during detector disassembly ({\sl right}).}
\end{figure}
\subsubsection{Tracker Layout and Electronics}
The layout of the tracker is displayed in
Figure~\ref{fig:tracker:scheme} ({\sl left}). The silicon was arranged
in 6 planes roughly 20\+cm apart (see
Table~\ref{table:trackerlayers}). Four of the planes were located
inside the magnet bore, enclosed in a carbon fiber cylindrical shell.
Each plane was supported by a 12\+mm thick aluminum honeycomb disk
dressed with 220\+$\mu$m carbon fiber skins.  Above and below the
magnet, two additional layers were mounted on stiffer disks (4\+cm
thick with 0.7\+mm cover) for mechanical stability.
\begin{table}[!h]
\begin{center}
\begin{tabular}{c|cccccc}
\hline
Layer no.&1&2&3&4&5&6\\
\hline
z position [cm] & 50.98 & 29.19 & 7.79 & -7.82 & -29.18 & -50.98 \\
\hline
\end{tabular}
\caption{\label{table:trackerlayers} The z position of the tracker layers
in the AMS coordinate system.}
\end{center}
\end{table}

The basic tracker elements were the 300\+$\mu$m thick silicon
microstrip sensors of size 40.14~$\times$~72.04\+mm$^2$. The sensors
were double sided with parallel strips implanted in the n$^+$ bulk
silicon on both sides. Strips on opposite sensor sides were
perpendicular to each other~\cite{burger02a}. On the
p-\mbox{(junction-)side}, which measures the y-coordinate, the
implants had a pitch of 27.5\+$\mu$m with every second implant covered
with an aluminum readout strip on the surface. On the ohmic n-side the
implantation pitch was 26\+$\mu$m with every implant aluminized. Every
fourth implant was read out on both sides, resulting in a readout
pitch of 110\+$\mu$m for the p-side and 208\+$\mu$m for the n-side.

\begin{figure}[!h]
\begin{center}
\includegraphics[width=12cm]{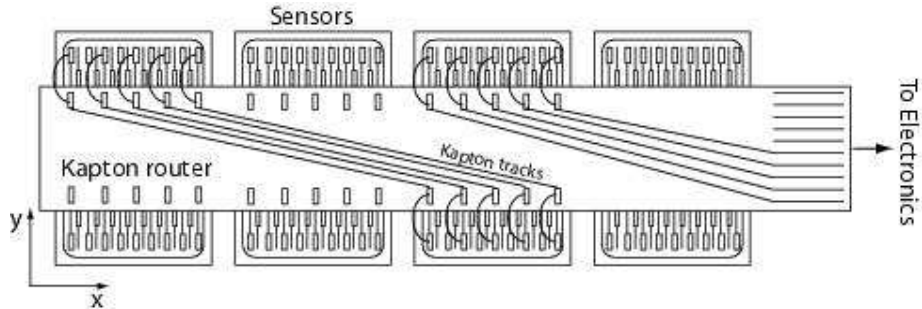}
\end{center} 
\caption{\label{fig:tracker:bonding} The principle of n-side strip connection
to the frontend electronics as a cause for ambiguities in the hit
reconstruction. Connection via one kapton track group only is
displayed.}
\end{figure}

Between 7 and 15 sensors were arranged in linear structures, the so
called {\sl ladders}, of different length to account for the overall
circular shape of the layers. Within a ladder, the p-side readout
strips of all sensors were daisy chained with wirebonds and connected
to the p-side frontend hybrid. The n-side strips were connected to the
n-side frontend hybrid through a metalized kapton foil, directly glued
to the sensors. The kapton foil incorporated 2 groups of 192 lines
each, with adjacent sensors connected to a different group. As a
consequence of this bonding scheme~\cite{ghete96a} (see
Figure~\ref{fig:tracker:bonding}) and depending on the number of
sensors in a ladder, between 4 and 8 strips on different sensors were
connected to the same readout channel, thus leading to an ambiguity in
the assignment of tracker hits to the n-side strips.
The two hybrids were attached back-to-back at each ladder's very end,
close to the magnet wall, and were perpendicular to the sensor planes.
They were equipped with 64-channel VA$_{\mathrm{hdr}}$ readout
chips~\cite{alpat00a}, which performed charge-sensitive signal
preamplification and shaping and also provided a sample-and-hold
stage, buffer and multiplexer for sequential analog data output. In
the onboard readout electronics, the output was digitized by fast
12-bit ADC after amplification. Further downstream, digital signal
processors performed pedestal subtraction, noise determination and
clustered the strip signals, so that compressed hit information was
transmitted to the data acquisition of the experiment (zero
suppression). For clustering, strips with signal amplitudes more than
three times above their noise level $\sigma_{\mathrm{ped}}$ were used
as seed strips and grouped with neighboring strips having amplitudes
above 1\+$\sigma_{\mathrm{ped}}$ to form clusters of at most 5
strips~\cite{alcaraz99b}. The cluster position was determined from the
center of gravity of the strip signal amplitudes.

As evident from Figure~\ref{fig:tracker:scheme} {\sl (right)\/}, the
tracker planes were not fully instrumented. From the full number of
1912 sensors, 778 were finally assembled~\cite{ionica03a}, covering a
total area of more than 2\+m$^2$. In this configuration, the tracker
had an acceptance of 0.31\+m$^2$sr for particles traversing four or
more layers of instrumentation~\cite{burger02a}.

\subsubsection{Tracker Characteristics and Performance}
\begin{figure}[!ht]
\begin{center}
\begin{tabular}{ll}
\begin{minipage}{7cm}
\includegraphics[width=6.6cm]{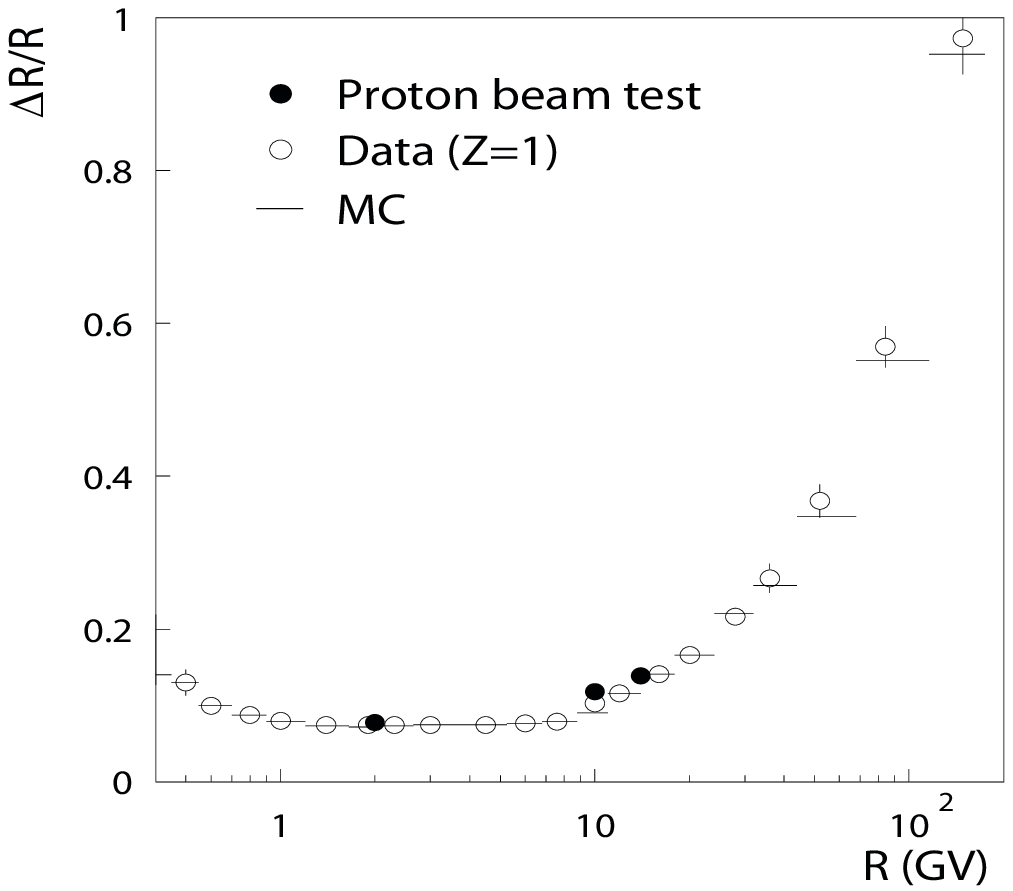}
\end{minipage}
&
\begin{minipage}{7cm}
\begin{center}
\includegraphics[width=6cm]{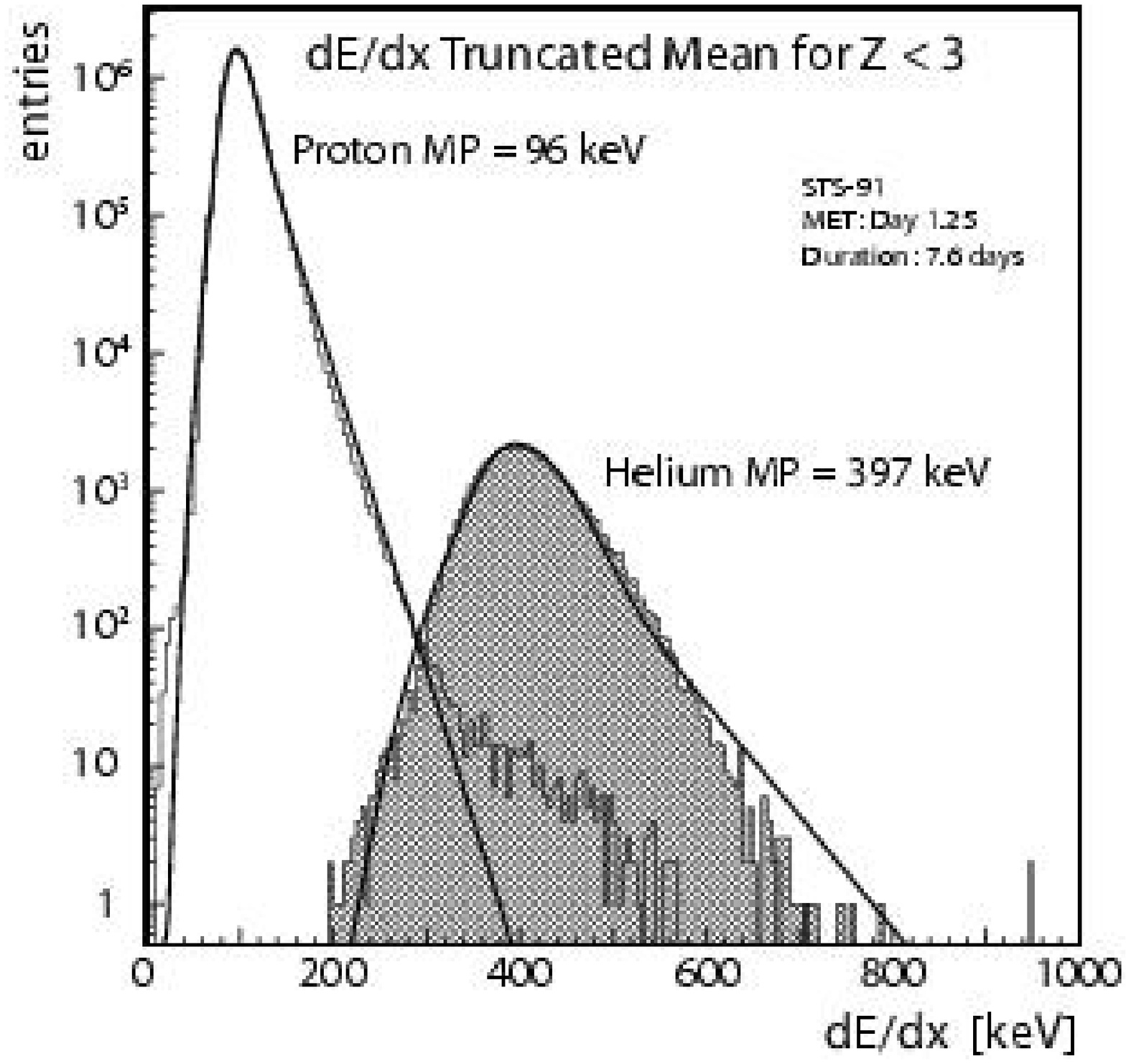}
\end{center}
\end{minipage}
\end{tabular} 
\end{center} 
\caption{\label{fig:tracker:performance} The tracker rigidity
  resolution as a function of rigidity~\cite{lamanna00a} determined
  for well defined tracks from flight and test beam data and compared
  with Monte Carlo simulations ({\sl left}); truncated mean of energy
  loss in the tracker planes~\cite{alcaraz99b} for protons and helium
  nuclei (hatched histogram) from data, based on p-side strip
  measurements ({\sl right}).}
\end{figure}

Using the framework of the offline software, the characteristics and
the performance of the tracker were determined with single-track
events from flight and from test beam data. Two different methods were
used for track fitting, based on an iterative Kalman filter approach
or on the numerical integration of the equation of motion of particles
in a magnetic field. For singly charged particles,
Figure~\ref{fig:tracker:performance} ({\sl left}) shows the rigidity
resolution $\Delta$R/R as a function of rigidity, for tracks with 4, 5
or 6 measured points from flight and CERN-PS test beam data, in
comparison with Monte Carlo results~\cite{lamanna00a}.  Multiple
scattering limits the rigidity resolution to a level of 9\+\% in the
low rigidity range up to approximately 10\+GV. For higher energies,
the resolution deteriorates rapidly as a consequence of the finite
magnetic field and the sensor spatial resolution.

On the right side of Figure~\ref{fig:tracker:performance}, the average
energy loss per tracker plane is displayed for protons and helium
nuclei from flight data~\cite{alcaraz99b}. The energy loss is derived
solely from p-side strip amplitudes and corrected for path length in
the silicon. Fits of a sum of Landau and Gauss functions to the
distributions give a most probable energy deposition of 96\+\keV{} for
protons and 397\+\keV{} for helium nuclei.

The detection efficiency for a single plane is determined by the
number of bad channels in the silicon, as well as by the fact that
adjacent ladders were mounted on the support structures without
overlap. The average total efficiency of the individual tracker planes
varied between 80\+\% and 85\+\%, depending on the signal to noise
ratio, which is a function of tracker temperature and thus of time.
The contribution from bad channels -- which may be due to high noise
level, defective readout electronics, or non-linear signal response --
sums up to approximately 12\+\% of the inefficiency.

During the Space Shuttle flight there were deviations of the tracker
support structure position from its nominal value in the order of
30\+$\mu$m due to temperature variations. These deformations were
monitored with an onboard infrared laser
system~\cite{vandenhirtz01a}. The tracker was finally aligned through
post-flight metrology and the analysis of high-rigidity tracks from
flight data~\cite{burger03a}.

\subsection{The Aerogel Threshold \v Cerenkov Detector}
Mounted underneath the lower time of flight scintillator layers, the
Aerogel Threshold \v Cerenkov Detector (ATC) provided separation of
antiprotons from electrons or protons from positrons for momenta below
3.5\+\GeV{}/c.
\begin{figure}[!h]
\begin{center}
\includegraphics[width=13cm]{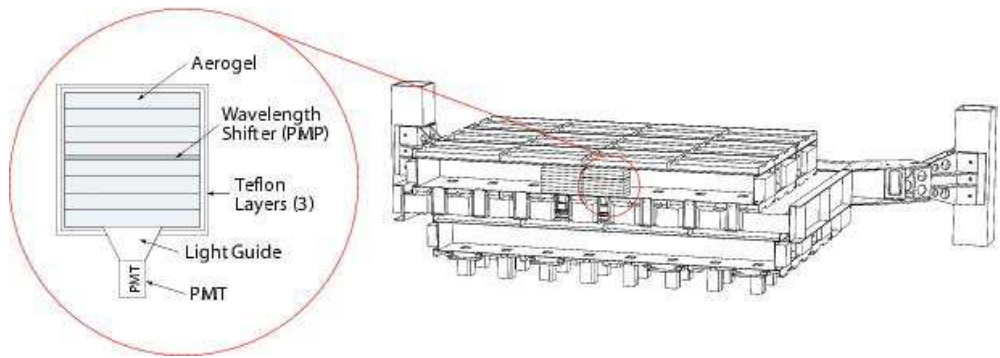}
\end{center} 
\caption{\label{fig:atc} Cross section of the Aerogel Threshold \v
  Cerenkov Detector and detail of the structure of an aerogel
  cell~\cite{barancourt01a}.}
\end{figure}

The ATC was composed of aerogel cells with a volume of
11$\times$11$\times$8.8\+cm$^3$, as shown in
Figure~\ref{fig:atc}. Each cell was filled with 8 slabs of 1.1\+cm
thick aerogel blocks and surrounded by 3 layers of teflon, each
250\+$\mu$m thick, to reflect \v Cerenkov photons back into the cell
volume. At the bottom end, a photomultiplier was attached to the cell
via a light guide. The efficiency of the photomultiplier reached a
maximum at wavelengths of approximately 420\+nm. In order to reduce
photon losses due to scattering and absorption, a 25\+$\mu$m
wavelength shifting layer was inserted between the aerogel slabs in
the middle of each cell.

The 168 aerogel cells were grouped into two layers of 8$\times$10 and
8$\times$11 cells respectively, and were enclosed in a thin carbon
fibre structure. Bolted to a 5\+cm thick honeycomb plate which was
glued into an aluminum frame, the layers were attached to the support
structure on which the \AMS{} experiment was mounted in the Space
Shuttle payload bay.

The data from the ATC are not used in this analysis, since they
provide no background suppression in the high momentum range above
3.5\+\GeV{}/c. A detailed description of the device is given
in~\cite{barancourt01a}.

\subsection{The Veto Counters}
The 16 veto or anti-coincidence counters (ACC) were mounted on the
inner magnet surface, thus surrounding the tracker. Their purpose was
to provide detection of particles which entered or exited the detector
volume through the magnet wall, thus typically indicating
multi-particle events involving $\delta$-ray creation or the presence
of low quality tracks. Each counter consisted of a 1\+cm thick
scintillator paddle 80\+cm high and 20\+cm wide with implanted
wavelength shifting fibers. At both ends of the paddles, the fibers
were viewed by a single photomultiplier.

\subsection{Material Budget}\label{sec:materialBudget}
Depending on direction of flight, particles had to traverse a certain
amount of material before they could reach the tracker. This fact has
important implications for this data analysis. Additional material
induces multiple scattering of particles and hence reduces the spatial
and angular resolution of reconstruction algorithms. On the other
hand, it increases the probability of processes such as bremsstrahlung
and gamma conversion, which are the focus of this work, and therefore
improves the reconstruction efficiency.
\begin{table}[!h]
\begin{center}
\begin{tabular}{lcl}
\hline
Component & \%$X_0$ & Ref. \STRUT \\ \hline
Thermal shielding & 3.2 & ~\cite{luebelsmeyer03a} \STRUT \\
LEPS & 3.9 & ~\cite{luebelsmeyer03a} \\ 
TOF double layer & 9.7 & ~\cite{luebelsmeyer03a} \\
TOF double layer support & 1.4 & ~\cite{luebelsmeyer03a} \\
Aerogel & 8.0 & ~\cite{henning03a} \\
\hline \\
\end{tabular}
\caption{\label{table:materialbudget} Contribution to the overall material budget
from the components of \AMS{} outside the tracker, in percent of a radiation length.}
\end{center}
\end{table}

Table~\ref{table:materialbudget} summarizes the dominant contributions
to the overall material budget from particular components and
subdetectors of \AMS{} outside the tracker. While the thermal
shielding and the LEPS were located on top of the experiment, the
aerogel was in the lowermost subdetector of \AMS{} and was located
below the tracker. As a consequence, for particles entering \AMS{}
from the top ({\sl downward}), the total amount of material to be
traversed before reaching the tracker was equivalent to 18.2\+\% of a
radiation length, while for particles entering from the bottom ({\sl
  upward}) it amounted to 19.2\+\%. Each tracker layer, including the
silicon and the support structure, contributed another 0.65\+\% of a
radiation length to the material budget~\cite{alcaraz99b}.

In addition to the components of the experiment itself, additional
objects have to be considered for the material budget for upward going
particles, namely the Space Shuttle and the \AMS{} support structure
in it. As shown in Figure~\ref{fig:sts91:shuttle}, the mounting frame
of \AMS{} {\sl (Unique Support Structure, USS)\/} consisted of
aluminum box beams and truss elements, some of which were within the
acceptance of the detector. The 4\+mm thick outer aluminum wall was
covered with low density silica fiber tiles~\cite{nasa04a}, forming
the heat shield.  Besides, the floor contained trusses, as well as
hydraulic and electric circuitry. Additionally, the inner surface of
the payload bay was lined with a protective blanket made of composite
fabric material.  Precise information about the structure of the Space
Shuttle's payload bay floor, which would allow an exact determination
of the material distribution underneath \AMS{}, is not available. The
amount of this additional material is estimated to be 4.5\+\% of a
radiation length from the specifications of dimensions and weight of
the Space Shuttle fuselage~\cite{nasa04a}.

\section{Trigger and DAQ Livetime}\label{sec:trigger} The trigger
chain of \AMS{} was divided into three successive stages: the
hardware-based {\sl Fast Trigger}, and the {\sl Level-1} and {\sl
  Level-3\/} stages, which were implemented in the data acquisition
software\footnote{A Level-2 stage was not implemented.}. Each trigger
stage definitely rejected an event if its particular requirements were
not fulfilled~\cite{casadei02a}. For trigger efficiency studies, a
small fraction of events -- approximately 1 in 1000 -- was accepted
after a positive decision of the Fast Trigger only {\sl (prescaled
  events)\/}.

The {\bf Fast Trigger} signal was derived solely from the
photomultiplier triplets (summed anode signals) viewing the TOF
scintillators. A Fast Trigger was generated if at least one
scintillator panel signaled a hit from one or both end triplets in
each of the four TOF planes. The typical time for the Fast Trigger
stage to generate the signal was 100\+ns. If after another 200\+ns none
of the following trigger requirements were fulfilled, the event was
rejected.

The first trigger condition of the {\bf Level-1 stage} was called {\sl
  matrix condition\/} and accounted for the fact that the tracker
layers were not fully equipped. By comparison with a correlation
matrix, those combinations of TOF scintillator panels on layers 1 and
4 which were incompatible with a particle traversing the tracker
instrumentation were identified and the corresponding events were
rejected. The second trigger condition of the first stage demanded the
absence of signals in any of the veto counters.

The {\bf Level-3 stage} assured the coincidence of the two
photomultiplier triplets on either side of the TOF panels used for the
fast trigger decision. Furthermore, it performed a clustering of the
TOF hits, forming clusters either from one hit in a single panel or
from two hits in two adjacent panels. A single cluster was then
required in each of the first and the last TOF planes in coincidence
with at least one additional cluster in the second and/or the third
plane. These TOF clusters defined a 6\+cm wide corridor in the bending
plane. The Level-3 stage retained the event, if at least three tracker
hits were found within this corridor with a signal to noise ratio
above~4~\cite{lamanna00a}. Events with multiple clusters on a single
TOF plane were rejected~\cite{choutko97a}. The latter trigger
criterion has a negative impact on this data analysis, since it
deteriorates the acceptance for the signal process towards lower
particle momentum (see \S~\ref{sec:acceptance}).

An additional feature of the Level-3 stage was active during the first
seven hours of data taking~\cite{choutko97a}. Its main purpose was the
rejection of low energy particles with \mbox{$\left|Z\right|<3$}. The
algorithm relied on the measurement of the average residual of tracker
clusters with respect to a straight line joining the uppermost and the
lowermost cluster. Events were accepted if the average residual was
below 300\+$\mu$m.

The total dead time per trigger operation amounted to
85\+$\mu$s~\cite{aguilar02a}. In case an event was rejected by the
Level-1 stage, it took 7\+$\mu$s until the data acquisition system
({\sl DAQ}) was again operational. Consequently, the fraction of time
during which the DAQ was active ({\sl DAQ livetime}) significantly
depended on the particle rate, which itself varied with the position
of the detector in the geomagnetic coordinate system. At low magnetic
latitude, particle rates were small and the readout rate reached up to
1.6\+kHz. Except over the South Atlantic, where the geomagnetic field
is exceptionally weak\footnote{A phenomenon known as the {\sl South
    Atlantic Anomaly} (SAA)} resulting in very low readout rates due
to a large amount of trigger dead time, the minimum readout rate was
100\+Hz.

\section{AMS-01 on the Space Shuttle}\label{section:shuttlemission}
\begin{figure}[h]
\begin{center}
\begin{tabular}{ll}
\begin{minipage}{7cm}
\begin{center}
\includegraphics[width=7cm]{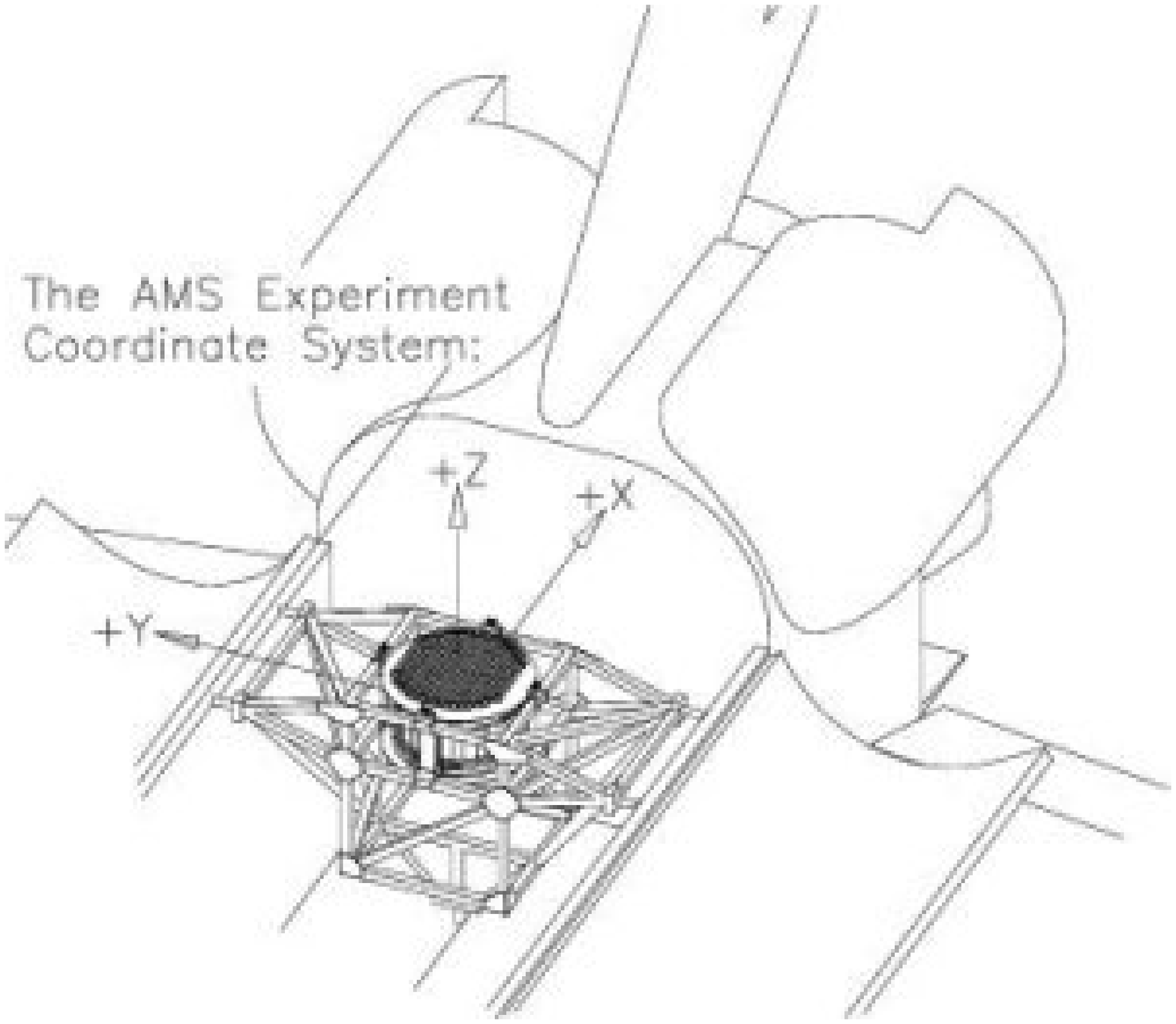}
\end{center}
\end{minipage}
&
\begin{minipage}{6cm}
\begin{center}
\includegraphics[width=5cm]{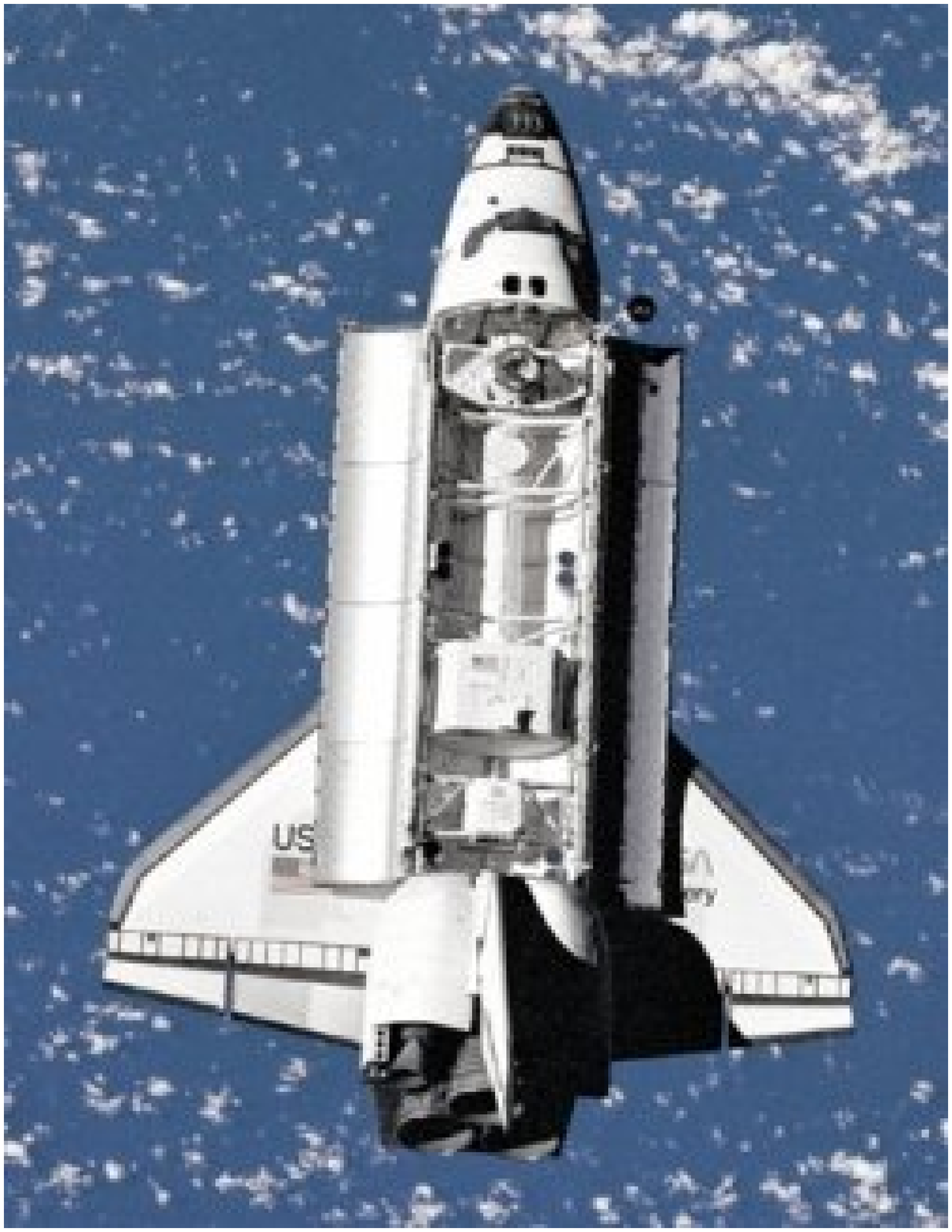}
\end{center}
\end{minipage}
\end{tabular} 
\end{center} 
\caption{\label{fig:sts91:shuttle} The mounting of \AMS{} in the Space
  Shuttle's cargo bay {\sl (left)\/}; photograph of the Space Shuttle
  taken from the Mir space station {\sl (right)\/}.}
\end{figure}

\begin{figure}[f]
\begin{center}
\includegraphics[width=10cm]{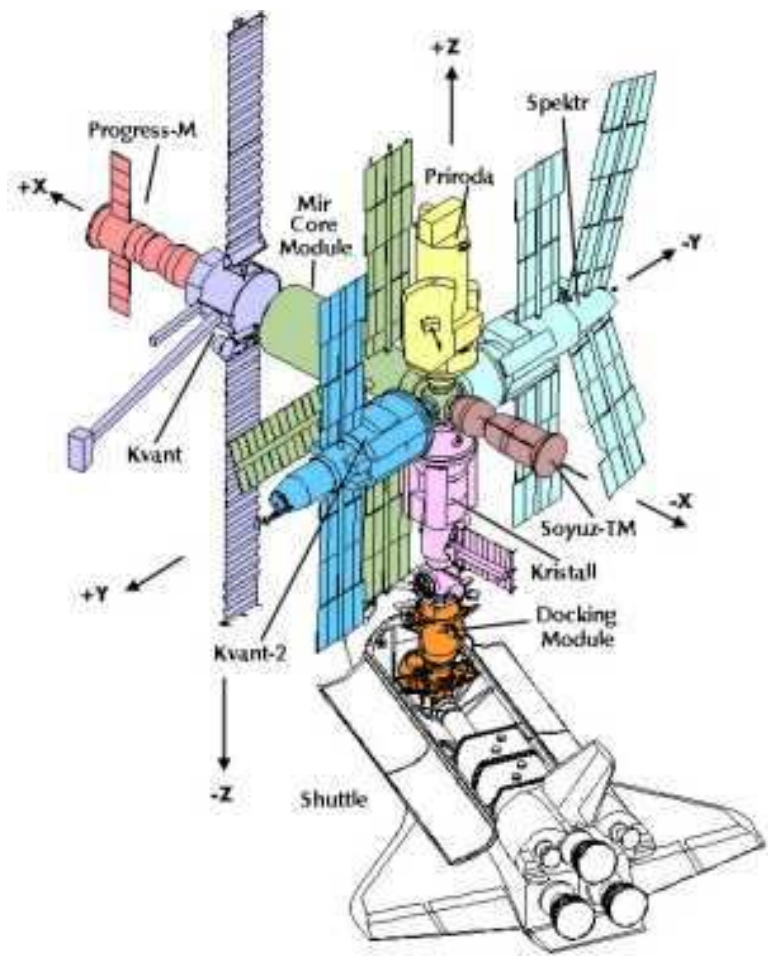}
\end{center} 
\caption{\label{fig:shuttlemirscheme} Schematic view of the Space
  Shuttle docked to the Mir space station~\cite{nasa99a}. The Soyuz-TM
  and the Priroda modules were within the acceptance of the \AMS{}
  experiment.}
\end{figure}

The \AMS{} experiment was flown on the Space Shuttle {\sl Discovery}
in the course of the STS-91 mission\footnote{STS: Space Transportation
  System} from June 2nd to 12th, 1998. Figure~\ref{fig:sts91:shuttle}
shows the mounting of \AMS{} in the spacecraft's cargo bay. During its
flight the Discovery orbited the Earth at altitudes between 320 and
390 km with an orbital period of 93\+min and an inclination of
$\pm$~51.7~degrees. From June 4th to 8th, for a total of 95~hours, the
Space Shuttle was docked to the {\sl Mir\/} space station. During
docking, parts of the Mir, such as the Soyuz-TM and the Priroda
modules, were within the acceptance solid angle of the \AMS{}
experiment~\cite{aguilar05a} (see Figure~\ref{fig:shuttlemirscheme}).

\begin{figure}[h]
\begin{center}
\includegraphics[width=12.6cm]{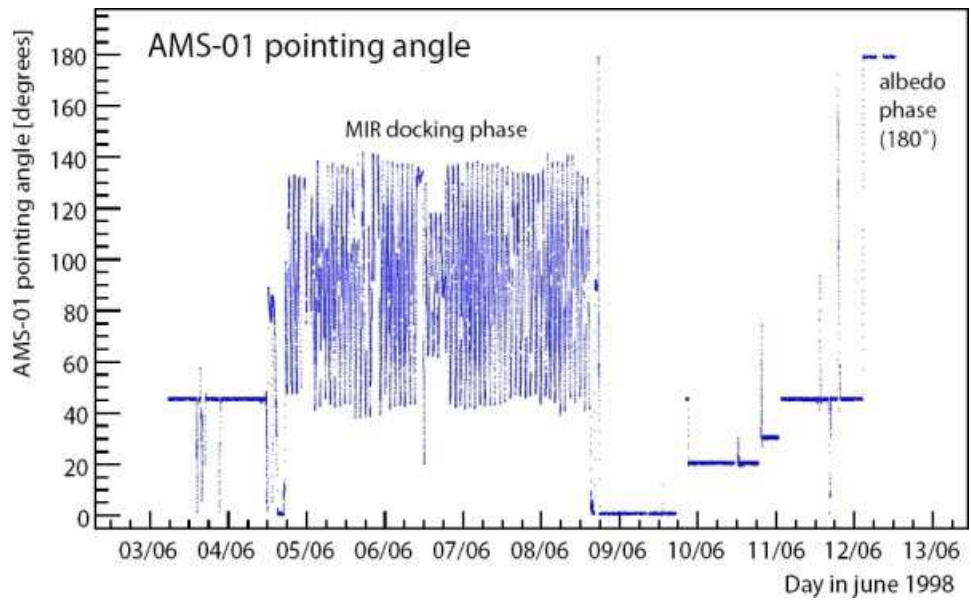}
\end{center} 
\caption{\label{fig:sts91:pointingangle} The pointing angle of \AMS{}
  as a function of time.}
\end{figure}

The 184~hours of data taking were divided into several periods. As
seen in Figure~\ref{fig:sts91:pointingangle}, the pointing angle of
\AMS{} -- the angle of the positive z-axis with respect to the zenith
direction -- varied with time. Before docking to the Mir, the pointing
angle was kept constant at 45 degrees. After decoupling from the space
station, the experiment was operated at consecutive angles of 0, 20,
30, 45 and 180 degrees. During the latter phase, the z-axis of \AMS{}
pointed towards the Earth for the measurement of outbound particles
emitted from the atmosphere {\sl (albedo phase)\/}. While the
Discovery was docked to the Mir, the pointing angle could not be kept
constant; it oscillated between roughly 40 and 140 degrees with the
Mir orbital frequency of about 0.65\+h$^{-1}$. Several short
loss-of-signal periods occured during the flight, partly because the
electronics had to be shut down due to overheating when the shuttle
was pointing towards the zenith~\cite{suter00a}.

\chapter{Positron Identification through Bremsstrahlung Conversion}
\section{The Challenge of Positron Measurements}
The main challenge of high energy cosmic ray positron measurements is
the suppression of the vast proton background. As already stated in
chapter \ref{chapter:cosmicrays}, the flux of cosmic ray protons
exceeds that of positrons by a factor of $10^4$ in the momentum range
of 1--50\+\GeV/c. Besides the fact that both particle species have the
same charge, their energy deposition in traversed material becomes
comparable at energies beyond a few GeV.  Thus, the identification of
positrons must rely either on techniques based on measurement of the
Lorentz factor, as realized for example in \v Cerenkov counters or
transition radiation detectors, or on high-resolution calorimetry.

However, as described in the last chapter, \AMS{} was not equipped
with a calorimeter and its ATC subdetector provided a sufficient
proton rejection only for momenta below 3.5\+\GeV/c. In order to extend
the energy range accessible to the experiment, a completely different
approach has been chosen. It relies on the identification of
bremsstrahlung emission through photon conversion, the latter also
referred to as pair production.

\section{Conversion of Bremsstrahlung Photons}
Both bremsstrahlung and photon conversion cannot occur in vacuum, but
require the presence of a Coulomb field, predominantly that of a heavy
object like a nucleus, so that momentum and energy conservation are
satisfied. They are closely related electromagnetic processes whose
energy-angle distributions can be calculated with the Bethe-Heitler
formalism. For bremsstrahlung from electrons or positrons and in the
limit of high energies this distribution is given by the cross section
\cite{schiff51a}
\begin{equation}\label{equation:bremsstrahlung}
\sigma_{\mathrm{b}}(k,x) \propto Z^2\, \frac{\mathrm{d}k\,\mathrm{d}x}{m^2k}\, x
\left\{ \frac{16x^2E}{A(x)} - \frac{(E_0+E)^2}{B(x)} 
+ \left[ \frac{E_0^2+E^2}{B(x)} - \frac{4x^2E}{A(x)} \right]\cdot \ln M(x)\right\},
\end{equation}
with
\begin{displaymath}
A(x)=(x^2+1)^4 E_0,\quad B(x)=(x^2+1)^2 E_0^2,\quad
\frac{1}{M(x)}=\left(\frac{mc^2k}{2E_0 E}\right)^2 + \left(\frac{Z^{1/3}}{C(x^2+1)}\right)^2, 
\end{displaymath}
where $E_0$ and $E$ denote the energy of the lepton before and after
the radiation process, $k=E_0-E$ the energy of the photon, $Z$ the
charge number of the scattering nucleus, $m$ the electron mass and
$c$ the speed of light.  $C$ is a constant of the order of $10^2$,
which expresses the screening effect of the atomic electrons bound to
the nucleus. The variable $x$ is equal to the reduced angle
$\gamma\theta$, $\gamma$ being the lepton's Lorentz factor and
$\theta$ the angle of the radiated photon with respect to the incident
lepton.

The bremsstrahlung matrix element is related to those of pair
production by the substitutions $k\leftrightarrow -k$ and
$p\leftrightarrow -p$, where $p$ is the four-momentum of the incident
particle in bremsstrahlung emission or the four-momentum of one of the
pair particles in pair production. Thus, the energy-angle distribution
for pair production has a mathematical form similar to that for
bremsstrahlung, with $\theta$ being the angle of one of the leptons
with respect to the incident photon
\cite{tsai74a}.

The above cross section (\ref{equation:bremsstrahlung}) depends
inversely on the square of the incident particle mass. Hence it
follows that bremsstrahlung emission is suppressed by a factor of more
than $3\cdot 10^{6}$ for protons with respect to positrons. This
important property of the bremsstrahlung process has a very beneficial
implication: identification of the radiated photon, in conjunction
with a measurement of the charge sign of the incident radiating
particle, determines the latter almost certainly as a positron (or
electron). This principle is the basis for the proton background
suppression as performed in this analysis.

\subsection{Event signature}\label{sec:eventsignature}
\begin{figure}[htb]
\begin{center}
\includegraphics[width=7cm]{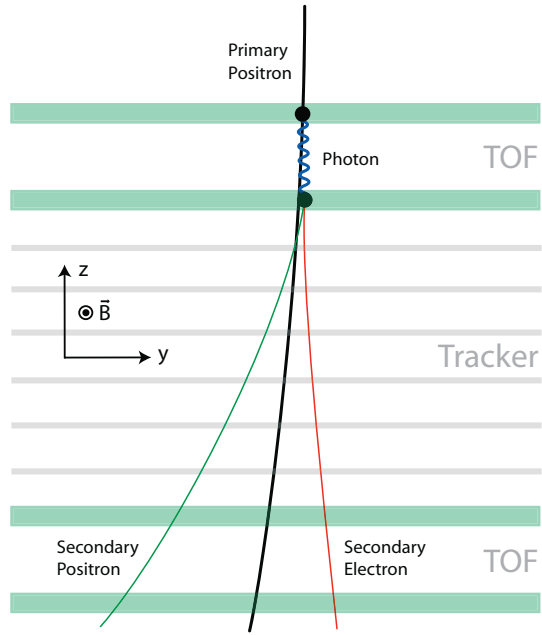}
\caption{\label{fig:convScheme}Schematic view of a converted bremsstrahlung event caused by a downward going positron.}
\end{center} 
\end{figure}
Figure~\ref{fig:convScheme} shows the topology of an event with a
converted bremsstrahlung photon. Here, a primary positron enters the
detector volume from above and emits a bremsstrahlung photon in the
first TOF scintillator layer. The photon then converts into a
secondary electron-positron pair in the second TOF layer in this
example. If the conversion process takes place above or in the upper
part of the tracker, the tracks of three particles can in principle be
reconstructed.

Eq.~(\ref{equation:bremsstrahlung}) can be integrated over $x$ to obtain
the root-mean-square angle $\theta_{rms}$ of the photon with respect
to the incident lepton, yielding $\theta_{rms} =
q(E_0,k,Z)/\gamma\cdot\ln\gamma$ for high energies
\cite{stearns49a}. Since $q(E_0,k,Z)$ depends primarily on $k/E_0$ 
and is always of the order of unity, it follows that
$\theta_{rms}\approx 1/\gamma$. As a consequence, in the GeV energy
range, bremsstrahlung photons are largely emitted under angles of
$10^{-3}$\+rad or smaller. The same applies to the opening angles of
the electron positron pairs from pair production. In a tracking
detector, these values fall below the typical limit of track
reconstruction accuracy given by multiple scattering and thus are
practically equal to zero.

Furthermore, it can be seen from eq.~(\ref{equation:bremsstrahlung})
that the spectrum of the emitted photons is soft, since the cross
section decreases with $1/k$. Because of the low fraction of momentum
which is typically carried away by the photon, the secondary particles
have lower momenta than the primary.  Therefore, in the bending plane
projection of the spectrometer, the secondaries tend to form the left
and right tracks, while the primary remains in the middle (see
Figure~\ref{fig:convScheme}). It must be pointed out that in the GeV
energy range the fraction of momentum given to each of the secondary
particles is approximately uniformly distributed between 0 and
$E_{\gamma}$. In the limit of high energies, the distribution of the
ratio of the two secondary momenta becomes slightly asymmetric, with
electron-positron pairs having considerably different momenta being
favored.

\subsection{Background}\label{sec:background}
The background to positron identification through bremsstrahlung
conversion is caused by protons undergoing hadronic reactions in the
material of the experiment, as well as by electrons with wrongly
reconstructed momentum sign.

Mesons produced in hadronic reactions involving protons can mimic the
three-track signature of converted bremsstrahlung events. These mesons
are almost exclusively pions; the cross section for the production of
kaons or other mesons is in comparison more than one order of magnitude
smaller~\cite{capiluppi74a}. Several processes contribute to this
background and can be grouped in two categories, where $p,n$ denote an
incident proton, or a proton or neutron in the scattering nucleus
(cross sections are taken from~\cite{nara99a}):
\begin{itemize} 
\item processes involving charged pion production, dominantly
$pp\to pp\pi^+\pi^-$ and $pn\to pn\pi^+\pi^-$, with cross sections of
approximately 2\+mb above a threshold of 3\+\GeV{} incident proton energy,
and
\item processes involving neutral pion production, such as
$pp\to pp\pi^0$ or $pn\to pn\pi^0$, with cross sections of
approximately 2\+mb or 4\+mb, respectively, above a
threshold of 1\+\GeV{} incident proton energy.
\end{itemize}
In case of neutral pion production, the $\pi^0$'s decay almost
exclusively into photon pairs after a mean flight path of
$c\tau=25$\+nm. If one of the photon remains undetected, while the
other one converts into an electron positron pair, the resulting
three-track event signature is that of a converted bremsstrahlung
event. Due to the procedure for vertex reconstruction, discussed in
\S~\ref{section:reconstructionquality}, the fact that -- in
contrast to the signal process -- only one vertex is present in the
above hadronic interaction cannot be used to effectively discriminate
against the hadronic background.

The material of the experiment, in which the relevant reactions take
place, mainly consists of carbon and hydrogen atoms. According to the
optical model \cite{carroll79a}, the proton-proton cross sections
$\sigma_{\mathrm{pp}}$ can be applied to proton-nucleus scattering
when appropriately scaled as
$\sigma_{\mathrm{pN}}=\sigma_{\mathrm{pp}} A^{\alpha}$, with $A$ being
the atomic weight of the target nucleus, and $\alpha=0.77$ for
carbon. Hence, for a scintillator of the TOF system with a thickness
of 1\+cm, the probability for a proton to undergo one of the above
reactions is of the order of 1\+\textperthousand\ for both charged and
neutral pion production. However, the probability for such a reaction
to give rise to a background event is much smaller due to several
reasons. The spectral distributions of the secondary pions show
distinct peaks at 100\+\MeV{} and decrease steeply towards higher energies
\cite{badhwar77a}. The resulting momenta are typically very low and
tend to prevent a hadronic event from exhibiting a three-track
signature due to the strong deflection in the magnetic field. On the
other hand, in contrast to converted bremsstrahlung events where the
secondary tracks emerge from the vertex at small angles and are thus
ordered according to their charge sign by the spectrometer magnetic
field, secondary particles from hadronic interactions are emitted
under large angles with respect to each other and may be any one of
the three tracks with almost equal probability. Background from
processes involving neutral pion production is further suppressed by
the requirement that exactly one of the two photons from pion decay
must convert. Let $p\approx 7/9 X_0$ denote the pair production
probability in a material with a thickness corresponding to a fraction
$X_0$ of a radiation length~\cite{rossi65a}. Then the suppression
factor is $\{p(1-p)\}^{-1}=8.33$ for $X_0=0.18$, in the case of
downward particles.

The average transverse momentum $\langle p_{\mathrm{T}}\rangle$ of
mesons stemming from proton-proton interactions is $\langle
p_{\mathrm{T}}\rangle\approx 0.35$\+\GeV{}/c, rather independent of the
collision energy~\cite{mccubbin81a}. As a consequence, the angles
$\theta$ between particles emerging from interactions in the GeV
energy range are large compared to those in bremsstrahlung
conversion. For example, $\langle\theta\rangle
\approx 33\cdot 10^{-3}$\+rad at 10\+\GeV{}. This important characteristic of
hadronic interactions allows to effectively discriminate against the
proton background.

Determining from first principles the background from electrons with
wrongly reconstructed charge sign is naturally harder due to the
rather complex behavior of track reconstruction algorithms. As
Figure~\ref{fig:tracker:performance} on
page~\pageref{fig:tracker:performance} shows, the rigidity resolution
of the tracker has a constant value of about 9\+\% in the range 1--9\+GV
and then rises linearly with increasing rigidity. For high energy
particles, the resolution rapidly worsens so that the probability of
bad reconstruction increases. If the resolution is approximated by the
width $\sigma$ of a normalized Gaussian distribution
$G(R,\hat{R},\sigma)$ of the rigidity $R$ around a true mean value
$\hat{R}$ (which is negative for electrons), then the probability
$p_{\mathrm{w}}$ of wrong charge sign reconstruction is given by the
integral
\begin{equation}\label{eq:misrecProb}
p_{\mathrm{w}} = \int^{\,\infty}_{0}\, G(R,\hat{R},\sigma)\, dR\, .
\end{equation}
Due to the quickly falling edges of the Gaussian distribution, the
value of $p_{\mathrm{w}}$ is practically zero for low rigidities. At
50\+GV it amounts to approximately 1\+\textperthousand\ and hence the
amount of misidentified electrons remains small compared to the
positron to electron ratio.

Apart from these rather coarse estimations, the properties of the
background sources, as well as their individual contributions to the
signal event sample, must be precisely determined by Monte Carlo
simulation. In \S~\ref{section:suppression} it will be shown that
in the course of the analysis the probability for a proton to create a
background event can be reduced to less than $10^{-7}$, and the
contamination of the positron sample from badly reconstructed
electrons remains below the percent level.

\section{Event Reconstruction}
The task of the event reconstruction process is to identify candidates
for converted bremsstrahlung events and reconstruct the event geometry
and particle momenta. In order to obtain the highest possible
selection efficiency, it is mandatory to apply sophisticated track and
vertex finding algorithms which are particularly customized for the
signature of converted bremsstrahlung. In a first step, a fairly
permissive preselection algorithm extracts signal candidates from the
\AMS{} data set. Two independent track finding algorithms, each of
them optimized for a particular event topology, are separately applied
to the preselected event sample. In case that three particle tracks
are found, their corresponding momenta are determined, and the
bremsstrahlung and conversion vertices are spatially reconstructed. As
argued in \S~\ref{sec:eventsignature}, the middle track is
considered to stem from the primary particle, while the left and right
tracks represent the secondary electron positron pair. To account for
the asymmetric geometry of the detector along its z-axis, event
reconstruction and analysis are performed separately for particles
traversing the detector from top to bottom {\sl (downward)\/} and from
bottom to top {\sl (upward)\/}.

Before the analysis, the raw data recorded by \AMS{} were
preprocessed~\cite{choutko03a}. This procedure included, amongst
others:
\begin{itemize}
\item reconstruction of the spatial coordinates of the tracker hits
and their errors,
\item reconstruction of the energy deposits and relative transit times 
of particles in the TOF scintillator layers, and
\item calculation of the {\sl trigger livetime\/}, i.e. the fraction
of time over which the trigger electronics of the experiment were
accepting signals (see \S~\ref{subsection:livetime}).
\end{itemize}

\subsection{Preselection}
The event reconstruction algorithms developed for this analysis are
rather complex and require substantial amounts of computing time
because of their combinatorial character. The task of the preselection
is to considerably reduce the size of the \AMS{} data, so that the
algorithms can operate on a reasonably small sample of event
candidates.

In a first step, all hits in the silicon strip detectors of the
tracker are projected into the bending plane for one dimensional
clustering. On a particular tracker layer, a cluster seed is defined
by a single arbitrary tracker hit. Subsequently, further hits are
added to an existing cluster if their distance from the
barycenter\footnote{The barycenter is the weighted mean of the hit
coordinates, with the hit amplitudes as weights.} of this cluster
along the y-axis is smaller than 100\+$\mu$m. If a hit cannot be added
to any existing cluster, it is used to start a new cluster. In this
procedure, the x-coordinates of the tracker hits are not regarded, and
the clusters are defined in the bending plane (y-z plane) only (see
Figure~\ref{fig:ghostHits}). Thus the hits in a given cluster may
spread over the whole length of the particular tracker layer along the
x-axis. Most of them are virtual hits caused by the tracker bonding
scheme (see \S~\ref{section:tracker}). This ambiguity is resolved
by the track finding algorithms in a subsequent step
(see \S~\ref{subsection:trackfinding}). Moreover, there may be
clusters comprised of a single hit only.

For further analysis, a minimum of 8 tracker clusters are
required. Events are selected in which at least two of the six layers
of silicon detectors contain exactly three clusters each {\sl
(triplets)\/}. By these means, a reduction of the data size by a
factor of 25 is achieved.

\subsection{Track Finding}\label{subsection:trackfinding}
\AMS{} was designed for single particle detection, thus the software 
package for data analysis provided by the \AMS{} collaboration is
optimized for the reconstruction of single-track events. The track
finding algorithms implemented in this package have a small
reconstruction efficiency for multi-track events, resulting in low
event statistics when applied to a rare process such as bremsstrahlung
conversion. As a consequence, alternative reconstruction methods had
to be developed for this analysis. Two different track finding
algorithms -- in the following referred to as the {\sl successive\/}
and the {\sl combinatorial\/} algorithm -- are independently applied
each to a subsample of the preselected events. Their purpose is to
establish if an event exhibits a three-track signature and, if so,
find the correct assignment of the tracker clusters to the three
tracks. Each algorithm is optimized for a particular cluster topology
and returns a set of parameterized track information.

With $t$ denoting time, and $\mathbf{p}$ and $\mathbf{v}$ being the
momentum and velocity of a particle of charge $e$, its equation of
motion in a magnetic field $\mathbf{B}$ is
\begin{equation}\label{equation:magfieldmotion}
\frac{\mathrm{d}\mathbf{p}}{\mathrm{d}t} = e\left(\mathbf{v}\times\mathbf{B}\right)\, .
\end{equation}
In a homogeneous field, the solution of this equation is a helix
trajectory. Since the magnetic field within \AMS{} is inhomogeneous,
eq.~(\ref{equation:magfieldmotion}) is preferably solved
numerically. Nevertheless, the resulting trajectories can still be
well approximated by a helix. The latter is fully parameterized with a
set $\mathbf{h}$ of five numbers, defined on any given plane $z=z_0$
as
\begin{equation}\label{equation:helixparameters}
\mathbf{h} = \left[x,\frac{\mathrm{d}x}{\mathrm{d}z},y,\frac{\mathrm{d}y}{\mathrm{d}z},\frac{1}{p}\right]\, .
\end{equation}
An iterative fitting algorithm~\cite{hart84a}, particularly adapted
for the \AMS{} experiment~\cite{choutko96a}, is used to determine
$\mathbf{h}$ from a set of point measurements $i$ of a particle's
trajectory. The algorithm is based on minimization of the quantity
\begin{equation}\label{equation:chisquare}
\chi^2 = \sum_i \left\{ \left(\frac{\hat{x}_i-x_i}{\sigma_{\hat{x}}}\right)^2 + \left(\frac{\hat{y}_i-y_i}{\sigma_{\hat{y}}}\right)^2\right\} \, ,
\end{equation}
with $x_i,y_i$ being helix parameters on the given z plane and
$\hat{x}_i,\hat{y}_i$ the measured hit coordinates on that plane with
the corresponding measurement errors
$\sigma_{\hat{x}},\sigma_{\hat{y}}$. The algorithm returns
$\mathbf{h}$ at the z-position of the first measurement in the
direction of flight. The minimum rigidity that can be reconstructed is
100\+MV.

\subsubsection{The successive track finding algorithm} 
\begin{figure}[b]
\begin{center}
\includegraphics[width=9cm]{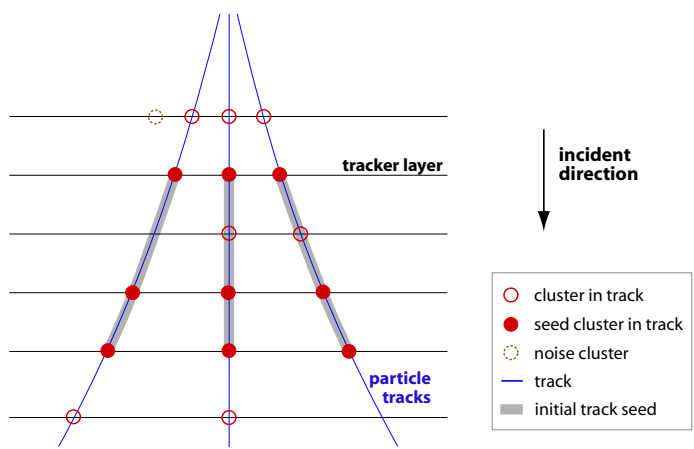}
\caption{\label{fig:trackFindScheme}Principle of the successive track finding 
algorithm in the bending plane projection. Initial track seeds are
created from seed clusters on layers with exactly three clusters.
Further clusters are successively added to the tracks, while noise
clusters, which do not fit to any track, remain isolated. Missing
clusters occur in parts of the tracker which are inefficient or not
instrumented.}
\end{center} 
\end{figure}

The successive track finding algorithm is customized for events in
which at least three tracker layers are found with exactly three
clusters on each of them. Its working principle is illustrated in
Figure~\ref{fig:trackFindScheme}. Since particle tracks diverge in the
magnetic field, the triplets -- clusters on layers with exactly three
clusters -- are required to have increasing cluster distances along
the z-axis in the direction of flight. With three particles having
traversed the tracker, the clusters in the triplets can directly be
assigned to a left, a middle and a right track segment {\sl (initial
track seeds)\/}.

Subsequently, starting from the uppermost layer, further clusters on
the other layers are tentatively added to each of the tracks. A
cluster is finally appended to the track which yields the lowest
$\chi^2$ according to eq.~(\ref{equation:chisquare}), if the value of
$\chi^2$ is below a threshold of 30 (the fitting procedure is
described in the next paragraph). In this case, the helix parameters
of this track, including the new cluster, are updated. As the tracks
grow, as many clusters as possible are attached to them. Clusters
caused by noise or the occurrence of $\delta$-ray electrons are mostly
incompatible with any of the tracks and remain isolated.

\begin{figure}[htb]
\begin{center}
\includegraphics[width=14cm]{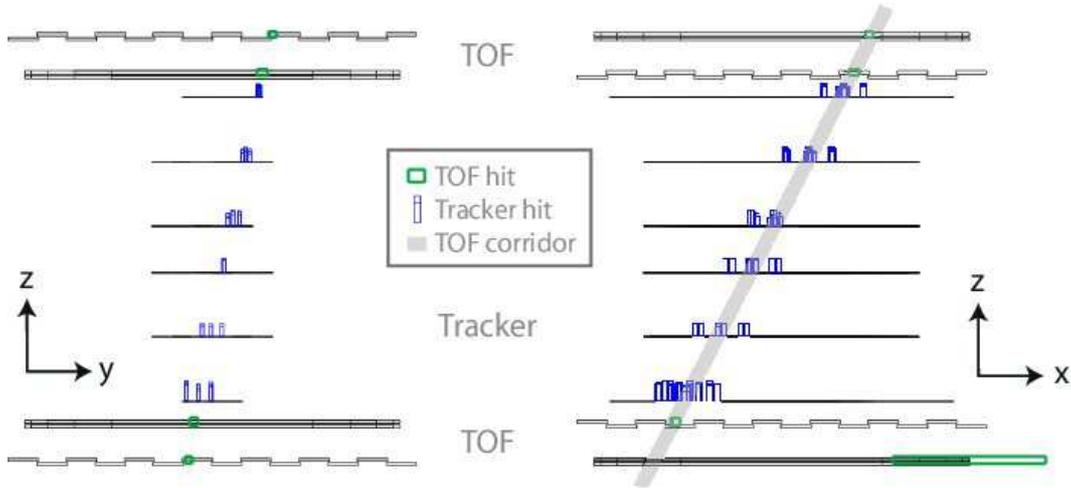}
\caption{\label{fig:ghostHits}Display of a three-track event, projected
in the bending plane {\sl (left)\/} and in the non-bending plane {\sl
(right)\/}. Only the TOF system and the tracker are shown. The height
of the hit boxes is proportional to the hit amplitudes, while their
width represents the position measurement errors. On the right side, the
hit selection corridor is plotted.}
\end{center} 
\end{figure}

For the fitting procedure mentioned above, it is necessary to resolve
the ambiguities in the x-coordinates of the clusters, caused by the
clustering only in the bending plane projection. For this, a linear
regression procedure~\cite{james94a} is carried out to obtain a
straight line through the hits in the TOF system
(see Figure~\ref{fig:ghostHits}). In the non-bending plane (x-z-plane), a
corridor of 10\+cm width is defined around the straight line and only
hits within this corridor are retained. All possible combinations of
remaining hits in the track -- with at most one hit from each cluster
-- are then tentatively fitted, and the helix parameters are updated
from the combination which yields the lowest value of $\chi^2$
according to eq.~(\ref{equation:chisquare}).

\subsubsection{The combinatorial track finding algorithm} 
In order to increase the efficiency for converted bremsstrahlung
events, a generalized algorithm has been developed for the treatment
of events that feature only two tracker layers with exactly three
clusters each~\cite{henning}. It is based on a combinatorial approach
to the track finding problem. The basic idea is to examine every
possibility {\sl (track hypothesis)\/} of arranging the tracker
clusters into three tracks and determine those combinations yielding
the lowest value of $\chi^2$ according to
eq.~(\ref{equation:chisquare}). These track hypotheses have to pass
several tests in order to reject wrongly assigned
tracks. Additionally, two tracks may share one single cluster. The
algorithm leads to an improvement of the efficiency by~40\+\%. A
detailed description of this method is given in~\cite{henning}.

\subsection{Vertex Reconstruction}
Vertex reconstruction is based on parallel backtracing of tracks
through the magnetic field of the detector using the functionality of
the GEANT3 package (routine GRKUTA)~\cite{brun87}. For a given
track of a particle $i$ with charge $q$ and momentum $\mathbf{p}$, a
virtual particle $\hat{\imath}$ with charge $-q$ and momentum
$-\mathbf{p}$ is started at the first tracker cluster found in the
flight direction of particle $i$. The propagation of $\hat{\imath}$ is
then performed by numerically solving the equation of motion in steps
of~100\+$\mu$m, so that $\hat{\imath}$ will follow the trajectory of
the original particle $i$ backward against its original flight
direction.

The vertex finding for the case of downward going particles is
performed as follows. The left and right tracks may have their
starting cluster on different tracker layers. As a first step, the
track starting at the lower layer is backtraced upward until it
reaches the starting layer of the other track. Both tracks are then
backtraced in parallel until their projected distance along the
y-axis reaches a minimum. The vertex is then defined as the
barycenter of the track points at the z-coordinate of closest approach
of the tracks. In case the tracks intersect in the bending plane
projection, the intersection point is taken as the vertex. The vertex
finding for upward going particles is exactly similar, but with all
directions reversed in z.

Finally, the four-vector of the photon is reconstructed from the
four-vectors of the left and right tracks at the vertex
position. Using the algorithm described above, the vertex of the
photon and the middle track is then computed. In this case, the photon
is simply propagated along a straight line.

\subsection{Reconstruction Quality and Monte Carlo}\label{section:reconstructionquality}
The quality of the reconstruction algorithms is verified with
$16.8\cdot 10^6$ electron and positron events from a complete Monte
Carlo simulation of the experiment using GEANT3. The incident momentum
spectrum of these Monte Carlo particles is proportional to $p^{-1}$
and ranges from 1 to 60\+\GeV{}$/$c. For background studies an additional
$170\cdot 10^6$ proton Monte Carlo events have been generated in the
momentum range \mbox{6--150\+\GeV{}/c}. All Monte Carlo particles are
injected both from above and below into the acceptance solid angle of
the experiment with a uniform spatial and angular distribution.
\begin{figure}[htb]
\begin{center}
\begin{tabular}{ll}
\begin{minipage}{7cm}
\includegraphics[width=7cm]{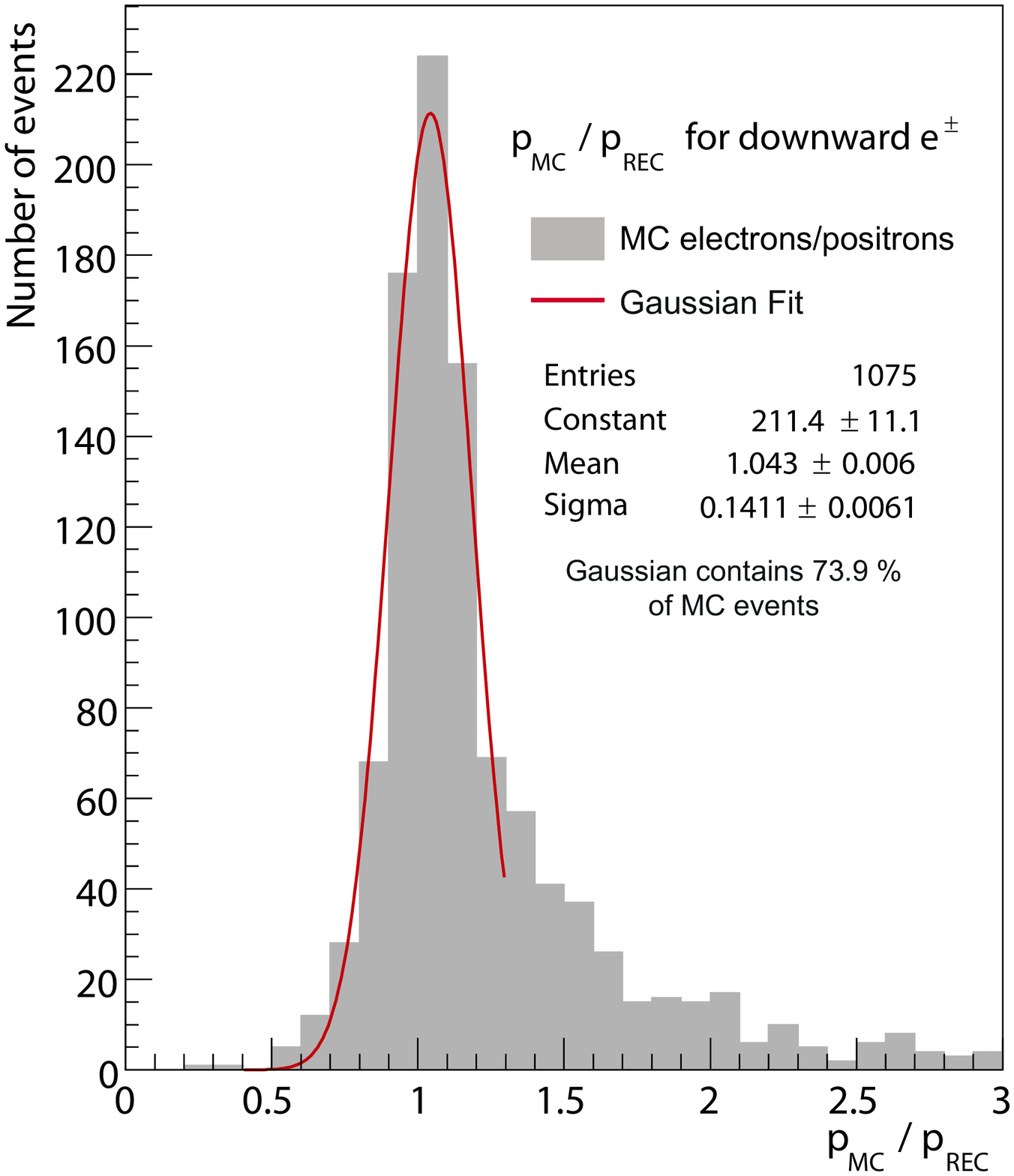}
\end{minipage}
&
\begin{minipage}{7.05cm}
\begin{center}
\includegraphics[width=7.05cm]{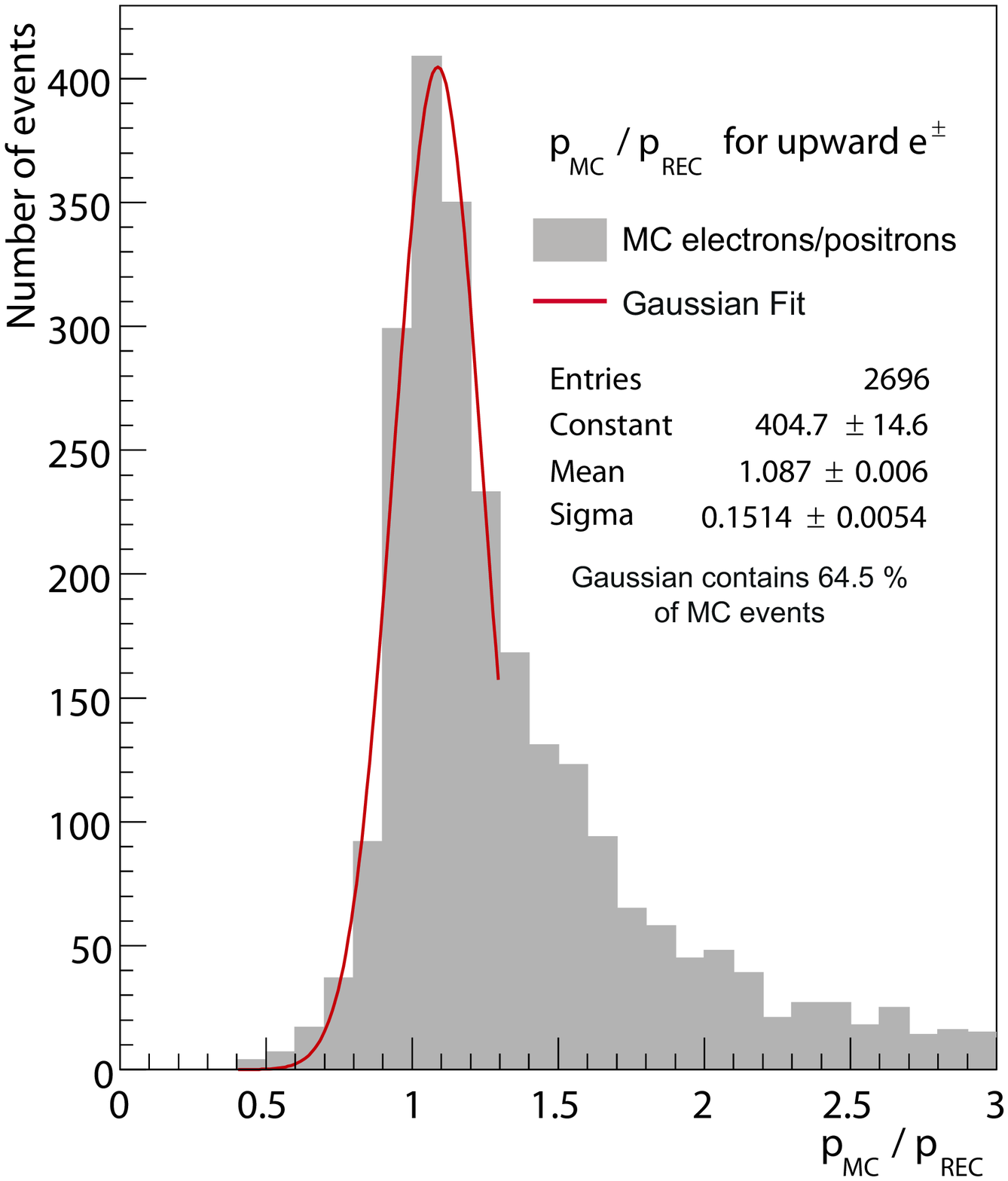}
\end{center}
\end{minipage}
\end{tabular} 
\end{center} 
\caption{\label{fig:relMom}Comparison of generated and reconstructed 
momentum (sum of three track momenta) for downward {\sl (left)\/} and
upward {\sl (right)\/} going positrons.}
\end{figure}
\begin{figure}[t]
\begin{center}
\begin{tabular}{ll}
\begin{minipage}{7cm}
\includegraphics[width=7cm]{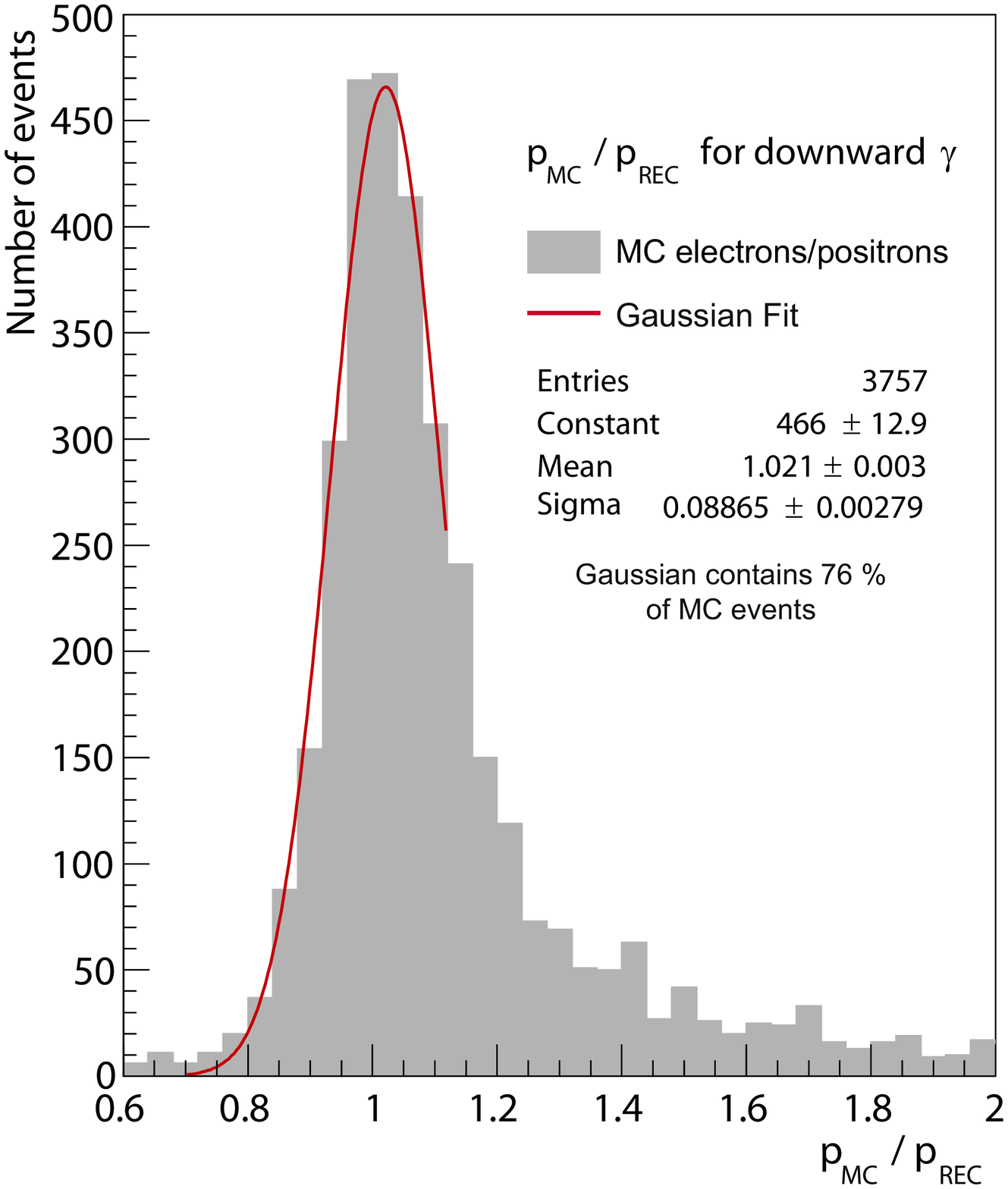}
\end{minipage}
&
\begin{minipage}{7.08cm}
\begin{center}
\includegraphics[width=7.08cm]{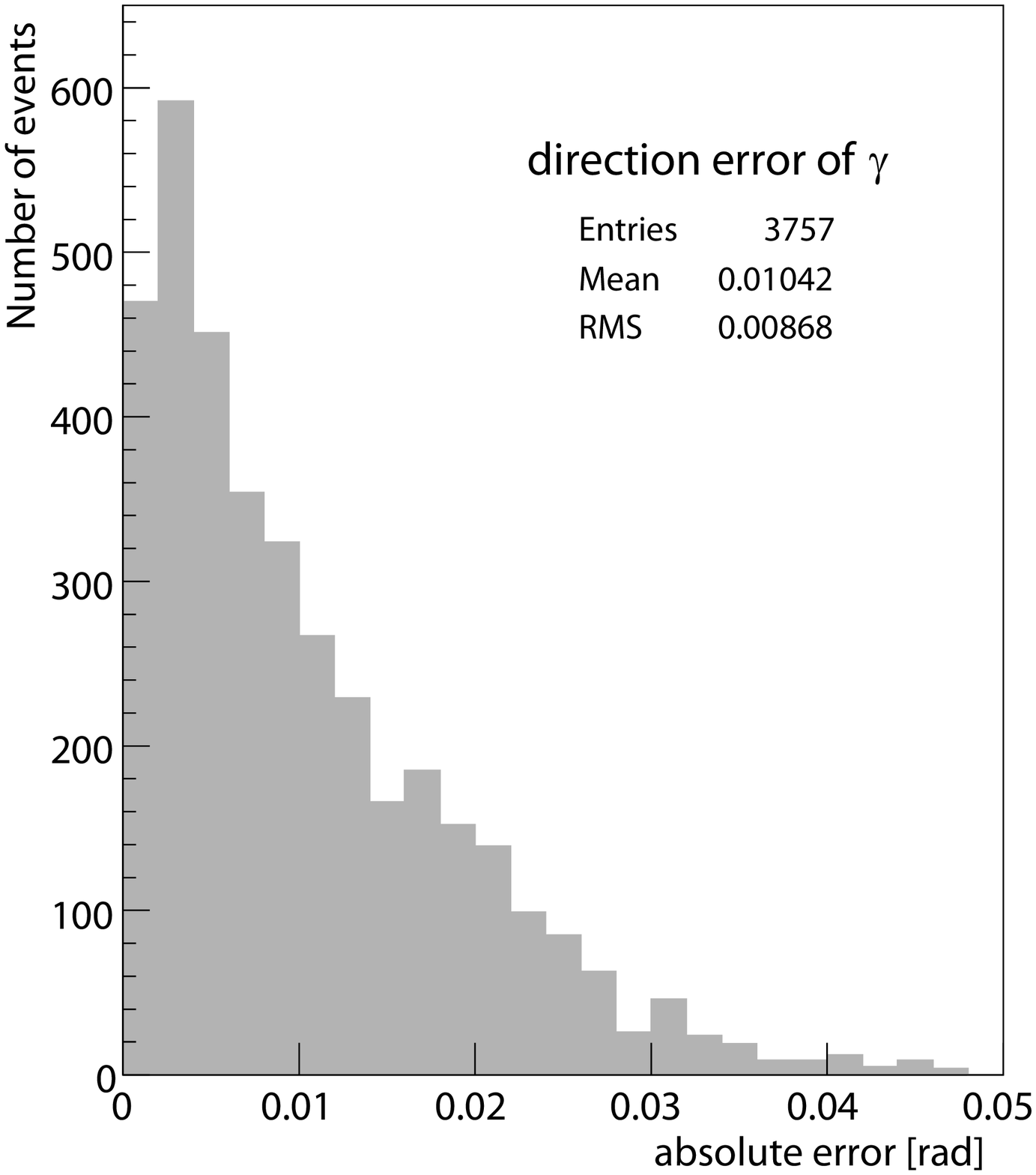}
\end{center}
\end{minipage}
\end{tabular} 
\end{center} 
\caption{\label{fig:relMomPhoton}Comparison of generated and 
reconstructed bremsstrahlung photon momentum {\sl(left)\/} and the
absolute error in its direction {\sl (right)\/} for downward and
upward going particles combined.}
\end{figure}

Figure~\ref{fig:relMom} shows the ratio of generated and reconstructed
momentum of converted bremsstrahlung events as the sum of the three
track momenta for downward {\sl (left)\/} and upward {\sl (right)\/}
going positrons. A Gaussian fit results in a momentum resolution of
approximately 14\+\% and 15\+\% for the downward and upward case,
respectively. This resolution is quite comparable to that for
single-track events in the energy range of 10\+\GeV{} and
above~\cite{alcaraz99b,chang01}, where the reconstruction algorithms
have their peak sensitivity (see \S~\ref{section:fluxcalculation}).

The tails towards higher relative momenta are caused by additional
bremsstrahlung photons which are emitted from the primary before it
enters the tracking volume. If such a photon remains undetected, the
measured momentum of the primary will be lower than the true initial
momentum. This effect is especially apparent in the case of upward
going particles due to the high amount of material they have to
traverse before reaching the tracker.

As displayed in Figure~\ref{fig:relMomPhoton}, the properties of the
bremsstrahlung photon can be particularly well reconstructed. A
Gaussian fit to the photon's relative momentum results in a momentum
resolution better than 9\+\%, while the absolute error in its direction
has a standard deviation below 9 mrad. The secondary tracks from which
the photon is reconstructed typically have lower momenta than the
primary particle, resulting in a better momentum resolution.
Additionally, in the case of the photon, the direction information of
both secondary tracks enters into the reconstruction, thus increasing
the resulting precision.
\begin{figure}[h] 
\begin{center}
\includegraphics[width=15cm]{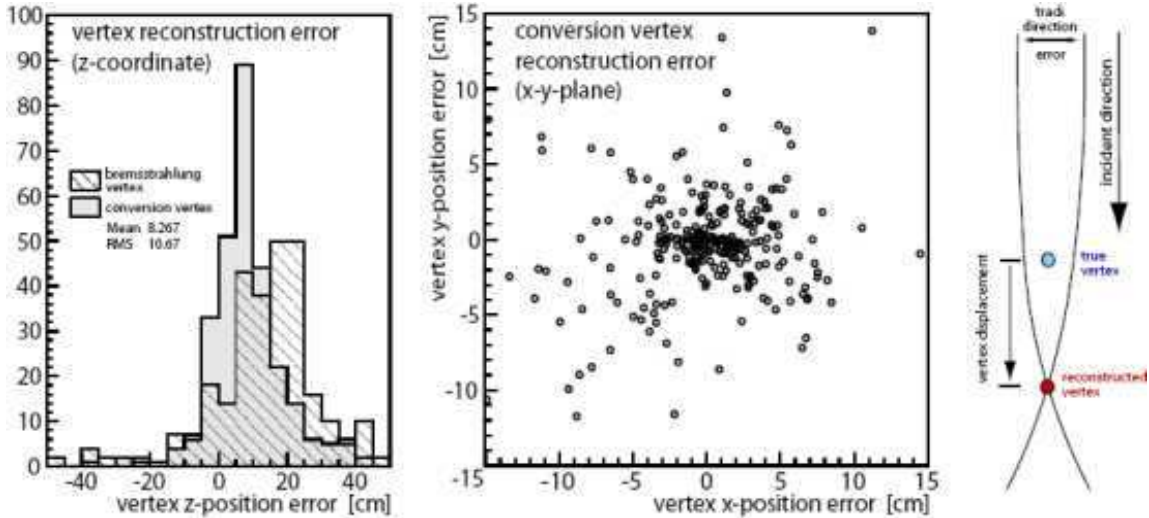}
\caption{\label{fig:vertexreconstruction}Absolute vertex position 
reconstruction error $z_{\mathrm{MC}} - z_{\mathrm{rec}}$ along the
z-coordinate {\sl (left)\/} and in the x-y-plane {\sl (middle)\/},
schematic view of the vertex finding in case of intersecting tracks
{\sl (right)\/}.}
\end{center} 
\end{figure}

The spatial reconstruction of the vertices is rather poor due to the
low angles under which the particles emerge from
them. Figure~\ref{fig:vertexreconstruction} {\sl (left, middle)\/} shows
the absolute error of the reconstructed vertex positions, calculated
from downward going Monte Carlo electrons and positrons. For the
conversion vertex, the resolution of the z-position is not better than
10\+cm. Furthermore, a mean displacement of 8\+cm is observed, i.e. the
vertex is typically reconstructed too close to the center of the
detector. The reason for this displacement lies in the geometrical
characteristics of the vertex finding algorithm. As displayed in
Figure~\ref{fig:vertexreconstruction} {\sl (right)\/}, small errors in
the track directions may lead to intersecting tracks in the bending
plane projection, always resulting in a vertex position which is
shifted in the flight direction of the particle. For the same reasons,
the bremsstrahlung vertex is reproduced with an accuracy of 16\+cm and
a displacement of about 13\+cm in the flight direction. For the
conversion vertex, the width of the error distribution in the
x-y-plane (Figure~\ref{fig:vertexreconstruction}, {\sl middle\/}) is
mostly a consequence of the z-errors when tracks are tilted with
respect to the \AMS{} z-axis. Since the opening angles of the tracks
vary only slightly with vertex position, the low quality of the vertex
position reconstruction has almost no implications for the analysis.

\section{Analysis}
\label{section:analysis}
Analysis and suppression of background mainly rely on the evaluation
of the topology and of the geometrical properties of the reconstructed
events and are therefore based on data from the tracker. Additionally,
cuts are applied on data from the TOF system. However, substantial
parts of the analysis deal with measures to account for the
environmental circumstances under which the \AMS{} experiment was
operated, especially the effect of the geomagnetic field.

\subsection{Basic Cuts}
\label{section:basiccuts}
Several simple cuts are applied to the data in order to eliminate
wrongly reconstructed events and a large portion of the background:
\begin{itemize}
\item As mentioned in \S~\ref{subsection:trackfinding}, the 
reconstruction of tracks with momenta lower than 100\+\MeV{} is not
reliable. Tracks reconstructed with momenta below this value are
tagged by the track fit algorithm and events containing them are
rejected.
\item Due to the deflection in the magnetic field and in conjunction 
with the small emission angles, the charge signs of the secondaries
are exactly constrained depending on the direction of incidence. For
downward going particles the left track is required to have positive
charge sign, while the right track must be reconstructed with negative
charge sign. For upward going particles, these requirements are
reversed. Thus the charge sum of all three tracks must be
$\pm$1\+e.
\item With higher energies, the track momentum resolution and
the signal over background ratio deteriorate. Thus the total
reconstructed momentum must not exceed~50\+\GeV{}/c.
\end{itemize}

\begin{figure}[btf] 
\begin{center}
\includegraphics[width=14cm]{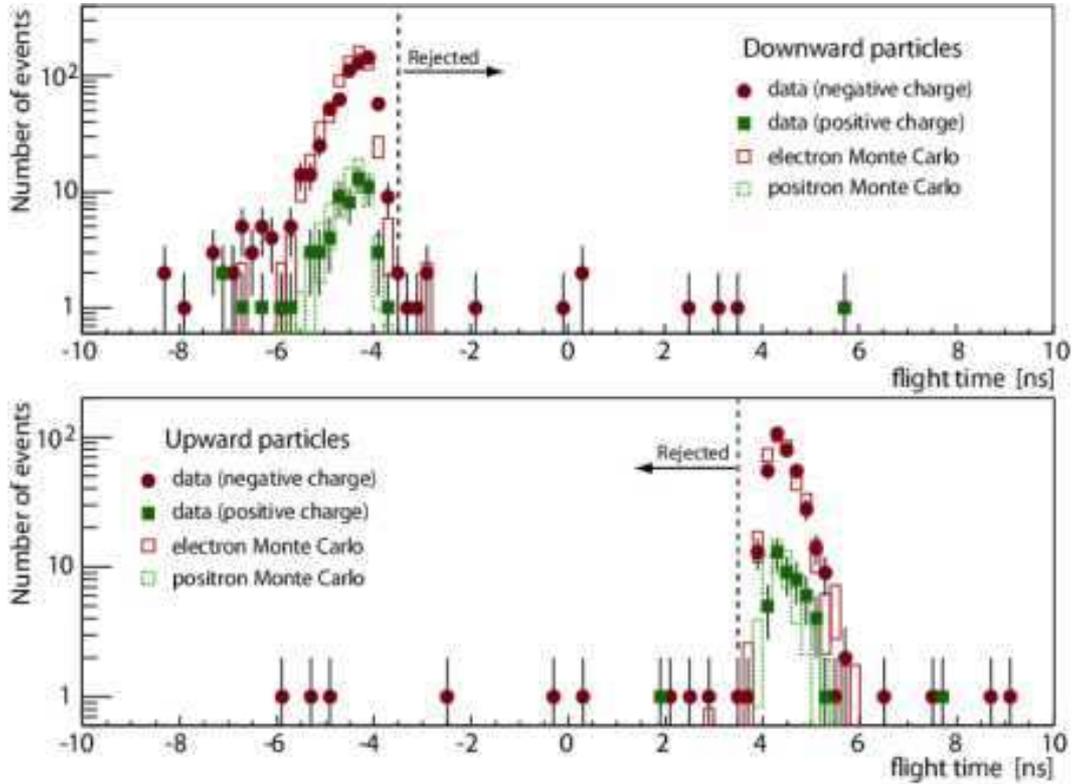}
\caption{\label{fig:toftime}Flight time along the z-axis measured with the TOF system
for downward {\sl (top)\/} and upward {\sl (bottom)\/} going electron
and positron candidates from data and Monte Carlo. The Monte Carlo distributions
are scaled to the data by integral.}
\end{center} 
\end{figure}
The requirement for increasing cluster distances within the seed
triplets along the flight path
(see \S~\ref{subsection:trackfinding}) largely distinguishes
between downward and upward going particles. To make sure the flight
direction is correctly recognized, timing information from the TOF
system is used. Figure~\ref{fig:toftime} shows the distributions of
flight times $t_f$ for downward and upward going particles calculated
according to
\begin{equation}\label{eq:timeofflight}
t_f = (t_1+t_2)/2 - t_3,
\end{equation}
where the $t_i$ denote the time of passage measured in TOF layer~$i$
($i$ is counted from top to bottom). Due to high voltage failures in
TOF layer~4~\cite{alvisi99a}, its timing information is not used. The
sign of $t_f$ in eq. (\ref{eq:timeofflight}) depends on the flight
direction. If it is not compatible with the flight direction derived
from the cluster topology, the event is rejected. The time needed by a
particle with $\beta\approx 1$ to pass the distance between the two
TOF stations is about 4.3\+ns. In rare cases particles are attributed
flight times considerably smaller than this. The reason for these
unphysical results is wrong reconstruction of the time of passage in
the scintillators. To account for the uncertainty of the time
measurements, $\left|t_f\right|$ is required to be larger than
3.5\+ns.

\begin{figure}[btf] 
\begin{center}
\includegraphics[width=14cm]{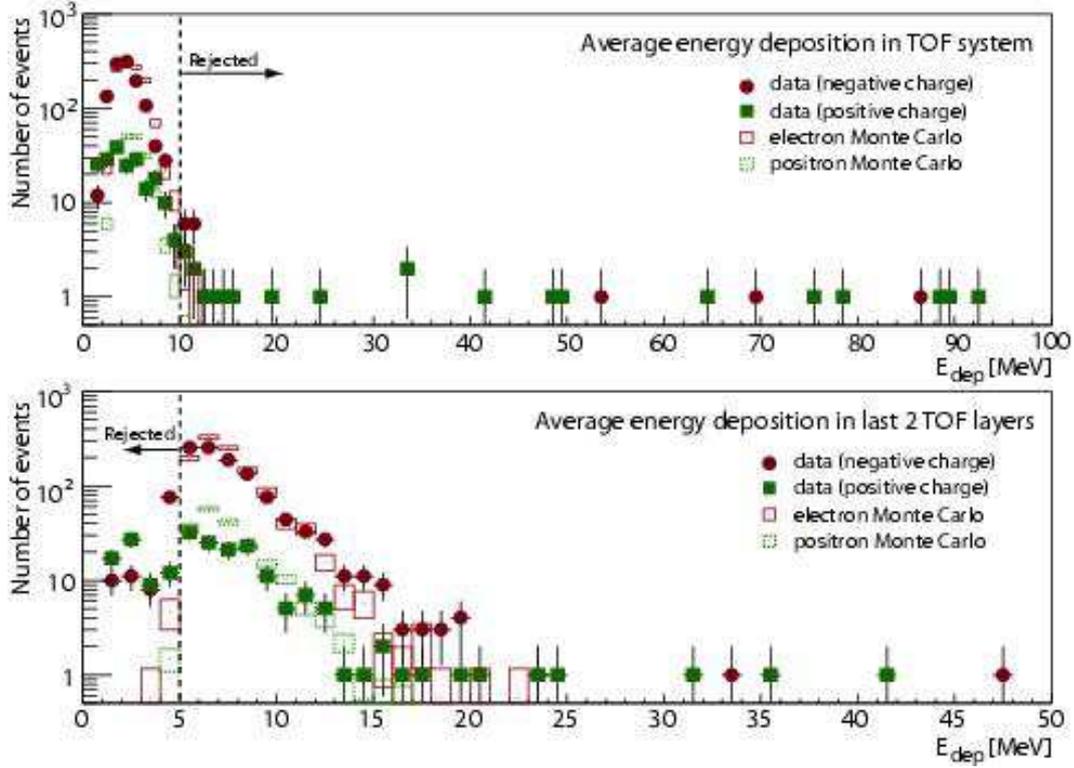}
\caption{\label{fig:tofedep} Average energy deposition in the TOF scintillators {\sl (top)\/}
and average energy deposition in the last two TOF layers in flight
direction {\sl (bottom)\/} for data and Monte Carlo electron and
positron candidates. The Monte Carlo distributions are scaled to the
data by integral.}
\end{center} 
\end{figure}
The energy loss $E_{\mathrm{dep}}$ in a TOF scintillator can be
calculated with the Bethe-Bloch equation~\cite{fano63a}. For an
electron or positron in the momentum range 1 to~50\+\GeV{}/c it
amounts to about 2.5\+\MeV{}, with fluctuations of about
0.7\+\MeV{}~\cite{bichsel88a}. Thus, to make sure that there are three
particles traversing the detector, consistent with the signature of a
converted bremsstrahlung photon, a minimum average energy loss is
required in the last two TOF layers in the direction of flight. The
distribution of this average is shown in Figure~\ref{fig:tofedep} {\sl
(bottom)}. To account for the fluctuations of the energy loss, events
with an average energy loss smaller than 5\+\MeV{} in the last two TOF
layers in the direction of flight are rejected.

Apart from protons, nuclei such as He or N have been observed to
induce background events through hadronic interactions in a few
cases. Such particles with $Z>1$ deposit significantly more energy in
the subdetectors than singly charged particles
(\mbox{$E_{\mathrm{dep}}>8$}\+\MeV{} for $^4$He in 1\+cm
scintillator). Thus they can be identified through the average energy
deposition in all TOF scintillators. The energy loss of an individual
particle, due to its stochastic nature, follows a Landau
distribution~\cite{yao06a}, which exhibits a long tail towards higher
values of $E_{\mathrm{dep}}$. To account for this effect, the
truncated mean of the energy deposition in the scintillators is
calculated by omitting the highest value. Figure~\ref{fig:tofedep}
{\sl (top)\/} shows the distribution of this truncated mean. Events
with a mean greater than 10\+\MeV{} are rejected.

\begin{figure}[hf] 
\begin{center}
\includegraphics[width=14cm]{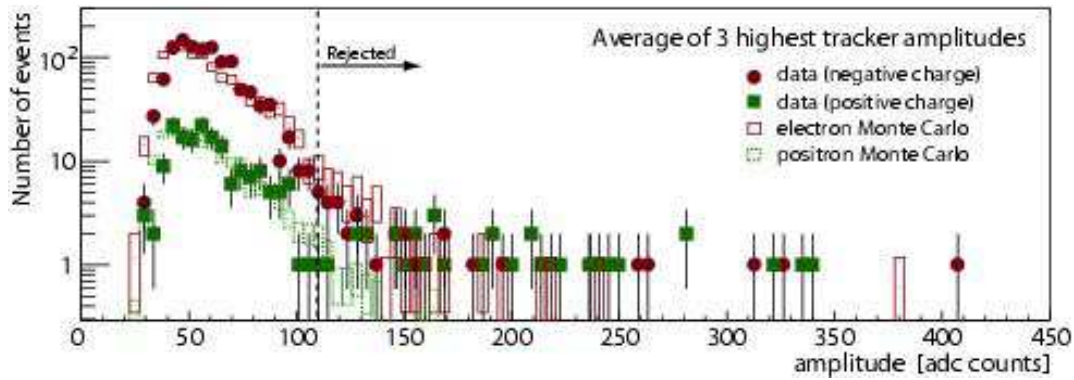}
\caption{\label{fig:trackeredep}Average of the three highest tracker
hit amplitudes for data and Monte Carlo electron and positron
candidates. The Monte Carlo distributions are scaled to the data by
integral.}
\end{center} 
\end{figure}
In addition to the TOF system, the energy deposition in the silicon
strip detectors can be used to discriminate against particles with
charge $q>1$. The amplitudes of the hits in the tracker are given in
ADC counts, thus the corresponding cut is determined from the electron
data sample. In Figure~\ref{fig:trackeredep} the distribution of the
average of the three highest tracker hit amplitudes, regardless of the
track to which they belong, is shown. The accumulation of positively
charged events at higher amplitude is caused mostly by helium
nuclei. Thus the average amplitude is required not to exceed 110 ADC
counts. In case of negative total charge, the events at higher
amplitude mostly result from wrongly reconstructed low momentum
background protons.

\subsection{Suppression of Background}\label{section:suppression}
For the suppression of background, use is made of the fact that
bremsstrahlung and photon conversion imply small opening angles of the
particles at the vertices (see \textsection~\ref{sec:eventsignature}),
while in hadronic reactions the particles are emitted under large
angles (see \textsection~\ref{sec:background}). However, the latter
are defined in the center-of-mass frame but observed in the laboratory
frame, whose relative speed varies with the particles' incident
momentum. Thus, in order to make the angles independent of the frame
of reference, the corresponding invariant mass is calculated according
to
\begin{equation}
m_{inv}^2 = 2 \cdot E_1 \cdot E_2 \cdot \left( 1 - \cos\theta \right),
\end{equation}
where $\theta$, $E_1$ and $E_2$ denote the opening angle and the
energies of the primary particle and the photon and of the conversion
pair, respectively.

\begin{figure}[t] 
\begin{center}
\includegraphics[width=10cm]{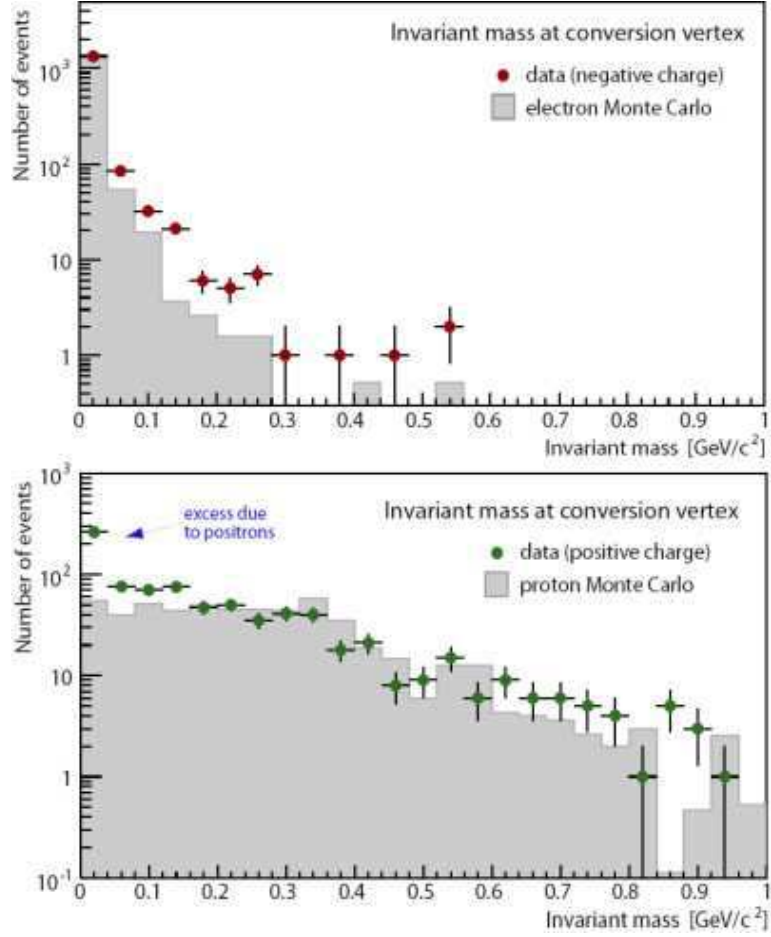}
\caption{\label{fig:conversionInvMass}The distribution of the
invariant mass at the conversion vertex for data {\sl (dots)\/} and Monte Carlo {\sl (shaded histograms)\/}:
data events with negative total charge compared to electron Monte Carlo {\sl (top)} and
data events with positive total charge compared to proton Monte Carlo {\sl (bottom)}.
The Monte Carlo distributions are scaled to the data by integral.} 
\end{center} 
\end{figure}
\begin{figure}[t]
\begin{center}
\includegraphics[width=10cm]{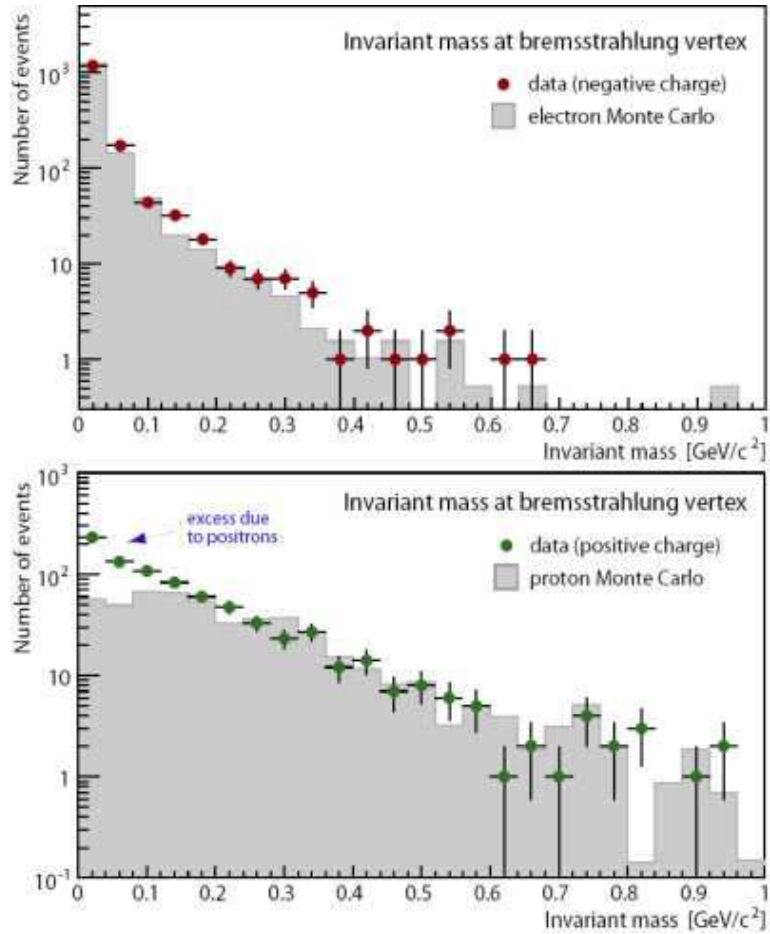}
\caption{\label{fig:bremsstrahlungInvMass}The distribution of the invariant mass at the 
bremsstrahlung vertex for data {\sl (dots)\/} and electron and proton Monte Carlo {\sl (shaded histograms)\/}:
data events with negative total charge compared to electron Monte Carlo {\sl (top)} and
data events with positive total charge compared to proton Monte Carlo {\sl (bottom)}.
The Monte Carlo distributions are scaled to the data by integral.}
\end{center} 
\end{figure}
The distributions of invariant mass at the conversion and
bremsstrahlung vertex are shown for data and electron and proton Monte
Carlo in Figures~\ref{fig:conversionInvMass} and
\ref{fig:bremsstrahlungInvMass}. For events with negative total 
charge, which represent a largely clean electron sample, they reveal a
narrow shape with a peak at zero. This is in agreement with
Monte Carlo results. In case of events with positive total charge,
consisting of positrons and background, the distributions also show a
peak at zero, but have a long tail towards higher invariant
masses which is largely caused by the proton background. The excess
due to the positrons is clearly visible near zero invariant mass.

\begin{figure}[t]
\begin{center}
\includegraphics[width=14cm]{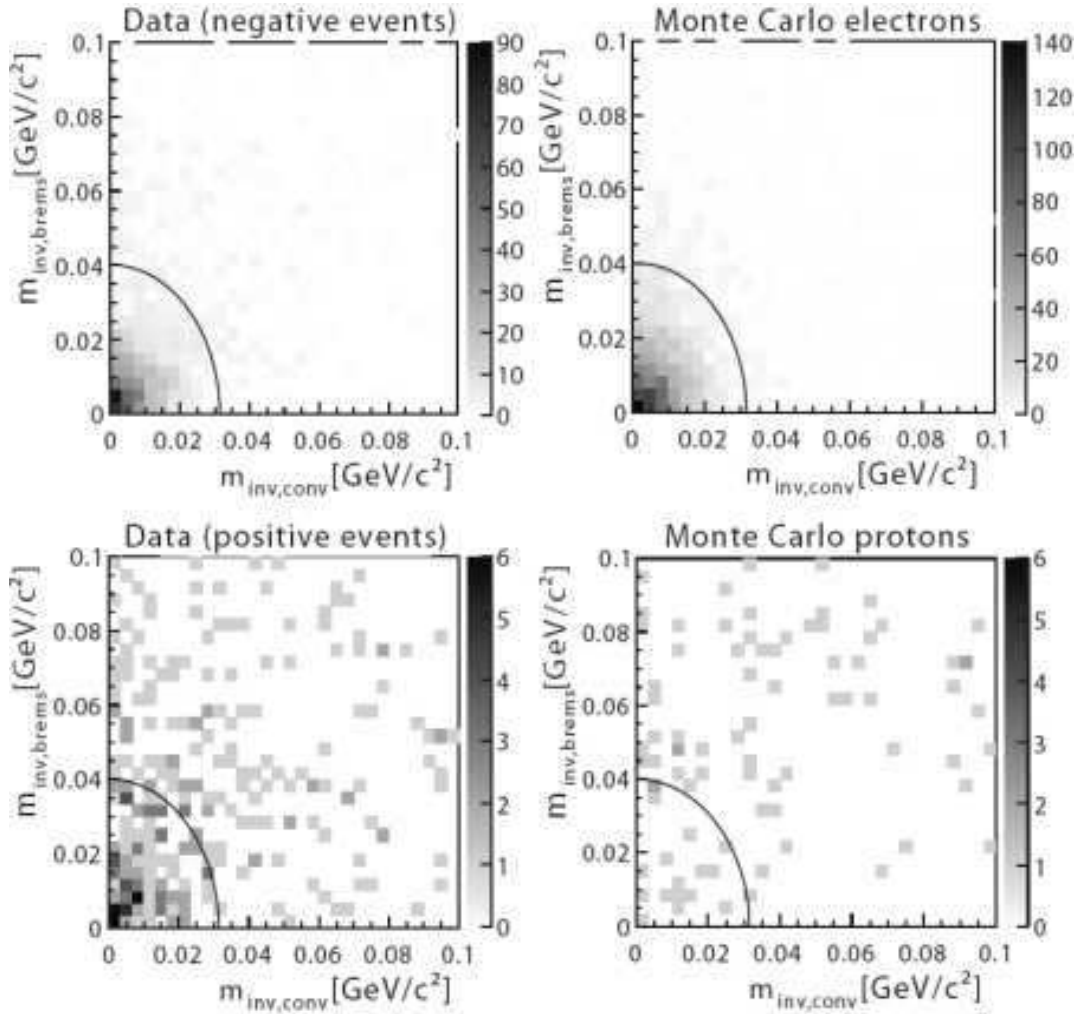}
\caption{\label{fig:invMassCanvases}Distributions in the invariant mass plane for data
events with negative ({\sl top left}\/) and positive ({\sl bottom
left}\/) total charge and for Monte Carlo electron ({\sl top right}\/)
and proton ({\sl bottom right}\/) events.}
\end{center} 
\end{figure}

Figure~\ref{fig:invMassCanvases} shows the distributions for data and
for Monte Carlo proton and electron events in the invariant mass
plane. In agreement with Monte Carlo, data events with negative total
charge (electrons) accumulate around zero invariant masses. Caused by
the positrons, such an accumulation is also observed in the
distribution of positively charged events, where the uniformly
distributed proton background indicated by the corresponding Monte
Carlo events is also seen.
In order to discriminate against background events, restrictions are
applied on the invariant masses. These cuts are parameterized as
ellipses in the invariant mass plane, centered at zero and with half
axes in units of the standard deviation of the respective distribution
in the electron data sample. Events outside the ellipses are
rejected. In order to keep the positron selection efficiency high,
both cut values have been set to $2\+\sigma$.

\subsection{Effect of the Mir Space Station}
As stated in \S~\ref{section:shuttlemission}, parts of the Mir
space station were within the \AMS{} field of view during the four day
Mir docking phase of the flight. Cosmic rays interacting with the Mir
generated secondary particles~\cite{aguilar05a}, some of which may
also have traversed the \AMS{} detector. Not being of cosmic origin,
these secondaries must be rejected in the analysis.
\begin{figure}[htb]
\begin{center}
\includegraphics[width=15.5cm]{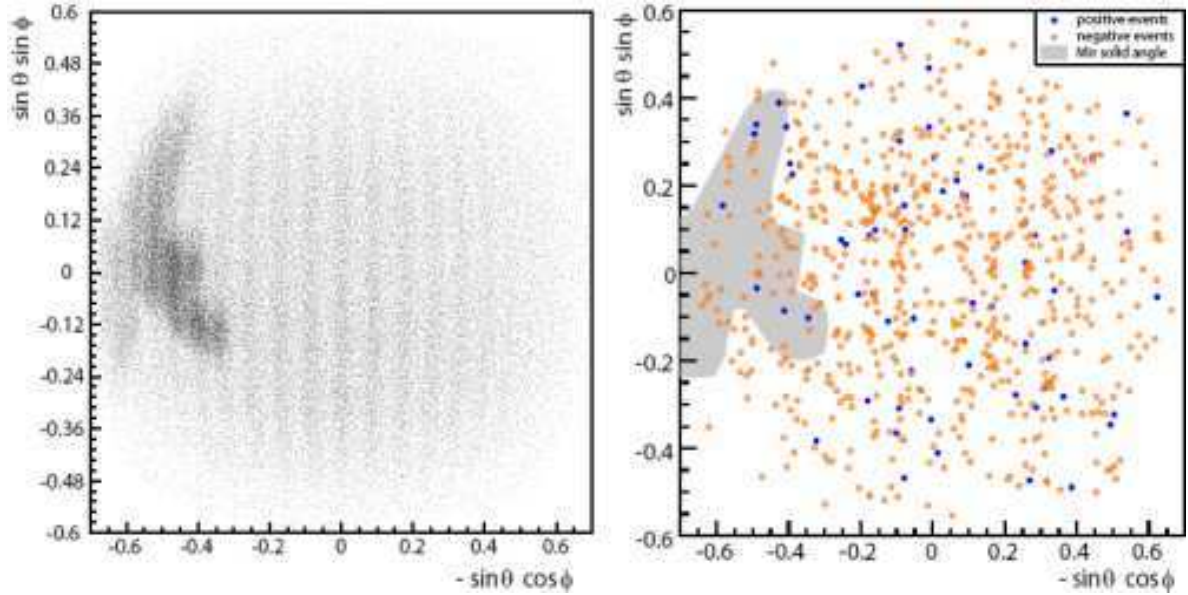}
\caption{\label{fig:mirProjection}Incoming directions projected 
on a x-y plane above the detector for downward going particles with
$Z=-1$ recorded during the Mir docking phase. The vertical stripes are
due to reconstruction artifacts {\sl (left)\/}~\cite{aguilar05a}; the
same display for downward going electron and positron candidates from
this analysis; also indicated is the solid angle obstructed by the Mir
{\sl (right)\/}.}
\end{center} 
\end{figure}

Figure~\ref{fig:mirProjection} {\sl (left)\/} displays the incoming
directions of downward going particles with $Z=-1$, projected on an
x-y-plane above the detector, which were recorded during the Mir
docking period~\cite{aguilar05a}. $\theta$ and $\phi$ denote the polar
and azimuthal angles of the incoming particles. There is an apparent
excess of particles in a clearly defined area on the left side of the
plot. It has been shown~\cite{aguilar05a} that these excess particles
originated in the Mir and were mostly spallation pions and muons from
their decay. Hence, since electrons and positrons can also result
from pion decay, the Mir must be regarded as a possible background
source.

In Figure~\ref{fig:mirProjection} {\sl (right)\/}, the same display is
shown for downward going electron and positron candidates from this
analysis, together with the solid angle obstructed by the Mir. No
significant excess of particles is observed in the affected solid
angle. However, particles in the Mir solid angle that were recorded
during the docking phase were not considered in this analysis.

\subsection{Geomagnetic Cutoff}\label{subsection:cutoff}
Energy spectra of cosmic rays are modulated by the geomagnetic field
(see \S~\ref{section:earthsVicinity}). Depending on the direction
of incidence and the geomagnetic coordinates of the entry point into
the magnetosphere, particles with momenta below a certain cutoff are
deflected by the magnetic field and cannot reach the Earth's
proximity. Hence, below geomagnetic cutoff the particles detected by
\AMS{} must originate from within the magnetosphere. They were mostly
produced as secondaries through hadronic interactions and trapped
inside the Earth's radiation belts.

To discriminate against these secondaries, particle trajectories were
individually traced back from their measured point of incidence, angle
and momentum through the geomagnetic field by numerical integration of
the equation of motion using an enhanced Runge-Kutta iteration
procedure~\cite{flueckiger90a}. For this purpose, a reference model of
the Earth's external field~\cite{tsyganenko89a} was used in
combination with the IGRF-6 coefficients~\cite{iaga06a} of the
internal field according to eq. (\ref{equation:sphericalharmonics}).

\begin{figure}[htb]
\begin{center}
\includegraphics[width=15cm]{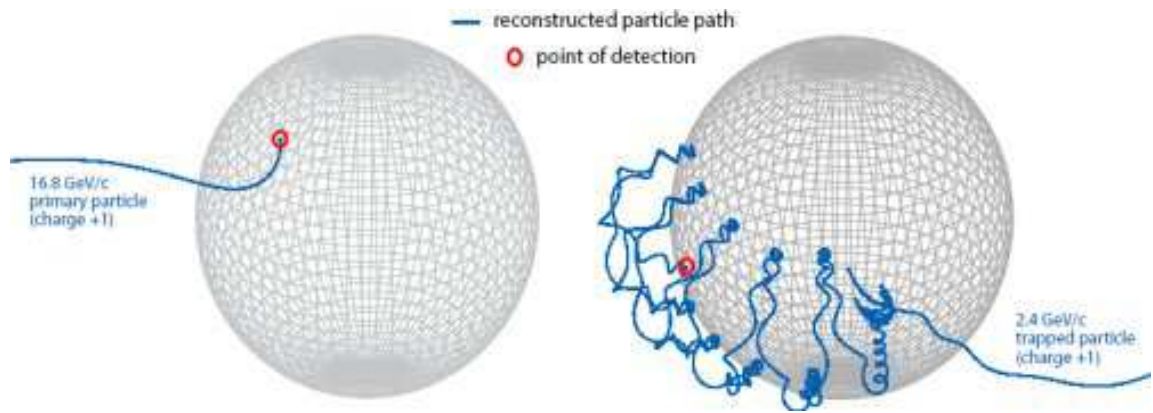}
\caption{\label{fig:geotracing}Reconstructed paths of positron candidates
in the Earth's magnetic field. The left display shows a 16.8\+\GeV{}/c 
primary particle, the right one a 2.4\+\GeV{}/c trapped particle which is
rejected as being of secondary origin.}
\end{center} 
\end{figure}
Figure~\ref{fig:geotracing} illustrates the flight paths of particles
with charge $+1e$ by means of two examples. The left display shows the
reconstructed path of a particle with a momentum of 16.8\+\GeV{}/c, as it
approached the Earth from outside the magnetosphere and was detected
by the \AMS{} experiment over central Northern America. Apparently,
this particle is of extraterrestrial origin and thus accepted for
further analysis. On the right side of Figure~\ref{fig:geotracing}, the
path of a trapped secondary particle with 2.4\+\GeV{}/c momentum is shown
which exhibits the typical superposition of a helix trajectory with
hemisphere bouncing and westward drift. It is certain that this
particle stems from an interaction with the atmosphere near one of the
numerous turning points of its trajectory, thus the early part of the
displayed path is purely hypothetical. Consequently, the particle is
not considered in the analysis. In general, particles are rejected as
secondaries if their trajectory touched the surface of the Earth at
least once. Particles always staying within a distance of 25 Earth
radii or not crossing the magnetopause are considered as trapped and
are also rejected.

\section{Correction for Irreducible Background}
\label{section:correction}
As seen in Figure~\ref{fig:invMassCanvases}, the distribution of protons
in the invariant mass space does not vanish in the signal region
around zero invariant mass. The same applies to the background from
wrongly reconstructed electrons. Consequently, a small fraction of
background events will not be rejected by the cut on the invariant
masses. This remaining irreducible background has to be corrected for,
which is accomplished by using Monte Carlo data.

In principle the approach is to run the analysis on an adequate number
of proton and electron Monte Carlo events as if they were data,
determine the amount and momentum distribution of particles that are
misidentified as positrons, and subtract these from the raw positron
counts obtained from data. However, such an implicit comparison of
Monte Carlo and data requires the adjustment of several properties of
the simulated events, which have not been affected by the geomagnetic
field and whose input spectrum is not exactly equal to the true
fluxes.

As discussed in \S~\ref{section:analysis}, the geomagnetic field
shields the Earth's vicinity from low energy particles. However, the
geomagnetic cutoff cannot be calculated individually for Monte Carlo
particles, since their four vector is not defined with respect to the
geomagnetic coordinates. To correct for the shielding effect, the
livetime function $T(p)$, described in
\S~\ref{section:fluxcalculation} and shown in
Figure~\ref{fig:livetime}, is used. The livetime function gives the
effective measurement time as a function of momentum for singly
charged particles. After normalization to a maximum value of 1,
resulting in the function $\hat{T}(p)$, its value at a given momentum
denotes the probability for a singly charged particle to penetrate the
geomagnetic field. Hence, it serves as a weight for distributions of
any event variable in Monte Carlo, particularly for the momentum
distribution of Monte Carlo background events. The livetime function
must be evaluated using the reconstructed momentum of the incident
particle to treat Monte Carlo and real data on an equal footing.

The incident momentum spectrum of the Monte Carlo particles follows a
distribution $\phi_{MC}(p) = p^{-1}$ and therefore differs
significantly from the true spectra which have spectral indices
between 2.7 and 3.4 and are affected by solar modulation. Therefore,
the Monte Carlo event variables have to be reweighted, since they are
correlated with the incident momentum. Using the parameterized flux
$\phi_D(p)$ of protons measured by \AMS{}~\cite{aguilar02a}, the
spectral reweighting function $w(p)$ is calculated as
$w(p)=\phi_D(p)/\phi_{MC}(p)$. In contrast to $\hat{T}(p)$, $w(p)$
must be evaluated using the incident particle's simulated momentum
and is not normalized.

The livetime function as well as the spectral reweighting function are
then applied as weights in histogramming an event variable $x$ from
the proton Monte Carlo sample, meaning that
$w(p_{\mathrm{s}})\cdot\hat{T}(p_{\mathrm{r}})\cdot x$ is histogrammed
instead of $x$, with the simulated momentum $p_{\mathrm{s}}$ and the
reconstructed momentum $p_{\mathrm{r}}$. In particular, the functions
correct for the shape of the momentum distribution of background
protons calculated from Monte Carlo. Subsequently, since $w(p)$ is not
normalized, the background distribution must be scaled to the data
to obtain the total number of background events to be subtracted from
the measured number of positrons.
\begin{figure}[!h]
\begin{center}
\includegraphics[width=12cm]{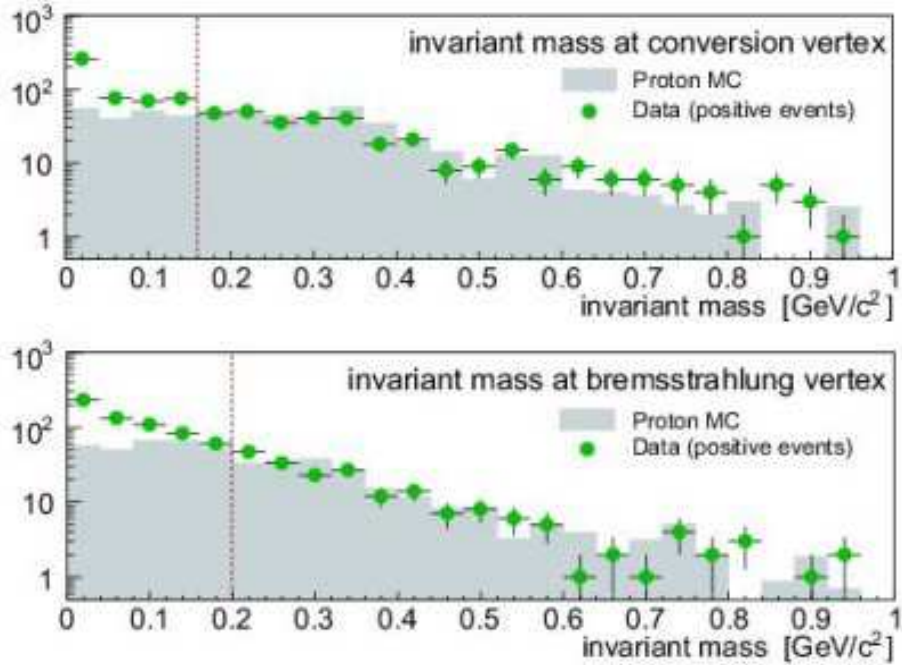}
\caption{\label{fig:sidebandScaling}Invariant mass distributions at the
conversion {\sl (top)\/} and bremsstrahlung vertex {\sl (bottom)\/}
for positively charged events from data and proton Monte Carlo. The
Monte Carlo distributions have been scaled to the data using the
sidebands whose borders are indicated by the vertical lines.}
\end{center} 
\end{figure}
\begin{figure}[!h]
\begin{center}
\includegraphics[width=10cm]{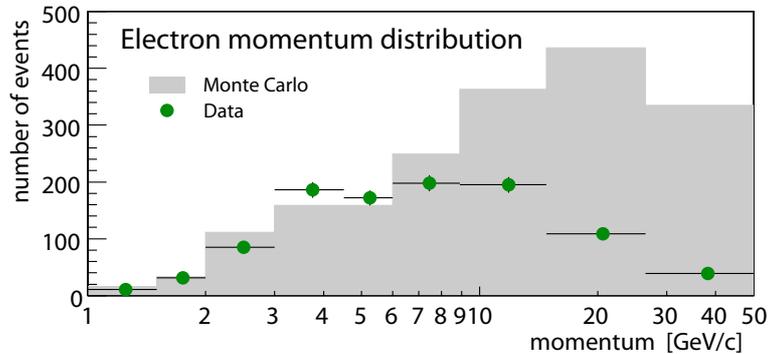}
\caption{\label{fig:electronMomDist}Distribution of the reconstructed momentum 
of electrons from data and Monte Carlo. The latter have been corrected
for the livetime $\hat{T}(p)$ but not for the distribution of momentum.}
\end{center} 
\end{figure}

Figure~\ref{fig:sidebandScaling} illustrates the scaling of the proton
Monte Carlo to the positively charged data events using the sidebands
of the invariant mass distributions. The sidebands are defined as the
ranges of invariant mass above certain thresholds in which the
positron contribution to the data is negligible. The thresholds are
determined from the electron distribution, which is identical to that
of positrons. They are set to 0.16\+\GeV{}/c$^2$ for the conversion
vertex and to 0.2\+\GeV{}/c$^2$ for the bremsstrahlung vertex. The
excess in the data due to the positron contribution is apparent below
threshold. Two scaling factors, one from each of the distributions at
the two vertices, are determined as the sideband ratios of the number
of events from data and Monte Carlo. The average of these two factors,
which differ less than 8\+\% from each other, is then applied to the
reweighted Monte Carlo proton background distribution, resulting in
the final proton background correction.

In contrast to protons, the electron data sample is clean, without
contamination from other particles. Therefore, it can be used directly
to reweight the spectrum of the electron Monte Carlo events in order
to estimate the background in the positron data resulting from wrongly
reconstructed electrons. Figure~\ref{fig:electronMomDist} shows the
distributions $D_{e^-\!,\,\mathrm{d}}(p)$ and
$D_{e^-\!,\,\mathrm{mc}}(p)$ of the measured momenta of electrons from
data and Monte Carlo respectively, where the latter have been weighted
with the normalized livetime function $\hat{T}(p)$. In a first step,
the electrons from Monte Carlo that have been wrongly reconstructed as
positrons are grouped in nine bins according to their simulated
momentum following the binning in Figure~\ref{fig:electronMomDist}. For
each group, the distribution of measured momenta is separately
histogrammed. Finally, each of these histograms is scaled with the
ratio $D_{e^-\!,\,\mathrm{d}}/D_{e^-\!,\,\mathrm{mc}}$ in the
respective bins. The sum of the nine resulting histograms then gives
the amount and distribution of electron background events to be
subtracted from the positron sample.

\begin{figure}[!h]
\begin{center}
\includegraphics[width=10cm]{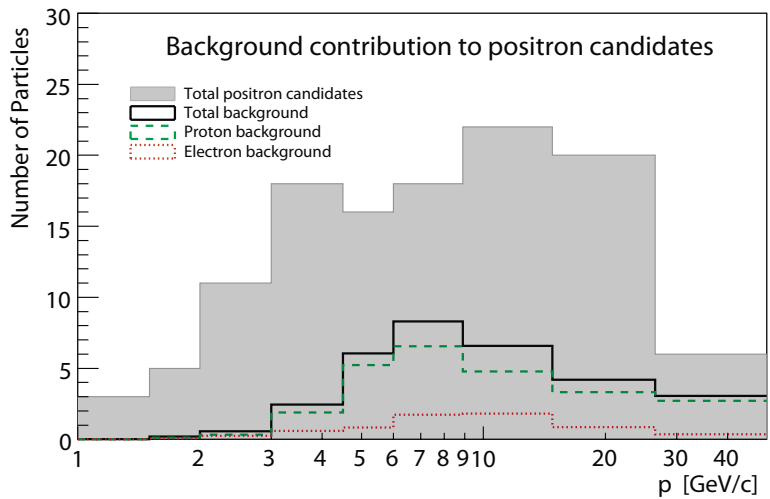}
\caption{\label{fig:subtractedBackground}Momentum distribution of positron candidates 
including background, and the background contribution from protons and
wrongly reconstructed electrons.}
\end{center} 
\end{figure}

In total, 1026 electrons and 119 positrons have been found prior to
the background correction in this
analysis. Figure~\ref{fig:subtractedBackground} shows the momentum
distribution of the uncorrected positron sample and the total
background correction as a function of momentum, itemized into
contributions from protons and wrongly reconstructed electrons, which
amount to 24.9 and 6.4 events, respectively.

\chapter{Results of the Positron Measurement}
\section{Positron Fraction}
\label{section:fraction}
\begin{figure}[t]
\begin{center}
\includegraphics[width=12cm]{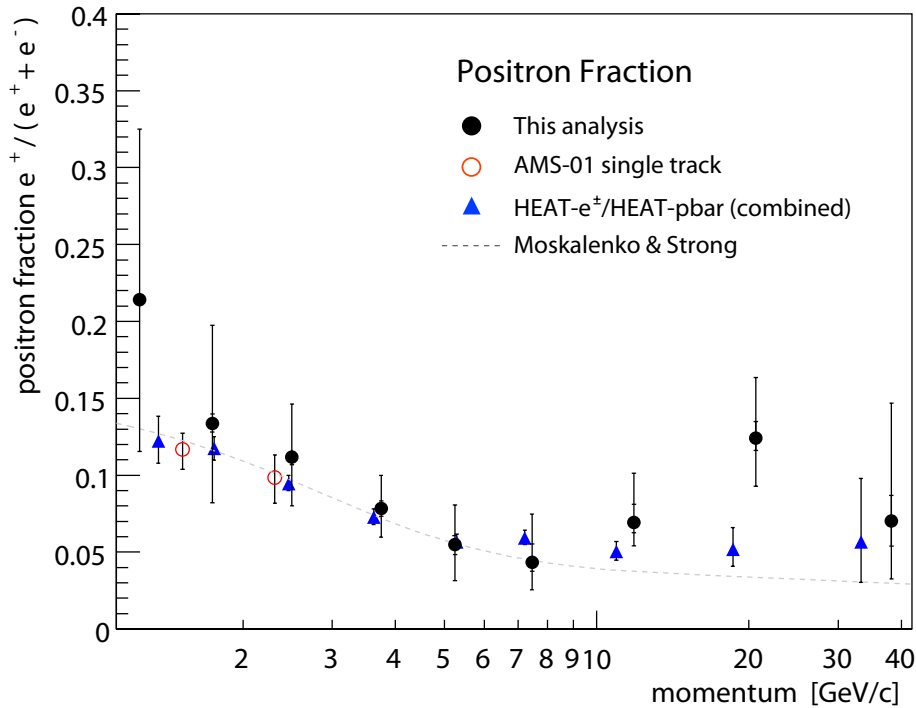}
\caption{\label{fig:fraction} The positron fraction $e^+/(e^{+}+e^-)$
measured in this analysis, compared with earlier results from
\AMS{}~\protect\cite{alcaraz00d}, the combined results from
HEAT-$e^{\pm}$ and HEAT-pbar~\protect\cite{beatty04a} and with a model
calculation for purely secondary positron
production~\protect\cite{moskalenko98a} (dashed line). The overall
errors are given by the outer error bars, while the inner bars
represent the systematic contributions $\sigma_{\mathrm{sys}}$ to the
error.}
\end{center} 
\end{figure}

The positron fraction $e^+/(e^{+}+e^-)$ is calculated from the
electron and corrected positron counts for each momentum bin. It is
shown in Figure~\ref{fig:fraction} and compared with earlier results and
with a model calculation based on purely secondary positron
production. Table~\ref{table:fraction} summarizes the results of this
analysis. The contributions to the error on the positron fraction are
discussed in the following.
\begin{table}[hb]
\begin{center}
\begin{tabular}{c c | c c c | c c c c}
\hline
\hline
momentum & range & \LL{$n_{e^-}$} & \LL{$n_{e^+\!,\,r}$} & \LL{$n_{e^+\!,\,c}$} & positron & \LL{$\sigma_{\mathrm{stat}}$} & \LL{$\sigma_{\mathrm{sys,b}}$} & \LL{$\sigma_{\mathrm{sys,a}}$}\STRUT  \\
 \GeV/c & \GeV/c & & & & fraction & & & \\ 
\hline
1.25 & 1--1.5      & 11  & 3  & 3    &	0.21    &    $_{-0.1} ^{+0.11}$    &   $\pm 0$               &   $\pm 0$\STRUT      \\
1.75 & 1.5--2      & 31  & 5  & 4.8  &	0.133   &    $_{-0.051}^{+0.064}$  &   $_{-0}^{+0.002}$      &   $\pm 0.006$\STRUT  \\
2.5  & 2--3        & 85  & 11 & 10.7 &	0.112   &    $_{-0.031}^{+0.034}$  &   $_{-0.003}^{+0.001}$  &   $\pm 0.004$\STRUT  \\
3.75 & 3--4.5      & 186 & 18 & 15.8 &	0.078   &    $_{-0.018}^{+0.021}$  &   $_{-0.003}^{+0.001}$  &   $\pm 0.004$\STRUT  \\
5.25 & 4.5--6      & 172 & 16 & 10   &	0.055   &    $_{-0.022}^{+0.025}$  &   $_{-0.007}^{+0.006}$  &   $\pm 0.001$\STRUT  \\
7.45 & 6--8.9      & 198 & 18 & 9    &	0.043   &    $_{-0.017}^{+0.029}$  &   $_{-0.004}^{+0.01}$   &   $\pm 0.004$\STRUT  \\
11.9 & 8.9--14.8   & 195 & 22 & 14.5 &	0.069   &    $_{-0.014}^{+0.03}$   &   $_{-0.002}^{+0.01}$   &   $\pm 0.006$\STRUT  \\
20.6 & 14.8--26.5  & 109 & 20 & 15.4 &	0.124   &    $_{-0.03}^{+0.038}$   &   $_{-0.003}^{+0.009}$  &   $\pm 0.007$\STRUT  \\
38.2 & 26.5--50    & 39  & 6  & 2.9  &	0.07    &    $_{-0.034}^{+0.075}$  &   $_{-0.01 }^{+0.01}$   &   $\pm 0.007$\STRUT  \\
\hline \\
\end{tabular}
\caption{\label{table:fraction}The positron fraction and its momentum dependence.
$n_{e^-}$, $n_{e^+\!,\,r}$ and $n_{e^+\!,\,c}$ denote the number of
electrons, raw (uncorrected) and corrected positrons,
respectively. The last three columns give the statistical errors
$\sigma_{\mathrm{stat}}$ and the systematic errors
$\sigma_{\mathrm{sys,b}}$ and $\sigma_{\mathrm{sys,a}}$ due to
background and acceptance correction.}
\end{center}
\end{table}

\subsection{Statistical Errors}
\begin{figure}[t]
\begin{center}
\includegraphics[width=15cm]{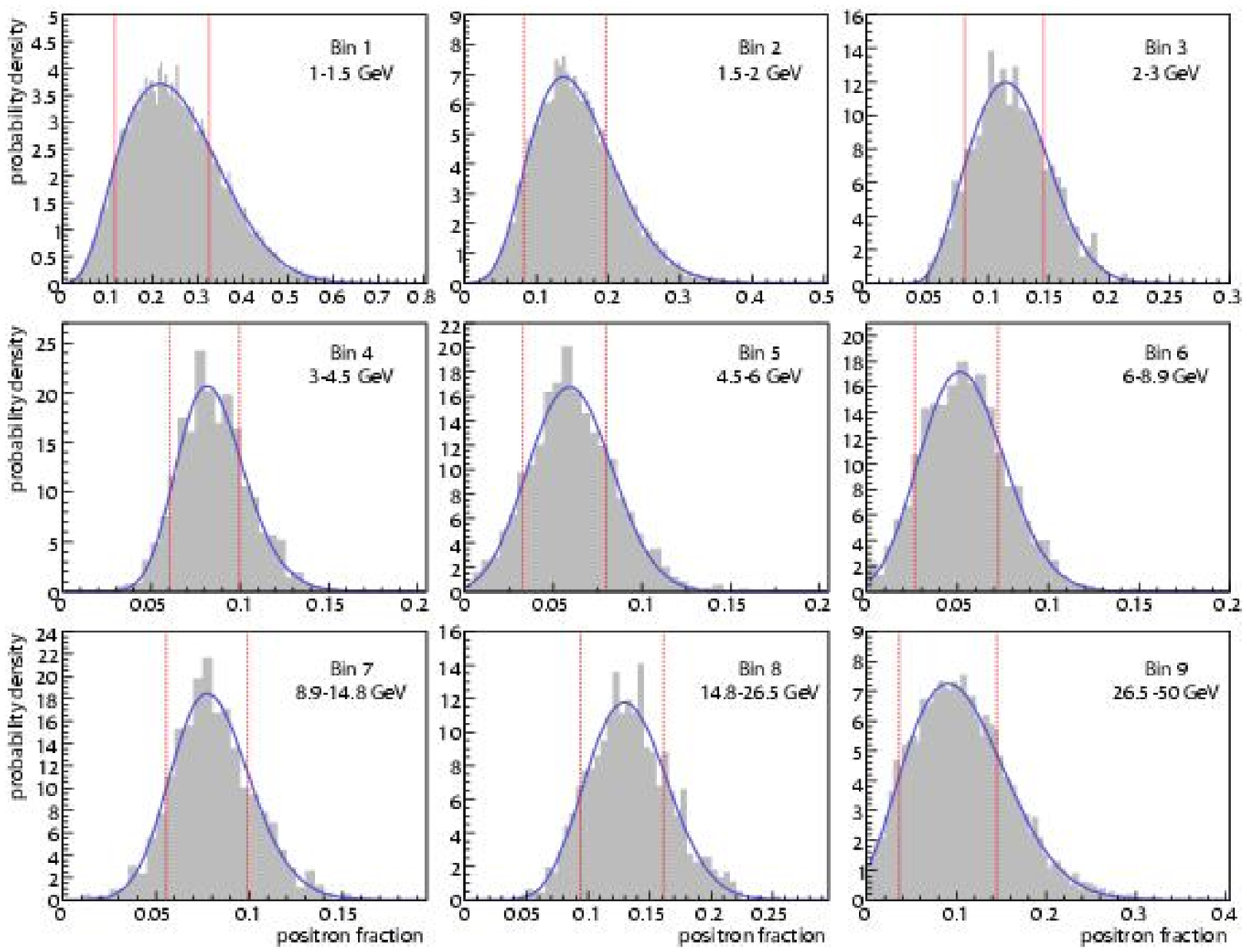}
\caption{\label{fig:statErrProbDensity}The normalized simulated distributions 
of the positron fraction $f_{\mathrm{true}}$ (shaded histograms), the
fitted functions $D_f$ (blue solid lines), and the resulting
confidence interval limits (red dashed lines) for each of the nine
momentum bins.}
\end{center} 
\end{figure}
Due to the complexity of the positron fraction analysis, having two
sources of irreducible background and low statistics, a Bayesian
approach based on Monte Carlo simulation has been chosen for the
determination of the statistical
errors~\cite{helene83a,helene84a}. The aim is to obtain the
probability distribution of all possible values of the positron
fraction which, together with the background, can lead to the observed
number of particle counts. Given this distribution, the
confidence intervals are derived by numerical integration. The Monte
Carlo procedure outlined below is carried out separately for each of
the momentum bins defined in Table~\ref{table:fraction}.

\begin{itemize}
\item In a first step, two random floating point numbers 
$S_{e^-\!,\,\mathrm{true}}$ and $S_{e^+\!,\,\mathrm{true}}$ are
generated by sampling a uniform distribution. They represent the
expectation values for the probability distributions of ``true''
numbers of electrons and positrons without background.
\item Subsequently, the number of background events are added to
$S_{e^+\!,\,\mathrm{true}}$. The corrected and scaled background
counts $B_p$ and $B_{e^-}$ from proton and electron Monte Carlo, as
displayed in Figure~\ref{fig:subtractedBackground}, are the result of a
counting experiment and therefore have statistical uncertainties,
which transform into systematic errors when added to the ``true''
signal. These errors follow Poisson distributions, whose expectation
values are, to best knowledge, assumed to be equal to $B_p$ and
$B_{e^-}$. Hence, to account for their uncertainty, $B_p$ and
$B_{e^-}$ are smeared according to the individual distributions before
adding them to $S_{e^+\!,\,\mathrm{true}}$. However, the width of the
Poisson distributions is determined by the statistics of the raw
number of Monte Carlo background events considered and not by $B_p$
and $B_{e^-}$, which are the result of correcting and scaling the raw
counts to the data. As a consequence, the raw Poisson width is altered
by the corresponding scale factor. To account for this effect, the raw
background counts are first smeared according to their respective
Poisson distribution and then multiplied with the ratio of corrected
to raw counts. The resulting numbers of background events are denoted
by $\hat{B}_p$ and $\hat{B}_{e^-}$.
\item The background counts thus obtained are added to the true number of
positrons, resulting in an expectation value
$T_+=S_{e^+\!,\,\mathrm{true}} + \hat{B}_p + \hat{B}_{e^-}$ for the
measured total number of positively charged particles.  On the other
hand, the electron sample is largely background-free and
$T_-=S_{e^-\!,\,\mathrm{true}}$.
\item Two integer random numbers $N_+$ and $N_-$ are finally generated 
according to Poisson distributions with expectation values $T_+$ and
$T_-$, respectively. These represent one simulated measurement of the
number of positron and electron candidates including background. If
$N_+$ and $N_-$ are exactly equal to the counts actually observed in
the experiment, the ``true'' positron fraction $f_{\mathrm{true}} =
S_{e^+\!,\,\mathrm{true}} / ( S_{e^+\!,\,\mathrm{true}} +
S_{e^-\!,\,\mathrm{true}})$ is accepted for further analysis, and the
whole procedure is repeated.
\end{itemize}

Figure~\ref{fig:statErrProbDensity} shows the distributions of the
simulated positron fraction $f_{\mathrm{true}}$ for each of the nine
momentum bins. Normalized to an integral of 1, they represent the
probability density functions of the positron fraction measured in
this analysis, taking into consideration the amount of background
calculated from Monte Carlo. As apparent in the figure, the
distributions are in all cases parameterized well by a function $D_f =
G(f)\cdot L(f)$, the product of a Gaussian and a Landau distribution
function $G(f)$ and $L(f)$, respectively. The slight asymmetries,
which become manifest in the tails towards higher values of
$f_{\mathrm{true}}$ and make the Landau function $L(f)$ necessary, are
due to the statistical uncertainty of the background correction.

By numerical integration, the smallest interval is found over which
the integral of the function $D_f$ equals 0.683, yielding the lower
and upper limits of the corresponding 1\+$\sigma$ Gaussian confidence
intervals for the appropriate momentum bin. These limits are stated as
the statistical errors on the positron fraction and are given in
Table~\ref{table:fraction}.

\subsection{Systematic Errors}
Since the positron fraction is a ratio of particle fluxes, most
sources of systematic error such as detector acceptance or trigger
efficiency naturally cancel out. The only sources of error that must
be considered are those which are asymmetric with respect to the
particle charge.

The background correction is applied to the sample of positron
candidates only and is therefore a source of systematic error. To a
certain degree, the description of the experimental setup may be
inaccurately implemented in the Monte Carlo program. Furthermore, in
contrast to the production of charged pions, background processes
involving neutral pion production imply photoconversion with typically
low angles between tracks emerging from the vertices. Hence, the
distribution of invariant masses depends on the angular cross sections
of charged and neutral pion production. Possible inaccuracies in the
implementation of the cross sections in the Monte Carlo program must
therefore be considered.

\begin{figure}[htb]
\begin{center}
\includegraphics[width=16cm]{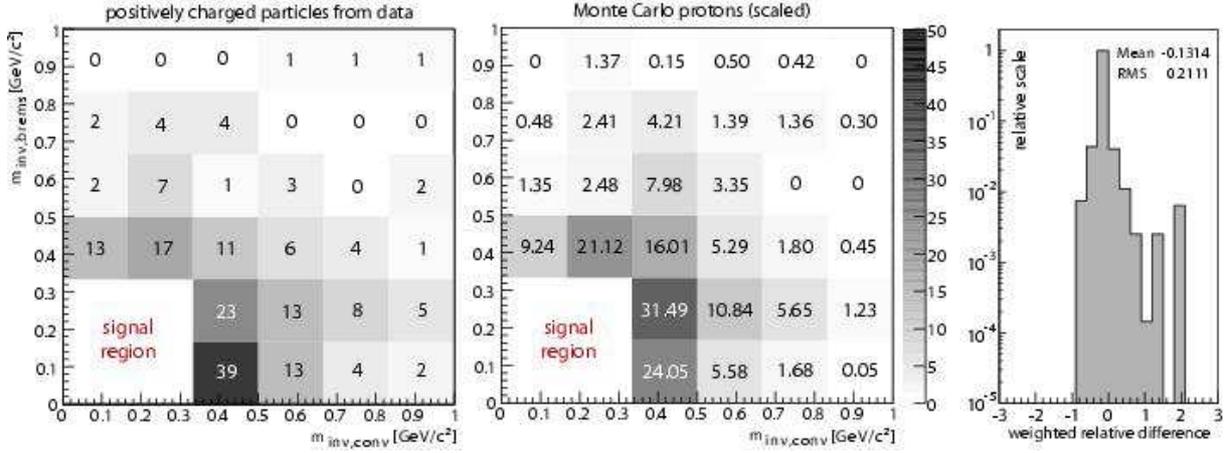}
\caption{\label{fig:protonSystematics}The distribution of particles
in the invariant mass space outside the positron signal region for
positively charged particles from data {\sl (left)\/} and scaled Monte
Carlo protons {\sl (middle)\/}. The numbers in the histograms denote
the particle counts in the respective bins; the distribution of the
weighted relative difference of the background in data and Monte Carlo
{\sl (right)\/}.}
\end{center} 
\end{figure}

The systematic error $\sigma_{\mathrm{sys,b}}$ from the background
correction can be estimated by evaluating the mean deviation of the
scaled Monte Carlo background from the data in the invariant mass
plane. Figure~\ref{fig:protonSystematics} shows the number density of
positively charged particles from data $n_{\mathrm{data}}$ {\sl
(left)\/} and Monte Carlo protons $n_{\mathrm{mc}}$ {\sl (middle)\/}
in the region of invariant masses $m_{inv}>0.33$\+\GeV/c$^2$ at both
vertices where the positron signal is negligible. The Monte Carlo
proton histogram has been scaled to the data using the ratio $\sum
n_{\mathrm{data}}/\sum n_{\mathrm{mc}}$ in the bins outside the signal
region. The histogram in the right panel of
Figure~\ref{fig:protonSystematics} shows the distribution of relative
differences $\delta_{\mathrm{r}}$ of data and Monte Carlo bin contents
given by
\begin{equation}
\delta_{\mathrm{r}} = w\cdot \frac{n_{\mathrm{data}} - n_{\mathrm{mc}}}{n_{\mathrm{mc}}}
\end{equation}
for $n_{\mathrm{mc}}\neq 0$ with weights $w$. Although the binning has
been chosen rather coarse to minimize statistical fluctuations, the
remaining statistical uncertainty is accounted for by weighting each
entry with the inverse of the relative statistical error
$w=n_{\mathrm{t}}/\sigma_{\mathrm{t}}$ with $w\approx
\sqrt{n_{\mathrm{t}}}$ in the Poisson limit and $n_{\mathrm{t}} =
(n_{\mathrm{data}} + n_{\mathrm{mc}})/2$. Finally, for display
purposes, $\delta_{\mathrm{r}}$ has been scaled to a maximum value
of~1. The RMS deviation of $\delta_{\mathrm{r}}$ leads to a systematic
error estimate of about 20\+\% of the background events. This value is
then propagated to the positron fraction for each momentum bin,
resulting in the asymmetric values of $\sigma_{\mathrm{sys,b}}$ as
given in Table~\ref{table:fraction}. No separate systematic error for
the background from wrongly reconstructed electrons is calculated,
since their contribution to the total background is small compared to
that from protons.

As a consequence of the East West Effect~\cite{johnson34a}, in
combination with the asymmetric layout of the \AMS{} tracker, the
product of the detector acceptance and the livetime as functions of
the direction of incidence of a particle may vary for positrons and
electrons, though no deviation of their average livetimes is apparent
(see section~\ref{section:fluxcalculation}). This effect is accounted
for with a second contribution $\sigma_{\mathrm{sys,a}}$ to the
systematic error of the positron fraction. It is estimated from the
mean variation of the difference in livetime of positrons and
electrons over the detector acceptance.  After propagation to the
positron fraction, the systematic error due to the East West Effect is
well below 10\+\% for all momentum bins, except for the highest momenta
above 26\+\GeV, where it amounts to approximately 14\+\% of the positron
fraction value.

The total errors $\sigma_{\mathrm{tot}}$ displayed in
Figure~\ref{fig:fraction} are calculated according to
$\sigma_{\mathrm{tot}}^2 =
\sigma_{\mathrm{stat}}^2+\sigma_{\mathrm{sys,b}}^2+\sigma_{\mathrm{sys,a}}^2$,
separately for the upper and lower ranges. Similarly, the systematic
contribution $\sigma_{\mathrm{sys}}$ to the total error is given by
$\sigma_{\mathrm{sys}}^2=\sigma_{\mathrm{sys,b}}^2+\sigma_{\mathrm{sys,a}}^2$.
Clearly, the accuracy of the positron fraction is statistically limited
by the small number of particle counts.

\subsection{Variation of the Positron Fraction with the Value of the Cut}
\begin{figure}[!h]
\begin{center}
\includegraphics[width=12cm]{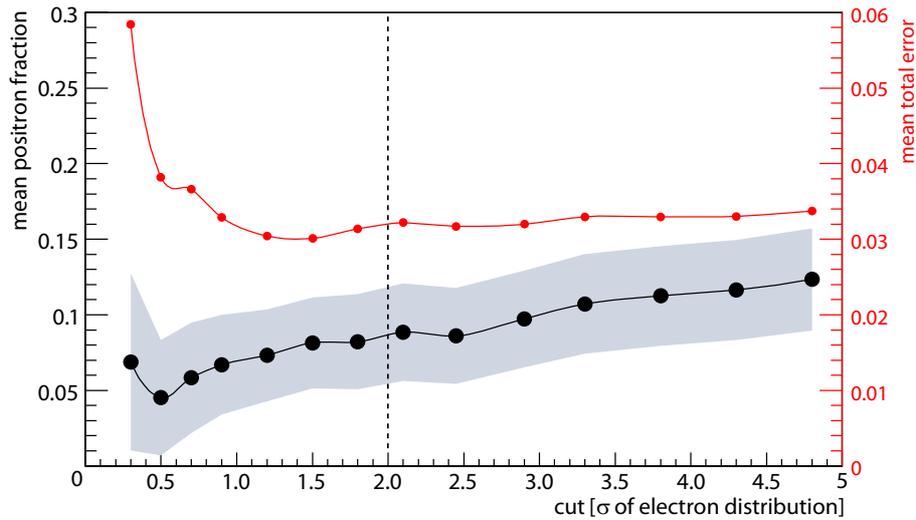}
\caption{\label{fig:meanFraction} Average value of the positron fraction
{\sl (large dots)} and its total error {\sl (small dots and shaded
area)} as functions of the cut parameter. The actual cut is indicated
by the vertical line.}
\end{center} 
\end{figure}

Figure~\ref{fig:meanFraction} shows the average positron fraction and
its total error $\sigma_{\mathrm{tot}}$ as a function of the invariant
mass cut parameter (see \S~\ref{section:analysis}). The fraction is
calculated as the weighted mean over the six momentum bins from 3 to
50\+\GeV/c, which have been corrected for a considerable amount of
background, with the total errors $\sigma_{\mathrm{tot}}$ as
weights. For values of the cut between approximately 1.5\+$\sigma$ and
2.5\+$\sigma$, the mean positron fraction remains largely stable,
showing that the background correction procedure compensates well for
the additional background events entering the positron sample when the
cut is loosened. Beyond $2.5\+\sigma$, the average positron fraction
rises constantly, probably due to systematic uncertainties of the
background correction. However, the cut of $2\+\sigma$ used in this
analysis, indicated by the vertical line in
Figure~\ref{fig:meanFraction}, is well within the region in which the
background correction procedure works reliably.

\section{Flux Calculation}
\label{section:fluxcalculation}
As a crosscheck to the measurement of the positron fraction, the
absolute incident fluxes of electrons and positrons are
calculated. The electron flux is then compared to measurements by
other experiments and to the results obtained previously by \AMS{}.
To calculate the fluxes it is necessary to determine
the acceptance as well as the livetime for positrons and electrons
in the context of this analysis.

The term {\sl livetime} means the effective amount of time during
which cosmic ray particles coming from outer space have the
opportunity to reach and trigger the experiment. The livetime may be
significantly lower than the total operation time of the experiment, a
fact for which four different effects are largely responsible. First,
due to a variety of reasons, there were several loss-of-signal periods
during the flight with average duration of roughly one hundred seconds
and these must be subtracted from the total operating
time. Furthermore, the total time during which the trigger system was
busy (see \S~\ref{sec:trigger}) must also be considered as dead
time.

Besides these rather simple effects, the influence of the Earth and
its magnetic field causes the livetime to be reduced even in a
momentum dependent way. The body of the Earth will obstruct particles
arriving from the ``wrong'' side, except if their momentum is low
enough so that their trajectories may be bent around the Earth by the
geomagnetic field. In addition the geomagnetic field will force the
trajectories of incoming particles on a helix and capture or shield
the Earth from particles with momentum below the geomagnetic
cutoff. These effects depend on the direction of incidence and on the
momentum of the particle, as well as on the orientation and position
of the Space Shuttle, and therefore on time. In principle, this means
that the experiment may be effectively shut down for the measurement
of particles in a certain momentum range at a given time.

From the particle count $N(p,\theta,\phi)$ in a particular momentum bin
$p$ of width $\Delta{}p$ and knowing the detector acceptance
$A(p,\theta,\phi)$, and the livetime $T(p,\theta,\phi)$, one can
calculate the differential flux as follows:
\begin{equation}\label{accurate_flux}
\diff{\Phi(p,\theta,\phi)}{p}=\frac{N(p,\theta,\phi)}{A(p,\theta,\phi)\cdot{}T(p,\theta,\phi)\cdot{}\Delta{}p}\, .
\end{equation}
If -- as is the case with the flux of downward going particles in
\AMS{} -- the livetime is only weakly dependent on direction, the
angular distribution of the particle count will follow that of the
acceptance. Then, one can approximate eq. (\ref{accurate_flux}) by
\begin{equation}\label{eq:approxFlux}
\diff{\Phi(p)}{p}=\frac{N(p)}{A(p)\cdot{}T(p)\cdot{}\Delta{}p}\, .
\end{equation}
The determination of the detector acceptance and the calculation of
the livetime is described in the following two sections.

\subsection{Detector Acceptance}\label{sec:acceptance}
The detector acceptance for the bremsstrahlung conversion process is
calculated from Monte Carlo, separately for downward and upward going
electrons and positrons. In the Monte Carlo simulation, particles are
emitted above or below the detector from a square surface $S$ with a
side length of 3.9 m. With $n_{t}$ the total number of particles
emitted from $S$ into the hemisphere facing the detector with an
isotropic angular distribution, and $n_{c}$ the number of
reconstructed events remaining after all cuts, the acceptance as a
function of incident momentum is~\cite{sullivan71a}
\begin{equation}
A(p) = S \cdot \pi \cdot \frac{n_{c}(p)}{n_{t}(p)}\, .
\end{equation}

\begin{figure}[bt]
\begin{center}
\includegraphics[width=11cm]{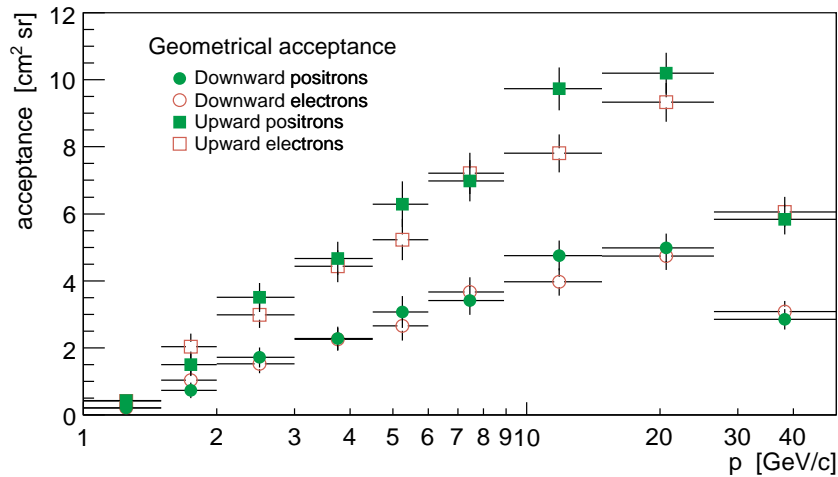}
\caption{\label{fig:acceptance} The acceptance $A(p)$ for
downward and upward going positrons and electrons, when identified
through bremsstrahlung conversion.}
\end{center} 
\end{figure}
Figure~\ref{fig:acceptance} shows $A(p)$ separately for upward or
downward going electrons and positrons. It is of the order of a few
cm$^2\cdot\,$sr and reaches a flat maximum at approximately 20\+\GeV/c. The decline towards higher momenta is caused by the detector
resolution. According to the cross section
eq.~(\ref{equation:bremsstrahlung}), high momentum incident positrons
or electrons radiate bremsstrahlung photons with higher energy,
resulting in higher momentum secondary particles which undergo a
smaller deflection in the magnetic field of the \AMS{}
tracker. Therefore, their trajectories are closer to each other and
the decreasing cluster separation approaches the resolution limit of
the silicon strip detectors. By contrast, at low momentum secondary
particles may be deflected so that they generate multiple separated
hits in single TOF planes. In this case an event is rejected by
the level 3 trigger logic of the experiment. Furthermore, for
secondary particles with too low momentum, the probability that they
will not be properly reconstructed increases, leading to a decrease of
the acceptance.

Additional material (the ATC, the Space Shuttle's payload bay floor
and the support structure of \AMS{}) is traversed by upward going
particles before they enter the detector
(see \S~\ref{sec:materialBudget}). Consequently, the acceptance for
upward going particles is generally higher than for downward going
ones. Except for the ATC, the amount of this additional material can
only be estimated, since precise information is not
available. Independent of their flight direction, no significant
difference in the acceptance for electrons and positrons is observed.

\subsection{Calculation of Livetime}\label{subsection:livetime}
For the calculation of the lifetime $T$ as a function of momentum and
of the direction of the incoming particles a method has been developed
which is derived from the determination of the geomagnetic cutoff as
described in sec.~\ref{subsection:cutoff}. The basic principle is to
determine the cutoff rigidity in short time steps for the whole
duration of the Space Shuttle flight and to count the livetime only
for the momentum range above the cutoff.

In a first step, the acceptance region of \AMS{} is divided into nine
bins of equal size in $\cos\theta$ in the interval $[0.7,1]$ or
$[-1,-0.7]$ for downward or upward going particles, respectively, and
into eight bins in $\phi$. Furthermore, the momentum range between
1\+\GeV/c and 50\+GeV/c is divided into eight bins. Then, for every four
seconds during the flight of the {\sl Discovery}, her recorded
position and attitude are obtained and for each of the 576
$(p,\mathrm{d}\Omega)$-bins, a virtual charged particle is started
with the corresponding values inside the aperture of the detector and
propagated backward through the geomagnetic field. If the virtual
particle fulfills the criteria of a primary cosmic ray particle as
described in section~\ref{section:analysis}, the four-second time
interval during which the particle has been detected is added to the
total livetime of the corresponding momentum and acceptance bin after
subtraction of the trigger dead time. This approach is as accurate as
it is CPU-intensive, with a total of approximately 5000 GHz-CPU days
spent for the calculations.

\begin{figure}[bt]
\begin{center}
\includegraphics[width=11cm]{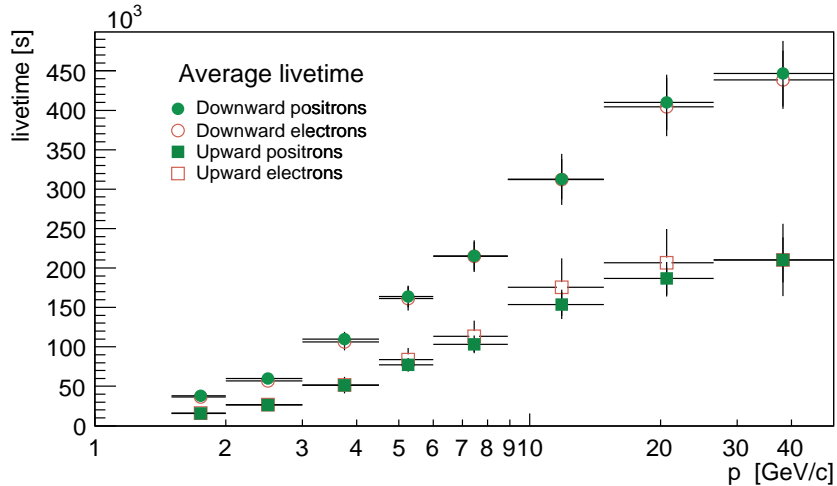}
\caption{\label{fig:livetime} The livetime, averaged over the detector acceptance, 
as a function of momentum for downward and upward going positrons and
electrons.}
\end{center} 
\end{figure}
The livetime, averaged over the detector acceptance, is displayed in
Figure~\ref{fig:livetime} for downward and upward going positrons and
electrons. As a consequence of the rigidity cutoff, it is a
monotonically increasing function of momentum and converges to
approximately 140 (55) hours for downward (upward) going
particles. These values correspond to 75\+\% (30\+\%) of the total data
taking time of about 184 hours. Since the Space Shuttle was mostly
oriented towards the zenith (see
section~\ref{section:shuttlemission}), and due to obstruction by the
Earth's body, the livetime for downward going particles exceeds that
of upward going ones. Concerning the average livetime, no significant
difference between positively and negatively charged particles is
apparent.

\subsection{Positron and Electron Fluxes}
Precise information about the thickness of the Space Shuttle's payload
bay floor is not accessible. Hence, the amount of material underneath
the detector, which strongly affects the acceptance for upward going
particles, cannot be ascertained to a degree which is suitable for
flux measurements, but only be estimated. As a consequence, in this
analysis particle fluxes are calculated solely for downward going particles.

\begin{figure}[hb]
\begin{center}
\includegraphics[width=10cm]{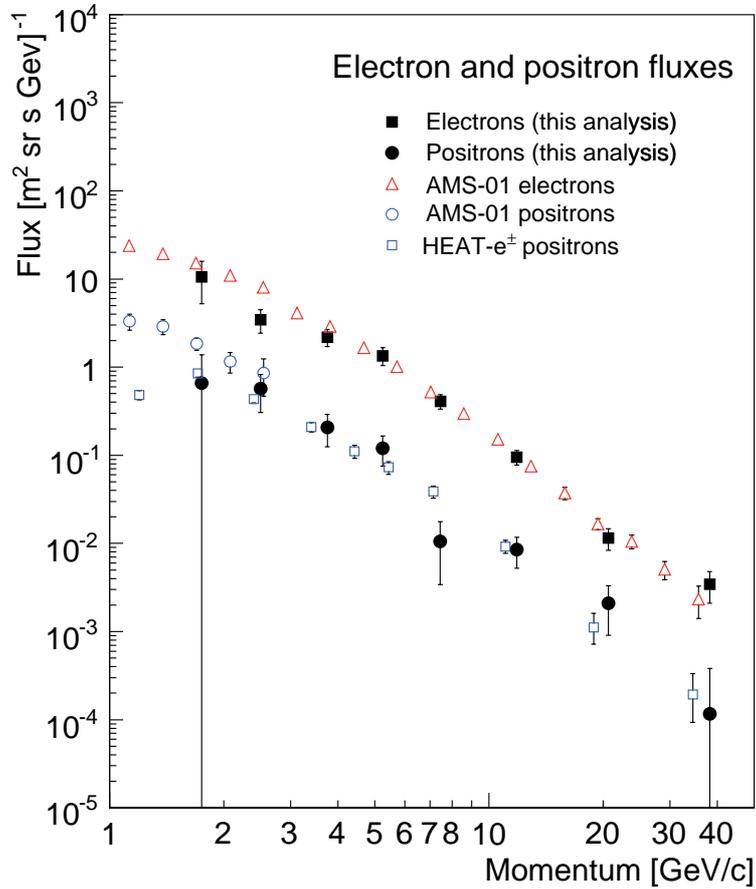}
\caption{\label{fig:fluxes} The fluxes of downward going positrons and electrons measured
in this analysis, compared with earlier results from
\AMS{}~\protect\cite{alcaraz00d} and HEAT-$e^{\pm}$~\protect\cite{duvernois01a}.
Error bars for results from this analysis denote statistical errors
only.}
\end{center} 
\end{figure}

In Figure~\ref{fig:fluxes} the fluxes of downward going positrons and
electrons, calculated according to eq. (\ref{eq:approxFlux}), together
with results published earlier by \AMS{}~\cite{alcaraz00d}, and
HEAT-e$^{\pm}$~\cite{duvernois01a}, are displayed together with their
statistical errors. The fluxes are in very good agreement with
previous measurements over the full momentum range, except for a
slight discrepancy in the electron fluxes between 2 and 3\+\GeV. Here,
at low momenta in combination with low statistics, additional
inaccuracies of the backtracing through the geomagnetic field are
expected to become the dominant source of systematic error to the
fluxes. However, this effect cancels out in the ratio giving the
positron fraction.

\chapter{Conclusions} 
Over the past decades it has become evident that luminous matter
amounts only to a small fraction of the energy density in the
universe. More than 75\+\% is accounted for by what is called the dark
energy, and about 20\+\% must exist in the form of some kind of
non-relativistic dark matter. At the same time new physics beyond the
Standard Model of particle physics is expected to appear when
present experimental limitations are exceeded. Among the many hypotheses
anticipating these new phenomena, supersymmetry yields the most
promising dark matter candidate, the neutralino. Thus cosmology,
astrophysics and particle physics may simultaneously approach the same
problem and their synergy could finally lead to a fundamentally new
understanding of basic physics.

The general purpose detectors, CMS and ATLAS, at the LHC will be able
to detect sparticles over large ranges of the supersymmetric
parameters. Measurements of cosmic ray particles, especially of
antiparticles such as positrons, can impose strong constraints on the
nature of new physics beyond the Standard Model. However, cosmic ray
positron measurements are experimentally very challenging due to the
vast proton background. While new cosmic ray experiments are under
construction or have recently started taking data, existing data from
past experiments can be re-analyzed using new analysis
techniques. This thesis describes a novel approach of positron
identification applied to the data of the \AMS{} experiment, namely
through the detection of bremsstrahlung conversion in a silicon
microstrip detector. In contrast to earlier single-track analyses,
this approach involves the selection and reconstruction of multi-track
events. As the discussion of the signal process shows, bremsstrahlung
from protons is suppressed by a factor of more than $3\cdot 10^6$ with
respect to positrons due to the dependence of the cross section on the
mass of the radiating particle. The background to the positron sample,
which dominantly stems from protons undergoing hadronic interactions
and from wrongly reconstructed electrons, can largely be suppressed
using the topological and geometrical properties of the events.

In order to obtain the highest positron selection efficiency possible,
novel combinatorial track finding algorithms were developed,
particularly optimized for the signature of converted
bremsstrahlung. These algorithms require no additional highly granular
subdetector for the determination of track seeds and their momentum
resolution is comparable to that of single-track events for momenta
above 10\+\GeV{}. By applying restrictions on the invariant mass of
particles the background to the positron sample can largely be
eliminated. The remaining background contamination was determined from
large samples of Monte Carlo data including the effects of the
geomagnetic field. It amounted to 26\+\% of the positron counts and
was corrected for. No indication was found for additional background
produced in interactions with the Mir space station. Finally, in order
to remove atmospheric secondaries from the positron and electron
samples, the precise method of trajectory backtracing was applied
individually to all positron and electron candidates.

The results of the positron measurement are presented in chapter
seven. It is shown that the bremsstrahlung approach extends the
sensitivity range of \AMS{} to positron momenta up to 50\+\GeV{}/c,
which is far beyond the original scope of the experiment. The
precision of the positron fraction measurement is clearly
statistically limited by the small number of particle counts. In the
momentum range 1 to 8\+\GeV{} the fraction is in good agreement with
model predictions for background from purely secondary positron
production, while at higher momenta there is indication for a positron
overabundance as already reported by other
experiments~\cite{beatty04a}. However, the results of the present
analysis are subject to different systematic errors, since they were
obtained with a different detector and a new method of
analysis. Furthermore, in contrast to the earlier balloon-borne
experiments, \AMS{} was operated in space well above the Earth's
atmosphere. Thus the
\AMS{} data lend further weight to the hints of a positron
overabundance seen in~\cite{beatty04a}.

In addition to the positron fraction, the absolute fluxes of positrons
and electrons were calculated from the event samples of this analysis.
For this purpose, a method was developed which allows the
determination of the livetime as a function of momentum and direction
of incidence with high accuracy. The results of the flux calculation
are found to be consistent with earlier
data~\cite{alcaraz00d,duvernois01a}. In particular, the electron flux
for momenta above 6\+\GeV{} is well in agreement with that obtained
from measurements of single-track events in the \AMS{} data,
confirming the good performance of particle selection with the
bremsstrahlung approach.

\begin{figure}[htb]
\begin{center}
\includegraphics[width=13.8cm]{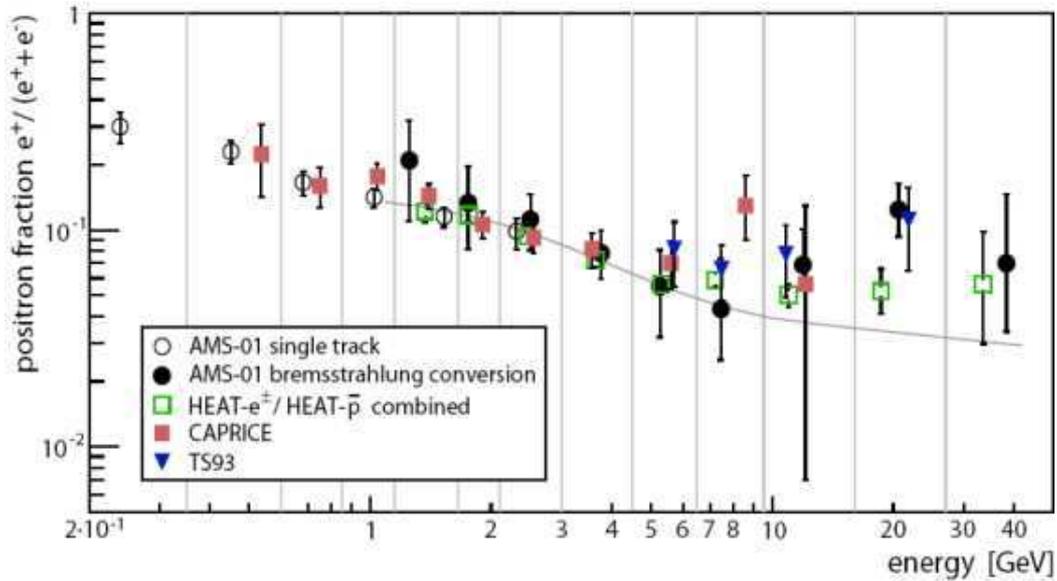}
\caption{\label{fig:allData}The data used for the calculation of the 
combined positron fraction from: this analysis,
\AMS{} (single tracks)~\protect\cite{alcaraz00d}, HEAT-$e^{\pm}$ and
HEAT-pbar~\cite{beatty04a}, CAPRICE~\cite{boezio00a} and
TS93~\cite{golden96a}. The data were grouped within intervals which
are denoted by vertical lines. The solid curve shows a model
prediction for purely secondary positron
production~\cite{moskalenko98a}. }
\end{center} 
\end{figure}

The positron fraction results from this analysis have been combined
with earlier results from \AMS{}~\protect\cite{alcaraz00d},
HEAT-$e^{\pm}$ and HEAT-pbar~\cite{beatty04a},
CAPRICE~\cite{boezio00a} and TS93~\cite{golden96a} (see
Figure~\ref{fig:allData}). For this, a procedure for combining data
with asymmetric errors was employed~\cite{barlow04a}. First, all data
were grouped in 13 momentum bins. For each data point $i$ the
corresponding likelihood function $\mathcal{L}_{i}$ was approximated
by a Gaussian function with a variable standard deviation
$\sigma_i$. The $\mathcal{L}_i$ were parameterized using the values of
the measurements and their upper and lower errors. The combined
likelihood function $\mathcal{L}$, from which the combined results and
errors are finally determined, is then defined by the product of the
$\mathcal{L}_i$ for a particular momentum bin. The central momentum
value for each bin is given by the mean momenta of the data in the
individual bins.

\begin{figure}[tb]
\begin{center}
\includegraphics[width=13cm]{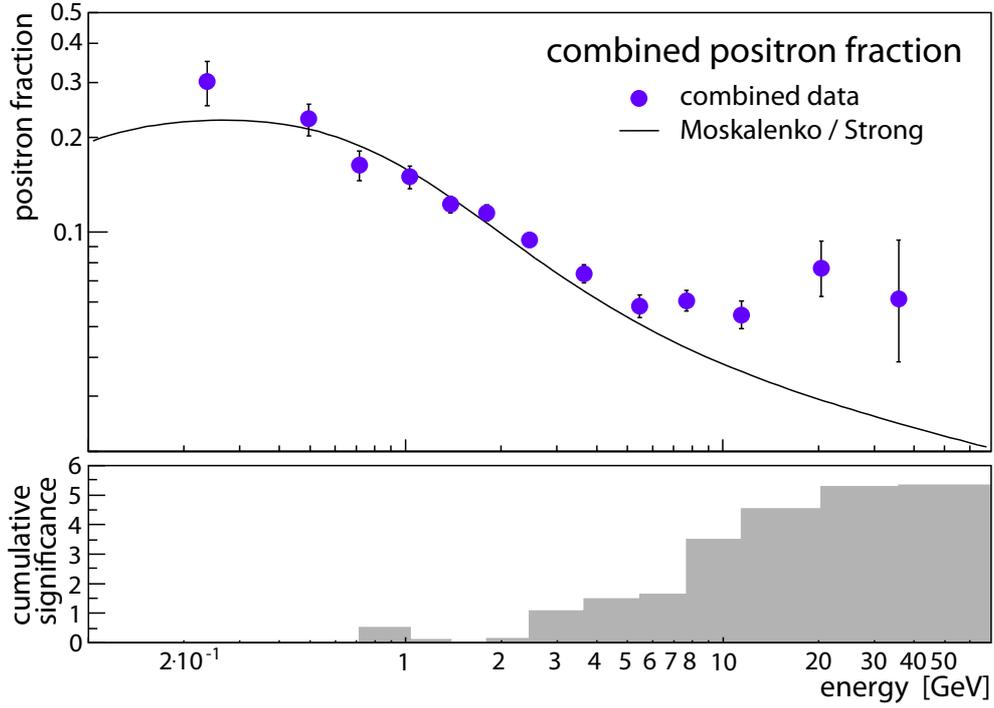}
\caption{\label{fig:superfraction}The combined positron fraction
of the results from this analysis and earlier results. The solid curve
shows a model prediction for purely secondary positron
production~\cite{moskalenko98a} {\sl (top)}; the cumulative
significance (the significance of all data for momenta $p<p'$) of the
positron overabundance in the data with respect to the background-only
prediction. The data point below 0.3\+\GeV{} was not considered in the
calculation {\sl (bottom)}.}
\end{center} 
\end{figure}

The combined positron fraction is shown in
Figure~\ref{fig:superfraction} {\sl (top)\/} and is compared to a
model calculation for the background expected from purely secondary
positron production (without diffuse
reacceleration)~\cite{moskalenko98a}. Table~\ref{tab:positronFractionAverage}
summarizes the results. The data indicate a positron overabundance for
momenta above 6\+\GeV{}. The bottom display of
Figure~\ref{fig:superfraction} shows the cumulative significance,
i.e. the significance of all data for momenta $p<p'$ with respect to
the background-only prediction. The significance was calculated from
the corresponding probability of the $\chi^2$ per degree of
freedom. Because of the possible effect of solar modulation on the
positron fraction at low momenta, the data point below 0.3\+\GeV{} was
not considered in the calculation. Apart from statistical fluctuations
on the 2 standard deviation level, the positron fraction is in good
agreement with purely secondary positron production for momenta below
6\+\GeV{}. However, when the data at higher momenta are considered,
the significance of the positron overabundance increases to 5.3
standard deviations, which would be reduced to 4.2 standard deviations
without the results of the present analysis. Therefore, a statistical
fluctuation causing the positron overabundance in the data with
respect to the background-only prediction can now be excluded.
\begin{sidewaystable}[!h]
\centering
\begin{tabular}{|r@{\,--}r|r|l|l|l|l|l|l|l|}
\hline
\multicolumn{2}{|c|}{$E$} & \multicolumn{1}{|c|}{$\bar{E}$} & \multicolumn{5}{|c|}{$e^{+}$ fraction data} &
        \multicolumn{1}{|c|}{Combined} & \multicolumn{1}{|c|}{$\chi^2$\,/$\,n_{\mathrm{dof}}$} \rule[-1ex]{0ex}{4ex}\\
\multicolumn{2}{|c|}{[GeV]\rule[-1.5ex]{0ex}{0ex}} & \multicolumn{1}{|c|}{[GeV]} & \multicolumn{1}{c}{This work} & 
        \multicolumn{1}{c}{\AMS{}} & \multicolumn{1}{c}{HEAT} & \multicolumn{1}{c}{CAPRICE} & 
        \multicolumn{1}{c}{TS93} & \multicolumn{1}{|c|}{$e^{+}$ fraction} & \multicolumn{1}{|c|}{of average} \\ \hline
0.20  & 0.35  & 0.24  & \makebox[32pt][r]{---}      & 0.300\,$^{+0.048}_{-0.049}$ & \makebox[35pt][r]{---}        & \makebox[34pt][r]{---}        & \makebox[34pt][r]{---}   & 0.300\,$^{+0.048}_{-0.049}$     & \makebox[27pt][c]{$\;$---}\,/\,1\rule[-1.2ex]{0ex}{3.7ex}\\
0.35  & 0.60  & 0.50  & \makebox[32pt][r]{---}      & 0.230\,$^{+0.028}_{-0.029}$ & \makebox[35pt][r]{---}        & 0.220\,$^{+ 0.081}_{- 0.081}$ & \makebox[34pt][r]{---}   & 0.230\,$^{+0.026}_{-0.027}$     & \makebox[27pt][r]{0.01}\,/\,2\rule[-1.2ex]{0ex}{3.7ex}\\
0.60  & 0.85  & 0.71  & \makebox[32pt][r]{---}      & 0.170\,$^{+0.02}_{-0.021}$  & \makebox[35pt][r]{---}        & 0.160\,$^{+ 0.034}_{- 0.034}$ & \makebox[34pt][r]{---}   & 0.160\,$^{+0.017}_{-0.018}$     & \makebox[27pt][r]{0.02}\,/\,2\rule[-1.2ex]{0ex}{3.7ex}\\
0.85  & 1.10  & 1.00  & \makebox[32pt][r]{---}      & 0.140\,$^{+0.015}_{-0.015}$ & \makebox[35pt][r]{---}        & 0.180\,$^{+ 0.025}_{- 0.025}$ & \makebox[34pt][r]{---}   & 0.150\,$^{+0.012}_{-0.013}$     & \makebox[27pt][r]{1.50}\,/\,2\rule[-1.2ex]{0ex}{3.7ex}\\
1.10  & 1.60  & 1.40  & 0.210\,$^{+0.11}_{-0.1}$    & 0.120\,$^{+0.01}_{-0.013}$  & 0.120\,$^{+0.016}_{-0.014}$   & 0.140\,$^{+ 0.019}_{- 0.019}$ & \makebox[34pt][r]{---}   & 0.120\,$^{+0.0071}_{-0.0076}$   & \makebox[27pt][r]{2.40}\,/\,4\rule[-1.2ex]{0ex}{3.7ex}\\
1.60  & 2.10  & 1.80  & 0.130\,$^{+0.064}_{-0.051}$ & \makebox[33pt][r]{---}      & 0.120\,$^{+0.0077}_{-0.0076}$ & 0.110\,$^{+ 0.015}_{- 0.015}$ & \makebox[34pt][r]{---}   & 0.120\,$^{+0.0067}_{-0.0067}$   & \makebox[27pt][r]{0.55}\,/\,3\rule[-1.2ex]{0ex}{3.7ex}\\
2.10  & 3.00  & 2.50  & 0.110\,$^{+0.034}_{-0.031}$ & 0.099\,$^{+0.015}_{-0.017}$ & 0.094\,$^{+0.0054}_{-0.0043}$ & 0.092\,$^{+ 0.014}_{- 0.014}$ & \makebox[34pt][r]{---}   & 0.094\,$^{+0.0049}_{-0.0041}$   & \makebox[27pt][r]{0.41}\,/\,4\rule[-1.2ex]{0ex}{3.7ex}\\
3.00  & 4.50  & 3.60  & 0.078\,$^{+0.021}_{-0.019}$ & \makebox[33pt][r]{---}      & 0.072\,$^{+0.0054}_{-0.0049}$ & 0.082\,$^{+ 0.015}_{- 0.015}$ & \makebox[34pt][r]{---}   & 0.074\,$^{+0.005}_{-0.0047}$    & \makebox[27pt][r]{0.43}\,/\,3\rule[-1.2ex]{0ex}{3.7ex}\\
4.50  & 6.50  & 5.50  & 0.055\,$^{+0.026}_{-0.023}$ & \makebox[33pt][r]{---}      & 0.056\,$^{+0.0052}_{-0.0048}$ & 0.070\,$^{+ 0.019}_{- 0.017}$ & 0.082\,$^{+0.027}_{-0.027}$& 0.058\,$^{+0.0049}_{-0.0046}$ & \makebox[27pt][r]{1.40}\,/\,4\rule[-1.2ex]{0ex}{3.7ex}\\
6.50  & 9.50  & 7.70  & 0.043\,$^{+0.031}_{-0.018}$ & \makebox[33pt][r]{---}      & 0.059\,$^{+0.0049}_{-0.0044}$ & 0.130\,$^{+ 0.049}_{- 0.039}$ & 0.066\,$^{+0.019}_{-0.019}$& 0.060\,$^{+0.0049}_{-0.0044}$ & \makebox[27pt][r]{4.30}\,/\,4\rule[-1.2ex]{0ex}{3.7ex}\\
9.50  & 16.00 & 11.00 & 0.069\,$^{+0.032}_{-0.015}$ & \makebox[33pt][r]{---}      & 0.050\,$^{+0.0065}_{-0.0057}$ & 0.056\,$^{+ 0.073}_{- 0.049}$ & 0.077\,$^{+0.028}_{-0.028}$& 0.054\,$^{+0.0059}_{-0.0052}$ & \makebox[27pt][r]{2.00}\,/\,4\rule[-1.2ex]{0ex}{3.7ex}\\
16.00 & 27.00 & 20.00 & 0.120\,$^{+0.04}_{-0.031}$  & \makebox[33pt][r]{---}      & 0.052\,$^{+0.014}_{-0.011}$   & \makebox[34pt][r]{---}        & 0.110\,$^{+0.046}_{-0.046}$& 0.077\,$^{+0.017}_{-0.014}$   & \makebox[27pt][r]{5.80}\,/\,3\rule[-1.2ex]{0ex}{3.7ex}\\
27.00 & 50.00 & 36.00 & 0.070\,$^{+0.076}_{-0.036}$ & \makebox[33pt][r]{---}      & 0.056\,$^{+0.04}_{-0.026}$    & \makebox[34pt][r]{---}        & \makebox[34pt][r]{---}   & 0.061\,$^{+0.033}_{-0.022}$     & \makebox[27pt][r]{0.06}\,/\,2\rule[-1.2ex]{0ex}{3.7ex}\\
\hline
\end{tabular}
\caption{\label{tab:positronFractionAverage}The combined 
positron fraction and the data used for its calculation from: this
work, \AMS{} (single tracks)~\cite{alcaraz00d}, HEAT-$e^{\pm}$
and HEAT-pbar~\cite{beatty04a}, CAPRICE~\cite{boezio00a} and
TS93~\cite{golden96a}. Also given is the average energy $\bar{E}$ of
the measurements and the energy interval $E$ that was used in grouping
them. The last column gives the $\chi^2$ and the number of degrees of
freedom $n_{\mathrm{dof}}$ of the combined values.}
\end{sidewaystable}

As stated in \S~\ref{sec:susyspectra}, besides neutralino annihilation
other possible contributions to the cosmic ray particle fluxes have been
proposed, such as hypothetical Kaluza-Klein particle states or galactic
pulsars. In order to establish through the analysis of cosmic ray
spectra if annihilations of dark matter neutralinos are in fact
responsible for the observed gamma and antiparticle overabundances and
to narrow down the parameters of supersymmetry, more precise
measurements remain necessary. Especially in the case of the cosmic
ray positron fraction, measurements up to about 300\+\GeV{} will allow
to precisely distinguish possible dark matter positron sources from
other contributions.

In June 2006 the PAMELA satellite experiment~\cite{picozza06a}, which
is designed to study charged particles in cosmic rays with a
particular focus on antiparticles, was launched for an at least three
year long mission. PAMELA features, amongst other subdetectors, a
silicon microstrip tracker in the 0.43\+T field of a permanent magnet,
a calorimeter and a neutron detector. However, with its small
geometrical acceptance of 21.5\+cm$^2$\+sr, it is unlikely that PAMELA
will be able to deliver precise data at highest energies relevant to a
possible dark matter signal. The successor to the \AMS{} experiment,
AMS-02~\cite{leluc05a}, is currently under construction and scheduled
for operation on the International Space Station {\sl (ISS)} from 2009
on. Developed on the basis of the \AMS{} design, AMS-02 will feature
strongly enhanced particle identification by means of a calorimeter, a
ring imaging \v Cerenkov detector and a transmission radiation
detector (TRD), as well as an improved silicon microstrip tracker in
the 0.87\+T field of a superconducting magnet. With a geometrical
acceptance of the tracking system of 4500\+cm$^2$\+sr and
900\+cm$^2$\+sr for the tracker in combination with the TRD and the
calorimeter, exceeding that of PAMELA by a factor of more than 40,
AMS-02 will conduct cosmic ray spectroscopy with unprecedentedly high
precision.

\chapter*{Acknowledgment} 
\addcontentsline{toc}{chapter}{Acknowledgment}
First, I want to thank my supervisor Prof. St. Schael for the
opportunity to work on this fascinating subject and his advice,
encouragement and thoroughgoing support I could always count on. I
also want to thank Prof. L. Baudis for her help and expert opinion.

The positron analysis would not have been successful without the
collaboration of H.~Gast who contributed very much valuable work and
countless excellent ideas for solving the many problems we were
encountering. It was really a marvelous and fruitful team
work. Besides, many other of my colleagues from the I. Physikalisches
Institut B have a part in the success of my work through their help,
advice, and the discussions I had with them: R.~Adolphi, C.~Berger,
W.~Braunschweig, C.~H.~Chung, L.~Feld, S.~K\"onig, K.~L\"ubelsmeyer,
A.~Ostaptchouk, D.~Pandoulas, A.~Schultz v. Dratzig, G.~Schwering,
T.~Siedenburg, W.~Wallraff and B.~Wittmer, only to name a
few. Furthermore, I enjoyed working with the members of the ARC group
of the III. Physikalisches Institut B, namely M.~Axer, F.~Bei\ss el,
T.~Franke, T.~Hermanns, S.~Kasselmann, J.~Mnich, J.~Niehusman and
M.~P\"ottgens.  Thank you for the excellent and straightforward
collaboration.

I benefited much from the aid and knowledge of many kind individuals
of the AMS and CMS collaborations, such as from J.~Alcaraz,
W.~Beaumont, H.~Breuker, M.~Bu\'enerd, M.~Capell, C.~Goy-Henningsen,
F.~Hartmann, G. Kim, J. Shin, H.-J.~Simonis, V.~Zhukov and
P.~Zuccon. Very special thanks go to V.~Choutko for his encouragement
and vital help and advice in many aspects of the positron analysis.

The smooth and powerful performance of the computing services of AMS
and especially here in Aachen were crucial for this work. I would like
to thank all the individuals who kept the machines running, and in
particular A.~Eline, A.~Klimentov, T.~Kre\ss\ and C.~Kukulies for
their kind support and advice. Many thanks to the staff of the
mechanical and electronics workshops in Aachen, particularly to
W.~Karpinski, G.~Kirchhoff, R.~Siedling$^{\dagger}$ and M.~Wlochal for
their help and guidance in the technical aspects of my work.

R.~Adolphi, H.~Gast, K.~Klein and especially D.~Pandoulas have read
the manuscript of this thesis very carefully, and I owe them many
substantial improvements. Thank you very much for this.

One good reason for spending more time at work rather than at home is
sharing the office with R.~Adolphi, G.~Anagnostou, H.~Gast, K.~Klein
and A.~Ostaptchouk. Thank you my colleagues and friends for the great
time we had during the past few years, and of course for all those
legendary {\it B\"uroabende}. Besides the abovementioned, a lot of
people from the Aachen physics institutes were responsible for the
amicable and comfortable atmosphere at work, such as R.~Brauer,
T.~Bruch, P.~v.~Doetinchem, M.~Fernandez Garcia, A.~Fischer, S.~Fopp,
S.~Funk, T.~Krynicki, J.~Hattenbach, A.~Heister, T.~Hennig,
S.~Hermann, T.~Kirn, M.~Kosbow, J.~Orboeck, F.~Raupach, G.~Roper,
M.~Thomas, M.~Weber, M.~Z\"oller, and many more. \\

\noindent Finally, I want to thank my family for making all this possible.
\clearpage\thispagestyle{plain}



\begin{thebibliography}{99999}
\addcontentsline{toc}{chapter}{Bibliography}
\setlength{\itemsep}{-1mm}
\bibitem{beckwith06a} R.~Beckwith et al., Astron. J. {\bf 132} (2006) 1729
\bibitem{alcaraz00d} J.~Alcaraz et al., Phys. Lett. B {\bf 484} (2000) 10
\bibitem{duvernois01a} M.~A.~DuVernois et al., ApJ {\bf 559}  (2001) 296
\bibitem{golden96a} R.~L.~Golden et al., ApJ {\bf 457} (1996) L103
\bibitem{boezio00a} M.~Boezio et al., ApJ {\bf 532} (2000) 653
\bibitem{alcaraz00a} J.~Alcaraz et al., Phys. Lett. B {\bf 472} (2000) 215
\bibitem{beatty04a} J.~J.~Beatty et al., Phys.~Rev.~Let. {\bf 93} (2004) 241102
\bibitem{friedman22a} A.~Friedman, Z. Phys. {\bf 10} (1922) 377
\bibitem{hubble29a} E.~Hubble, Proc. Nat. Acad. {\bf 15} (1929) 168
\bibitem{gamow46a} G.~Gamow, Phys. Rev. {\bf 70} (1946) 572
\bibitem{gamow48a} G.~Gamow, Phys. Rev. {\bf 74} (1948) 505
\bibitem{penzias65a} A.~A.~Penzias, R.~W.~Wilson, ApJ {\bf 142} (1965) 419
\bibitem{alpher49a} R.~A.~Alpher, R.~C.~Herman, Phys.~Rev. {\bf 75} (1949) 1089
\bibitem{spergel03a} D.~N.~Spergel et al., ApJS {\bf 148} (2003) 175 
\bibitem{delapparent86a} V.~de~Lapparent, M.~J.~Geller, J.~P.~Huchra, ApJ {\bf 302} (1986) L1
\bibitem{tonry03a} J.~L.~Tonry et al., ApJ {\bf 594} (2003) 1
\bibitem{spergel06a} D.~N.~Spergel et al., submitted article, {\tt astro-ph/0603449} (2006)
\bibitem{gott05a} J.~R.~Gott et al., ApJ {\bf 624} (2005) 463
\bibitem{guth05a} A.~H.~Guth, D.~I.~Kaiser, Science {\bf 307} (2005) 884
\bibitem{turner99a} M.~S.~Turner, J.~A.~Tyson, Rev. Mod. Phys. {\bf 71} (1999) S145
\bibitem{sakharov67a} A.~D.~Sakharov, J. Exp. Theor. Phys. Lett. {\bf 5} (1967) 24
\bibitem{tyson98a} J.~A.~Tyson, G.~P.~Kochanski, I.~P.~Dell'Antonio, ApJ {\bf 498} (1998) L107
\bibitem{begeman91a} K.~G.~Begeman, A.~H.~Broeils, R.~H.~Sanders, \\ Mon. Not. R. Astron. Soc. {\bf 249} (1991) 523
\bibitem{zwicky33a} F.~Zwicky. Helv. Phys. Acta {\bf 6} (1933) 110
\bibitem{bertone04a} G.~Bertone, D.~Hooper, J.~Silk, Phys. Rep. {\bf 405} (2004) 279
\bibitem{carlberg98a} R.~G.~Carlberg et al., ApJ {\bf 516} (1998) 552
\bibitem{sofue01a} Y.~Sofue, V.~Rubin, Annu. Rev. Astron. Astrophys. {\bf 39} (2001) 137
\bibitem{hinshaw06a} G.~Hinshaw et al., submitted article, {\tt astro-ph/0603451} (2006)
\bibitem{hu03a} W.~Hu, Ann. Phys. {\bf 303} (2003) 203
\bibitem{cole05a} S.~Cole et al., Mon. Not. R. Astr. Soc. {\bf 362} (2005) 505
\bibitem{tegmark04b} M.~Tegmark et al., ApJ {\bf 606} (2004) 702
\bibitem{yao06a} W.-M.~Yao et al., J. Phys. G {\bf 33} (2006) 1, and references therein
\bibitem{higgs64a} P.~W.~Higgs, Phys. Lett. {\bf 12} (1964) 132; \\
                   P.~W.~Higgs, Phys. Rev. {\bf 145} (1966) 1156
\bibitem{cabibbo63a} N.~Cabibbo, Phys. Rev. Lett. {\bf 10} (1963) 531
\bibitem{kobayashi73a} M.~Kobayashi, T.~Maskawa, Prog. Theor. Phys. {\bf 49} (1973) 652
\bibitem{overduin04a} J.~Overduin, P.~Wesson, Phys. Rep. {\bf 402} (2004) 267
\bibitem{paczynski86a} B.~Paczynski, ApJ {\bf 304} (1986) 1
\bibitem{martin97a} S.~Martin, {\tt hep-ph/9709356} (1997)
\bibitem{amaldi91a} U.~Amaldi, W.~de~Boer, H.~F\"urstenau, Phys. Lett. B {\bf 260} (1991) 447
\bibitem{chung05a} D.~Chung et al., Phys. Rep. {\bf 407} (2005) 1
\bibitem{chung07a} C.~H.~Chung, PhD thesis in preparation, RWTH Aachen (2007)
\bibitem{hooper05a} D.~Hooper, J.~Silk, Phys. Rev. D {\bf 71} (2005) 083503
\bibitem{jungman96a} G.~Jungman, M.~Kamionkowski, K.~Griest, Phys. Rep. {\bf 267} (1996) 195
\bibitem{baer04a} H.~Baer et al., J. Cosmol. Astropart. Phys. {\bf 08} (2004) 005
\bibitem{gaitskell04a} R.~J.~Gaitskell, Annu. Rev. Nucl. Part. Sci. {\bf 54} (2004) 315
\bibitem{aguilar02a} M.~Aguilar et al., Phys.~Rep. {\bf 366} (2002) 331
\bibitem{deboer06a} W.~de Boer et al., Phys. Lett. B {\bf 636} (2006) 13
\bibitem{bruening04a} O.~Br\"uning et al. (ed.), CERN Yellow Report 2004-003-V-1 (2004)
\bibitem{bruening04b} O.~Br\"uning et al. (ed.), CERN Yellow Report 2004-003-V-2 (2004)
\bibitem{benedikt04a} M.~Benedikt et al. (ed.),  CERN Yellow Report 2004-003-V-3 (2004)
\bibitem{brandt00a} D.~Brandt et al., Rep. Prog. Phys. {\bf 63} (2000) 939
\bibitem{aleph06a} The ALEPH Collaboration et al., Phys. Rep. {\bf 427} (2006) 257
\bibitem{lamont06a} M.~Lamont et al., CERN LHC Project Report {\bf 949} (2006) 
\bibitem{lamont05a} M.~Lamont, CERN LHC Project Note {\bf 375} (2005)
\bibitem{atlas99} The ATLAS Collaboration, CERN-LHCC-99-014, CERN-LHCC 99-015 (1999)
\bibitem{dellanegra06a} M.~Della~Negra et al., CERN-LHCC-2006-001 (2006)
\bibitem{dellanegra06b} M.~Della~Negra et al., CERN-LHCC-2006-021 (2006)
\bibitem{lhcb98a} The LHCb Collaboration, CERN-LHCC-98-004 (1998)
\bibitem{antunesnobrega03a} R.~Antunes-Nobrega et al., CERN-LHCC-2003-030 (2003)
\bibitem{alice95a} The ALICE Collaboration, CERN-LHCC-95-071 (1995)
\bibitem{abdullin02a} S.~Abdullin et al., J. Phys. G {\bf 28} (2002) 469
\bibitem{gladyshev06a} A.~V.~Gladyshev, D.~I.~Kazakov, {\tt hep-ph/0606288} (2006)
\bibitem{hinchliffe97a} I.~Hinchliffe et al., Phys. Rev. D {\bf 55} (1997) 5520
\bibitem{kitano06a} R.~Kitano, Y.~Nomura, Phys. Rev. D {\bf 73} (2006) 095004
\bibitem{denegri05a} D.~Denegri, Nuovo Cim. B {\bf 120} (2005) 687
\bibitem{cms94a} The CMS Collaboration, CERN-LHCC-94-38 (1994)
\bibitem{cms97b} The CMS Collaboration, CERN-LHCC-97-10 (1997)
\bibitem{cms97a} The CMS Collaboration, CERN-LHCC-97-32 (1997)
\bibitem{cms97c} The CMS Collaboration, CERN-LHCC-97-33 (1997)
\bibitem{cms97d} The CMS Collaboration, CERN-LHCC-97-31 (1997)
\bibitem{cms98a} The CMS Collaboration, CERN-LHCC-98-6 (1998)
\bibitem{brauer05a} R.~Brauer et al., CERN CMS Note 2005/025 (2005)
\bibitem{cms00a} The CMS Collaboration, CERN-LHCC-2000-016 (2000)
\bibitem{cucciarelli05a} S.~Cucciarelli, Nucl. Inst. Meth. A {\bf 549} (2005) 49
\bibitem{franke05a} T.~Franke, PhD Thesis, RWTH Aachen (2005)
\bibitem{french01a} M.~J.~French et al., Nucl. Inst. Meth. A {\bf 466} (2001) 359
\bibitem{axer03a} M.~Axer, PhD Thesis, RWTH Aachen (2003)
\bibitem{braibant02a} S.~Braibant et al., Nucl. Inst. Meth. A {\bf 485} (2002) 343
\bibitem{scholze00a} F.~Scholze et al., Nucl. Inst. Meth. A {\bf 439} (2000) 208
\bibitem{abt99a} I.~Abt et al., Nucl. Inst. Meth. A {\bf 423} (1999) 303
\bibitem{dierlamm03a} A.~Dierlamm, Nucl. Inst. Meth. A {\bf 514} (2003) 167
\bibitem{meschini04a} M.~Meschini et al., CMS Internal Note 2004/018 (2004)
\bibitem{axer02a} M.~Axer et al., Nucl. Inst. Meth A {\bf 485} (2002) 73
\bibitem{wittmer02a} B.~Wittmer, PhD Thesis, RWTH Aachen (2002)
\bibitem{axer04a} M.~Axer et al., Nucl. Inst. Meth. A {\bf 518} (2004) 321, and references therein
\bibitem{hess12a} V.~Hess, Phys.~Zeit. {\bf 13} (1912) 1084
\bibitem{anderson33a} C.~D.~Anderson, Phys. Rev. {\bf 43} (1933) 491
\bibitem{neddermeyer37a} S.~H.~Neddermeyer, C.~D.~Anderson, Phys. Rev. {\bf 51} (1937) 884
\bibitem{lattes47a} C.~M.~G.~Lattes et al., Nature {\bf 159} (1947) 694;\\
        C.~M.~G.~Lattes, G.~P.~S.~Occhialini, C.~F.~Powell, Nature {\bf 160} (1947) 453;
\bibitem{rochester47a} G.~D.~Rochester, C.~C.~Butler, Nature {\bf 160} (1947) 855
\bibitem{cronin97a} J.~Cronin, T.~K.~Gaisser, S.~P.~Swordy, Sci. Amer. {\bf v276} (1997) 44
\bibitem{alcaraz00b} J.~Alcaraz et al., Phys. Lett. B {\bf 490} (2000) 27
\bibitem{alcaraz00c} J.~Alcaraz et al., Phys. Lett. B {\bf 494} (2000) 193
\bibitem{bird94a} D.~J.~Bird et al., ApJ {\bf 424} (1994) 491
\bibitem{cronin99a} J.~Cronin, Rev. Mod. Phys. {\bf 71} (1999) S165
\bibitem{alcaraz99a} J.~Alcaraz et al., Phys. Lett. B {\bf 461} (1999) 387
\bibitem{orito00a} S.~Orito et al., Phys. Rev. Lett. {\bf 84} (2000) 1078
\bibitem{lezniak78a} J.~A.~Lezniak, W.~R.~Webber, ApJ {\bf 223} (1978) 676
\bibitem{fermi49a} E.~Fermi, Phys. Rev. {\bf 75} (1949) 1169
\bibitem{gaisser90a} T.~K.~Gaisser, Cosmic Rays and Particle Physics, Camb. Univ. Press (1990)
\bibitem{blandford78a} R.~D.~Blandford, J.~P.~Ostriker, ApJ {\bf 221} (1978) L29
\bibitem{gloeckler84a} G.~Gloeckler, Adv.~Space~Res. {\bf 4} (2--3) (1984) 127
\bibitem{steinacker89a} J.~Steinacker, R.~Schlickeiser, Astron. Astrophys. {\bf 224} (1989) 259
\bibitem{blandford87a} R.~D.~Blandford, D.~Eichler, Phys. Rep. {\bf 154} (1987) 1
\bibitem{uchiyama02a} Y.~Uchiyama, T.~Takahashi, F.~Aharonian, Publ. Astron. Soc. Jap. {\bf 54} (2002) L73
\bibitem{aharonian04a} F.~Aharonian et al., Nature {\bf 432} (2004) 75
\bibitem{moskalenko04a} I.~Moskalenko, Frascati Phys. Ser. {\bf 35} (2004) 115
\bibitem{stephens98a} S.~Stephens, R.~Streitmatter, ApJ {\bf 505} (1998) 266
\bibitem{maurin01a} D.~Maurin et al., ApJ {\bf 555} (2001) 585
\bibitem{vallee94a} J.~P.~Vall\'ee, ApJ {\bf 437} (1994) 179
\bibitem{moskalenko98a} I.~V.~Moskalenko, A.~W.~Strong, ApJ {\bf 493} (1998) 694
\bibitem{moskalenko98b} I.~Moskalenko, A.~Strong, O.~Reimer, Astron. Astrophys. {\bf 338} (1998) L75
\bibitem{murphy87a} R.~Murphy, C.~Dermer, R.~Ramaty, ApJS {\bf 63} (1987) 721
\bibitem{gleeson68a} L.~Gleeson, W.~Axford, ApJ {\bf 154} (1968) 1011
\bibitem{clem96a} J.~M.~Clem et al., ApJ {\bf 464} (1996) 507
\bibitem{usoskin05a} I.~Usoskin et al., J. Geophys. Res. {\bf 110} (2005) A12108
\bibitem{olsen00a} N.~Olsen et al., Geophys. Res. Lett. {\bf 27} (2000) 3607
\bibitem{russell91a} C.~Russell, Ann. Rev. Earth Planet Sci. {\bf 19} (1991) 169
\bibitem{li01a} X.~Li, M.~Temerin, Space Sci. Rev {\bf 95} (2001) 569
\bibitem{smart99a} D.~Smart, M.~Shea, E.~Fl\"uckiger, Proc. 26$^{th}$ ICRC {\bf 7} (1999) 337
\bibitem{kane02a} G.~Kane, L.~Wang, T.~Wang, Phys. Lett. B {\bf 536} (2002) 263
\bibitem{deboer05a} W.~de~Boer et al., Astron. Astrophys. {\bf 444} (2005) 51
\bibitem{coutu01a} S.~Coutu et al., Proc 27$^{th}$ ICRC (2001) 1687
\bibitem{deboer04a} W.~de~Boer et al., Eur. Phys. J. C {\bf 33} (2004) S981
\bibitem{bergstrom06a} L.~Bergstr\"om et al., J. Cosmol. Astropart. Phys. {\bf 05} (2006) 006
\bibitem{zhang01a} L.~Zhang, K.~S.~Cheng, Astron. Astrophys. {\bf 368} (2001) 1063
\bibitem{nasa99b} NASA -- National Aeronautics and Space Administration, \\NASA/TP--1999--209260 (1999)
\bibitem{suter98a} H.~Suter, AMS Internal Note 98/29 (1998)
\bibitem{alvisi99a} D.~Alvisi et al., Nucl.~Inst.~Meth. A {\bf 437} (1999) 212
\bibitem{baldini01a} L.~Baldini et al., Proc. 27$^{th}$ ICRC (2001) 2211
\bibitem{alcaraz99b} J.~Alcaraz et al., Nuovo~Cim. A {\bf 112} (1999) 1325
\bibitem{burger02a} W.~J.~Burger, Nucl. Inst. Meth. A {\bf 435} (1999) 202
\bibitem{ghete96a} V.~M.~Ghete, UBPub. EPPG/Phys. {\bf 33} (1996)
\bibitem{alpat00a} B.~Alpat et al., Nucl. Inst. Meth. A {\bf 439} (2000) 53, and references therein
\bibitem{ionica03a} M.~Ionica, Nucl. Inst. Meth. A {\bf 513} (2003) 222
\bibitem{lamanna00a} G.~Lamanna, PhD Thesis, University of Perugia (2000)
\bibitem{vandenhirtz01a} J.~Vandenhirtz, W.~Wallraff, M.~Weisgerber, Proc 27$^{th}$ ICRC (2001) 2197
\bibitem{burger03a} W.~J.~Burger et al., Nucl. Inst. Meth. A {\bf 512} (2003) 517
\bibitem{barancourt01a} D.~Barancourt et al., Nucl. Inst. Meth. A {\bf 465} (2001) 306
\bibitem{luebelsmeyer03a} K.~Luebelsmeyer, private communication
\bibitem{henning03a} R.~Henning, PhD Thesis, Massachusetts Institute of Technology (2003)
\bibitem{nasa04a} NASA -- National Aeronautics and Space Administration (2004)\\ {\tt http://spaceflight.nasa.gov/shuttle/reference}
\bibitem{casadei02a} D.~Casadei, AMS Bologna Internal Note 2000-03-02, updated (2002)
\bibitem{choutko97a} V.~Choutko, AMS Internal Note 97/95a.
\bibitem{aguilar05a} M.~Aguilar et al., Nucl.~Inst.~Meth. B {\bf 234} (2005) 321
\bibitem{suter00a} H.~Suter, AMS Internal Note 2000-07-03 (2000)
\bibitem{nasa99a} NASA -- National Aeronautics and Space Administration (1999)\\ {\tt http://spaceflight.nasa.gov/history/shuttle-mir}
\bibitem{schiff51a} L.~Schiff, Phys. Rev. {\bf 83} (1951) 252
\bibitem{tsai74a} Y.~Tsai, Rev. Mod. Phys. {\bf 46} (1974) 815, Rev. Mod. Phys. {\bf 49} (1977) 421
\bibitem{stearns49a} M.~Stearns, Phys. Rev. {\bf 76} (1949) 836
\bibitem{capiluppi74a} P.~Capiluppi et al., Nucl. Phys. B {\bf 70} (1974) 1
\bibitem{nara99a} Y.~Nara et al., Phys. Rev. C {\bf 61} (1999) 024901, and references therein
\bibitem{carroll79a} A.~S.~Carroll et al., Phys. Lett. B {\bf 80} (1979) 319
\bibitem{badhwar77a} G.~D.~Badhwar, S.~A.~Stephens, R.~L.~Golden, Phys. Rev. D {\bf 15} (1977) 820
\bibitem{rossi65a} B.~Rossi, High-Energy Particles. Prentice-Hall (1965)
\bibitem{mccubbin81a} N.~A.~McCubbin, Rep. Prog. Phys. {\bf 44} (1981) 1027
\bibitem{choutko03a} V.~Choutko, Unpublished AMS Note (2003)\\{\tt http://ams.cern.ch/AMS/Analysis/hpl3itp1/ams\_rec\_sel.ps}
\bibitem{hart84a} J.~C.~Hart, D.~H.~Saxon, Nucl. Inst. Meth. {\bf 220} (1984) 309
\bibitem{choutko96a} V.~Choutko, Unpublished AMS Technical Note (1996)\\{\tt http://ams.cern.ch/AMS/Analysis/hpl3itp1/amsnote.ps}
\bibitem{james94a} F.~James, M.~Roos, Comput. Phys. Commun. {\bf 10} (1975) 343
\bibitem{henning} H. Gast, Diploma Thesis, RWTH Aachen (2004)
\bibitem{brun87} R.~Brun et al., CERN DD/EE/84-1 (1987)
\bibitem{chang01} Y.-H.~Chang, Nucl.~Inst.~Meth. A {\bf 466} (2001) 282
\bibitem{fano63a} U.~Fano, Annu. Rev. Nucl. Sci. {\bf 13} (1963) 1
\bibitem{bichsel88a} H.~Bichsel, Rev. Mod. Phys. {\bf 60} (1988) 663
\bibitem{flueckiger90a} E.~Fl\"uckiger, E.~Kobel, J.~Geomag.~Geoelectr. {\bf 42} (1990) 1123
\bibitem{tsyganenko89a} N.~A.~Tsyganenko, Planet. Space Sci. {\bf 37} (1989) 5
\bibitem{iaga06a} IAGA -- International Association of Geomagnetism and Aeronomy (2006)\\ {\tt http://www.ngdc.noaa.gov/IAGA/vmod }
\bibitem{helene83a} O.~Helene, Nucl. Inst. Meth. {\bf 212} (1983) 319
\bibitem{helene84a} O.~Helene, Nucl. Inst. Meth. A {\bf 228} (1984) 120
\bibitem{johnson34a} T.~Johnson, Phys. Rev. {\bf 45} (1934) 569
\bibitem{sullivan71a} J.~D.~Sullivan, Nucl.~Inst.~Meth. {\bf 95} (1971) 5
\bibitem{barlow04a} R.~Barlow, {\tt physics/0603449} (2004)
\bibitem{picozza06a} P.~Picozza et al., submitted article, {\tt astro-ph/0608697} (2006)
\bibitem{leluc05a} C.~Lechanoine-Leluc, Proc 29$^{th}$ ICRC {\bf 3} (2005) 381
\end{thebibliography}
\end{document}